\documentclass[12pt,twoside]{scrreprt}

\pdfoutput=1

\usepackage{amsmath,amssymb}
\usepackage{cmap}
\usepackage{braket}
\usepackage{cancel}
\usepackage{wrapfig}
\usepackage{slashed}
\usepackage[stable]{footmisc}
\usepackage{graphicx}
\usepackage{aeguill}
\usepackage[romanian,french,english]{babel}
\usepackage[utf8]{inputenc}
\usepackage[T1]{fontenc}
\usepackage{microtype}
\usepackage{hyperref}

\long\def\beginpgfgraphicnamed#1#2\endpgfgraphicnamed{\includegraphics{#1}}

\DeclareMathOperator{\tr}{Tr}
\DeclareMathOperator{\Li}{Li}
\DeclareMathOperator{\Gl}{Gl}

\subject{PhD Thesis}
\title{Twistors, Strings and Supersymmetric Gauge Theories}
\author{Cristian Vergu}

\begin{document}

\extratitle{\centering
\includegraphics[width=.4\textwidth]{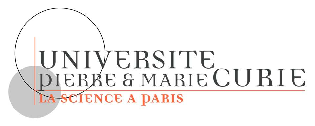}
\vfill
\selectlanguage{french}%
{\LARGE Universit\'e Paris VI---Pierre et Marie Curie}\\*[.5cm]
{\LARGE Institut de Physique Th\'eorique---CEA/Saclay}\\
\vfill
{\Large Th\`ese de physique th\'eorique}
\vfill
{\Large Cristian Vergu}
\vfill
{\huge Twisteurs, cordes et th\'eories de jauge supersym\'etriques}
\vfill
{\Large Th\`ese dirig\'ee par~David Kosower}\\*[2mm]
{\Large Soutenue le: 15 juillet 2008}\\
\vfill
{\large\begin{tabular}{lll}
\multicolumn{3}{l}{\textbf{Jury :}} \\*[2mm]

Iosif Bena & IPhT, CEA/Saclay & (invit\'e)\\
Bernard Julia & LPT, ENS & (pr\'esident) \\
Gregory Korchemsky & Paris XI, Orsay & (rapporteur)\\
David Kosower & IPhT, CEA/Saclay & (directeur)\\
Fabio Maltoni & Catholic University of Louvain & (rapporteur)\\
Pierre Vanhove & IPhT, CEA/Saclay & (examinateur)
\end{tabular}}
\vfill
}

\selectlanguage{english}
\title{Twistors, strings and supersymmetric gauge theories}
\author{Cristian Vergu}
\date{}
\publishers{




\selectlanguage{english}
\textbf{Abstract}\\[.7\baselineskip]
\begin{minipage}{\linewidth}
  \small This thesis is concerned with the study of scattering amplitudes in four-dimensional conformal field theories, more particularly the $\mathcal{N}=4$ super-Yang-Mills theory.  We study this theory first at tree level by using twistor space techniques and review the twistor string models that were proposed to describe it.

  Then, we turn to the issue of iteration relations and all loop ansatze for scattering amplitudes.  We review the unitarity method for computing scattering amplitudes and discuss the Wilson loop--scattering amplitude duality that was inspired by the strong-coupling prescription of Alday and Maldacena for scattering amplitudes.

  We describe in some detail the computation of a two-loop six-point scattering amplitude and its surprising equality to the polygonal Wilson loop.
\end{minipage}

\bigskip\bigskip


\selectlanguage{english}
}








\dedication{{\selectlanguage{romanian}\emph{Familiei mele}}}

\maketitle

\tableofcontents

\chapter{Acknowledgements}

First I would like to thank my thesis adviser, David Kosower, for setting such high standards of quality of research.  I benefited a lot from his advice and expertise.

I am very grateful to the members of my thesis committee: to Gregory Korchemsky and Fabio Maltoni for having accepted to report on my thesis manuscript, to David Kosower and Pierre Vanhove for their careful reading and the numerous improvements they suggested, to Iosif Bena for his continuous encouragement and to Bernard Julia for making place for me in his very busy schedule.

Then, I would like to thank all those with whom I shared an office, however briefly: Gerhard Goetz, Jérôme Rech, Yann Michel, Loïc Estève, Clément Gombeaud, Emmanuel Schenck, Arunansu Sil and Clément Ruef.  Thank you for the pleasant time we spent together.

Next, I would like to thank all the students I met during my PhD: Alexey Andreanov, Adel Benlagra, Guillaume Beuf, Micha\"el Bon, Jean-Emile Bourgine, Tristan Brunier, Candu Constantin, Cédric Delaunay, Jérôme Dubail, Clément Gombeaud, Pierre Hosteins, Nicolas Orantin, Jeanne Parmentier, Sylvain Prolhac, Clément Ruef, Emmanuel Schenck, Dmitro Volyn.  Thank you, I will miss having lunch with you!

I have benefited a lot from the contacts I had with my colleagues from the DEA class: Davide Cassani, Adel Dayarian, Benoît Estienne, Romain Gicquaud, R\u azvan Gur\u au, Pedro Vieira and the other friends I made at ENS: Plamen Bokov, Alina Grigorescu, Radu Ignat, Oana Ivanovici, Piotr Karwasz, Patrick Labatut, Tristan Machado, Kenneth Maussang, Nicolae Mihalache, Yashonidhi Pandey, Preden Roulleau, Ovidiu Sferlea, Sorin T\u anase-Nicola, Nicolas Templier, Félix Werner.

During my PhD I taught LP206 (Mathematical methods for physicists) together with: Claude Aslangul, Eli Ben-Haïm, Raphaël Benichou, Bruno Deremble, Dominique Mouhanna, Philippe Sindzingre, Julien Tailleur, Jean-Bernard Zuber.  It was a great experience to work with all of them.  I also thank my colleagues from LP203 (Electromagnetism) and especially Jacques Chauveau and Bernard Clerjaud.

I learned a lot from my scientific collaborators: Zvi Bern, Lance Dixon, Radu Roiban, Marcus Spradlin, Anastasia Volovich.  I have also benefited from discussions with Fernando Alday, John Joseph Carrasco, James Drummond, Johannes Henn, Gregory Korchemsky, Daniel Ma\^\i tre and Emery Sokatchev.

I also thank the other members of my laboratory and especially: Simon Badger, Marco Cirelli, Michele Frigerio, Riccardo Guida, Edmond Iancu, Jean-Marc Luck, Jean-Yves Ollitrault and Henri Orland.

I am indebted to the secretary of our laboratory, Sylvie Zaffanella, who helped me navigate the treacherous paths of French administration.  I also thank Bruno Savelli for helping me whenever the paper got stuck in the photocopier (which happened a lot).

During my PhD there were a lot of interesting lectures organised by the laboratory.  Many thanks to the lecturers: Larry Schulman, Daniel Estève, Pierre Binétruy, Bertrand Eynard, Olivier Babelon, François David, Didina Serban, Xavier Viennot, Francis Bernardeau, Pierre Fayet for their efforts.

I was very lucky to be able to attend to very interesting conferences like the Claude Itzykson Meetings, the celebration of Jean-Bernard Zuber sixtieth anniversary, the conference ``Wonders of Gauge theory and Supergravity'' and also the seminars ``Rencontres Théoriciennes'' organised at the Institut Henri Poincaré.  Many thanks to all the people who contributed to creating such a stimulating atmosphere.

I am indebted to Roland Combescot and Jean-Bernard Zuber for helping me meet all the deadlines and obtain all the necessary approvals from the Paris VI University, before the date of my thesis defence.  I am especially grateful to Jean-Bernard Zuber for helping me constitute a thesis committee, for his interest in my research and for hosting me in his laboratory during the difficult days of the strike of public transportation system.  His energy and generosity were an inspiration for me.

Last but not least, I would like to thank my family for their unconditional support.

\selectlanguage{english}%
\chapter{Introduction}

The main subject of this thesis is scattering amplitudes in $\mathcal{N}=4$ super-Yang-Mills theory and in conformal field theories in general.  We will also briefly discuss scattering amplitudes in supergravity and conformal supergravity.

In a conformal field theory (CFT), one usually considers correlation functions of gauge invariant operators.  Operators of definite scaling dimension have simple correlation functions whose form is constrained by conformal symmetry.  All the information needed to compute these correlation functions is in the anomalous dimensions and the fusion coefficients in the OPE (Operator Product Expansion).

Studying scattering amplitudes in a CFT may seem strange for several reasons.  One reason is that these scattering amplitudes are not really well defined because of infrared divergences and therefore are not good observables.  They can be used however as building blocks for physical observables.

Another reason is that the single particle states used for computing scattering amplitudes are unnatural from the point of view of conformal symmetry.  The way to see this is as follows: the single particle states transform in irreducible representations of the Poincar\'e group.  The (super-)conformal group, however, is a bigger group that contains the Poincar\'e group as a subgroup.  It is therefore natural to consider states that transform irreducibly under the (super-)conformal group.  The irreducible representations of the (super-)conformal group are in general reducible with respect to the Poincar\'e subgroup, so there is a tension between the conformal symmetry and the particle interpretation which is necessary for computing scattering amplitudes.  We will describe below how the twistor-space constructions solve this problem.

In the framework of AdS/CFT correspondence (see refs.~\cite{Maldacena:1997re, Gubser:1998bc, Witten:1998qj}), at first the dual gravitational interpretation of correlation functions in the CFT was understood and studied.  The dual gravitational interpretation of the scattering amplitude was proposed only recently and it has several surprising features that we will discuss.

We will consider the scattering amplitudes from two very different points of view.  In part I we will discuss twistor-space constructions that emphasise (super-)conformal symmetry but are mostly restricted to tree-level amplitudes.  In part II we will emphasise higher-loop computations by using the unitarity method and discuss all-loop ansatze.  At this point there are no firm conclusions that we can present since much remains to be understood, but we will compile a list of open questions.

The principal motivation for studying scattering amplitudes is of course phenomenological.  Precise results for scattering amplitudes are of great importance for extracting new physics from data collected at colliders.  Many years of study have produced results and have resulted in techniques that tame the complexity of the computations and of the final results.

It is conceivable however that further advances are possible; indeed, recent findings seem to confirm this.  Even though great advances have made feasible computations that were once considered out of reach, phenomenologically relevant theories like QCD remain pretty complicated.  It is then useful to apply these techniques to simpler theories.  One obvious choice is the maximally supersymmetric $\mathcal{N}=4$ Yang-Mills theory.  This theory has the maximal number of supersymmetries compatible with helicity $\pm 1$ (theories with more supersymmetry must include gravity).  It turns out, however, that this theory has even more symmetry than is apparent at first.  It has a classical conformal symmetry that survives quantisation and is enlarged to superconformal symmetry.

Another reason to be interested in supersymmetric $\mathcal{N}=4$ Yang-Mills is that it provides the simplest incarnation of the celebrated AdS/CFT correspondence.  It is believed that the supersymmetric $\mathcal{N}=4$ Yang-Mills in four dimensions is \emph{equivalent} to Type IIB string theory on a $AdS_5 \times \mathbb{S}^5$ background.  There is by now a fairly detailed dictionary between observables on both sides of the correspondence but computations of dual quantities can usually only be performed in non-overlapping regions of the parameter space.  Integrability techniques yield exact solutions and have afforded non-trivial tests of the AdS/CFT correspondence.  It is important to note here that perturbative computations at weak and strong coupling played a decisive role in finding solutions for the integrable models which appeared in studying the dilatation operator for the supersymmetric $\mathcal{N}=4$ theory.

There have been many hints that the scattering amplitudes in gauge theories are much simpler that what one could naively expect.  One striking example is the simplicity of the MHV amplitudes (the MHV, or maximally helicity violating amplitudes, are amplitudes with two external lines of helicity minus and the rest of helicity plus).  In fact, we will see that the structure of MHV amplitudes fits perfectly with the $\mathcal{N}=4$ supersymmetry.  Of course, spinor techniques played a crucial role in uncovering this simplicity.

Another example is the simple structure of the self-dual Yang-Mills theory, whose scattering amplitudes only receive corrections at one-loop order.  This has been argued to result from anomaly arguments and integrability in ref.~\cite{Bardeen:1995gk}.

In ref.~\cite{Nair:1988bq} Nair proposed that MHV scattering amplitudes are computable from a Wess-Zumino-Witten model.  All these early advances, together with twistor ideas of Penrose~\cite{Penrose:1968me}, enabled Witten~\cite{Witten:2003nn} to formulate conjectures about geometric interpretations of the scattering amplitudes in twistor-space and to put forward a string theory proposal that computes these scattering amplitudes.

Further advances along these lines include the formulation of MHV rules, where the amplitudes are computed from vertices composed of MHV amplitudes, and tree level recurrence relations.  It has also become obvious that there are multiple twistor string prescriptions and, after a proposal by Berkovits (see ref.~\cite{Berkovits:2004hg}), several twistor string theories for computing the scattering amplitudes.

The MHV rules are the simplest known way to compute tree amplitudes in gauge theories but they have the awkward feature that Lorentz invariance, though present, is not manifest.  Because of this simplicity, it is natural to try to extend the the construction to loop level.  This was done in ref.~\cite{Brandhuber:2004yw} (see also refs.~\cite{Brandhuber:2005kd, Brandhuber:2006bf} for more details and applications to theories with less supersymmetry).  The Lagrangian origin of the MHV rules was investigated in refs.~\cite{Mansfield:2005yd, Ettle:2006bw, Ettle:2007qc, Gorsky:2005sf}.

There are reasons to suspect that the various twistor string theories that were formulated (and maybe others that remain to be found) will have important lessons to teach us in the future.  In particular, the relations between the twistor strings and the usual string theories that appear in the AdS/CFT correspondence are not at all clear at present.

A point that we will not discuss in detail in this thesis but which still deserves mention, is the issue of finiteness of maximal $\mathcal{N}=8$ supergravity.  There is by now a growing body of evidence that supports the perturbative finiteness of $\mathcal{N}=8$ supergravity.  There is by now a growing body of evidence that supports the perturbative finiteness of $\mathcal{N}=8$ supergravity.  This question can be approached in different ways: by explicit computations (see refs.~\cite{Bern:1998ug, Bern:2006kd, Bern:2007hh, Bern:2007xj}), by using the constraints imposed by string theory dualities on the low energy limit of four-dimensional compactifications (see refs.~\cite{Green:2006gt, Green:2008bf}).  Supersymmetry arguments (see~\cite{Howe:2002ui, Howe:2004pn}) yield predictions for the onset of ultraviolet divergences.  The most restrictive constraints were presented in ref.~\cite{Green:2006yu}, which uses non-renormalisation theorems proved in ref.~\cite{Berkovits:2006vc} by using the pure spinor formalism.  See also refs.~\cite{BjerrumBohr:2008vc, BjerrumBohr:2008ji, Bern:2008qj} for recent developments.

The supergravity scattering amplitudes are computed by the unitarity method and the tree amplitudes that are needed as ingredients are computed by Kawai-Lewellen-Tye (KLT) relations (see ref.~\cite{Kawai:1985xq}) that were proved in string theory (see ref.~\cite{Bern:2002kj} for a review of the computational techniques).

Let us now turn to the second center of focus of this thesis, namely the iteration relations~\cite{Anastasiou:2003kj} and all-loop ansatze~\cite{Bern:2005iz} for scattering amplitudes.

Anastasiou, Bern, Dixon and Kosower (ABDK) observed in ref.~\cite{Anastasiou:2003kj} that the splitting function, an universal quantity which characterises the collinear limit, obeys an iteration relation.  Namely, the splitting function at two loops can be expressed purely in terms of its value at one loop.  From this, they conjectured an iteration relation for the MHV amplitudes that is compatible with the iteration relation for the splitting function.

Later, Bern, Dixon and Smirnov (BDS) proved~\cite{Bern:2005iz} an extension of the ABDK ansatz to three loops and proposed an all-loop ansatz for MHV amplitudes.  These ansatze were further tested for five-point amplitude at two loops in refs.~\cite{Cachazo:2006tj, Bern:2006vw}.

Even though BDS proposed an all-loop ansatz, all the available evidence for it came from weak coupling perturbative computations. This changed after Alday and Maldacena~\cite{Alday:2007hr} found a prescription for computing scattering amplitudes at strong coupling. Their explicit four-point computation matched the BDS ansatz perfectly.

At strong coupling, after a $T$-duality transformation along the space-time directions which maps the $AdS$ space into itself, the computation of scattering amplitudes is identical to the computation to a light-like polygonal Wilson loop, whose sides are constructed from the on-shell momenta of the scattered particles.

The interesting question of whether this similarity between scattering amplitudes and Wilson loops is restricted to strong coupling, was addressed in ref.~\cite{Drummond:2007aua} for one-loop four-point case and in ref.~\cite{Brandhuber:2007yx} for one-loop and an arbitrary number of points.  Then Drummond, Henn, Korchemsky and Sokatchev computed the two-loop corrections to the four-~\cite{Drummond:2007cf} and five-point~\cite{Drummond:2007au} Wilson loops.

Drummond, Henn, Korchemsky and Sokatchev also proved that the polygonal Wilson loops with light-like sides obey anomalous Ward identities whose origin is the conformal symmetry of the $\mathcal{N}=4$ Yang-Mills.  These Ward identities fix the finite part of the logarithm of Wilson loop up to additive constants.  It turns out that the finite parts of the four- and five-point one loop amplitudes also satisfy these Ward identities, which hints that the scattering amplitudes also have a conformal symmetry.  This conformal symmetry is \emph{different} from the usual conformal symmetry of the $\mathcal{N}=4$ Yang-Mills and was called `dual conformal symmetry.'

It was also observed in ref.~\cite{Drummond:2006rz} that the scattering amplitudes can be written in terms of `pseudo-conformal' integrals, which are integrals which have a conformal symmetry in momentum space.

Starting at six points, the anomalous Ward identities for the Wilson loop do not completely fix the finite part of the logarithm of the Wilson loop, because their general solution allows an arbitrary function of three conformal cross-ratios.  This function was computed numerically in ref.~\cite{Drummond:2007bm}.  This computation motivated a hypothesis for why the Wilson loop and scattering amplitudes are identical at four and five points; if one accepts the conjecture that the scattering amplitudes exhibit a `dual conformal symmetry' similar to the conformal symmetry of the Wilson loop, then the two are equal simply because they have the same symmetry which constrains them completely.

As mentioned above, starting at six points the symmetry does not constrain the results completely.  In order to test this hypothesis, and also to check if the BDS conjecture is correct at six points (some arguments that the BDS conjecture must fail for a sufficiently large number of external legs were put forward in ref.~\cite{Alday:2007he}) it became necessary to compute the six-point two-loop MHV scattering amplitude.  This computation was announced in ref.~\cite{Bern:2008ap} and compared numerically to the results of refs.~\cite{Drummond:2007bm, Drummond:2008aq} for the six-point Wilson loop.

Remarkably, the results for the Wilson loop and the scattering amplitudes agree at the six points as well.  There is as yet no argument for why this equality holds.  At strong coupling the situation is a bit better because there the equality can be explained by $T$-duality, but only to first order in $\frac 1 {\sqrt{\lambda}}$.  No such understanding is available at weak coupling.

In ref.~\cite{Bern:2008ap} it was also shown that the BDS ansatz fails at six points for MHV amplitudes.  An important question is whether the BDS ansatz can be fixed to take into account the remainder appearing at six points.

\part{General Introduction}


\chapter{Short Review of Computational Techniques}
\label{ch:short_review_of_computational_techniques}

Here we review some aspects of techniques used for computing scattering amplitudes in gauge theories.  This will be useful in the subsequent chapters.  The techniques reviewed are quite standard and are treated in a number of review articles~\cite{Mangano:1990by, Dixon:1996wi, Bern:1996je, Bern:2007dw}, but are not textbook material yet.  See also~\cite{Chalmers:1997ui, Chalmers:1998jb, Chalmers:2001cy}.

One of the central ideas is that, due to gauge invariance, there are a lot of cancellations that take place when one sums all the Feynman graphs that contribute to a given amplitude.  One can simplify the intermediate results somewhat by a clever gauge choice, but a better way is to decompose an amplitude in a sum of gauge invariant pieces and then compute each piece with a gauge choice that is convenient. One gains the possibility of choosing \emph{different} gauges for different gauge invariant pieces, which is an advantage over the traditional approach.  The decomposition in gauge invariant pieces will be discussed in sec.~\ref{sec:colour_decomposition} below.

Another technique that proves very powerful in computing scattering amplitudes in four dimensions is the spinor-helicity method.  This is discussed in sec.~\ref{sec:spinor_helicity}.  The spinor-helicity method is useful for computing on-shell scattering amplitudes in four dimensions.  For loop amplitudes, it is useful in combination with a variant of dimensional reduction regularisation, the four dimensional helicity scheme, which is a supersymmetry-preserving regularisation scheme where the polarisations of external lines are kept in four dimensions.

The tree-level scattering amplitudes for $n$ gluons in a non-supersymmetric gauge theory are the same as those for a supersymmetric gauge theory.  Moreover, for a supersymmetric gauge theory one can use supersymmetry Ward identities (also called SWI) to relate scattering amplitudes with only gluons to scattering amplitudes containing fermions.  For extended supersymmetry one can also relate gluon amplitudes to amplitudes containing scalars.  Note that these Ward identities are exact for a supersymmetric theory but also hold at tree level for non-supersymmetric theories. One can then compute the scattering amplitudes in the supersymmetric theory by computing the amplitudes containing fermions or scalars, which are often easier to compute.  The SWI are also very useful when computing loop amplitudes using the unitarity method.  A discussion of the supersymmetric Ward identities can be found in sec.~\ref{sec:swi}.

\section{Colour Decomposition}
\label{sec:colour_decomposition}

The Feynman rules for a gauge theory have the structure of a product between a kinematic part, containing momenta and coupling constants and a colour part, containing colour factors, represented by the structure constants $f_{a b c}$ of the gauge algebra\footnote{Sometimes in the literature the following defining relation is used instead $\left[T^a, T^b\right] = i \sqrt{2} f_{a b c} T^c$.  For our purposes this is not needed because we will express the results of our computations independently of $f$.  The convention above is used in order to be able to use the usual Feynman rules, where the normalisation is fixed by $\tr\left(T^a T^b\right) = \tfrac 1 2 \delta^{a b}$.}
\[
  \left[T^a, T^b\right] = i f_{a b c} T^c,
\] where we fix the normalisation of the generators by
\[
  \tr\left(T^a T^b\right) = \delta^{a b},
\] (with this normalisation there is no need to distinguish between upper and lower indices for the structure constants $f$).

Now, take a tree level Feynman diagram, pick a vertex and replace the colour structure function using
\[
  f_{a b c} = - i \tr\left(T^a T^b T^c - T^c T^b T^a\right).
\]  If the vertex is a four-gluon vertex\footnote{In fact, as far as only colour factors are considered, one can split a four-gluon vertex into two three-gluon vertices.}, whose Feynman rules contain a product of type $f_{a b e} f_{c d e}$ use the commutation relations to eliminate the remaining $f$ factor together with a matrix in the trace
\[
  f_{c d e} T^e = - i \left[T^c, T^d\right].
\]

Then, one can traverse the tree starting at this vertex and using the rule above to eliminate the structure constants $f$.  In the end, if the external lines are all in the adjoint representation, the colour factor for a $n$-particle scattering amplitude can be decomposed on a basis
\[
  \tr\left(T^{a_1} \cdots T^{a_n}\right),
\] where $(a_1, \ldots, a_n)$ is a permutation of the colour indices of external gluons (obviously the cyclic permutations don't yield new elements).  It is easy to see that only external colour indices survive because the internal colour indices appear in pairs and are also annihilated in pairs by the above procedure.  In order to complete the proof one must apply the same procedure to all the Feynman diagrams of a given process.

In the case of loop amplitudes this procedure does not work as described above.  In this case, one can also find contributions like\footnote{If the two gauge algebra generators are neighbours, their product can be computed by using the fact that the quadratic Casimir operator is a multiple of identity.}
\[
  \sum_c \tr\left( \cdots T^c \cdots T^c \cdots\right),
\] arising from the use of the above replacement rules in the expression,
\[
  \tr\left(\cdots T^a \cdots T^b \cdots\right) f_{a b c}.
\]

Alternatively, one can see how this kind of structure arises in the case of loop amplitudes by using the result above for tree amplitudes and sewing some external legs to form loop amplitudes.  In the case of an $L$ loop amplitude there will be a sum over $L$ colour indices, each appearing twice inside the trace.  This is the generic case, as sometimes a pair of indices can be eliminated by using the fact that the quadratic Casimir is a multiple of identity.

Using the fact that a basis in the colour space for tree amplitudes is formed of elements $\tr\left(T^{a_1} \cdots T^{a_n}\right)$, where $(a_1, \ldots, a_n)$ are permutations of the external colour indices, one can write the general tree level $n$-gluon amplitude in the following form
\begin{equation}
  \label{eq:colour_decomposition}
  \mathcal{A}_n^{\text{tree}} = \sum_{\sigma \in     \mathcal{S}_n/\mathbb{Z}_n} \tr\left(T^{a_{\sigma(1)}} \cdots     T^{a_{\sigma(n)}}\right) A^{\text{tree}}(p_{\sigma(1)},   \epsilon_{\sigma(1)}; \ldots ; p_{\sigma(n)}, \epsilon_{\sigma(n)}),
\end{equation} where $\mathcal{S}_n$ is the permutation group of $n$ elements and the sum runs over all permutations modulo cyclic permutations.  One can also sum over all permutations and divide the result by $n$.

The colour stripped amplitudes $A^{\text{tree}}$ can be taken to be cyclically symmetric; clearly there is nothing to be gained by allowing a more general symmetry.

Another very important property of the sub-amplitudes $A^\text{tree}$ is their gauge invariance.  This can be seen as follows: in the asymptotic region where the interactions are supposed to turn off, the non-abelian gauge theory can be seen as a product of abelian gauge theories.  Therefore, the gauge invariance means the invariance of the scattering amplitude under transformations $\epsilon_i^\mu \rightarrow \epsilon_i^\mu + \alpha k_i^\mu$, where $\epsilon_i$ is the polarisation and $k_i$ is the momentum of the $i^\text{th}$ gluon.  As the scattering amplitude is gauge invariant and the gauge transformations considered above do not act on the colour structure, it follows that the sub-amplitudes $A^\text{tree}$ are also gauge invariant.

At this point it looks like there are $(n-1)!$ different sub-amplitudes to consider.  These sub-amplitudes are however linked by some identities, so the set of independent sub-amplitudes one needs to compute is actually smaller.  See~\cite{Kleiss:1988ne} for a discussion of the complexity of the computation.  See also ref.~\cite{DelDuca:1999rs} for an alternative colour decomposition where only $(n-2)!$ sub-amplitudes need to be computed.

Let us list all the properties of the amplitudes below:
\begin{itemize}
\item $A^{\text{tree}}(1, \ldots, n)$ is gauge invariant,
\item $A^{\text{tree}}(1, \ldots, n)$ is invariant under cyclic permutations of $(1, \ldots, n)$,
\item $A^{\text{tree}}(1, 2, \ldots, n) = (-1)^n A^{\text{tree}}(n,   \ldots, 2, 1)$ (the reflection identity),
\item the dual Ward identity
\begin{multline*}
  A^{\text{tree}}(1, 2, 3, \ldots, n) + A^{\text{tree}}(2, 1, 3,   \ldots, n) + A^{\text{tree}}(2, 3, 1, \ldots, n) +\\+   A^{\text{tree}}(2, 3, \ldots, 1, n) = 0.
\end{multline*}
\end{itemize}  See ref.~\cite{Berends:1987me} for proofs of these results.

\section{\texorpdfstring{Large $N_c$}{Large Nc}}
\label{sec:large_N}

Up to this point, the discussion about colour decomposition has been completely general.  Here we will specialise to $SU(N_c)$ and sometimes to $U(N_c)$ gauge group.  In these cases one can take the 't Hooft limit~\cite{'t Hooft:1973jz} $N_c \rightarrow \infty$ with $\lambda = g^2 N_c = \text{constant}$.

In usual (non-conformal) gauge theories the coupling constant runs with the energy scale so it is not a parameter of the theory.  There is a dimensionful parameter $\Lambda$ which is the scale where the coupling constant becomes large but, as it is dimensionful, it can't be used as an expansion parameter in a perturbative expansion.  't Hooft observed that for gauge theories with gauge group $SU(N_c)$ there is another dimensionless parameter, $\tfrac 1 {N_c}$ and that one can construct an expansion in this parameter.

This limit has been very useful in studies of AdS/CFT correspondence~\cite{Maldacena:1997re, Gubser:1998bc, Witten:1998qj} (see refs.~\cite{Aharony:1999ti, D'Hoker:2002aw, Nastase:2007kj} for reviews) which conjectures an equivalence between conformal four dimensional\footnote{The correspondence is conjectured to hold for $M$-theory backgrounds and for non-conformal gauge theories.  Also, the background does not necessarily have to be of type $AdS_5 \times X$, but its isometry group must contain $SO(2,4)$. However, these generalisations are less understood.} gauge theories and string theories in backgrounds $AdS_5 \times X$.

The most famous incarnation of the correspondence is the one involving the $\mathcal{N}=4$ supersymmetric gauge theory with gauge group $SU(N_c)$ on the gauge side and Type IIB string theory on $AdS_5 \times \mathbb{S}^5$ on the string side.

In the case of $SU(N_c)$ gauge groups one can use the following formula (where $T^a$ are the gauge group generators in the fundamental representation with our conventional normalisation $\tr\left(T^a T^a\right) = \delta^{a b}$)
\begin{equation}
  \label{eq:su(n)_completeness}
  \sum_{a=1}^{N_c^2-1} \left(T^a\right)_i^{~j} \left(T^a\right)_k^{~l} = \delta_i^l \delta_k^j - \frac 1 {N_c} \delta_i^j \delta_k^l.
\end{equation} This is just the statement that the generators $T^a$ form a complete set of traceless hermitian matrices.

This identity suggests a new way to prove the results in sec.~\ref{sec:colour_decomposition} by recursion over the number of external gluons.  Assume that eq.~\eqref{eq:colour_decomposition} holds for a number of gluons less than $n$.  Then, by constructing a tree out of two sub-trees and using eq.~\eqref{eq:su(n)_completeness} one obtains the expected colour structure up a sub-dominant in $\tfrac 1 {N_c}$ contribution.

The following arguments show that this $\tfrac 1 {N_c}$ contribution cancels upon summing over all the permutations.  In fact, this sub-dominant contribution corresponds to the subtraction of a $U(1)$ part from a $U(N_c)$ gauge theory.  This $U(1)$ factor does not couple to the $SU(N_c)$ gluons, so it can't contribute to the tree level scattering amplitude of gluons (the $U(1)$ factor does couple to fermions or scalars and so can contribute to the scattering amplitude of gluons at loop level in a theory containing fermions and scalars).  This can be used to prove the dual Ward identity mentioned above.

Yet another way to prove some of the identities discussed above is to use the antisymmetry of colour-stripped Feynman rules under exchange of two lines.  The usual Feynman rules are symmetric in permutation of external legs but the colour part is completely antisymmetric for three point interaction vertices.  Therefore, the colour-stripped Feynman rules are also antisymmetric.  Using these antisymmetry properties the diagrams contributing to some amplitudes can be grouped in pairs that cancel one another.

One can generalise this by using the subgroup $U(P) \times U(N_c-P)$ of the gauge group $U(N_c)$.  By observing that the gluons in the first gauge group do not interact with the gluons in the second gauge group, we conclude that the tree-level scattering amplitude containing  gluons of both types is zero.  Using this vanishing one can write a generalised dual Ward identity.

Let us briefly discuss the colour structure of one-loop amplitudes~\cite{Bern:1990ux}.  By using the completeness relation for the generators $T^a$ (eq.~\eqref{eq:su(n)_completeness}) and the results in sec.~\ref{sec:colour_decomposition} one can decompose the one-loop amplitudes over a colour basis of at most two traces.  The general form is as follows
\begin{align*}
  \label{eq:1loop_colour_decomposition}
  \mathcal{A}_n^{\text{1-loop}} =& \sum_{\sigma \in     \mathcal{S}_n/\mathbb{Z}_n} N_c \tr\left(T^{a_{\sigma(1)}} \cdots     T^{a_{\sigma(n)}}\right) A^{\text{1-loop}}_{n;1}(\sigma(1),   \ldots, \sigma(n)) +\\ &+\sum_{c=2}^{\lfloor \frac n 2\rfloor + 1}   \sum_{\sigma \in \mathcal{S}_n/\mathcal{S}_{n;c}}   \tr\left(T^{a_{\sigma(1)}} \cdots T^{a_{\sigma(c-1)}}\right)   \tr\left(T^{a_{\sigma(c)}} \cdots T^{a_{\sigma(n)}}\right)   A_{n;c}^{\text{1-loop}}(\sigma(1), \ldots, \sigma(n))
\end{align*}  Observe here that the leading contribution as $N_c \rightarrow \infty$ is the single trace contribution.  This is completely general and can be proven by using the double line notation for adjoint fields.  The dominant contribution comes from planar diagrams and each loop yields a factor of $N_c$.

Amusingly, some of the properties of the colour decomposition and of the colour-stripped sub-amplitudes are more easily understood in string theory.  Then, by taking the zero slope limit $\alpha' \rightarrow 0$ and the compactification radii to zero in a correlated manner as detailed in ref.~\cite{Green:1982sw}, one obtains the on-shell scattering amplitudes of the light states in the field theory limit.  Also, the properties of the string theory amplitudes survive when taking this limit.  See ref.~\cite{Bern:1990ux} for a discussion of the colour decomposition in a string theory setting.

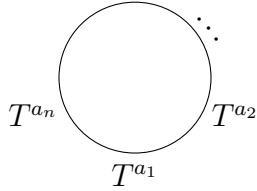
\begin{figure}
  \centering
  \beginpgfgraphicnamed{1}
  \begin{tikzpicture}
    \draw (0,0) circle (1);
    \draw (0,0) ++(-90:1) node [below] {$T^{a_1}$};
    \draw (0,0) ++(-30:1) node [right] {$T^{a_2}$};
    \draw (0,0) ++(210:1) node [left] {$T^{a_n}$};
    \draw (0,0) ++(30:1) node [right] {$\cdot$};
    \draw (0,0) ++(40:1) node [right] {$\cdot$};
    \draw (0,0) ++(50:1) node [right] {$\cdot$};
    \end{tikzpicture}
    \endpgfgraphicnamed
    \caption{Colour decomposition for an $n$-gluon tree amplitude       scattering in open string theory.  The colour factor is       $\tr\left(T^{a_1} \cdots T^{a_n}\right)$.}
\label{fig:string_tree}
\end{figure}

For example, the tree amplitude for gluon scattering in open string theory is obtained by computing the disk correlation function of gluon vertex operators inserted on the boundary of the disk (see fig.~\ref{fig:string_tree}).  Each gluon vertex operator contains a generator $T^a$ of the gauge algebra and the boundary of the disk (worldsheet) has a Chan-Paton index.  The summation over Chan-Paton indices is equivalent to contracting the indices of the gauge algebra generators into a trace.  The colour-stripped amplitudes are given by the Koba-Nielsen formula.  Some properties of the tree amplitude are easier to establish starting from this formula than from field theory.  This is exactly the same structure as above (see eq.~\eqref{eq:colour_decomposition}).

It is a remarkable result that the double trace colour stripped amplitudes $A_{n;c}^{\text{1-loop}}(\sigma(1), \ldots, \sigma(n))$ can be expressed in terms of the single trace colour stripped amplitudes (see~\cite{Bern:1994zx}).  The relation between the two partial amplitudes is as follows
\begin{equation}
  \label{eq:sub-dominant_colour}
  A_{n;c}(1, 2, \ldots, c-1; c, c+1, \ldots, n) = \sum_{\sigma \in COP (1, 2, \ldots, c-1) (c, c+1, \ldots, n)} A_{n;1} (\sigma),
\end{equation} where the sum is over permutations preserving the cyclic ordering of $(1, 2, \ldots, c-1)$ and of $(c, c+1, \ldots, n)$.

\begin{figure}
  \centering
  \beginpgfgraphicnamed{2}
  \begin{tikzpicture}
    \draw (0,0) circle (3);
    \draw (0,0) ++(-90:3) node [below] {$T^{a_1}$};
    \draw (0,0) ++(-30:3) node [right] {$T^{a_2}$};
    \draw (0,0) ++(210:3) node [left] {$T^{a_{c-1}}$};
    \draw (0,0) ++(30:3) node [right] {$\cdot$};
    \draw (0,0) ++(40:3) node [right] {$\cdot$};
    \draw (0,0) ++(50:3) node [right] {$\cdot$};
    \draw (0,0) circle (2);
    \draw (0,0) ++(-90:2) node [above] {$T^{a_c}$};
    \draw (0,0) ++(-30:2) node [left] {$T^{a_{c+1}}$};
    \draw (0,0) ++(210:2) node [right] {$T^{a_n}$};
    \draw (0,0) ++(30:2) node [left] {$\cdot$};
    \draw (0,0) ++(40:2) node [left] {$\cdot$};
    \draw (0,0) ++(50:2) node [left] {$\cdot$};
  \end{tikzpicture}
  \endpgfgraphicnamed
  \centering
  \caption{Colour decomposition for an $n$-gluon one loop amplitude     scattering in open string theory (We restrict to the case of an     orientable worldsheet, which is the only possibility for an     $U(N_c)$ gauge group.     See~\cite{Green:1987sp,Green:1987mn,Polchinski:1998rq,Polchinski:1998rr}.)     The colour factor is $\tr\left(T^{a_1} \cdots T^{a_{c-1}}\right)     \tr\left(T^{a_c} \cdots T^{a_n}\right)$.}
\label{fig:string_loop}
\end{figure}
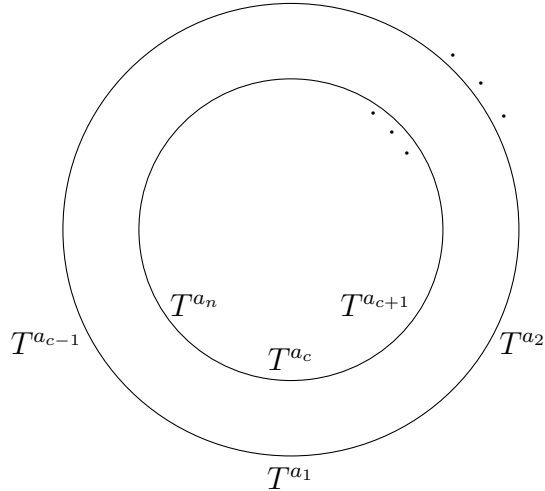

One can get an intuitive understanding of this formula from the string rules.  In string theory the vertex operators are distributed on the two boundaries of the annulus diagram (see fig.~\ref{fig:string_loop}).  When performing the worldsheet path integral the cyclic ordering of the vertices remains the same, while one sums over all the relative orderings of the vertices on different boundaries.  Moreover, when performing the zero slope limit $\alpha' \rightarrow 0$, the distinction between the two boundaries disappears so, in this limit, the amplitudes with vertex operators inserted on both boundaries are equal to the amplitudes where the vertex operators are inserted on only one of the boundaries.  This then implies eq.~\eqref{eq:sub-dominant_colour}.

Of course, one can also give a field theory argument for this identity.  See~\cite{Bern:1994zx} for more details.

\section{Spinor Helicity Method}
\label{sec:spinor_helicity}

Let $x_\mu$ be a quadrivector in a Minkowski space.  We can associate to $x_\mu$ a $2 \times 2$ complex matrix $x_\mu \sigma^\mu$, where $\sigma^\mu = (1, \sigma^i)$ and $\sigma^i$ are the Pauli matrices.

This matrix is explicitly given by:
\begin{equation}
  X_{\alpha \dot{\alpha}} =
  \begin{pmatrix}
    x_0 + x_3 & x_1 - i x_2\\
    x_1 + i x_2 & x_0 - x_3
  \end{pmatrix}.
\end{equation}

The indices $\alpha$ and $\dot{\alpha}$ take the values $1, 2$ and $\dot{1}, \dot{2}$ respectively.  If the vector $x_\mu$ is real (all its components $x_\mu$ are real), $X_{\alpha \dot{\alpha}}$ is hermitian and we have a one-to-one correspondence between quadrivectors and hermitian $2 \times 2$ matrices. Furthermore, the determinant of $X_{\alpha \dot{\alpha}}$ is $\det X_{\alpha \dot{\alpha}} = (x_0)^2 - (x_1)^2 - (x_2)^2 - (x_3)^2 = \eta_{\mu \nu} x^\mu x^\nu$.

As the quadrivectors are the representation space of the Lorentz transformations (the group $O(3,1)$) and there is a linear one-to-one correspondence between the quadrivectors and the $2 \times 2$ hermitian matrices, it is obvious that there must be a linear action of $O(3,1)$ on the hermitian matrices.  Moreover, this action must preserve the determinant of the matrices, because it preserves the norm of the quadrivectors on which it acts.

In order to keep the hermiticity of the matrices and to have a linear action, the transformation must act as $X \rightarrow X' = c M X M^\dagger$, where $M^\dagger$ is the hermitian conjugate of $M$ and $c$ is a real constant.  If we make the redefinition $M \rightarrow |c|^{1/2} M$, the condition that $\det X' = \det X$ gives $|\det M| = 1$.  We can restrict our attention to $\det M = 1$ without loss of generality (if $\det M = -1$ we can do a transformation $M \rightarrow i M$).  The transformation of $X$ is therefore $X \rightarrow X' = \pm M X M^\dagger$, where $M \in SL(2, \mathbb{C})$.

Apart from the continuous transformations described above, there is a \emph{discrete} transformation, namely $X \rightarrow X' = X^*$, where $X^*$ is the complex-conjugate matrix.  This transformation preserves the hermiticity of $X$ and also its determinant, as all hermitian matrices have real determinant.  It has the effect of reversing the direction of the $x^2$ axis, therefore it is a parity transformation (the transformation $x^i \rightarrow - x^i$ can be obtained by performing an additional rotation in the $x^1 x^3$ plane).

The time reversal transformation is obtained by $X \rightarrow X' = -
M X^* M^\dagger$, where $M =\begin{pmatrix}0 & 1\\-1 & 0\end{pmatrix}$.

We shall concentrate on the transformation $X \rightarrow X' = M X M^\dagger$ below.  The Lorentz transformations which act this way are orthochronous.  We can see that from $x_0' = \frac 1 2 \tr X' = \frac 1 2 \tr(M X M^\dagger) = \frac 1 2 x_0 \tr(M M^\dagger) + \cdots$ (this gives $\Lambda^0_{~0} = \frac 1 2 \tr(M M^\dagger) > 0$).

$O(3,1)$ is not connected, but has four connected components.  These are usually denoted by $L_+^\uparrow, L_+^\downarrow, L_-^\uparrow,
L_-^\downarrow$ (the $+, -$ indices represent the sign of $\det \Lambda$ and $\uparrow, \downarrow$ correspond to $+$ and $-$ sign of $\Lambda^0_{~0}$ respectively, where $\Lambda$ is an element of $O(3,1)$).  Sometimes $L_+^\uparrow$ is denoted by $SO(1,3)_0$.

As the action of $O(3,1)$ on the $2 \times 2$ hermitian matrices $V$ is given by $X \rightarrow X' = \pm M X M^\dagger$ and $X \rightarrow X' = X^*$, it is clear that $L_+^\uparrow$, the connected component of $O(3,1)$ which contains the identity, is in correspondence with $X \rightarrow X' = M X M^\dagger$ (which also contains the identity $X \rightarrow X' = X$).  This correspondence is one-to-two, meaning that for each $\Lambda \in L_+^\uparrow$ there are two elements $M$ and $-M$ in $SL(2, \mathbb{C})$.

It can be proven\footnote{To see this we can calculate the kernel of   the correspondence $SL(2, \mathbb{C}) \rightarrow L_+^\uparrow$.   This amounts to finding all matrices $M \in SL(2, \mathbb{C})$ which   satisfy $M X M^\dagger = X$ for all hermitian matrices $X$.  In the   particular case $X = 1$ we find $M M^\dagger = 1$.  Using this in $M X M^\dagger = X$ we find that $M$ commutes with all hermitian   matrices, which is possible only if $M$ is proportional to the   identity matrix.  The condition $\det M = 1$ only leaves $M = \pm   1$.} that there are always two and only two elements of $SL(2, \mathbb{C})$ corresponding to one element of $L_+^\uparrow$.  This means that $SL(2,\mathbb{C})$ is a double cover\footnote{Let us prove   that $SL(2, \mathbb{C})$ is the universal covering group of   $L_+^\uparrow$.  This means that we have to show that $SL(2,   \mathbb{C})$ is simply connected.  Let $M$ be a matrix of   determinant one.  Any invertible matrix has a polar decomposition $M   = H U$ where $H$ is hermitian and $U$ is unitary.  We can see this   as follows: $M^\dagger = U^\dagger H$, therefore $M M^\dagger = H^2$   and $H = (M M^\dagger)^{1/2}$.  $U = H^{-1} M$, $U^\dagger =   M^\dagger H^{-1}$.  $U$ is therefore unitary because $U^\dagger U =   M^\dagger H^{-2} M = M^\dagger (M M^\dagger)^{-1} M = 1$.

  The condition $\det M = 1$ gives $\det H \det U = 1$ and, using   $|\det U| = 1$ and $\det H \in \mathbb{R}$, $\det U = 1$ and $\det H   = 1$.  If $H = \begin{pmatrix}\alpha & \beta + i \gamma\\ \beta - i     \gamma & \delta \end{pmatrix}$ and $U = \begin{pmatrix}a + i b & c     + i d\\ -c + i d & a - i b \end{pmatrix}$, the conditions are   $\alpha \delta - \beta^2 - \gamma^2 = 1$ and $a^2 + b^2 + c^2 + d^2   = 1$.  This means that the topology of $SL(2, \mathbb{C})$ is   identical to $\mathbb{R}^3 \times \mathbb{S}^3$, where   $\mathbb{S}^3$ is the three dimensional sphere (the topology of   $L_+^\uparrow$ is $\mathbb{R}^3 \times \mathbb{S}^3 /   \mathbb{Z}_2$).  This proves that $SL(2, \mathbb{C})$ is simply   connected.} of $L_+^\uparrow$.

We will now consider a two-dimensional space, called the space of spinors. This is the space on which $SL(2, \mathbb{C})$ acts.  Its importance resides in the fact that the irreducible representations of the Poincar\' e group give the transformations of one-particle states.  As $SL(2, \mathbb{C})$ is the universal covering group of $L_+^\uparrow$, its representations are important in classifying the one particle states, especially those of half-integer spin.

We can formulate a tensorial calculus on the spinor space.  For example, the transformation law $X'_{\alpha \dot{\alpha}} = M_\alpha^{~\beta} X_{\beta \dot{\beta}} \smash[t]{M^\dagger}^{\dot{\beta}}_{~\dot{\alpha}}$, where ${M^\dagger}^{\dot{\beta}}_{~\dot{\alpha}} = (M_\alpha^{~\beta})^*$. One rule that immediately apparent is that complex conjugation ``puts a dot on the index.''  (More precisely, $SL(2, \mathbb{C})$ has two complex representations, $\mathbf{2}$ and $\bar{\mathbf{2}}$ which are complex conjugate of one another.  The undotted spinors transform in $\mathbf{2}$ and the dotted spinors transform in $\bar{\mathbf{2}}$.)

If the spinors are to be interpreted as Grassmann anti-commuting variables it is convenient to define a spinor scalar product by contracting the undotted spinor indices from North-West to South-East ($\psi^\alpha \chi_\alpha$) and the dotted indices from South-West to North-East ($\bar{\psi}_{\dot{\alpha}} \bar{\chi}^{\dot{\alpha}}$). This fits well with the fact that, in this case, the scalar product of a spinor with itself is generally non-zero and with the convention that the hermitian conjugation of a product of two Grassmannian numbers changes their order ($(\psi^\alpha \chi_\alpha)^\dagger = \bar{\chi}_{\dot{\alpha}} \bar{\psi}^{\dot{\alpha}}$).  However, here we are more interested in classifying the representations and it proves to be inconvenient to work with anti-commuting spinors, so all the spinors we will use are commuting.

The role of the invariant metric is played by $\epsilon_{\alpha \beta} = \begin{pmatrix} 0 & 1\\ -1& 0\end{pmatrix}$.  This is because $\epsilon'_{\alpha \gamma} = M_\alpha^{~\beta} M_\gamma^{~\delta} \epsilon_{\beta \delta} = \epsilon_{\alpha \gamma} \det M = \epsilon_{\alpha \gamma}$.

We can use this metric to lower indices by $T_\alpha = T^\beta \epsilon_{\beta \alpha}$.  We also have an ``inverse'' metric, which has upper indices $\epsilon^{\alpha \beta} \epsilon_{\alpha \gamma} \epsilon_{\beta \delta} = \epsilon_{\gamma \delta}$.  The consequence is that $\epsilon^{\alpha \beta} = \begin{pmatrix}0 & 1\\ -1 & 0 \end{pmatrix}$.  We raise the index with $T^\beta = \epsilon^{\beta   \alpha} T_\alpha$ and we use similar formulae for dotted indices. One awkward consequence is that $\epsilon^{\alpha \beta} \epsilon_{\beta \gamma} = - \delta^\alpha_\gamma$ so that the upper index metric is not the inverse of the lower index metric (this is the reason for the quotation marks above).  This will only be slightly awkward however, because we will not have to contract the indices on two $\epsilon$ tensors very often.

There are several conventions in use in the literature concerning the definition of the metric $\epsilon$, of its ``inverse,'' of the contraction rules, etc.  One convention \emph{defines} the $\epsilon$ with upper indices to be the inverse of the one with lower indices.  We do not choose this convention as it implies that $\epsilon$ is not a tensor (more precisely, the epsilon tensor with upper indices is \emph{not} equal to the tensor obtained by raising  the indices of the epsilon tensor with lower indices).  Also, sometimes the metric is given by $C_{\alpha \beta} = i \epsilon_{\alpha \beta}$.

Let us introduce some notation\footnote{The conventions defined here and in Appendix~\ref{ch:spinor_conventions} are only used in this part.  In part~\ref{part:twistor_string} we will use other conventions that are more convenient.  The conventions we use here are closer to the ones used for explicit computations while the conventions used in part~\ref{part:twistor_string} are closer to the ones used in the twistor literature.} (see also Appendix~\ref{ch:spinor_conventions}).  We define two types of contractions, for dotted and undotted indices:
\begin{align}
  \langle \xi \zeta\rangle =& \xi^\alpha   \zeta^\beta \epsilon_{\beta \alpha},\\
  \left[\overline{\xi} \overline{\zeta}\right] =&   \epsilon_{\dot{\alpha} \dot{\beta}} \overline{\xi}^{\dot{\beta}}   \overline{\zeta}^{\dot{\alpha}}.
\end{align}
These spinor products are $SL(2, \mathbb{C})$ and therefore Lorentz invariant.  They will be extensively used in the following.  These conventions are the ones used in the QCD literature and we will use them in this part and in part~\ref{part:scattering}.  In part~\ref{part:twistor_string} we use twistor-literature conventions which are different.  The two usages can be distinguished by the fact that, in twistor conventions, the spinors inside the angle brackets or square brackets are separated by a comma.

The spinor products satisfy an important identity, called Schouten identity.  For arbitrary spinors $\psi$, $\chi$, $\rho$ and $\mu$ we have
\begin{equation}
  \langle \psi \chi\rangle \langle \rho \mu\rangle + \langle \psi \rho\rangle \langle \mu \chi\rangle + \langle \psi \mu\rangle \langle \chi \rho\rangle = 0.
\end{equation}  This identity is the consequence of the fact that there is no rank-three completely symmetric tensor in two dimensions (see also Appendix~\ref{ch:spinor_conventions}).

After this intermezzo we are ready to apply the spinor formalism to computation of scattering amplitudes in four dimensions.

Take a null momentum $p^\mu$, such that $p^2=0$.  This implies that the determinant of $p_{\alpha \dot{\alpha}}$ is zero, so the rank of the $2 \times 2$ matrix $p_{\alpha \dot{\alpha}}$ is less or equal to one.  This in turn implies that one can find $\lambda_\alpha$ and $\tilde{\lambda}_{\dot{\alpha}}$ such that
\begin{equation}
  \label{eq:null_decomposition}
  p_{\alpha \dot{\alpha}} = \lambda_\alpha \tilde{\lambda}_{\dot{\alpha}}.
\end{equation}  If we impose the reality of momentum $p^\mu$, then the two spinors $\lambda$ and $\tilde{\lambda}$ can be normalised such that $\tilde{\lambda}^* = \pm \lambda$, where the minus signs is for negative energy and the plus sign is for positive energy.

Note that in Minkowski signature and with $p^\mu$ real, the spinors $\lambda$ and $\tilde{\lambda}$ are not uniquely determined.  If we rotate them by a phase
\begin{gather}
  \label{eq:lambda_ambiguity}
  \lambda \rightarrow e^{i \phi} \lambda,\\
  \tilde{\lambda} \rightarrow e^{-i \phi} \tilde{\lambda},
\end{gather} $p$ is unchanged.

The central idea of the spinor helicity method is to express the amplitudes using spinor language rather than vector language (see Appendix~\ref{ch:spinor_conventions} for a discussion of our choice of conventions and Appendix~\ref{ch:wavefunctions} for a discussion of wavefunctions in this language).  Of course, the description in terms of spinors can be done as described above only for on-shell scattering amplitudes of massless particles.

There are at least two reasons why the spinor helicity techniques lead to very simple expressions.  One reason is that the spinor products capture soft and collinear singularities in the amplitudes more naturally than the dot products (see Appendix.~\ref{ch:ir_divergencies}).  The second reason is that the supersymmetry Ward identities are naturally expressed in spinor language, as detailed in the following section.

\section{Supersymmetry Ward Identities}
\label{sec:swi}

Here we study the constraints the supersymmetry imposes on S-matrix elements.  Because supersymmetry links bosonic and fermionic states, we expect some implications for the $S$-matrix elements.  This was first discussed in refs.~\cite{Grisaru:1977px, Parke:1985pn}.

Consider an $N$-extended SUSY algebra
\begin{equation}
  \label{eq:susy_algebra}
  \left\{\mathbf{Q}_\alpha^i, \bar{\mathbf{Q}}_{\dot{\alpha}             j}\right\} = 2 \delta^i_j \mathbf{P}_{\alpha \dot{\alpha}},
\end{equation} where $i, j = 1, \ldots, N$ and where we have used a boldface font for operators.

Consider on-shell representations of the SUSY super-algebra in the massless case ($\mathbf{P}^2 = 0$ on these representations).  Take a one particle state\footnote{Note that we have labelled the state by its four dimensional momentum $p$, not by $\vec{p}$ as it is usually done.  The constraint $p^2 = 0$ is assumed.  One could label the states is by using $(\lambda, \overline{\lambda})$.} $\ket{p, h}$ characterised by momentum $p$ and helicity $h$, where $p^2 = 0$ and $p_{\alpha \dot{\alpha}} = \lambda_\alpha \bar{\lambda}_{\dot{\alpha}}$ and such that $\mathbf{Q}_\alpha^i \ket{p, h} = 0$.  This kind of state always exists (Clifford vacuum).

Define\footnote{Though we don't indicate it explicitly, it is   important to bear in mind that $\tilde{\mathbf{Q}}^i$ (and   $\mathbf{Q}^i$ defined below) depend on $\lambda$.} $\tilde{\mathbf{Q}}^i \equiv \lambda^\alpha \mathbf{Q}_\alpha^i$ and $\bar{\tilde{\mathbf{Q}}}_j \equiv \bar{\lambda}^{\dot{\alpha}} \bar{\mathbf{Q}}_{\dot{\alpha} j}$.  We also have $\bar{\tilde{\mathbf{Q}}}_j = \left(\tilde{\mathbf{Q}}^j\right)^\dagger$ and $\left\{\tilde{\mathbf{Q}}^i, \bar{\tilde{\mathbf{Q}}}_j\right\} = 0$.

Then (no summation on $i$),
\begin{equation}
  0 = \braket{p, h| \left\{\tilde{\mathbf{Q}}^i,
      \bar{\tilde{\mathbf{Q}}}_i\right\}| p, h} = \| \tilde{\mathbf{Q}}^i
  \ket{p, h} \|^2,
\end{equation} implies that $\tilde{\mathbf{Q}}^i = 0$.

Now define\footnote{Usually, when one discusses the on-shell representation of SUSY on massless particles, one goes to a reference frame where the momentum is $(k, 0, 0, k)$.  Using this one can prove that half of the supercharges are zero, count the number of degrees of freedom, etc.  This choice breaks Lorentz invariance, however.  In our discussion there is no need to choose a preferred reference frame.  This kind of parametrisation differs from the one usually presented in the literature.}
\begin{equation}
  \mathbf{Q}^i \equiv \frac {\mathbf{Q}_\alpha^i (\sigma^0)^{\alpha
      \dot{\alpha}} \bar{\lambda}_{\dot{\alpha}}}{\sqrt{2}     \tr(p_{\beta \dot{\beta}})}
\end{equation} and
\begin{equation}
  \bar{\mathbf{Q}}_j \equiv \frac {\bar{\mathbf{Q}}_{\dot{\alpha} j}
    (\bar{\sigma}^0)^{\dot{\alpha} \alpha} \lambda_\alpha}{\sqrt{2}         \tr(p_{\beta \dot{\beta}})}.
\end{equation}  The only \emph{raison d'\^ etre} for $\sigma^0$ is to convert from dotted to undotted indices in order to allow the contractions ($\sigma^0 = \mathbf{1}$ so it really has no other influence).  The traces above are also defined with its help.

$\mathbf{Q}$ and $\bar{\mathbf{Q}}$ have a simple algebra
\begin{equation}
  \left\{\mathbf{Q}^i, \bar{\mathbf{Q}}_j\right\} = \delta^i_j.
\end{equation}

Now, for the Clifford vacuum $\ket{p, h}$ we have $\mathbf{Q}^i \ket{p, h} = 0$ by construction and $\tilde{\mathbf{Q}}^i \ket{p, h} = \bar{\tilde{\mathbf{Q}}}_j \ket{p, h} = 0$.

It follows that the states of the super-multiplet are constructed by action of the lowering operators $\bar{\mathbf{Q}}_i$ on the highest weight $\ket{p, h}$.

We have

\begin{center}
\begin{tabular}{r|c|c}
  state(s) & helicity & multiplicity\\
  \hline
  $\ket{p, h}$ & $h$ & $1$\\
  $\bar{\mathbf{Q}}_i \ket{p, h}$ & $h - \frac 1 2$ & $N$\\
  $\bar{\mathbf{Q}}_i \bar{\mathbf{Q}}_j \ket{p, h}$ & $h - 1$ & $\frac {N
    (N - 1)} 2$\\
  $\vdots$ & $\vdots$ & $\vdots$\\
  $\bar{\mathbf{Q}}^N \ket{p, h}$ & $h - \frac N 2$ & 1
\end{tabular}
\end{center}

Strictly speaking, the states should also bear a label indicating their $R$-symmetry transformation properties.  The states above are already completely antisymmetric in indices $i, j, \ldots$, therefore they are irreducible tensors under $R$ symmetry transformations.

We will study the scattering amplitude of $n$ in-going particles (we take no particles to be out-going which is possible because of crossing symmetry\footnote{When using crossing symmetry one has negative energy particles (this is necessary by momentum conservation).  When considering negative energy states, some of the formulae above need to be modified by signs.  However, for our purposes these signs are irrelevant as they cancel in the final result.}).  Therefore, the $S$ matrix element will be
\begin{equation}
  \label{eq:s-matrix}
  {}_{\text{out}} \braket{\text{vac} | p_1, h_1; \cdots; p_n,
    h_n}_{\text{in}}.
\end{equation}

Before finding the action of a SUSY transformation on a $n$ particle
sector, study the action on a one particle sector.
\begin{align}
  \ket{p, h} &= a_p^\dagger(h) \ket{\text{vac}},\\
  \ket{p, h - \tfrac 1 2, i} &= \bar{\mathbf{Q}}_i \ket{p, h} =
  \bar{\mathbf{Q}}_i a_p^\dagger(h) \ket{\text{vac}} =\\
  &= a_p^\dagger(h - \tfrac 1 2, i) \ket{\text{vac}}.
\end{align}

From these formulae we deduce that the $\bar{\mathbf{Q}}_i$ can be
expressed in terms of creation and annihilation operators as follows
\begin{equation}
  \bar{\mathbf{Q}}_i = \int d^3 \vec{p} \left[a_p^\dagger(h - \tfrac 1         2, i) a_p(h) + \sum_j a_p^\dagger(h - 1, i j) a_p(h - \tfrac 1 2, j) +     \cdots\right].
\end{equation}

Now use the (anti-)commutation relations $[a_p(h), a_q^\dagger(h')]_{\pm} = \delta_{h h'} \delta^3(\vec{p} - \vec{q})$ and $[A B, C]_{\pm} = A [B, C]_{\pm} \mp [A, C] B$ we get
\begin{equation}
  a_p^\dagger(h' - \tfrac 1 2, i \ldots) = [\bar{\mathbf{Q}}_i,
  a_p^\dagger(h', \ldots)]_\pm.
\end{equation}  In fact, one can \emph{define} the lower helicity creation operators through the above relations.

On a multiparticle Fock space, in the $n$ particles sector, we have
\begin{multline}
  \bar{\zeta}^{\dot{\alpha}} \bar{\mathbf{Q}}_{\dot{\alpha} i}
  a_{p_1}^\dagger(h_1, i_1, \cdots) a_{p_2}^\dagger(h_2, i_2, \cdots)
  \cdots \ket{\text{vac}} =\\= [\bar{\zeta}^{\dot{\alpha}}
  \bar{\mathbf{Q}}_{\dot{\alpha} i},\; a_{p_1}^\dagger(h_1, i_1, \cdots)]_\pm a_{p_2}^\dagger(h_2, i_2, \cdots) \cdots \ket{\text{vac}}
  \mp\\ a_{p_1}^\dagger(h_1, i_1, \cdots) [\bar{\zeta}^{\dot{\alpha}}
  \bar{\mathbf{Q}}_{\dot{\alpha} i},\; a_{p_2}^\dagger(h_2, i_2, \cdots)]_\pm \cdots \ket{\text{vac}} \cdots
\end{multline} Inserting this operator between the in and out states above
  we get zero when acting to the left because the vacuum is invariant under
  SUSY, but we get a non-trivial expression when acting to the right.

Now, what is the action of $\bar{\zeta}^{\dot{\alpha}}
\bar{\mathbf{Q}}_{\dot{\alpha} i}$ when acting on a one particle state
$\bar{\mathbf{Q}}_k \cdots \bar{\mathbf{Q}}_l \ket{p, h}$?  We need to
project onto $\bar{\mathbf{Q}}_i$.

Decompose $\bar{\zeta}^{\dot{\alpha}}$ on the basis $\bar{\lambda}^{\dot{\alpha}}$, $(\bar{\sigma}^0)^{\dot{\alpha}   \alpha} \lambda_\alpha$ (these vectors are the eigenvectors of $p_{\alpha \dot{\alpha}}$ corresponding to the eigenvalues $0$ and $\tr (p_{\alpha \dot{\alpha}})$; since $p_{\alpha \dot{\alpha}}$ is hermitian, they are also orthogonal).  We have
\begin{equation}
  \bar{\zeta}^{\dot{\alpha}} = A \bar{\lambda}^{\dot{\alpha}} + B
  (\bar{\sigma}^0)^{\dot{\alpha} \alpha} \lambda_\alpha,
\end{equation} so $B = \frac {[\lambda, \zeta]}{\tr{p}}$.  The part which contains $A$ doesn't contribute because it projects to $\tilde{\mathbf{Q}}$
which is zero.

Therefore, when acting on a one particle state with momentum $p_{\alpha \dot{\alpha}} = \lambda_{\alpha} \bar{\lambda}_{\dot{\alpha}}$, we can make the replacement
\begin{equation}
  \bar{\zeta}^{\dot{\alpha}} \bar{\mathbf{Q}}_{\dot{\alpha} i} \equiv
  \sqrt{2} [\lambda, \zeta] \bar{\mathbf{Q}}_i.
\end{equation}

This finally gives
\begin{equation}
  \sum_{a = 1}^n \pm [\lambda_a, \zeta] A_n(p_1, h_1; \cdots; p_a, h_a - \frac 1 2; \cdots ; p_n, h_n) = 0,
\end{equation} where $\pm$ arises because when rearranging the operators $\bar{\mathbf{Q}}_i$ we may pick up some signs.  There is also a sign given by the parity of the number of fermions at the left of the particle where we subtract $\tfrac 1 2$.

Note that the formula above is an \emph{exact} result when the supersymmetry is not spontaneously broken.  As such, it holds order by order in perturbation theory.  Here we have only used the transformations under SUSY and the fact that, because it is a symmetry, the \emph{same} operator performs the SUSY transformations on the `in' and on the `out' states.

If the super-multiplets are not self-conjugate, not all possible helicities can be generated from the above Clifford vacuum (for example, the case of gauge multiplet in $N=2$ SUSY).  In this case, one must also start with an alternative vacuum $\ket{p, -h}$ which is such that $\bar{\mathbf{Q}}_i \ket{p, -h} = 0$ and $\tilde{\mathbf{Q}}^i \ket{p, -h} = \bar{\tilde{\mathbf{Q}}}_j \ket{p,   -h} = 0$.

Let us compute some relations between scattering amplitudes.  Each time we start with a list of helicities and apply the operator $\bar{\zeta}^{\dot{\alpha}} \bar{\mathbf{Q}}_{\dot{\alpha} i}$ keeping track of the signs generated by the permutation of this operator with the fermionic creation operators.  In order to obtain equations with a minimum number of terms, we take a maximum number of terms, we take most of the helicities be $-1$ (these states are annihilated by $\bar{\zeta}^{\dot{\alpha}} \bar{\mathbf{Q}}_{\dot{\alpha} i}$).

By starting with helicities $(- \tfrac 1 2, -1, -1, \ldots, -1)$ and applying the helicity lowering operator we get
\begin{equation}
  \left[\lambda_1\ \zeta\right] A(-1, -1, \ldots, -1) = 0.
\end{equation}  This proves that the all-minus helicity amplitude is zero.

By starting with helicities $(- \tfrac 1 2, 1, -1, \ldots, -1)$ and applying the helicity lowering operator we get
\begin{equation}
  \left[\lambda_1\ \zeta\right] A(-1, 1, -1, \ldots, -1) - \left[\lambda_2\ \zeta\right] A(- \tfrac 1 2, \tfrac 1 2, -1, \ldots, -1) = 0.
\end{equation}  Substituting $\zeta = \lambda_2$ and $\zeta = \lambda_1$ we get
\begin{align*}
  A(-1, 1, -1, \ldots, -1) =& 0,\\
  A(- \tfrac 1 2, \tfrac 1 2, -1, \ldots, -1) =& 0.
\end{align*}

An important result is that the helicity amplitude with one helicity plus and the remaining helicities minus is also identically zero.  Of course, all these relations between helicity amplitudes are also valid when the arguments are permuted\footnote{When working with colour ordered amplitudes the permutation symmetry is lost and only a cyclic symmetry survives.  One can also consider the case where no colour decomposition was performed.} (except maybe for signs from permutation of fermions).

We have shown above that the helicity amplitudes with all external legs with helicity minus or with all external legs helicity minus except for one leg with helicity plus are zero by SUSY.  At tree level, this is also true for non-supersymmetric theories because the Feynman diagrams contributing to gluon scattering amplitudes are the same in supersymmetric and non-supersymmetric theories.

It turns out that the simplest amplitudes (and the first that are not constrained to be zero by supersymmetry) are the so-called MHV amplitudes.  For $n$ external legs, these amplitudes have $n-2$ gluons of helicity plus and two gluons of helicity minus.  The helicity flipped amplitudes with $n-2$ gluons of helicity minus and two gluons of helicity plus are called $\overline{\text{MHV}}$ amplitudes.

Let us now obtain a relation linking two helicity amplitudes.  We start with helicities $(1, 1, - \tfrac 1 2, -1, \ldots -1)$ and apply the helicity lowering operator to get
\begin{multline}
  \left[\lambda_1 \ \zeta\right] A(\tfrac 1 2, 1, - \tfrac 1 2, -1, \ldots, -1) + \left[\lambda_2 \ \zeta\right] A(1, \tfrac 1 2, - \tfrac 1 2, -1, \ldots, -1) +\\+ \left[\lambda_3 \ \zeta\right] A(1, 1, -1, \ldots, -1) = 0.
\end{multline}

From this we obtain
\begin{equation}
  \label{eq:swi_step1}
  A(1, \tfrac 1 2, -\tfrac 1 2, -1, \ldots, -1) = - \frac {[1\ 3]}{[1\ 2]} A(1, 1, -1, \ldots, -1),
\end{equation} where we abbreviated $\left[\lambda_i\ \lambda_j\right] = [i\ j]$.  This tells us that amplitudes with external fermions are related to amplitudes with external gluons by supersymmetry.

Let us now prove an important formula for $\overline{\text{MHV}}$ amplitudes (the analogous formula for MHV amplitudes is obtained by parity conjugation).  Consider helicities $(1, \tfrac 1 2, 0, -1, \ldots, -1)$ and proceed as above to obtain
\begin{multline}
  \left[\lambda_1 \ \zeta\right] A(\tfrac 1 2, \tfrac 1 2, 0, -1, \ldots, -1) + \left[\lambda_2 \ \zeta\right] A(1, 0, 0, -1, \ldots, -1) -\\- \left[\lambda_3 \ \zeta\right] A(1, \tfrac 1 2, -\tfrac 1 2, -1, \ldots, -1) = 0,
\end{multline} which, for $\zeta = \lambda_1$ yields
\begin{equation}
  \label{eq:swi_step2}
  A(1, 0, 0, -1, \ldots, -1) = \frac {[1\ 3]}{[1\ 2]} A(1, \tfrac 1 2, -\tfrac 1 2, -1, \ldots, -1).
\end{equation}  In the next step, consider helicities $(1, 0, \tfrac 1 2, -1, \ldots, -1)$ and get
\begin{multline}
  \left[\lambda_1 \ \zeta\right] A(\tfrac 1 2, 0, \tfrac 1 2, -1, \ldots, -1) + \left[\lambda_2 \ \zeta\right] A(1, -\tfrac 1 2, \tfrac 1 2, -1, \ldots, -1) +\\+ \left[\lambda_3 \ \zeta\right] A(1, 0, 0, -1, \ldots, -1) = 0,
\end{multline} which, for $\zeta = \lambda_1$ yields
\begin{equation}
  \label{eq:swi_step3}
  A(1, -\tfrac 1 2, \tfrac 1 2, -1, \ldots, -1) = - \frac {[1\ 3]}{[1\ 2]} A(1, 0, 0, -1, \ldots, -1).
\end{equation}

Finally, in the last step, start with helicities $(1, -\tfrac 1 2, 1, -1, \ldots, -1)$ and get
\begin{multline}
  \left[\lambda_1 \ \zeta\right] A(\tfrac 1 2, -\tfrac 1 2, 1, -1 \ldots, -1) + \left[\lambda_2 \ \zeta\right] A(1, -1, 1, -1, \ldots, -1) -\\- \left[\lambda_3 \ \zeta\right] A(1, -\tfrac 1 2, \tfrac 1 2, -1, \ldots, -1) = 0,
\end{multline} which, for $\zeta = \lambda_1$ yields
\begin{equation}
  \label{eq:swi_step4}
  A(1, -1, 1, -1, \ldots, -1) = \frac {[1\ 3]}{[1\ 2]} A(1, -\tfrac 1 2, \tfrac 1 2, -1, \ldots, -1).
\end{equation}

Putting together eqns.~\eqref{eq:swi_step1}, \eqref{eq:swi_step2}, \eqref{eq:swi_step3}, \eqref{eq:swi_step4} we get
\begin{equation}
  \frac 1 {[1\ 2]^4} A(1, 1, -1, -1,\ldots, -1) = \frac 1 {[1\ 3]^4} A(1, -1, 1, -1, \ldots, -1).
\end{equation}  It is now obvious that, for an MHV amplitude where legs $i$ and $j$ have helicity minus and all the others have helicity plus, the ratio
\begin{equation}
  \frac 1 {\langle i\ j\rangle^4} A(1^+, \ldots, i^-, \ldots, j^-, \ldots, n^+)
\end{equation} is independent of $i$ and $j$.  This formula is true for $\mathcal{N}=4$ supersymmetry to all orders or for any amount of supersymmetry at tree level.  This observation is very useful when computing helicity amplitudes, because it tells us that, for MHV amplitudes, there is essentially only one helicity structure one needs to compute (all the others follow by supersymmetry).

Let us now study the six-point case.  Here, for the first time, more complicated amplitudes (that are not MHV or $\overline{\text{MHV}}$) appear.  The amplitudes with three gluons of helicity plus and three gluons of helicity minus are more complicated (see~\cite{Mangano:1990by}).  We will be interested here in how much information one can extract using supersymmetry Ward identities.

Start with helicities $(1,1,1,-\tfrac 1 2,-1,-1)$ and apply the helicity lowering operator to get
\begin{multline}
  [1\ \zeta] A(\tfrac 1 2, 1, 1, -\tfrac 1 2, -1, -1) + [2\ \zeta] A(1, \tfrac 1 2, 1, -\tfrac 1 2, -1, -1) +\\+ [3\ \zeta] A(1, 1, \tfrac 1 2, -\tfrac 1 2, -1, -1) + [4\ \zeta] A(1,1,1,-1,-1,-1) = 0.
\end{multline}  Letting $\zeta$ be $\overline{\lambda}_1$, $\overline{\lambda}_2$, $\overline{\lambda}_3$, we get
\begin{equation}
  \begin{pmatrix}
    0 & [2\ 1] & [3\ 1]\\
    [1\ 2] & 0 & [3\ 2]\\
    [1\ 3] & [2\ 3] & 0
  \end{pmatrix}
  \begin{pmatrix}
    A(\tfrac 1 2, 1, 1, -\tfrac 1 2, -1, -1)\\
    A(1, \tfrac 1 2, 1, -\tfrac 1 2, -1, -1)\\
    A(1, 1, \tfrac 1 2, -\tfrac 1 2, -1, -1)
  \end{pmatrix} =
  \begin{pmatrix}
    [1\ 4]\\ [2\ 4]\\ [3\ 4]
  \end{pmatrix} A(1,1,1,-1,-1,-1).
\end{equation}  If we could solve this linear system of equations it would be possible to express the amplitudes with fermions in terms of amplitudes with gluons only.  Unfortunately we can't solve the system as the determinant of the matrix is zero.\footnote{To see that this is indeed so it suffices to observe that this is an odd dimension antisymmetric matrix ($\det M = \det M^t = \det (- M) = (-1)^{2 d + 1} \det M = - \det M$, if $M$ is a $(2 d + 1) \times (2 d + 1)$ matrix).}

\section{MHV amplitudes}
\label{sec:MHV}

We have seen in the section~\ref{sec:swi} that the amplitudes with all plus helicity gluons and the amplitudes with all but one gluon of helicity plus and the remaining gluon of helicity minus are zero by supersymmetry.  We emphasize again that this result is correct at tree level even when the theory is not supersymmetric since the tree-level all gluon amplitudes receive contributions only from gluon vertices.

So the first non-trivial amplitudes are those with two helicity minus gluons and the rest with helicity plus.  These are called MHV (maximally helicity violating) amplitudes.  Remember that, when labelling helicities, we consider all particles to be outgoing.  By crossing symmetry, when we transform an outgoing particle to an ingoing one we have to change the sign of the helicity.

The origin of the name `maximally helicity violating' is the following: consider a scattering process of $n$ gluons.  As we recalled above, when taking into account supersymmetric Ward identities the first non-vanishing amplitude is the one with two gluons of helicity minus and $n-2$ of helicity plus.

Now, in collision experiments there usually are two incoming particles and the rest are outgoing.  In order to make contact with this experimental situation we have to take two gluons to be incoming by using crossing symmetry and we need to flip their helicity in the process.

If these gluons have helicity plus in the formulation with all the particles outgoing, they will have helicity minus after crossing symmetry and the MHV amplitude will be the scattering of two negative helicity gluons into $n-2$ gluons with $2$ of helicity minus and $n-4$ of helicity plus.  So this amplitude is maximally helicity violating in the sense that, given two helicity minus gluons in the initial state, one can't have more than $n-4$ gluons of opposite helicity in the final state of an $n$ particle tree-level scattering process.

It is remarkable that these MHV amplitudes have very simple form at tree level.  In fact, closed expressions for MHV amplitudes with arbitrary number of external legs are known (for a study of MHV amplitudes in string theory see ref.~\cite{Stieberger:2006te}).  This is remarkable since the tree level amplitudes for large numbers of external legs are obtained by summing a large number of Feynman diagrams and the complexity grows very rapidly with the number of external legs (see Table~1 in ref.~\cite{Mangano:1990by}; the number of diagrams one has to sum for a ten-gluon tree amplitude is larger than ten million!).

The expression for a colour-stripped $n$-point MHV amplitude with legs $i$ and $j$ of helicity minus is
\begin{equation}
  \label{eq:MHV}
  A(1^+, \ldots i^-, \ldots j^-, \ldots, n) = i g^{n-2} \frac {\langle i\ j\rangle^4}{\langle 1\ 2\rangle \langle 2\ 3\rangle \cdots \langle n\ 1\rangle},
\end{equation} where $g$ is the Yang-Mills coupling constant, $\langle\ \rangle$ is the spinor product defined in sec.~\ref{sec:spinor_helicity} and Appendix~\ref{ch:spinor_conventions}.

Note that, after dividing the amplitude by $\langle i\ j\rangle^4$, the legs $i$ and $j$ do not play a special role anymore, in agreement with the implications of $\mathcal{N}=4$ supersymmetry Ward identities.  One can also verify that the properties enumerated in sec.~\ref{sec:colour_decomposition} hold.  The cyclic symmetry and the reflection symmetry are obvious.  The dual Ward identity is not obvious and its proof necessitates some non-trivial spinor manipulations.

Let us show that, for MHV tree-level amplitudes,
\begin{equation}
  A(q, 1, 2, \ldots, n) + A(1, q, 2, \ldots, n) + \cdots A(1, \ldots, q, n) = 0.
\end{equation}  After using the expression for MHV amplitudes \ref{eq:MHV}, this reduces to proving that
\begin{equation}
  \sum_{l=1}^n \frac {\langle l\ (l+1)\rangle}{\langle l q\rangle \langle q\ (l+1)\rangle} = 0,
\end{equation} where we take $l=n+1 \equiv 1$ for the summation index $l$.  By multiplying the above sum by $\langle a\ q\rangle$ where $a$ is an arbitrary spinor and using Schouten identity we get
\begin{align}
  \langle a\ q\rangle \sum_{l=1}^n \frac {\langle l\ (l + 1)\rangle}{\langle l q\rangle \langle q\ (l + 1)\rangle} =& - \sum_{l=1}^n \frac {\langle a\ l\rangle \langle (l + 1)\ q\rangle + \langle (l + 1)\ a\rangle \langle l\ q\rangle}{\langle l q\rangle \langle q\ (l + 1)\rangle}\\ =& \sum_{l=1}^n \left(- \frac {\langle l\ a\rangle}{\langle l\ q\rangle} + \frac {\langle (l + 1)\ a\rangle}{\langle (l + 1)\ q\rangle}\right) = 0.
\end{align}

Another feature that will be important later is that the MHV amplitude depends only on the left-handed spinors $\lambda_i$ and does not depend on $\tilde{\lambda}_i$.  In $+---$ signature, where $\tilde{\lambda} = \pm \lambda^*$, we say that the MHV amplitude is holomorphic.

Let us now discuss the nature of the amplitudes in some examples of low number of external legs.

The $n=3$ case only admits degenerate kinematics in signature $+---$ and for real on-shell momenta.  The momentum conservation $p_1 + p_2 + p_3 = 0$ and on-shell conditions $p_1^2 = p_2^2 = p_3^2 = 0$ imply that $p_i \cdot p_j = 0$ for all $i, j$ form $1$ to $3$.  In signature $+---$ and for real on-shell momenta, this implies that $p_1, p_2, p_3$ are all collinear.  Therefore, the MHV amplitude is zero since all the spinor products are zero and the numerator has a higher power (four) than the denominator (three).  However, in other signatures or for complex on-shell momenta the kinematics is non-degenerate and the amplitude does not vanish.

If $n=4$, only MHV amplitudes are non-vanishing.  In this case, the amplitude can be equally considered as MHV or as $\overline{\text{MHV}}$.

For example, the four-point MHV amplitude $A(1^-, 2^-, 3^+, 4^+)$ with helicities indicated by a superscript can be written in two ways
\begin{align}
  A(1^-, 2^-, 3^+, 4^+) &= i g^2 \frac {\langle 1\ 2\rangle^3}{\langle 2\ 3\rangle \langle 3\ 4\rangle \langle 4\ 1\rangle},  \label{eq:MHV-4pt}\\
  A(1^-, 2^-, 3^+, 4^+) &= i g^2 \frac {\left[3\ 4\right]^3}{\left[1\ 2\right] \left[2\ 3\right] \left[4\ 1\right]}.   \label{eq:MHVBar-4pt}
\end{align}  Above we used the short-hand notation $|\lambda_i\rangle \to | i\rangle$ and $|\tilde{\lambda}_i] \to | i]$, where $p_i^{\alpha \dot{\alpha}} = \lambda_i^{\alpha} \tilde{\lambda}_i^{\dot{\alpha}}$.

In fact, the two expressions are identical as can be seen by using momentum conservation.  Momentum conservation in spinor language can be written
\begin{equation}
  |1\rangle [1| +  |2\rangle [2| +  |3\rangle [3| +  |4\rangle [4| = 0.
\end{equation}  Multiplying with $\langle 2|$ at left and with $|4]$ at right we get $\langle 2\ 1\rangle [1\ 4] + \langle 2\ 3\rangle [3\ 4] = 0$, or $\frac {\langle 1\ 2\rangle}{\langle 2\ 3\rangle} = - \frac {[3\ 4]}{[4\ 1]}$.  Other identities can be found in the same way and used to prove the identity of eq.~\eqref{eq:MHV-4pt} and eq.~\eqref{eq:MHVBar-4pt}.

It is a general feature of the spinor language computations that the same result can be written in very different ways.  So, while the spinor notation leads to the most compact results, it has to be kept in mind that simplifications are often very difficult to perform.  There are many examples in the literature where an analytical proof of identity of two expressions in spinor language is not known, but when evaluated numerically they always yield the same result.

For the case $n=5$ all non-vanishing amplitudes are MHV or $\overline{\text{MHV}}$.

The last case we will discuss is the case $n=6$.  Here, for the first time one has NMHV amplitudes, which have three helicity minus gluons.  By using cyclic and reflection symmetries we find there are only three independent amplitudes (the helicity distribution in these three cases can be taken to be $+++---$, $++--+-$ and $+-+-+-$).  These amplitudes were first computed in refs.~\cite{Mangano:1987xk, Berends:1987cv}.

One has to keep in mind that these amplitudes are not completely independent, as the dual Ward identity links them.  Using the dual Ward identity and the cyclic and reflection symmetries
\begin{multline}
  A(1^+, 2^+, 3^+, 4^-, 5^-, 6^-) + A(2^+, 1^+, 3^+, 4^-, 5^-, 6^-) + A(2^+, 3^+, 1^+, 4^-, 5^-, 6^-) +\\  A(3^+, 2^+, 6^-, 5^-, 1^+, 4^-) + A(2^+, 3^+, 4^-, 5^-, 1^+, 6^-) = 0.
\end{multline}  This is a link between $+++---$ and $++--+-$ tree-level amplitudes.

\section{Regularisation Schemes}
\label{sec:regularisation_schemes}

When computing loop amplitudes in field theories, one usually encounters divergences.  In this thesis we will be concerned with (super-)conformal field theories which are free of ultraviolet divergences.  However, the regularisation is needed even for UV finite theories for several reasons.

One reason is that often the UV divergences do not cancel diagram by diagram but between different diagrams, so that only the final results are finite. (Sometimes the cancellations can be made explicit by working in some kind of superspace formalism where the cancellations due to supersymmetry are manifest.  However, for some extended supersymmetry theories a superspace construction does not exist, or when it does it is very complicated.)  Moreover, when computing scattering amplitudes for on-shell massless particles one encounters infrared divergences which must also be regulated.  Also, even if the theory is UV-finite, composite operators have UV divergences which have to be regulated.  Ultimately, these UV divergences lead to non-trivial renormalisation properties for these operators, encoded in their anomalous dimensions.

The IR divergences prevent, in a strict textbook sense, the definition of an $S$-matrix for massless theories in low enough dimensions.  One can construct some other observables which are free of IR divergences (infrared safe observables).  Examples are inclusive cross-sections and jet observables.  The jet observables depend on the details of the detectors and the precise definitions of jets but do not depend on the regulator and are free of IR divergences.  In the following we will not try to construct IR safe observables and the amplitudes we will compute at loop level will explicitly depend on the IR regulator (that is, in dimensional regularisation they will contain poles in $\tfrac 1 \epsilon$).

There are several prescriptions for dimensional regularisation (and even for a given prescription there are several subtraction schemes).  We will describe in what follows the most widely used and their characteristics (see~\cite{Jack:1997sr} for a more detailed discussion).

The simplest regularisation prescription is conventional dimensional regularisation (CDR).  In this prescription the momenta and also the vector boson indices are continued to $D$ dimensions, where $D$ is complex.  The $\gamma$ matrices are kept four dimensional but their vector index takes values form $0$ to $D-1$ $(\gamma^0, \ldots \gamma^{D-1})$ The four dimensional divergences manifest themselves as poles in the complex variable $D$, when $D=4$.  It is customary to express the results as functions of $\epsilon$, where $D = 4 - 2 \epsilon$.

The 't Hooft-Veltman (HV) scheme is similar to CDR except that the external polarisations of the particles are kept in four dimensions.  This scheme is better suited to computing helicity amplitudes (amplitudes where one has well defined helicities for all the external legs), because helicity is only defined in four dimensions.

These regularisation prescriptions break the supersymmetry.  This is easy to see because a necessary condition for supersymmetry is the equality of the number of fermionic and bosonic states.  This equality is only valid in some particular dimensions and it is not respected for arbitrary dimension $D$.

Several SUSY preserving regularisation prescriptions were constructed, inspired by the idea of dimensional regularisation.  One widely used such regularisation prescription is the dimensional reduction scheme (DR) (see~\cite{Siegel:1979wq} for the original paper and~\cite{Capper:1979ns} for a pedagogical introduction).  The idea here is to mimic the dimensional reduction construction that was used to construct some SUSY theories (see~\cite{Brink:1976bc}).  The regularised theory is obtained by dimensional reduction from four dimensions to $D<4$ dimensions (note that at first we define the regularised theory for real $D$ but once the integrals are evaluated we can continue $D$ to complex values).

As in all the other prescriptions based on the idea of dimensional regularisation, the loop momenta are $D$-dimensional.  This implies that all the Kronecker $\delta$'s resulting from the loop integrals are also $D$-dimensional.

All the indices on the fields and on the $\gamma$ matrices have their four dimensional values.  There appears therefore a four-dimensional Kronecker $\delta_{(4)}$.  In the end, we have both four-dimensional and $D$-dimensional vectors and tensors.  All the contractions between four and $D$-dimensional tensors are performed as if the $D$-dimensional quantities were embedded in a four dimensional space, by adding $4 - D$ zero components.

In the DR regularisation prescription, the four-dimensional vector boson decomposes into a $D = 4 - 2 \epsilon$ dimensional vector boson and $4 - D = 2 \epsilon$ scalars.  Note that the scalars and vector bosons have different renormalisation properties so the separation above in vectors and scalars is important.

The DR regularisation prescription is plagued by an inconsistency first pointed out by Siegel in~\cite{Siegel:1980qs}.  The inconsistency comes from the inability to properly define a completely antisymmetric tensor of rank-four in $D$ dimensions, where $D < 4$ as is the case for dimensional reduction.

Let us briefly discuss this inconsistency.  Denote the $D$-dimensional tensors by hatted symbols and the four-dimensional ones by unhatted symbols.  Take the $D$-dimensional indices to be $\mu, \nu, \rho, \ldots$ and the four-dimensional ones to be $\alpha, \beta, \gamma, \ldots$.

Then, one can try to define a completely antisymmetric rank-four $D$-dimensional tensor by
\begin{equation*}
  \hat{\epsilon}^{\alpha \beta \gamma \delta} = \hat{\eta}^{\alpha \mu} \hat{\eta}^{\beta \nu} \hat{\eta}^{\gamma \rho} \hat{\eta}^{\delta \sigma} \epsilon_{\mu \nu \rho \sigma}.
\end{equation*}

The four-dimensional completely antisymmetric tensor $\epsilon$ satisfies the following identity
\begin{equation*}
  \epsilon^{\mu \nu \rho \sigma} \epsilon_{\mu' \nu' \rho' \sigma'} = - \delta_{\left[\mu'\right.}^\mu \delta_{\nu'}^\nu \delta_{\rho'}^\rho \delta_{\left.\sigma'\right]}^\sigma
\end{equation*} and, using the definition above for $\hat{\epsilon}$ we can prove a similar relation
\begin{equation*}
  \hat{\epsilon}^{\alpha \beta \gamma \delta} \hat{\epsilon}_{\alpha' \beta' \gamma' \delta'} = - \hat{\delta}_{\left[\alpha'\right.}^\alpha \hat{\delta}_{\beta'}^\beta \hat{\delta}_{\gamma'}^\gamma \hat{\delta}_{\left.\delta'\right]}^\delta.
\end{equation*}

Now try to compute the following quantity
\begin{equation*}
  \hat{\epsilon}^{\alpha' \beta' \gamma' \delta'} \hat{\epsilon}^{\alpha \beta \gamma \delta} \hat{\epsilon}_{\alpha \beta \gamma \delta},
\end{equation*} in two different ways.  One can first compute the contraction of the second and third $\hat{\epsilon}$ tensors
\begin{equation*}
  \hat{\epsilon}^{\alpha \beta \gamma \delta} \hat{\epsilon}_{\alpha \beta \gamma \delta} = - D (D - 1) (D - 2) (D - 3),
\end{equation*} or one can first compute the product of the first and third $\hat{\epsilon}$ tensors
\begin{equation*}
  \hat{\epsilon}^{\alpha' \beta' \gamma' \delta'} \hat{\epsilon}_{\alpha \beta \gamma \delta} = - \hat{\delta}_{\left[\alpha\right.}^{\alpha'} \hat{\delta}_\beta^{\beta'} \hat{\delta}_\gamma^{\gamma'} \hat{\delta}_{\left.\delta\right]}^{\delta'}.
\end{equation*}

One finally obtains two different results for this computation
\begin{equation*}
  D (D - 1) (D - 2) (D - 3) \hat{\epsilon}^{\alpha' \beta' \gamma' \delta'} = 4 ! \hat{\epsilon}^{\alpha' \beta' \gamma' \delta'}.
\end{equation*}  The relation above forbids analytical continuation in $D$.

Note that the kind of manipulations that gave rise to this ambiguity do not arise at one loop so the DR regularisation is generally considered to be safe at one loop.

The DR regularisation is not suited to computing helicity amplitudes.  One way to see this is to observe that the little group is $SO(D-2) = SO(2 - 2 \epsilon)$.  This group is smaller than $SO(2) \equiv U(1)$ which is the helicity group.  In other words, there is no plane in which one can perform a rotation which is necessary to define the helicity.

In order to surpass the inability to compute helicity amplitudes using the DR regularisation, the four-dimensional helicity scheme (FDH) was devised (see refs.~\cite{Bern:1991aq, Bern:2002zk}).  This scheme is a kind of hybrid between the HV and DR schemes.

Since this is more involved than the other prescriptions and perhaps less widely known, let us describe it in more detail.

\begin{itemize}
\item As in all variants of dimensional regularisation the internal momenta are $D$-dimensional.  The tensors such as Kronecker $\delta$ arising from the integrals are also $D$-dimensional.
\item All the external polarisations and momenta are kept in four dimensions.  This allows us to compute helicity amplitudes.
\item All the internal states are treated as they are in $D_s$-dimensions\footnote{But with $D$-dimensional momenta.}($D_s$ is called the spin dimension).  All the sums over polarisations and fermion states should be performed as if the internal states were in $D_s$-dimensions.
\item All the index contractions are performed \emph{as if} $D_s > D > 4$.
\item The rules for $\gamma_5$ are that $\gamma_5$ \emph{commutes} with $\gamma_\mu$ if the index $\mu$ is outside four dimensions.
\item After all the computations are performed the result is a function of $D_s$ and $D$ which can be analytically continued outside the region $D_s > D > 4$ where we initially defined it.
\item Finally, after performing all computations, set $D_s = 4$.  This final step is necessary for preserving supersymmetry.  In fact, when $D_s = D$ we obtain the HV scheme and when $D_s = 4$ we obtain the FDH scheme.
\end{itemize}

That the FDH scheme obeys supersymmetry was checked in~\cite{Kunszt:1993sd, Bern:1994zx, Bern:1994cg} to one loop and in~\cite{Bern:2002zk} to two loops.  At this point there is no known inconsistency in the FDH regularisation.

\part{Twistor String}
\label{part:twistor_string}

\chapter{More on spinors}

In this part, we change the conventions a bit since, with the old conventions a lot of awkward factors and signs appear frequently.  Our new conventions will be the same for dotted and undotted indices
\begin{equation}
  \begin{alignedat}{3}
    \psi^\alpha =& \epsilon^{\alpha \beta} \psi_\beta, \qquad&     \psi_\beta =& \psi^\alpha \epsilon_{\alpha \beta},\\
    \overline{\psi}^{\dot{\alpha}} =& \epsilon^{\dot{\alpha} \dot{\beta}} \overline{\psi}_{\dot{\beta}}, \qquad& \overline{\psi}_{\dot{\beta}} =& \overline{\psi}^{\dot{\alpha}} \epsilon_{\dot{\alpha} \dot{\beta}}.
  \end{alignedat}
\end{equation}

We also define the $2 \times 2$ matrix corresponding to a quadrivector with a factor of $\tfrac 1 {\sqrt{2}}$
\begin{equation}
  p_{\alpha \dot{\alpha}} = \frac 1 {\sqrt{2}} p_\mu \sigma^\mu_{\alpha \dot{\alpha}}.
\end{equation}

We can define the notion of tensor with spinorial indices.  There are four types of indices: upper dotted and undotted and lower dotted and undotted.  Once we have the metric $\epsilon$ we can connect an upper index spinor to a lower index spinor for dotted and undotted components (we can do the same for multi-index tensors).

In order to find the tensors with spinorial indices which correspond to tensors in Minkowski space, we have to generalise the notion of hermiticity to tensors of several indices.  We first define the conjugate of a tensor $\xi^{\alpha \beta \ldots \dot{\gamma}   \dot{\delta} \ldots}$ by $\smash[t]{\bar{\xi}}^{\dot{\alpha}   \dot{\beta} \ldots \gamma \delta \ldots} = (\xi^{\alpha \beta \ldots   \dot{\gamma} \dot{\delta} \ldots})^*$.  Obviously the number of dotted and undotted indices must be the same in order to represent a Minkowski space tensor (this is obvious if we recall that that a tensor with $n$ indices can be represented as a sum of terms of the form $v_1 \otimes \cdots \otimes v_n$ and each $v$ is represented by a tensor with a dotted and an undotted index).  The way to generalise the hermiticity condition is: $\bar{\xi}^{\alpha \beta \ldots   \dot{\gamma} \dot{\delta} \ldots} = \xi^{\alpha \beta \ldots   \dot{\gamma} \dot{\delta} \ldots}$.

Let us find the corresponding spinorial tensors for the Minkowski metric $\eta_{\mu \nu}$ and the totally antisymmetric tensor $\epsilon_{\mu \nu \rho \sigma}$.  These tensors are invariant tensors, therefore they have to be constructed from the only invariant spinorial tensors, $\epsilon_{\alpha \beta}$ and $\epsilon_{\smash[t]{\dot{\alpha}} \smash[t]{\dot{\beta}}}$.  To the tensor $\eta_{\mu \nu}$ corresponds $\eta_{\alpha \dot{\alpha} \beta   \dot{\beta}}$.  The only way to write a tensor of this kind by using $\epsilon$ is $\epsilon_{\alpha \beta} \epsilon_{\smash[t]{\dot{\alpha}} \smash[t]{\dot{\beta}}}$.  This means that $\eta_{\alpha \dot{\alpha} \beta \dot{\beta}} \propto \epsilon_{\alpha \beta} \epsilon_{\smash[t]{\dot{\alpha}}   \smash[t]{\dot{\beta}}}$.

In order to find the proportionality constant we use $\eta_{\alpha   \dot{\alpha} \beta \dot{\beta}} = \tfrac 1 2 \eta_{\mu \nu} (\sigma^\mu)_{\alpha \dot{\alpha}} (\sigma^\nu)_{\beta \dot{\beta}}$ and we calculate $\eta_{1 \dot{1} 2 \dot{2}} = 1$.  This implies $\eta_{\alpha \dot{\alpha} \beta \dot{\beta}} = \epsilon_{\alpha   \beta} \epsilon_{\smash[t]{\dot{\alpha}} \smash[t]{\dot{\beta}}}$. The upper index metric is given by $\eta^{\alpha \dot{\alpha} \beta   \dot{\beta}} = \epsilon^{\alpha \beta} \epsilon^{\smash[t]{\dot{\alpha}} \smash[t]{\dot{\beta}}}$.  This is pertinent because the translation of $p^\mu = \eta^{\mu \nu} p_\nu$ in spinor language is $p^{\alpha \dot{\alpha}} = \epsilon^{\alpha \beta} \epsilon^{\smash[t]{\dot{\alpha}} \smash[t]{\dot{\beta}}} p_{\beta   \dot{\beta}}$ which is just the raising of the indices in spinor language.  One unexpected conclusion is that in order for this to work we need $(+,-,-,-)$ signature.  Indeed, for the $(-,+,+,+)$ signature we get an extra $-$ sign.  So in order for the the raising or lowering of spinor and vector indices to be compatible, one has to use the mostly minus metric signature.  See refs.~\cite{Penrose:1985jw,Penrose:1986ca,Wald:1984rg} for a related discussion.

The scalar product can be written in spinor language $p \cdot q = \eta^{\mu \nu} p_\mu q_\nu = \epsilon^{\alpha \beta} \epsilon^{\dot{\alpha} \dot{\beta}} p_{\alpha \dot{\alpha}} q_{\beta \dot{\beta}} = p^{\alpha \dot{\alpha}} q_{\alpha \dot{\alpha}}$.  The case of light-like vectors $p^2 = 0$ is particularly interesting.  In that case, $\det p_{\alpha \dot{\alpha}} = 0$ so we can find two spinors $\lambda_\alpha$ and $\tilde{\lambda}_{\dot{\alpha}}$ such that $p_{\alpha \dot{\alpha}} = \lambda_\alpha \tilde{\lambda}_{\dot{\alpha}}$.  For two light-like quadrivectors $p_{\alpha \dot{\alpha}} = \lambda_\alpha \tilde{\lambda}_{\dot{\alpha}}$ and $q_{\alpha \dot{\alpha}} = \mu_\alpha \tilde{\mu}_{\dot{\alpha}}$, we have $p \cdot q = \langle \lambda,\ \mu\rangle [\tilde{\lambda},\ \tilde{\mu}]$. (Note that in the conventions of Part~\ref{part:scattering}  we would have $2 p \cdot q = \langle\lambda\ \mu\rangle [\tilde{\mu}\ \tilde{\lambda}]$.)

Let us now try to construct the tensor corresponding to the completely antisymmetric tensor $\epsilon_{\mu \nu \rho \sigma}$.  This tensor has the following index structure $\epsilon_{\alpha \dot{\alpha} \beta   \dot{\beta} \gamma \dot{\gamma} \delta \dot{\delta}}$.  We will have to construct this tensor from $\epsilon_{\alpha \beta}, \epsilon_{\alpha \gamma}, \epsilon_{\alpha \delta}, \epsilon_{\beta   \gamma}, \epsilon_{\beta \delta}, \epsilon_{\gamma \delta}$ and the corresponding expressions with dotted indices.  In order to use all undotted indices we will have to use $\epsilon_{\alpha \beta} \epsilon_{\gamma \delta}, \epsilon_{\alpha \gamma} \epsilon_{\beta   \delta}, \epsilon_{\alpha \delta} \epsilon_{\beta \gamma}$.  The expression we look for must be antisymmetric in any pairs of undotted-dotted indices: $\epsilon_{\alpha \dot{\alpha} \beta   \dot{\beta} \gamma \dot{\gamma} \delta \dot{\delta}} = -\epsilon_{\beta \dot{\beta} \alpha \dot{\alpha} \gamma \dot{\gamma}   \delta \dot{\delta}}$, \ldots These conditions restrict the most general form of $\epsilon_{\alpha \dot{\alpha} \beta   \dot{\beta} \gamma \dot{\gamma} \delta \dot{\delta}}$ to be proportional to
\begin{equation*}
  \epsilon_{\alpha \beta} \epsilon_{\gamma \delta}      (\epsilon_{\dot{\alpha} \dot{\gamma}} \epsilon_{\dot{\beta}
    \dot{\delta}} + \epsilon_{\dot{\beta} \dot{\gamma}}      \epsilon_{\dot{\alpha} \dot{\delta}}) - \epsilon_{\alpha \gamma}      \epsilon_{\beta \delta} (\epsilon_{\dot{\alpha} \dot{\beta}}      \epsilon_{\dot{\gamma} \dot{\delta}} + \epsilon_{\dot{\alpha}
    \dot{\delta}} \epsilon_{\dot{\gamma} \dot{\beta}}) -      \epsilon_{\alpha \delta} \epsilon_{\beta \gamma}      (\epsilon_{\dot{\alpha} \dot{\gamma}} \epsilon_{\dot{\beta}
    \dot{\delta}} + \epsilon_{\dot{\alpha} \dot{\beta}}      \epsilon_{\dot{\gamma} \dot{\delta}}).
\end{equation*}

This expression can be simplified however, by using $\epsilon_{\alpha   \beta} \epsilon_{\gamma \delta} + \epsilon_{\alpha \gamma} \epsilon_{\delta \beta} + \epsilon_{\alpha \delta} \epsilon_{\beta   \gamma} = 0$ (this formula is the consequence of the fact that the left-hand side is completely antisymmetric in $\beta, \gamma, \delta$).  Applying this identity repeatedly we can simplify the above general form to $-3 (\epsilon_{\alpha \delta} \epsilon_{\beta \gamma} \epsilon_{\dot{\alpha} \dot{\gamma}} \epsilon_{\dot{\beta} \dot{\delta}} - \epsilon_{\alpha \gamma} \epsilon_{\beta \delta} \epsilon_{\dot{\alpha} \dot{\delta}} \epsilon_{\dot{\beta} \dot{\gamma}})$.  We have therefore the general form of the completely antisymmetric tensor apart a normalisation. In order to fix the normalisation we calculate a non-zero component of the tensor, $\epsilon_{1 \dot{1} 2 \dot{2} 1 \dot{2} 2 \dot{1}}$ for example, by using the formula $\epsilon_{\alpha \dot{\alpha} \beta \dot{\beta} \gamma \dot{\gamma} \delta \dot{\delta}} = \tfrac 1 4 \epsilon_{\mu \nu \rho \tau} (\sigma^\mu)_{\alpha \dot{\alpha}} (\sigma^\nu)_{\beta \dot{\beta}} (\sigma^\rho)_{\gamma \dot{\gamma}} (\sigma^\tau)_{\delta \dot{\delta}}$.  This calculation fixes the normalisation and gives
\begin{equation}
  \label{eq:spinorial_epsilon}
  \epsilon_{\alpha \dot{\alpha} \beta \dot{\beta} \gamma \dot{\gamma}
    \delta \dot{\delta}} = i (\epsilon_{\alpha \delta} \epsilon_{\beta
    \gamma} \epsilon_{\dot{\alpha} \dot{\gamma}} \epsilon_{\dot{\beta}
    \dot{\delta}} - \epsilon_{\alpha \gamma} \epsilon_{\beta \delta}    \epsilon_{\dot{\alpha} \dot{\delta}} \epsilon_{\dot{\beta}                 \dot{\gamma}}).
\end{equation}  The appearance of $i$ might seem surprising but it is in fact necessary to insure hermiticity.

The irreducible representations of $SL(2,\mathbb{C})$ can be formed from tensorial products of spinors with dotted and undotted indices. For this it is important to classify the tensors by their symmetry properties.  It is obvious that these symmetry properties are preserved by transformations.  It will be sufficient to restrict our analysis to tensors with all upper dotted and all lower undotted indices (this is because we can raise or lower the indices at will).

An antisymmetric tensor in two indices $T_{\ldots \alpha \ldots \beta
  \ldots} = - T_{\ldots \beta \ldots \alpha \ldots}$ has the property $T_{\ldots \alpha \ldots \beta \ldots} = \frac 1 2 \epsilon_{\alpha
  \beta} T_{\ldots \gamma \ldots \phantom{\alpha}
  \ldots}^{\phantom{\ldots \alpha \ldots} \gamma}$ (the same is obviously true for dotted indices).  This has the effect that an antisymmetric tensor effectively transforms as a tensor with two fewer indices.  There remain only the totally symmetric tensors in the undotted and dotted indices which are irreducible.  We can therefore label the irreducible transformations of $SL(2, \mathbb{C})$ by the couple $(\mathbf{\tfrac k 2}, \mathbf{\tfrac l 2})$.  The dimension of the $(\mathbf{\tfrac k 2}, \mathbf{\tfrac l 2})$ representation is $(k + 1)(l + 1)$.

For example, the $(\mathbf{\tfrac 1 2}, \mathbf{0})$ representation is given by the action of $M_\alpha^{~\beta}$ on the spinors $\xi_\beta$, the $(\mathbf{0}, \mathbf{\tfrac 1 2})$ representation given by the action of $(M^\dagger)^{\dot{\alpha}}_{~\dot{\beta}}$ on $\zeta^{\dot{\beta}}$ (this can also be described as the action of $(M^*)_{\dot{\alpha}}^{~\smash[t]{\dot{\beta}}}$ on $\zeta_{\dot{\beta}}$).

As we have already seen, the $(\mathbf{\tfrac 1 2}, \mathbf{\tfrac 1 2})$ representation corresponds to the quadrivectors.  In what concerns the rank-two tensors, we can reduce this representation to the symmetric traceless, antisymmetric, and identity representations of dimensions $9$, $6$ and $1$.  The symmetric rank-two tensors correspond to $(\mathbf{1}, \mathbf{1})$ and the identity part corresponds to $(\mathbf{0}, \mathbf{0})$.  What does the antisymmetric part corresponds to?  The quickest way to answer is to observe that a rank-two tensor transforms as a tensor product of vectors.  As the vectors are of type $(\mathbf{\tfrac 1 2}, \mathbf{\tfrac 1 2})$, this means that the rank-two tensors $(\mathbf{\tfrac 1 2}, \mathbf{\tfrac 1 2}) \otimes (\mathbf{\tfrac 1 2}, \mathbf{\tfrac 1 2})$ reduces to $(\mathbf{1}, \mathbf{1}) \oplus (\mathbf{1}, \mathbf{0}) \oplus (\mathbf{0}, \mathbf{1}) \oplus (\mathbf{0}, \mathbf{0})$.  This means that the antisymmetric tensors are given by $(\mathbf{1}, \mathbf{0}) \oplus (\mathbf{0}, \mathbf{1})$.

What are the tensors which transform under $(\mathbf{0}, \mathbf{1})$ and $(\mathbf{1}, \mathbf{0})$?  An antisymmetric rank-two tensor can be decomposed in a self-dual and an anti-self-dual part corresponding to the irreducible representations $(\mathbf{1}, \mathbf{0})$ and $(\mathbf{0}, \mathbf{1})$ respectively of the of complexified Lorentz group $SL(2, \mathbb{C})$.

The projection on the self-dual and anti-self-dual parts can done using the following $2 \times 2$ matrices
\begin{gather}
  \left(\sigma^{\mu \nu}\right)_\alpha^{~\beta} = \frac 1 4 \left(\sigma^\mu \overline{\sigma}^\nu - \sigma^\nu \overline{\sigma}^\mu\right)_\alpha^{~\beta},\\
  \left(\overline{\sigma}^{\mu \nu}\right)^{\dot{\alpha}}_{~\dot{\beta}} = \frac 1 4 \left(\overline{\sigma}^\mu \sigma^\nu - \overline{\sigma}^\nu \sigma^\mu\right)^{\dot{\alpha}}_{~\dot{\beta}}.
\end{gather}

Then, by using the duality properties\footnote{Here we use the convention $\epsilon_{0123} = 1$.}
\begin{equation}
   \sigma^{\mu \nu} = \frac i 2 \epsilon^{\mu \nu \rho \tau} \sigma_{\rho \tau}, \quad \overline{\sigma}^{\mu \nu} = - \frac i 2 \epsilon^{\mu \nu \rho \tau} \overline{\sigma}_{\rho \tau},
\end{equation} and the symmetry properties $\sigma_{\alpha \beta}^{\mu \nu} = \sigma_{\beta \alpha}^{\mu \nu}$ and $\overline{\sigma}_{\dot{\alpha} \dot{\beta}}^{\mu \nu} = \overline{\sigma}_{\dot{\beta} \dot{\alpha}}^{\mu \nu}$ one can easily establish the duality properties and the transformations under the complexified Lorentz group of the following quantities  \begin{equation}
   f_{\alpha \beta} = \frac 1 2 \sigma_{\alpha \beta}^{\mu \nu} F_{\mu \nu}, \quad \overline{f}_{\alpha \beta} = \frac 1 2 \overline{\sigma}_{\alpha \beta}^{\mu \nu} F_{\mu \nu}.
\end{equation} $f_{\alpha \beta}$ transforms as $(\mathbf{1}, \mathbf{0})$ and is self-dual, while $\overline{f}_{\dot{\alpha} \dot{\beta}}$ transforms as $(\mathbf{0}, \mathbf{1})$ and is anti-self-dual.

\chapter{Conformal group and twistor space}
\label{ch:conformal_group_and_twistor_space}

Field theories with conformal symmetry are very important in modern physics.  They describe the physics of fixed points of renormalisation group.  The conformal symmetry is especially powerful in two dimensions.  We will be interested here in  (super-)conformal symmetry in four dimensions.

The conformal group can be seen as an extension of the Poincar\' e group by adding dilatation and special conformal transformations (or conformal boosts).
\begin{align}
  {x'}^\mu &= x^\mu + a^\mu, \qquad \text{translations},\\
  {x'}^\mu &= M^\mu_{\hphantom{\mu} \nu} x^\nu, \qquad \text{Lorentz     transformations},\\
  {x'}^\mu &= \alpha x^\mu, \qquad \text{dilatation},\\
  {x'}^\mu &= \frac {x^\mu - b^\mu x^2}{1 - 2 b \cdot x + b^2 x^2}, \qquad \text{special conformal transformations}
\end{align}

The Lie algebra corresponding to the conformal group is
\begin{align}
  \left[\mathbf{D}, \mathbf{P}_\mu\right] &= -i \mathbf{P}_\mu, \qquad   \left[\mathbf{D}, \mathbf{K}_\mu\right] = i \mathbf{K}_\mu,\\
  \left[\mathbf{K}_\mu, \mathbf{P}_\nu\right] &= -2 i \left(\eta_{\mu       \nu} \mathbf{D} + \mathbf{L}_{\mu \nu}\right)\\
  \left[\mathbf{P}_\rho, \mathbf{L}_{\mu \nu}\right] &= -i \left(\mathbf{P}_\mu \eta_{\rho \nu} - \mathbf{P}_\nu \eta_{\rho \mu}\right),\\
  \left[\mathbf{K}_\rho, \mathbf{L}_{\mu \nu}\right] &= -i \left(\eta_{\rho \mu} \mathbf{K}_\nu - \eta_{\rho \nu}   \mathbf{K}_\mu\right),\\
  \left[\mathbf{L}_{\mu \nu}, \mathbf{L}_{\rho \sigma}\right] &= -i \left(\eta_{\mu \rho} \mathbf{L}_{\nu \sigma} + \cdots\right).
\end{align}  The remaining commutators are zero.  These generators can be repackaged into a $SO(2, 4)$ algebra.

The representation on the coordinates is
\begin{align}
  \mathbf{P}_\mu &= i \partial_\mu, \qquad \mathbf{D} = i x^\mu   \partial_\mu,\\
  \mathbf{L}_{\mu \nu} &= i(x_\mu \partial_\nu - x_\nu   \partial_\mu),\\
  \mathbf{K}_\mu &= i (2 x_\mu x^\nu \partial_\nu - x^2 \partial_\mu).
\end{align}

There is another representation of the conformal group on massless, on-shell, one-particle states where the momentum is represented by $\mathbf{P}_{\alpha \dot{\alpha}} = \lambda_\alpha \tilde{\lambda}_{\dot{\alpha}}$.  The momentum has dimension one under the dilatation so we attribute dimension one-half to $\lambda$ and $\tilde{\lambda}$.  The discussion below closely follows the one in ref.~\cite{Witten:2003nn}.

The Lorentz generators $\mathbf{L}_{\mu \nu}$ can be expressed in spinor language and they decompose in a self-dual and an anti-self-dual part $\mathbf{L}_{\alpha \dot{\alpha} \beta \dot{\alpha}} = \epsilon_{\alpha \beta} \mathbf{L}_{\dot{\alpha} \dot{\beta}} + \epsilon_{\dot{\alpha} \dot{\beta}} \mathbf{L}_{\alpha \beta}$, where $\mathbf{L}_{\alpha \beta}$ and $\mathbf{L}_{\dot{\alpha} \dot{\beta}}$ are symmetric.  In order to guide us to an explicit expression for the generators $\mathbf{L}_{\alpha \beta}$ and $\mathbf{L}_{\dot{\alpha} \dot{\beta}}$ we use the symmetry properties and the fact that the generators $\mathbf{L}$ have dimension zero.  This fixes the form completely apart from a global constant,
\begin{align}
  \mathbf{L}_{\alpha \beta} &= \frac i 2 \left(\lambda_\alpha \frac \partial {\partial \lambda^\beta} + \lambda_\beta \frac \partial {\partial \lambda^\alpha}\right),\\
  \mathbf{L}_{\dot{\alpha} \dot{\beta}} &= \frac i 2 \left(\tilde{\lambda}_{\dot{\alpha}} \frac \partial {\partial \tilde{\lambda}^{\dot{\beta}}} + \tilde{\lambda}_{\dot{\beta}} \frac \partial {\partial \tilde{\lambda}^{\dot{\alpha}}}\right).
\end{align}  The normalisation of the generators $\mathbf{L}$ can be found by computing $\left[\mathbf{P}_\rho, \mathbf{L}_{\mu \nu}\right] = -i \left(\eta_{\rho \mu} \mathbf{P}_\nu - \eta_{\rho \nu}   \mathbf{P}_\mu\right)$, using the decomposition $\mathbf{L}_{\alpha \dot{\alpha} \beta \dot{\alpha}} = \epsilon_{\alpha \beta} \mathbf{L}_{\dot{\alpha} \dot{\beta}} + \epsilon_{\dot{\alpha} \dot{\beta}} \mathbf{L}_{\alpha \beta}$ and the translation of $\eta_{\mu \nu}$ to spinor language.  This computation is not completely trivial, one necessary ingredient being identities like $\epsilon_{\alpha \beta} \lambda_\gamma + \epsilon_{\beta \gamma} \lambda_\alpha + \epsilon_{\gamma \alpha} \lambda_\beta = 0$.  Another possibility for normalising the $\mathbf{L}$ is using $\left[\mathbf{L}_{\mu \nu}, \mathbf{L}_{\rho \sigma}\right] = i \left(\eta_{\mu \rho} \mathbf{L}_{\nu \sigma} + \cdots\right)$, but this is more difficult to compute, as it demands some more intricate spinor manipulations than above.

The generator $\mathbf{K}_{\alpha \dot{\alpha}}$ has dimension minus one so the simplest possibility is to represent it as a second order derivative
\begin{equation}
  \mathbf{K}_{\alpha \dot{\alpha}} = \frac {\partial^2}{\partial \lambda^\alpha \partial \tilde{\lambda}^{\dot{\alpha}}}.
\end{equation}

Finally, by using $\left[\mathbf{K}_\mu, \mathbf{P}_\nu\right] = -2 i \left(\eta_{\mu \nu} \mathbf{D} + \mathbf{L}_{\mu \nu}\right)$, we find that we need to take
\begin{equation}
  \mathbf{D} = -\frac i 2 \left(\lambda^\alpha \frac \partial {\partial \lambda^\alpha} + \tilde{\lambda}^{\dot{\alpha}} \frac \partial {\partial \tilde{\lambda}^{\dot{\alpha}}} + 2\right).
\end{equation}

An important symmetry transformation (which is not in the component connected to the identity of the conformal group) is the inversion.  In fact, the symmetry under inversion is sufficient to insure the symmetry under the full conformal group. (The composition of the inversion, translation and inversion gives a special conformal transformation.  Then, the invariance under special conformal transformations and under Poincar\' e transformations insures the invariance under dilatation.)

The inversion induces an automorphism of the Lie algebra.  The action of this inversion automorphism is
\begin{align}
  \mathcal{I}(\mathbf{K}_\mu) &= \mathbf{P}_\mu,\\
  \mathcal{I}(\mathbf{P}_\mu) &= \mathbf{K}_\mu,\\
  \mathcal{I}(\mathbf{D}) &= -\mathbf{D}.
\end{align}

The representation we found above is a bit complicated, having non-homogeneous operators like $\mathbf{D}$ and a mix of zero order, first order and second order differential operators.

A better way to represent the conformal group is to perform a transformation to twistor space~\cite{Penrose:1968me}.  For this, we make the following replacements
\begin{align}
  \tilde{\lambda}_{\dot{\alpha}} & \rightarrow i \frac \partial   {\partial \mu^{\dot{\alpha}}},\\
    -i \frac \partial {\partial \tilde{\lambda}^{\dot{\alpha}}} & \rightarrow \mu_{\dot{\alpha}}.
\end{align}

By using this in the expressions for the representation on $(\lambda, \tilde{\lambda})$, we get
\begin{align}
  \mathbf{P}_{\alpha \dot{\alpha}} & = i \lambda_\alpha \frac \partial {\partial \mu^{\dot{\alpha}}}, \qquad \mathbf{K}_{\alpha \dot{\alpha}}  = i \mu_{\dot{\alpha}} \frac \partial {\partial \lambda^\alpha},\\
  \mathbf{L}_{\alpha \beta} & = \frac i 2 \left(\lambda_\alpha \frac \partial {\partial \lambda^\beta} + \lambda_\beta \frac \partial {\partial \lambda^\alpha}\right),\\
  \mathbf{L}_{\dot{\alpha} \dot{\beta}} & = \frac i 2 \left(\mu_{\dot{\alpha}} \frac \partial {\partial \mu^{\dot{\beta}}} + \mu_{\dot{\beta}} \frac \partial {\partial   \mu^{\dot{\alpha}}}\right),\\
  \mathbf{D} & = \frac i 2 \left(-\lambda^\alpha \frac \partial {\partial \lambda^\alpha} + \mu^{\dot{\alpha}} \frac \partial {\partial \mu^{\dot{\alpha}}}\right).
\end{align}  Note that the choice to transform $\tilde{\lambda}$ rather than $\lambda$ breaks parity.

After transforming to twistor space, the representation of the conformal group becomes simpler (all the generators are represented by homogeneous, first order derivation operators) and more symmetric.  The space $(\lambda_\alpha, \mu_{\dot{\alpha}})$ is called twistor space, $\mathbb{T}$. (More precisely, what Penrose calls twistor space is a complexified version of this space.  See below for more details.)

We have already commented on the fact that, given a massless on-shell momentum $p$, the values for the spinors $\lambda$ and $\tilde{\lambda}$ are not uniquely defined.  For example, in $+---$ signature where $\tilde{\lambda} = \pm \lambda^*$, one can change $\lambda$ and $\tilde{\lambda}$ by phases, while keeping $p$ unchanged
\begin{equation}
  \lambda \rightarrow e^{i \phi} \lambda, \qquad \tilde{\lambda} \rightarrow e^{-i \phi} \tilde{\lambda}.
\end{equation}  It is easy to see that $\mu$ should transform in the same way as $\lambda$: $\mu \rightarrow e^{i \phi} \mu$.  All the generators defined above are invariant under this transformation.

\chapter{Super-twistor space}
\label{ch:super-twistor_space}

Let us now extend the construction in the previous section to the case of super-conformal symmetry.  This construction was first done by Ferber in ref.~\cite{Ferber:1977qx} (see also ref.~\cite{Witten:2003nn}).

We will introduce $\eta_A$ with $A=1, \ldots, 4$, a Grassmann variable transforming in the $\overline{\mathbf{4}}$ of $SU(4)$ $R$-symmetry.  The particles will be described by spinors $\lambda$ and $\tilde{\lambda}$ as before but also by a polynomial in the Grassmann variables $\eta_A$.

\begin{center}
  \begin{tabular}{r|l|l}
    helicity & Grassmann factor & $SU(4)_R$ representation\\
    \hline
    $1$ & $1$ & $\mathbf{1}$\\
    $\tfrac 1 2$ & $\eta_A$ & $\overline{\mathbf{4}}$\\
    $0$ & $\eta_A \eta_B$ & $\mathbf{6}$\\
    $-\tfrac 1 2$ & $\tfrac 1 {3 !} \epsilon^{A B C D} \eta_A \eta_B     \eta_C$ & $\mathbf{4}$\\
    $-1$ & $\eta_1 \eta_2 \eta_3 \eta_4$ & $\mathbf{1}$
  \end{tabular}
\end{center}
We emphasize that this choice also breaks parity.

It turns out that the MHV amplitudes have a supersymmetrised version that is also very simple.  If we define
\begin{align}
  P_{\alpha \dot{\alpha}} & = \sum_i \lambda_{i \alpha}   \tilde{\lambda}_{i \dot{\alpha}},\\
  \Theta_{A \alpha} & = \sum_i \lambda_{i \alpha} \eta_{i A},
\end{align} then the supersymmetrised version of MHV amplitudes is
\begin{equation}
  \label{eq:susy_mhv}
  A = i g^{n-2} \delta^4(P) \delta^8(\Theta) \frac 1 {\langle 1,\ 2\rangle \cdots \langle n,\ 1\rangle}.
\end{equation} (Remember that for a Grassmann variable $\psi$, $\delta(\psi) = \psi$ by definition.  For $\delta^8(\Theta)$ we choose the ordering $\prod_{A=1}^4 \Theta_{A 1} \Theta_{A 2}$.)

In order to get the MHV amplitude with gluons $i$ and $j$ of helicity minus, we have to pick out the coefficient of $\eta_{i 1} \eta_{i 2} \eta_{i 3} \eta_{i 4} \eta_{j 1} \eta_{j 2} \eta_{j 3} \eta_{j 4}$ in the expansion of the amplitude in eq.~\eqref{eq:susy_mhv}.  As we are interested only in the terms containing $\eta_i$ and $\eta_j$ we can ignore the Grassmann variables corresponding to other particles and the expansion of $\delta^8(\Theta)$ yields
\begin{multline}
  \delta^8(\Theta) = \prod_{A=1}^4 \left(\lambda_{i 1} \eta_{i A} + \lambda_{j 1} \eta_{j A}\right) \left(\lambda_{i 2} \eta_{i A} + \lambda_{j 2} \eta_{j A}\right) =\\= \prod_{A=1}^4 \left(\lambda_{i 1} \lambda_{j 2} - \lambda_{i 2} \lambda_{j 1}\right) \eta_{i A} \eta_{j A} = \langle i,\ j\rangle^4 \prod_{A=1}^4 \eta_{i A} \eta_{j A}.
\end{multline}

The super-conformal algebra of the super-conformal group $PSU(2,2|4)$ can be represented as follows:
\begin{itemize}
\item the conformal group does not act on the $\eta$'s and is represented in the same way as in Chap.~\ref{ch:conformal_group_and_twistor_space}.
\item the $SU(4)$ $R$-symmetry is represented by
  \begin{equation}
    \eta_A \frac \partial {\partial \eta_B} - \frac 1 4 \delta_B^A \eta_C \frac \partial {\partial \eta_C},
  \end{equation} where we subtracted the trace
\item the 16 supercharges $Q$ which have dimension $\tfrac 1 2$ are represented by\footnote{It is useful to recall here that there is an ambiguity in the definition of $\lambda$ and $\tilde{\lambda}$ (see eq.~\eqref{eq:lambda_ambiguity}).  In signature $+---$ one can transform $\lambda$ by a phase factor and $\tilde{\lambda}$ by the complex conjugate of this phase factor and the momentum remains unchanged.  For $++--$ signature, both $\lambda$ and $\tilde{\lambda}$ are real and one can transform $\lambda$ by a real, non-zero factor $t$ and the momentum remains unchanged if we also transform $\tilde{\lambda}$ by the factor $t^{-1}$.  As the generators have to be invariant under this transformation, we must transform $\eta$ in the same way as $\tilde{\lambda}$ and this restricts the form of the generators to be the one given here.}
\begin{equation}
  \tilde{\lambda}^{\dot{\alpha}} \frac \partial {\partial \eta_A}, \qquad \lambda^\alpha \eta_A,
\end{equation} and the 16 supercharges $S$ which have dimension $- \tfrac 1 2$ are represented by
\begin{equation}
  \eta_A \frac \partial {\partial \tilde{\lambda}^\alpha}, \qquad \frac {\partial^2} {\partial \lambda^\alpha \partial \eta_A}.
\end{equation}
\end{itemize}

This representation also has the unwanted feature that the operators appearing are differential operators of different degrees.  In order to obtain a simpler representation we transform to the super-twistor space, in way which is analogous to the transformation to the twistor space\footnote{Note that after the transform to the super-twistor space the all the coordinates $\lambda$, $\mu$, $\psi$ have the same scaling under the transformations discussed in the previous footnote.}
\begin{alignat}{3}
  \tilde{\lambda}_{\dot{\alpha}} & \rightarrow i \frac \partial {\partial \mu^{\dot{\alpha}}}, \qquad & -i \frac \partial {\partial     \tilde{\lambda}^{\dot{\alpha}}} & \rightarrow \mu_{\dot{\alpha}},\\
  \eta_A & \rightarrow i \frac \partial {\partial \psi^A}, \qquad & -i \frac \partial {\partial \eta_A} & \rightarrow \psi^A.
\end{alignat}

We now introduce a space on which the representation defined above acts.  It is $\hat{\mathbb{T}} = \mathbb{C}^{4|4}$ (the hat serves to distinguish this version of the twistor space which is supersymmetric from the one defined above).  The space $\hat{\mathbb{T}}$ is parametrised by four bosonic coordinates $Z^I = (\lambda^\alpha, \mu^{\dot{\alpha}})$ and four fermionic coordinates $\psi^A$.

The projective twistor space is parametrised by $(Z^I, \psi^A) = (\lambda^\alpha, \mu^{\dot{\alpha}}, \psi^A)$, subject to the equivalence $(Z^I, \psi^A) \sim (t Z^I, t \psi^A)$ for $t$ a complex non-zero number.  This space is the same as the projective super-space $\mathbb{CP}^{3|4}$.

There is another version of the twistor space that one can construct for $++--$ signature.  For this signature $\lambda$ and $\mu$ can be taken to be real ($\psi$ can't be real since this would be incompatible with the $SU(4)$ $R$-symmetry, but its conjugate field $\bar{\psi}$ never appears).  So we can denote this `real' version of the super-twistor space by $\mathbb{RP}^{3|4}$.

Our discussion about the transformation to (super-)twistor space has been formal and, in practice, the transformation to twistor space is difficult to do.  Let us take a function defined on the space parametrised by $(\lambda, \tilde{\lambda})$ and transform it into a function defined on twistor space $Z = (\lambda, \mu)$.  In the case of signature $++--$, where the variables $(\lambda, \tilde{\lambda})$ are real, we can interpret the transformation to the twistor space as a Fourier transform in $\tilde{\lambda}$:
\begin{equation}
  f \rightarrow \tilde{f}, \qquad \text{where}\ \tilde{f}(\lambda, \mu) = \int \frac {d^2 \tilde{\lambda}}{(2 \pi)^2} e^{i \mu^{\dot{\alpha}} \tilde{\lambda}_{\dot{\alpha}}} f(\lambda, \tilde{\lambda}).
\end{equation}

In general, $\tilde{f}$ or $f$ are fairly complicated so finding one from the other by using the Fourier transform or its inverse is difficult.  Also, the integrals will often not exist in the usual sense and the answer will have to be interpreted in the language of distributions (we will see some examples below where this happens).

In $+ - - -$ signature the twistor variables are complex so one can try to extend the prescription that works in the real case by choosing an integration contour for the `Fourier transform.'  It is not guaranteed that this prescription works, however.  There is an alternative and more systematic approach used by Penrose (see ref.~\cite{Penrose:1968me}) which uses $\overline{\partial}$ cohomology or sheaf cohomology.  We will not use this language here.

In the following, we will mostly use the $+ + - -$ signature and the transform to twistor space will be a Fourier transform.  The fact that we use signature $++--$ will have no adverse implications for the tree-level amplitudes we compute, though it might become important for amplitudes at loop level.  See ref.~\cite{Witten:2003nn} for a construction that applies to Euclidean case ($+ + + +$ signature).

What is the interpretation of the scattering amplitude transformed to twistor space?  Consider the scattering amplitude $A(p_1, \ldots, p_n)$ of $n$ particles with on-shell momenta $p_i^2 = 0$.  The scattering amplitude for the same particles which are in the states characterised by wavefunctions
\begin{equation}
  \label{eq:wavefunction}
  \phi_i(x) = \int d^4 p \delta(p^2) a_i(p) e^{i p \cdot x},
\end{equation} can be found from the amplitude in momentum space as follows
\begin{equation}
  A(\phi_1, \ldots, \phi_n) = \int A(p_1, \ldots, p_n) \prod_{i=1}^n d^4 p_i \delta(p_i^2) a_i(p_i).
\end{equation}

Recall that the scattering amplitude in twistor space is similar in form to the above formula, where we replace the measure of integration $d^4 p \delta(p^2) a_i(p)$ by $\frac {d^2 \tilde{\lambda}}{(2 \pi)^2} e^{i \tilde{\lambda}^{\dot{\alpha}} \mu_{\dot{\alpha}}}$.

If we now make the same replacement inside the equation for the wavefunction (see eq.~\eqref{eq:wavefunction}) we get
\begin{equation}
  \phi_{\lambda, \mu}(x) = \int \frac {d^2 \tilde{\lambda}}{(2 \pi)^2} e^{i \tilde{\lambda}^{\dot{\alpha}} \mu_{\dot{\alpha}}} e^{i \lambda^\alpha \tilde{\lambda}^{\dot{\alpha}} x_{\alpha \dot{\alpha}}} = \delta^2(\mu_{\dot{\alpha}} + \lambda^\alpha x_{\alpha \dot{\alpha}}).
\end{equation}

We therefore see that we can consider the amplitude transformed to twistor space as the scattering of particles with wavefunctions given by the above expression.  This unusual wavefunction is supported on the points in Minkowski space which satisfy the equation
\begin{equation}
  \label{eq:twistor_eq}
  \mu_{\dot{\alpha}} + \lambda^\alpha x_{\alpha \dot{\alpha}} = 0.
\end{equation}

It is important to notice here that the ambiguity affecting $\lambda$ and $\mu$ is inconsequential in $+---$ signature as it modifies the wavefunction by a phase (we need to remember, though, that all the arguments leading to the form of the wavefunction are correct only in $++--$ signature).  This equation for the support of the wavefunction can be read in two ways: given $(\lambda, \mu)$, what is $x$? and given $x$, what are $(\lambda, \mu)$?  It establishes a link between space-time and twistor space and is thus central to all twistor constructions.

If we consider $(\lambda, \mu)$ as given and fixed, we have two equations and four components of $x$, so the solution will be a two-dimensional manifold.  If $x$ and $y$ are both solutions of this equation, then $\lambda^\alpha (x - y)_{\alpha \dot{\alpha}} = 0$.  This implies that $(x - y)^2 = 0$ because $(x - y)_{\alpha \dot{\alpha}}$ has a right eigenvector with eigenvalue zero so its determinant must be zero.  Any two solutions are therefore separated by a light-like interval.  As the equations are linear they lead to a linear manifold which is a two-dimensional light-like plane.

We can also see that, for $\lambda = 0$ and any finite $x$, $\mu =0$ also.  So $\lambda = 0$ corresponds to wavefunctions which are localised at infinity.\footnote{This is so because if $\lambda = 0$, then $\mu \neq 0$ because otherwise one can't define the associated projective twistor space.}  As we are interested in scattering of wavefunctions, we will omit the points in twistor space where $\lambda = 0$.  This space is usually denoted by $\mathbb{T}'$.

Let us return now to a subtle point that has not been emphasized in the literature.  We have found a nice representation of the (super-)conformal group acting on spinors $(\lambda, \tilde{\lambda})$ and on twistors $(\lambda, \mu)$, but is this the same as the conformal symmetry in space-time?  This is not obvious and, in fact, there seems to be another conformal symmetry acting in momentum space as discovered in ref.~\cite{Drummond:2006rz} (see also sec.~\ref{ch:pseudo-conformal}).

We will show that, after transforming both the position $x$ and the twistor space coordinates $(\lambda, \mu)$, the wavefunction $\psi_{\lambda, \mu}(x) = \delta^2(\mu_{\dot{\alpha}} + \lambda^\alpha x_{\alpha \dot{\alpha}})$ remains unchanged up to a Jacobian factor that compensates the transformation of the measure in the relation for the normalisation of the wavefunction.

We will only discuss the invariance under translations and dilatation here.  The invariance under conformal boosts is a bit more complicated to establish, but follows the same logic.

The translations are performed by the operator $\exp (i a \cdot \mathbf{P})$ which, for infinitesimal $a$ produces the transformations
\begin{align}
  x_\mu & \rightarrow x_\mu - a_\mu,\\
  \mu_{\dot{\alpha}} & \rightarrow \mu_{\dot{\alpha}} + \lambda^\alpha   a_{\alpha \dot{\alpha}},\\
  \lambda_\alpha & \rightarrow \lambda_\alpha,
\end{align} where we have used the following expressions for the representation of $\mathbf{P}$:
\begin{equation}
  \mathbf{P}_\mu = i \partial_\mu, \qquad \mathbf{P}_{\alpha \dot{\alpha}} = i \lambda_\alpha \frac \partial {\partial \mu^{\dot{\alpha}}}.
\end{equation}  Under these transformations $\mu_{\dot{\alpha}} + \lambda^\alpha x_{\alpha \dot{\alpha}}$ remains unchanged.

Let us now study the dilatation symmetry.  Dilatation transformations are performed by $\exp (i \rho \mathbf{D})$ and, for infinitesimal $\rho$, we have the following transformations
\begin{align}
  x_\mu & \rightarrow (1 - \rho) x_\mu,\\
  \lambda^\alpha & \rightarrow \left(1 + \frac \rho 2\right) \lambda^\alpha,\\
  \mu^{\dot{\alpha}} & \rightarrow \left(1 - \frac \rho 2\right) \mu^{\dot{\alpha}},
\end{align} where we have used the following expressions for the representation of $\mathbf{D}$:
\begin{equation}
  \mathbf{D} = i x^\mu \partial_\mu, \qquad \mathbf{D} = \frac i 2 \left(- \lambda^\alpha \frac \partial {\partial \lambda^\alpha} + \mu^{\dot{\alpha}} \frac \partial {\partial \mu^{\dot{\alpha}}}\right).
\end{equation}  Again, $\mu_{\dot{\alpha}} + \lambda^\alpha x_{\alpha \dot{\alpha}}$ gets multiplied by a factor $1 - \tfrac \rho 2$ which should be absorbed in the normalisation of the wavefunction.

\chapter{Geometric interpretation and Witten's conjecture}

The scattering amplitudes can be transformed to twistor space and studied in this setting.  As discussed above, this can be interpreted as the scattering of particles in some peculiar states.  These states are characterised by their homogeneous coordinates $Z=(\lambda, \mu)$ and $\psi$ and can be represented by a point in the projective super-twistor space.  An $n$-point amplitude then is a function that associates a number to a collection of $n$ points in the projective super-twistor space.

The fact that the conformal group has a simple representation when acting on the twistor space leads us to suspect that the amplitudes might have a simple geometric representation in twistor space.

In ref.~\cite{Witten:2003nn}, Witten formulated the following conjecture: the $n$-point scattering amplitude in twistor space is non-zero only if the points $P_i$ representing the states are supported on an algebraic curve inside the projective twistor space with the following characteristics
\begin{itemize}
\item it is not necessarily connected,
\item it has degree $d$ given by $d = q + l - 1$, where $q$ is the number of helicity minus and $l$ is the number of loops,
\item its genus $g$ is bounded by the number of loops $g \leq l$.
\end{itemize}

At tree level, the gluon amplitudes are the same in supersymmetric and non-supersymmetric theories.  We will therefore discuss the geometric interpretation in twistor and in super-twistor space.  For studies at loop level see refs.~\cite{Cachazo:2004zb, Bena:2004xu}.  The simplest non-vanishing amplitudes are MHV amplitudes so it is natural to study them first.

We only need the fact that the MHV amplitudes are holomorphic, i.e.
\begin{equation}
  A(\lambda_i, \tilde{\lambda}_i) = (2 \pi)^4 i g^{n-2} \delta^4\left(\sum_{i=1}^n \lambda_{i \alpha} \tilde{\lambda}_{i \dot{\alpha}}\right) f(\lambda_i).
\end{equation}  Using a representation of the delta function, we rewrite the amplitude in a way that facilitates the transformation to twistor space.
\begin{equation}
  A(\lambda_i, \tilde{\lambda}_i) = i g^{n-2} \int d^4 x e^{i x_{\alpha \dot{\alpha}} \sum_{i=1}^n \lambda_i^\alpha \tilde{\lambda}_i^{\dot{\alpha}}} f(\lambda_i).
\end{equation}  Then,
\begin{align*}
  \label{eq:mhv_twistor}
  \tilde{A}(\lambda, \mu) & = i g^{n-2} \int d^4 x \prod_{i=1}^n \int \frac {d^2 \tilde{\lambda}_i}{(2 \pi)^2} e^{i \sum_{i=1}^n \mu_{i \dot{\alpha}} \tilde{\lambda}^{\dot{\alpha}}} e^{i x_{\alpha \dot{\alpha}} \sum_{i=1}^n \lambda_i^\alpha \tilde{\lambda}_i^{\dot{\alpha}}} f(\lambda_i) \\
& = i g^{n-2} \int d^4 x \prod_{i=1}^n \delta^2(\mu_{i \dot{\alpha}} + x_{\alpha \dot{\alpha}} \lambda_i^\alpha) f(\lambda_i).
\end{align*}

The interpretation of this result is the following: the two equations (for $\dot{\alpha} = \dot{1}, \dot{2}$) $\mu_{\dot{\alpha}} + x_{\alpha \dot{\alpha}} \lambda^\alpha = 0$ define a plane in twistor space and a line in the projective twistor space.  This line is, of course, a degree one, genus zero algebraic curve.  If the $(\lambda_i, \mu_i)$ are not collinear, the equations $\mu_{i \dot{\alpha}} + x_{\alpha \dot{\alpha}} \lambda_i^\alpha = 0$ cannot be satisfied simultaneously and the amplitude is zero.  This satisfies Witten's conjecture for $l=0$, $g=0$, $d=1$ and $q=2$.

It is easy to see that all the lines in the real version of the projective twistor space, $\mathbb{RP}^3$ are of the form $\mu_{i \dot{\alpha}} + x_{\alpha \dot{\alpha}} \lambda_i^\alpha = 0$.  Then, the integral $\int d^4 x$ can be interpreted as an integral over the moduli space of degree one, genus zero algebraic curves in $\mathbb{RP}^3$.  This interpretation will be useful when we will study higher degree curves.

Let us now study the supersymmetric version of the MHV amplitudes.  Our starting point will be eq.~\ref{eq:susy_mhv}.  Here also we rewrite the bosonic delta function as in the case of ordinary twistor space.  The fermionic delta function can be rewritten as
\begin{equation}
  \delta^8(\Theta) = \int d^8 \theta_\alpha^A \exp\left(i \theta_\alpha^A \sum_{i=1}^n \eta_{i A} \lambda_i^\alpha\right),
\end{equation} after which the amplitude becomes
\begin{equation}
  A = i g^{n-2} \int d^4 x d^8 \theta \exp\left(i x_{\alpha \dot{\alpha}} \sum_{i=1}^n \lambda_i^\alpha \tilde{\lambda}_i^{\dot{\alpha}}\right) \exp\left(i \theta_\alpha^A \sum_{i=1}^n \eta_{i A} \lambda_i^\alpha\right) \prod_{i=1}^n \frac 1 {\langle i,\ (i+1)\rangle}.
\end{equation}

After this rewriting, the transformation to super-twistor space is easy to do
\begin{align*}
  \label{eq:susy_mhv_twistor}
  \tilde{A}(\lambda_i, \mu_i, \psi_i) & = \int \prod_{i=1}^n \frac {d^2 \tilde{\lambda}_i d^4 \eta_i}{(2 \pi)^2} \exp\left(i \sum_{i=1}^n \mu_i^{\dot{\alpha}} \tilde{\lambda}_{\dot{\alpha}} + i \sum_{i=1}^n \psi_i^A \eta_{i A}\right) A(\lambda_i, \tilde{\lambda}_i, \eta_i)\\
& = i g^{n-2} \int d^4 x d^8 \theta \prod_{i=1}^n \frac {\delta^2(\mu_{i \dot{\alpha}} + x_{\alpha \dot{\alpha}} \lambda_i ^\alpha) \delta^4(\psi_i^A + \theta_\alpha^A \lambda_i^\alpha)}{\langle i,\ (i + 1)\rangle}.
\end{align*}

The result is very similar to the one in the bosonic case.  Here, given $x$ and $\theta$ we have a curve in the projective super-twistor space defined by equations
\begin{align}
  \mu_{\dot{\alpha}} + x_{\alpha \dot{\alpha}} \lambda^\alpha & = 0,\\
  \psi^A + \theta_\alpha^A \lambda^\alpha & = 0.
\end{align}  The interpretation is also similar to the bosonic case: the MHV amplitude vanishes unless the points representing the external states are collinear in the projective super-twistor space.  In this case, the integrals $\int d^4 x d^8 \theta$ are integrals over the super-moduli space of lines in the projective super-twistor space.

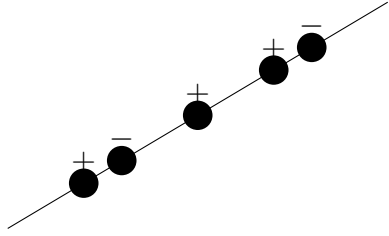
\begin{figure}
  \centering
  \beginpgfgraphicnamed{3}
  \begin{tikzpicture}
    \draw (-2,-1)--(3,2)
    node[fill,circle,minimum size=3pt, pos=.2]{} node[pos=.2, above]{$+$}
    node[fill,circle,minimum size=3pt, pos=.3]{} node[pos=.3, above]{$-$}
    node[fill,circle,minimum size=3pt, pos=.5]{} node[pos=.5, above]{$+$}
    node[fill,circle,minimum size=3pt, pos=.7]{} node[pos=.7, above]{$+$}
    node[fill,circle,minimum size=3pt, pos=.8]{} node[pos=.8, above]{$-$};
  \end{tikzpicture}
  \endpgfgraphicnamed
  \caption{Collinear distribution of points in twistor space corresponding to an MHV amplitude.}
  \label{fig:twistor_collinear}
\end{figure}

Witten's conjecture also works for amplitudes with $n$ positive helicities or $n-1$ positive helicities and one negative helicity.  In the first case, at tree level, $q=0$ and $l=0$ so the degree is $d=-1$.  As there are no algebraic curves of degree $-1$ the amplitude vanishes.

In the second case, of amplitudes with $n-1$ helicity plus and one helicity minus gluons, $q=1$ and $l=0$ so the degree is $d=0$.  A degree zero curve is a point so, unless all the external states are represented by the same point, the amplitude vanishes.  In fact, when proving the vanishing of the amplitudes by the supersymmetry Ward identities, one also needs the $\lambda_i$ (or $\tilde{\lambda}_i$) to be different (see sec.~\ref{sec:swi}).

The three-point amplitude is special, because of the exceptional kinematics.  The on-shell conditions $p_i^2=0$ and the momentum conservation imply $p_i \cdot p_j = 0$ for all $i, j = 1, 2, 3$.  Since $p_i \cdot p_j = \langle i,\ j\rangle [i,\ j]$, this implies that either $\lambda_i$ and $\lambda_j$ or $\tilde{\lambda}_i$ and $\tilde{\lambda}_j$ are proportional.  Taking all possible combinations, it follows that either all $\lambda_i$ or all $\tilde{\lambda}_i$ are proportional (in signature $+---$, where $\lambda_i$ and $\tilde{\lambda}_i$ are not independent, both sets are proportional).

In momentum space the three-point $-++$ amplitude is:\footnote{This can be regarded as an $\overline{\text{MHV}}$ amplitude.}
\begin{equation}
  \label{eq:3point_MHVbar}
  A = i g \frac {[2,\ 3]^3}{[1,\ 2] [3,\ 1]}.
\end{equation}  If all the $\tilde{\lambda}_i$ are proportional this vanishes (we actually have a ratio $\tfrac 0 0$ but the numerator has a higher exponent) so the amplitude is supported on configurations for which all the $\lambda_i$ are proportional.  However, we want to show that all the $Z_i = (\lambda_i, \mu_i)$, for $i=1, 2, 3$ are proportional.  It does not seem possible to prove this proportionality by transforming to twistor space the amplitude in eq.~\ref{eq:3point_MHVbar}.

One other test that should be discussed (and which does not seem to have been studied in the literature) is the vanishing of amplitude with $n$ negative helicities or with $n-1$ negative helicities and one positive helicity at loop level (this vanishing is a consequence of supersymmetry Ward identities so it is an exact statement).  To prove this it would be sufficient to show that there are no algebraic curves in projective twistor space such that $g \leq d$.\footnote{The simplest type of algebraic curves that can be embedded in $\mathbb{RP}^3$ can be described by the common zeros of two homogeneous polynomials $P(Z^I)$ and $Q(Z^I)$.  This kind of curve is a called complete intersection.  The simplest example of complete intersection is the line supporting MHV amplitudes.  Not all curves are of this type; in some cases one needs three or more polynomials.  If the curve is a complete intersection and the polynomials have degrees $d_1$ and $d_2$, the degree of the resulting algebraic curve is $d = d_1 d_2$.  I don't know if the result $g \leq d$ is true or not.}

The non-MHV amplitudes are more complicated and the transform to twistor space is very difficult to compute.  However, one can translate the geometrical information encoded in the twistor space amplitude into differential equations satisfied by the momentum space amplitudes.  The idea is the following: suppose we can find a polynomial expression $O(Z_i^I)$ which vanishes on the support of the scattering amplitude $\tilde{A}$ in twistor space (this kind of expressions can be obtained by considering the equations of the curves which support the amplitude).  Then, in twistor space we have
\begin{equation}
  O(Z_i^I) \tilde{A}(Z_i^I) = 0.
\end{equation}  It follows that
\begin{align*}
  O(Z_i^I) \tilde{A}(Z_i^I) & = \int \prod_{i=1}^n \frac {d^2 \tilde{\lambda}_i}{(2 \pi)^2} \left(O\bigg(\lambda_i^\alpha, -i \frac \partial {\partial \tilde{\lambda}_i^{\dot{\alpha}}}\bigg) \exp\bigg(i \sum_{i=1}^n \mu_{i \dot{\alpha}} \tilde{\lambda}_i^{\dot{\alpha}}\bigg)\right) A(\lambda_i, \tilde{\lambda}_i) \\ & = \int \prod_{i=1}^n \frac {d^2 \tilde{\lambda}_i}{(2 \pi)^2} \exp\bigg(i \sum_{i=1}^n \mu_{i \dot{\alpha}} \tilde{\lambda}_i^{\dot{\alpha}}\bigg) \left( O\bigg(\lambda_i^\alpha, i \frac \partial {\partial \tilde{\lambda}_i^{\dot{\alpha}}}\bigg) A(\lambda_i, \tilde{\lambda}_i) \right),
\end{align*} where in the second line we have done an integration by parts.

The conclusion is that $O(Z_i^I) \tilde{A}(Z_i^I) = 0$ implies a differential equation in momentum space
\begin{equation}
  O\bigg(\lambda_i^\alpha, i \frac \partial {\partial \tilde{\lambda}_i^{\dot{\alpha}}}\bigg) A(\lambda_i, \tilde{\lambda}_i) = 0.
\end{equation}

There are some obvious candidates for $O$ that one can consider.  In ref.~\cite{Witten:2003nn}, Witten introduced the following quantities
\begin{gather}
  K_{i j k l} = \epsilon_{I J K L} Z_i^I Z_j^J Z_k^K Z_l^L,\\
  F_{I;i j k} = \epsilon_{I J K L} Z_i^J Z_j^K Z_k^L.
\end{gather}  The first quantity is zero when the points $P_i, P_j, P_k$ and $P_l$ are contained in an $\mathbb{RP}^2$ inside the $\mathbb{RP}^3$, while the second quantity vanishes if the points $P_i, P_j, P_k$ are collinear.

These operators have been used to explore the twistor space properties of scattering amplitudes in refs.~\cite{Witten:2003nn, Cachazo:2004zb}.  A subtlety in the interpretation of these results, termed `holomorphic anomaly', was discussed in ref.~\cite{Cachazo:2004by}.

\chapter{Yang-Mills from twistor string}

In this chapter we briefly review how the $\mathcal{N}=4$ super-Yang-Mills theory arises from a string theory in twistor space.

In ref.~\cite{Witten:2003nn}, Witten considered the topological $B$-model on $\mathbb{CP}^{3|4}$, which is a Calabi-Yau super-manifold.  One could consider $\mathbb{CP}^{3|\mathcal{N}}$ instead for a theory with $\mathcal{N}$ supersymmetries, but such super-manifold is Calabi-Yau only for $\mathcal{N}=4$.  The target-space manifold has to be Calabi-Yau in order for the $B$-model to be consistent (see Appendix~\ref{ch:topological}).

Let us describe first the case of a purely bosonic Calabi-Yau threefold $X$.  In that case, the model we are interested in is an open-string $B$-model where the open strings end on space-filling $D5$-branes.  For a stack of $N$ $D5$-branes we have a gauge group $U(N)$.

The low energy effective action of the $D5-D5$ strings can be described in terms of a single $(0, 1)$-form field $A$, while the BRST operator $Q$ acts as $\overline{\partial}$ on $A$.  The low energy effective action is a holomorphic Chern-Simons theory
\begin{equation}
  \mathcal{S} = \frac 1 2 \int_X \Omega \wedge \tr \left(A \wedge \overline{\partial} A + \frac 2 3 A \wedge A \wedge A\right),
\end{equation} where $\Omega$ is the holomorphic three-form of the Calabi-Yau manifold.  The field $A$ has a gauge invariance
\begin{equation}
  \delta A = \overline{\partial} \epsilon + [A, \epsilon],
\end{equation} where $\epsilon$ is a zero-form.

The path integral is a bit subtle to compute since the gauge field $A$ is complex; one should interpret the integrals over the modes of $A$ as contour integrals.

The extension to the case of a Calabi-Yau super-manifold is done as follows: we also consider space-filling $D$-branes but only in the bosonic and $\psi$ directions and we take them to be placed at $\overline{\psi}=0$.  The world-volume $Y$ of these branes is parametrised by $Z, \overline{Z}, \psi$ and $\overline{\psi}=0$.  The low-energy theory is described by a field $\mathcal{A} = \mathcal{A}_{\overline{I}} d \overline{Z}^{\overline{I}}$, where we can expand $\mathcal{A}_{\overline{I}}$ in powers of $\psi$
\begin{multline}
  \mathcal{A}(Z, \overline{Z}, \psi) = d \overline{Z}^{\overline{I}} \bigg(A_{\overline{I}}(Z, \overline{Z}) + \psi^A \chi_{{\overline{I}} A} + \frac 1 2 \psi^A \psi^B \phi_{A B \overline{I}} (Z, \overline{Z}) +\\ \frac 1 {3!} \epsilon_{A B C D} \psi^A \psi^B \psi^C \tilde{\chi}_{\overline{I}}^D(Z, \overline{Z}) + \frac 1 {4!} \epsilon_{A B C D} \psi^A \psi^B \psi^C \psi^D G_{\overline{I}} (Z, \overline{Z})\bigg).
\end{multline}  The twistor-space fields $A$, $\chi$, $\phi$, $\tilde{\chi}$ and $G$ correspond to space-time fields after Penrose transform from twistor-space to space-time (the correspondence is described in more detail below; the Penrose transform is described in Appendix~\ref{ch:penrose_transform})).

The action is
\begin{equation}
  S = \frac 1 2 \int_Y \Omega \wedge \tr \left(\mathcal{A} \wedge \overline{\partial} \mathcal{A} + \frac 2 3 \mathcal{A} \wedge \mathcal{A} \wedge \mathcal{A}\right),
\end{equation} with
\begin{equation}
  \Omega = \frac 1 {(4!)^2} \epsilon_{I J K L} Z^I d Z^J d Z^K d Z^L \epsilon_{A B C D} d \psi^A d \psi^B d \psi^C d \psi^D.
\end{equation}  Note that this measure is invariant under scalings of coordinates (as needed for a measure on a projective space) because the bosonic and fermionic parts transform with opposite weights.

After integrating the fermionic coordinates we get
\begin{multline}
  \label{eq:twistor_action}
  \mathcal{S} = \int_{\mathbb{CP}^3} \omega \wedge \tr \bigg(G \wedge (\overline{\partial} A + A \wedge A) + \tilde{\chi}^A \wedge \overline{D} \chi_A +\\+ \frac 1 4 \epsilon^{A B C D} \phi_{A B} \wedge \overline{D} \phi_{C D} + \frac 1 2 \epsilon^{A B C D} \chi_A \wedge \chi_B \wedge \phi_{C D}\bigg),
\end{multline} where $\overline{D} \phi = \overline{\partial} \phi + A \phi$ and $\omega = \frac 1 {4!} \epsilon_{I J K L} Z^I d Z^J d Z^K d Z^L$.

The classical equations of motion are
\begin{equation}
  \overline{\partial} \mathcal{A} + \mathcal{A} \wedge \mathcal{A} = 0.
\end{equation}  By linearisation around $\mathcal{A} = 0$ we have $\overline{\partial} \Phi = 0$, where $\Phi$ is any of the component fields obtained by expanding $\mathcal{A}$ in powers of $\psi$.  The linearised gauge invariance is $\delta \Phi = \overline{\partial} \alpha$, so each of these fields are elements of cohomology groups.

Following Witten, we associate to the component fields $(A, \chi, \phi, \tilde{\chi}, G)$ a charge called $S$-charge of $k = (0, -1, -2, -3, -4)$ respectively.  So these fields are elements of cohomology groups $H^1(\mathbb{PT}', \mathcal{O}(-k))$ which, by Penrose transform (see ref.~\cite{Penrose:1968me} and Appendix~\ref{ch:penrose_transform}), map to solutions of the massless wave equation for fields of helicity $1 - \tfrac k 2$. ($\mathcal{O}(-k)$ denotes fields of homogeneity $-k$ and $\mathbb{PT}'$ denotes the projective twistor space without the points where $\lambda = 0$.)

So by the Penrose transform the fields $(A, \chi, \phi, \tilde{\chi}, G)$ in twistor space map to to fields\footnote{We denote the space-time fields by the same letters as the twistor space ones and hope that this will not provoke any confusion.} $(A, \chi, \phi, \tilde{\chi}, G)$ in space-times with helicities $(1, \tfrac 1 2, 0, -\tfrac 1 2, -1)$.

It is easy to see that the the action obtained by a Penrose transform from eq.~\eqref{eq:twistor_action} has an $S$-charge equal to $-4$.  Therefore, it can't be the full $\mathcal{N}=4$ action which also has $S=-8$ terms (the four-scalar interactions in $\mathcal{N}=4$ super-Yang-Mills have $S=-8$ because each scalar has $S=-2$).  In fact, what we get is a supersymmetrised version (see ref.~\cite{Siegel:1992wd, Siegel:1992xp}) of the self-dual Yang-Mills Theory (see ref.~\cite{Chalmers:1996rq}).

Following Witten (see ref.~\cite{Witten:2003nn}) let us describe how one can arrive at the Yang-Mills action starting with the self-dual action.  In string language, the missing terms come from $D$-instantons, but we will not describe that in detail.

It is illuminating however to describe the passage from self-dual Yang-Mills to the conventional Yang-Mills theory.  This is the non-supersymmetric version of what needs to be done in order to get the full $\mathcal{N}=4$ theory.

The self-dual action is (see ref.~\cite{Siegel:1992wd, Siegel:1992xp})
\begin{equation}
  \mathcal{S} = \int_{\mathbb{R}^{1,3}} \tr (G \wedge F) = \int_{\mathbb{R}^{1,3}} \tr (G \wedge F^+),
\end{equation} where $G$ is a self-dual $2$-form in the adjoint representation (more precisely $* G = i G$ where $*$ is the Hodge star), $F = d A + A \wedge A$ is the conventional field strength and $F^{\pm} = \tfrac 1 2 (F \pm i * F)$ (recall that in Minkowski signature $** = -1$ so $* F^{\pm} = \pm i F^{\pm}$).  The second equality follows from the fact that the wedge product of a self-dual and anti-self-dual $2$-forms is zero.\footnote{For two $r$-forms $\omega$ and $\eta$ we have $\omega \wedge * \eta = \eta \wedge * \omega$.  This is important when proving the symmetry of the inner product $(\omega, \eta) = \int \omega \wedge * \eta$.  Using this symmetry property it is easy to prove that the wedge product of self-dual and anti-self-dual $2$-forms is zero.}

The equations of motion obtained by varying $G$ are $F^+ = 0$, so the only non-trivial part in $F$ is its anti-self-dual part $F^-$.  The self-dual field $G$ describes a field of helicity $-1$ and the its anti-self-dual field $F^-$ describes a field of helicity $+1$.  This theory differs from the conventional Yang-Mills theory in that it has only a $A A G$ vertex, describing a $--+$ interaction, but no $G G A$ vertex and no four-gluon vertex.

However, by adding a $G \wedge G$ term to the action
\begin{equation}
  \label{eq:chiral_ym}
  \mathcal{S}_1 = \int_{\mathbb{R}^{1,3}} \tr (G \wedge F - \frac \epsilon 2 G \wedge G),
\end{equation} where $G$ is still self-dual and integrating out $G$ we get
\begin{equation}
  \mathcal{S}_2 = \frac 1 {2 \epsilon} \int_{\mathbb{R}^{1,3}} \tr (F^+ \wedge F^+).
\end{equation}

The Yang-Mills action
\begin{equation}
  \mathcal{S}_{\text{YM}} = \frac 1 {g^2} \int_{\mathbb{R}^{1,3}} \tr (F \wedge * F) = \frac 1 {g^2} \int_{\mathbb{R}^{1,3}} i \tr (F^+ \wedge F^+ - F^- \wedge F^-),
\end{equation} can be obtained from the above action $\mathcal{S}_2$ by adding a multiple of the topological term (which does not matter in perturbation theory)
\begin{equation}
  \int \tr (F \wedge F) = \int_{\mathbb{R}^{1,3}} \tr (F^+ \wedge F^+ + F^- \wedge F^-).
\end{equation}  It follows that the action described in eq.~\eqref{eq:chiral_ym} is equivalent to the conventional Yang-Mill action in perturbation theory.

The perturbation theory (in the limit $\epsilon \rightarrow 0$, which is the same as $g^2 \rightarrow 0$) derived from the action in eq.~\eqref{eq:chiral_ym} is very interesting and has been studied first in ref.~\cite{Chalmers:1996rq}.

When $\epsilon = 0$, the only non-vanishing two-point function is $\langle A G\rangle$ so, when using Wick's theorem the only allowed contractions are between fields $A$ and $G$.  Consider Feynman graphs in this theory with $v$ vertices, $e$ internal lines, $l$ loops and $n$ external lines.  This means there are $2 v$ fields $A$ and $v$ fields $G$, $e - n$ contractions and $2 v - (e - n)$ external fields $A$ and $v - (e - n)$ external fields $G$.  Denote the number of external fields $G$ by $k$ ($k = v - (e - n)$).

The topological constraints
\begin{align}
  2 (e - n) + n & = 3 v,\\
  (v + n) - e + l & = 1,
\end{align} can be solved to yield
\begin{align}
  v & = n + 2 (l - 1),\\
  e & = 2 n + 3 (l - 1), 
\end{align} and finally $k = 1 - l$.  We see here that if $k > 1$ the amplitude is zero, if $k = 1$ the amplitude receives contributions only at tree level\footnote{It turns out that at tree level this amplitude is also zero since there exists a supersymmetric theory in which the graphs contributing to the amplitude are the same as in this non-supersymmetric theory.  By invoking the supersymmetry Ward identities the result follows.} and for $k = 0$ only the one-loop graphs contribute.

If $\epsilon$ is non-vanishing, the vertices remain unchanged but the two-point function $\langle A A\rangle$ is non-vanishing.  The analysis is very easy to adapt to this case.  Suppose there are $d$ $A$-$A$ contractions and $e - d$ $A$-$G$ contractions.  In this case there will be $2 v - (e - d) - 2 d$ external $A$ fields and $v - (e - d)$ external $G$ fields.  By replacing $k \rightarrow k - d$ we can reuse the results above and we get
\begin{equation}
  k = d + 1 - l.
\end{equation}  This is the same as Witten's conjecture but this time in a field theory setting.  Indeed, it was this formula that in part motivated Witten's conjecture.

A similar construction was done for (super-)gravity in refs.~\cite{AbouZeid:2005dg, AbouZeid:2006wu}.  See ref.~\cite{Mason:2007ct} for a construction of $\mathcal{N}=8$ supergravity from twistor space.

The self-dual Yang-Mills theory described above was also the starting in point for understanding the MHV rules (see refs.~\cite{Cachazo:2004kj, Mansfield:2005yd, Ettle:2006bw, Ettle:2007qc}).

\chapter{Connected and disconnected prescriptions}

The exploration of geometrical properties of scattering amplitudes in twistor space revealed that the support of amplitudes is on connected \emph{and} disconnected curves, so the initial prescription described in ref.~\cite{Witten:2003nn} was that one should sum over all these contributions.  Later, Roiban, Spradlin and Volovich showed that the sum over connected curves only also gives a result that is proportional\footnote{The twistor string results were only defined up to a multiplicative constant.} to the known gauge theory answers at tree level (see refs.~\cite{Roiban:2004vt, Roiban:2004ka, Roiban:2004yf}).

Also, the completely disconnected prescription, where one sums over disconnected lines led to the MHV rules (which appeared first in ref.~\cite{Cachazo:2004kj}), where one uses the MHV amplitudes as a kind of elementary interaction vertex (later some mixed prescriptions appeared; see refs.~\cite{Bena:2004ry, Gukov:2004ei}).

The MHV rules are very convenient for performing computations but they lack manifest Lorentz symmetry (the internal legs in an MHV diagram have to be taken off-shell and the prescription for doing this involves an arbitrary light-like vector $\mu$).  In the end, the amplitudes can be proven to be Lorentz invariant~\cite{Cachazo:2004kj}.

In the disconnected prescription the factorisation properties in multiparticle invariants of the amplitude are obvious.  (Note that it is only because the MHV amplitude does not have any multiparticle poles that we are allowed to consider it as an elementary vertex.)  The right factorisation properties and the fact that the MHV rules yield the right results for some low-point tree-level amplitudes insure that the MHV rules yield the right results for tree amplitudes with arbitrary number of external legs.

It is more difficult to prove that the connected prescription is correct.  For instance, the factorisation properties in multiparticle invariants are not obvious.  Nevertheless, the connected prescription has passed a number of tests: it yields the right expressions for the MHV amplitudes, it has the right soft and collinear limits, it is parity symmetric (though this is not immediately obvious), satisfies the dual Ward identity and has also been tested numerically (see refs.~\cite{Roiban:2004vt, Roiban:2004ka, Roiban:2004yf} for more details).  Arguments that the connected prescription has the right factorisation properties have appeared in ref.~\cite{Vergu:2006np}.  In ref.~\cite{Gukov:2004ei}, Gukov, Motl and Neitzke also presented arguments that the connected and disconnected prescriptions yield the same result.

Let us describe in more detail the connected prescription and present a sample computation.  Start with the space of maps of degree $d$ and genus zero from $\mathbb{CP}^1$ to $\mathbb{CP}^{3|4}$.  If we parametrise $\mathbb{CP}^1$ by homogeneous coordinates $(\sigma^1, \sigma^2)$, then the degree $d$ genus zero maps from $\mathbb{CP}^1$ to $\mathbb{CP}^{3|4}$ can be described by
\begin{subequations}
  \label{eq:moduli_param}
  \begin{align}
    Z^I(\sigma^1, \sigma^2) & = \sum_{k=0}^d a_k^I (\sigma^1)^k (\sigma^2)^{d - k}, \\
    \psi^A(\sigma^1, \sigma^2) & = \sum_{k=0}^d \beta_k^A (\sigma^1)^k (\sigma^2)^{d - k}.
  \end{align}
\end{subequations}

We will try to parametrise the moduli space of the holomorphic curves of degree $d$ and genus zero by the coordinates $(a_k^I, \beta_k^A)$. A curve described by eq.~\eqref{eq:moduli_param} and by some fixed coefficients $(a_k^I, \beta_k^A)$ remains unchanged under reparametrisations of $\mathbb{CP}^1$
\begin{equation}
  \label{eq:sl2c}
  \begin{pmatrix}
    (\sigma^1)' \\ (\sigma^2)'
  \end{pmatrix} = M
  \begin{pmatrix}
    \sigma^1 \\ \sigma^2
  \end{pmatrix},
\end{equation} where $M \in SL(2, \mathbb{C})$.  In fact, as a global
rescaling of the super-twistor space coordinates $(Z^I, \psi^A)$ does
not matter, we can take $M \in \Gl(2, \mathbb{C})$.  Also because of
this scaling symmetry we can pass from homogeneous coordinates
$(\sigma^1, \sigma^2)$ on $\mathbb{CP}^1$ to local coordinates $\sigma
= \tfrac {\sigma^1}{\sigma^2}$.

Therefore, the moduli space we need to integrate over is $\mathbb{C}^{4 d + 4|4 d + 4}/\Gl(2, \mathbb{C})$, where the $\Gl(1, \mathbb{C})$ group in $\Gl(2, \mathbb{C}) \simeq \Gl(1, \mathbb{C}) \times SL(2, \mathbb{C})$ serves to cancel the global scaling symmetry of the projective space (the action of $\Gl(2, \mathbb{C})$ on the moduli $(a_k^I, \beta_k^A)$ is easy to infer from the eqns.~\ref{eq:moduli_param} and \ref{eq:sl2c}).

After the passage to local coordinates on $\mathbb{CP}^1$, the expressions of $\lambda$, $\mu$ and $\psi$ are
\begin{subequations}
  \begin{align}
    \lambda^\alpha(\sigma) = & \sum_{k=0}^d a_k^\alpha \sigma^k,\\
    \mu^{\dot{\alpha}}(\sigma) = & \sum_{k=0}^d a^{\dot{\alpha}}_k \sigma^k,\\
    \psi^A(\sigma) = & \sum_{k=0}^d \beta^A_k \sigma^k,
  \end{align}
\end{subequations} and the connected prescription for the twistor space amplitude is
\begin{multline}
  \label{eq:rsv_formula}
  \tilde{A}(\lambda_i, \mu_i, \psi_i) = \int \frac {d^{4 d + 4} a d^{4 d +4} \beta}{\text{Vol}(\Gl(2, \mathbb{C}))} \prod_{i=1}^n \frac {d \sigma_i}{\sigma_i - \sigma_{i+1}} \\ \delta^3\left(\frac {Z_i^I}{Z_i^1} - \frac {Z^I(\sigma_i)}{Z^1(\sigma_i)}\right) \delta^4\left(\frac {\psi_i^A}{Z_i^1} - \frac {\psi^A(\sigma_i)}{Z^1(\sigma_i)}\right).
\end{multline}

Let us now transform this amplitude from twistor space variables $(\lambda, \mu, \psi)$ to momentum space variables $(\lambda, \tilde{\lambda}, \eta)$
\begin{equation}
  A(\lambda_i, \tilde{\lambda}_i, \eta_i) = \int \prod_{i=1}^n d^2 \mu_i d^4 \psi_i  \exp\left(- i \sum_{i=1}^n \mu_i^{\dot{\alpha}} \tilde{\lambda}_{\dot{\alpha}} - i \sum_{i=1}^n \psi_i^A \eta_{i A}\right) \tilde{A}(\lambda_i, \mu_i, \psi_i)
\end{equation}

After integration over $\mu_i^{\dot{\alpha}}$ and then over the coefficients $a_k^{\dot{\alpha}}$, we get (leaving aside the fermionic part for now)
\begin{multline}
  (2 \pi)^{2 d + 2} \int \frac {d^{2 d + 2} a}{\text{Vol}(\Gl(2, \mathbb{C}))} \prod_{i=1}^n \frac {(\lambda_i^1)^2  d \sigma_i}{\sigma_i - \sigma_{i+1}} \delta\left(\frac {\lambda_i^2}{\lambda_i^1} - \frac {\lambda^2(\sigma_i)}{\lambda^1(\sigma_i)}\right) \prod_{k=0}^d \delta^2\left(\sum_{j=1}^n \frac {\lambda_j^1 \tilde{\lambda}_j^{\dot{\alpha}} \sigma_j^k}{\lambda^1(\sigma_j)}\right).
\end{multline}

Calculate the fermionic Fourier transform\footnote{It is easily seen   that, for $\psi$, $\chi$ Grassmann variables, $\int d \psi \delta(\psi - \chi) e^{i \eta \psi} = \int d \psi (\psi - \chi) (1 +   i \eta \psi) = 1 - i \chi \eta = e^{i \eta \chi}$.}
\begin{multline}
  \int \prod_{i=1}^n d^4 \psi_i^A \delta^4\left(\frac {\psi_i}{\lambda_i^1} - \frac {\psi^A(\sigma_i)}{\lambda^1(\sigma_i)}\right) \exp\left(-i \sum_{i=1}^n \psi_i^A \eta_{i A}\right) =\\= \prod_{i=1}^n (\lambda_i^1)^{-4} \exp\left( -i \sum_{i=1}^n \frac {\lambda_i^1 \psi^A(\sigma_i) \eta_{i A}}{\lambda^1(\sigma_i)}\right).
\end{multline}

The integrals over the fermionic moduli $\beta$ are now easy to perform and the result is:
\begin{equation}
  \prod_{k=0}^d \delta^4\left(\sum_{i=1}^n \frac {\lambda_i^1 \eta_{i A} \sigma_i^k}{\lambda^1(\sigma_i)}\right).
\end{equation}  The degree of homogeneity in $\eta$ will be linked to the helicity by the formula $h = 1 - \frac 1 2 \sum_A \eta_A \frac \partial {\partial \eta_A}$.  By expanding the amplitude $A(\lambda, \tilde{\lambda}, \eta)$ in series of $\eta$ we get the scattering amplitudes for all helicity combinations as coefficients of the expansion.

We need to construct the delta functions that impose momentum conservation.  We can do this by using the following formula
\begin{multline}
  \delta\left(\frac {\lambda_1^2}{\lambda_1^1} - \frac {\lambda^2(\sigma_1)}{\lambda^1(\sigma_1)}\right) \delta\left(\frac {\lambda_2^2}{\lambda_2^1} - \frac {\lambda^2(\sigma_2)}{\lambda^1(\sigma_2)}\right) \delta^2\left(\sum_{i=1}^n \frac {\lambda_i^1 \tilde{\lambda}_i^{\dot{\alpha}}} {\lambda^1(\sigma_i)}\right) =\\= \lambda_1^1 \lambda_2^1 (a_0^1)^2 [1,\ 2] \delta^4\left(\sum_{i=1}^n \lambda_i^\alpha \tilde{\lambda}_i^{\dot{\alpha}}\right),
\end{multline} where the factor $\lambda_1^1 \lambda_2^1 (a_0^1)^2 [1,\ 2]$ is the Jacobian of the change of variables.  Strictly speaking the formula above is only valid when multiplied by the remaining delta functions.

We also have to fix the $\Gl(2, \mathbb{C})$ gauge.  One way to do this is to set the variables $\sigma_1$, $\sigma_2$, $\sigma_3$ and $a_0^1$ to some fixed values.  The resulting Faddeev-Popov determinant is $a_0^1 (\sigma_1 - \sigma_2) (\sigma_2 - \sigma_3) (\sigma_3 - \sigma_1)$.

For the purpose of making the symmetry manifest it is convenient to use the following identity
\begin{equation}
  1 = \int \prod_{i=1}^n d \xi_i \delta\left(\xi_i - \frac {\lambda_i^1}{P^1(\sigma_i)}\right)
\end{equation} in the expression for the tree amplitude.

For $i = 3, \ldots, n$ we transform the integrand
\begin{multline}
  (\lambda_i^1)^{-2} \delta\left(\xi_i - \frac {\lambda_i^1}{P^1(\sigma_i)}\right) \delta\left(\frac {\lambda_i^2}{\lambda_i^1} - \frac {P^2(\sigma_i)}{P^1(\sigma_i)}\right) =\\ (\lambda_i^1)^{-1} \delta\left(\xi_i - \frac {\lambda_i^1}{P^1(\sigma_i)}\right) \delta(\lambda_i^2 - \xi_i P^2(\sigma_i)) =\\ \frac 1 {\xi_i} \delta(\lambda_i^1 - \xi_i P^1(\sigma_i)) \delta(\lambda_i^2 - \xi_i P^2(\sigma_i)).
\end{multline}

The delta functions with $i=1,2$ and $k=0$ are used for pulling out momentum conservation, so $i=1,2$ will benefit from a special treatment.  For $i=1$, $i=2$
\begin{equation}
  \frac 1 {\lambda_i^1} \delta\left(\xi_i - \frac {\lambda_i^1}{P^1(\sigma_i)}\right) = \frac 1 {\xi_i} \delta(\lambda_i^1 - \xi_i P^1(\sigma_i)).
\end{equation}

Putting all the results together
\begin{multline}
  A(\lambda_i, \tilde{\lambda}_i, \eta_i) = (2 \pi)^{2 d + 2} [1,\ 2] (a_0^1)^3 \delta^4\left(\sum_{i=1}^n \lambda_i^\alpha \tilde{\lambda}_i^{\dot{\alpha}}\right) \int d^{2 d + 1} a d^{n - 3} \sigma d^n \xi \\ \frac {(\sigma_1 - \sigma_2)(\sigma_2 - \sigma_3)(\sigma_3 - \sigma_1)}{\prod_{i=1}^n \xi_i (\sigma_i - \sigma_{i+1})} \prod_{i=1}^n \delta(\lambda_i^1 - \xi_i P^1(\sigma_i)) \prod_{i=3}^n \delta(\lambda_i^2 - \xi_i P^2(\sigma_i))\\ \prod_{k=1}^d \delta^2\left(\sum_{j=1}^n \tilde{\lambda}_j^{\dot{\alpha}} \xi_j \sigma_j^k\right) \prod_{k=0}^d \delta^4\left(\sum_{i=1}^n \eta_{i A} \xi_i \sigma_i^k\right)
\end{multline}

In order to express the gluon scattering amplitude we have to extract from the product of fermionic delta functions the factors which have degree of homogeneity in $\eta$ equal to zero for helicity $+$ and four for helicity $-$.  In order to separate the relevant contribution we use:
\begin{equation}
  \label{eq:fermionic_coefficient}
  \prod_{i=1}^n \sum_{j=1}^n a_{i j} \eta_j = \det a \prod_{i=1}^n \eta_i.
\end{equation}

Suppose we want the $n$-point amplitude with $q = d + 1$ helicity $-$ gluons and the rest with helicity $+$.  Suppose also that the negative helicity gluons are in positions $i_1, \ldots, i_{q}$.  In order to get this amplitude we have to compute the coefficient of $\prod_{k=1}^q \prod_{A=1}^4 \eta_{i_k A}$.  By using eq.~(\ref{eq:fermionic_coefficient}) we easily see that the coefficient is $(\det F)^4$, where $F$ is obtained from $\tilde{F}$ (which is $n \times (d + 1)$ matrix whose elements are $\tilde{F}_{i   k} = \xi_i \sigma_i^k$) by taking only the lines $i_1, \ldots i_q$ of $\tilde{F}$.

The general formula for a $n$-point tree-level amplitude with $q = d + 1$ negative helicity gluons is,
\begin{multline}
  \label{eq:general_formula}
  A(\lambda_i, \tilde{\lambda}_i, h_i) = (2 \pi)^{2 d + 2} [1,\ 2] (a_0^1)^3 \delta^4\left(\sum_{i=1}^n \lambda_i^\alpha \tilde{\lambda}_i^{\dot{\alpha}}\right)\\ \sum_{\text{solutions of $\{B_r(q_s)=0, \forall r,s\}$}} \frac {(\sigma_1 - \sigma_2)(\sigma_2 \sigma_3)(\sigma_3 - \sigma_1)}{\prod_{i=1}^n \xi_i (\sigma_i - \sigma_{i+1})} \frac {(\det F)^4} {\det \left(\frac {\partial B_r}{\partial q_s}\right)}
\end{multline} where $q_s = (\sigma_{i=4, \ldots, n}, \xi_{i=1,   \ldots, n}, a^1_{k=1, \ldots, d}, a^2_{k=0, \ldots, d})$,
\begin{equation}
  B_r =
  \begin{cases}
    \lambda_i^\alpha - \xi_i \sum_{k=0}^d a_k^\alpha \sigma_i^k,& \text{for $i=1, \ldots, n$ if $\alpha=1$ and $i=3, \ldots, n$ if $\alpha=2$}\\
    \sum_{i=1}^n \tilde{\lambda}_i^{\dot{\alpha}} \xi_i \sigma_i^k,& \text{for $k=1, \ldots, d$ and $\dot{\alpha}=\dot{1},\dot{2}$}
  \end{cases}
\end{equation} and the $(d + 1) \times (d + 1)$ matrix $F$ is obtained
by selecting the lines $i_1, \ldots, i_q$ corresponding the the
positions of negative helicity gluons from the matrix
$\tilde{F}_{i k} = \xi_i \sigma_i^k$.

The final result is a bit complicated, but the surprising outcome is that, in order to compute the tree-level amplitudes, one needs to find the solutions to a system of algebraic equations and them sum a certain Jacobian evaluated for the values of these solutions (the number of equations is equal to the number of unknowns).  Unfortunately, solving such systems of algebraic equations is a complicated task.  However, numerical comparisons to the known values of scattering amplitudes yield excellent agreement.

\chapter{Factorisation of the connected prescription}

In this section we discuss the factorisation of the connected prescription~\cite{Roiban:2004vt, Roiban:2004ka, Roiban:2004yf} of Roiban, Spradlin and Volovich, as detailed in ref.~\cite{Vergu:2006np}.

One drawback of the connected prescription is the lack of manifest factorisation properties.  The factorisation properties of tree amplitudes suffice to determine the amplitude for an arbitrary number of external legs.  Therefore, they are a crucial test for the correctness of various ansatze for the amplitudes.

Let us quickly review the factorisation properties at tree level.  As we already explained, the tree-level scattering amplitude decomposes on a colour basis formed of single traces of colour factors of external particles,
\begin{equation}
  \mathcal{A}_n(\{k_i, h_i, a_i\}) = g^{n-2} \sum_{\sigma \in
    S_n/\mathbb{Z}_n} Tr(T^{a_{\sigma(1)}} \cdots T^{a_{\sigma(1)}})
  A_n(\sigma(1), h_{\sigma(1)}; \ldots; \sigma(n), h_{\sigma(n)}),
\end{equation} where $k_i$ is the momentum of the $i$-th particle, $h_i$ is its helicity (we consider all particles to be out-going), $a_i$ label the generators $T^{a_i}$ of the colour algebra, $g$ is the Yang-Mills coupling constant, and $\sigma \in S_n/\mathbb{Z}_n$ instructs us to sum only over cyclically inequivalent permutations $\sigma$.

The colour-ordered amplitudes $A_n$ satisfy the following factorisation property: in the limit where the sum of more than two adjacent particles goes on-shell ($P = p_1 + \cdots + p_m$ and $P^2 \to 0$), the amplitude has the following behaviour
\begin{equation}
  \label{eq:fact}
  A_n(p_1, \ldots, p_n) \sim \sum_{h=\pm} A_{m+1}(p_1, \ldots, p_m, P^h) \frac i {P^2} A_{n-m+1}(P^{-h}, p_{m+1}, \ldots, p_n),
\end{equation} where $h = \pm$ represents the sum over the two
helicities of an particle going on-shell.

Note that, in contrast to the case of the full amplitudes, the colour ordered amplitudes can only have poles when the sum of \emph{adjacent} external momenta goes on-shell.  This means they have a simpler singularity structure than the full amplitude.

One issue that remains mysterious in the prescription of Roiban, Spradlin and Volovich is the interpretation of the delta functions and of the holomorphic Jacobian.  It was observed in~\cite{Roiban:2004ka,   Roiban:2004yf} that, after accounting for the delta functions corresponding to momentum conservation, the number of integrals equals the number of delta functions.  This means that, in the end, computing the integrals is equivalent to summing a Jacobian factor for each solution of the delta functions inside the integrals.

There are two puzzling issues, however.  The first is that, while the prescription of Roiban, Spradlin and Volovich was formulated in a real version of twistor space and the integrals appearing are real integrals, some of the solutions of the equations imposed by the delta functions are complex.  In order to get the right result one needs to sum over all solutions, be they real or complex.

The second issue concerns the Jacobian factor.  The formulae for the integrals involve the absolute value of the Jacobian but using this in the connected prescription does not give the right results.  The prescription that gives correct results involves the Jacobian without the absolute value.

In ref.~\cite{Vergu:2006np} we proposed interpreting the integrals as contour integrals.  How should the delta functions be interpreted when the integrals are contour integrals?  The defining property of the delta function is its property
\begin{equation}
  \label{eq:delta_def}
  f(a) = \int d x \delta(x - a) f(x),
\end{equation} for all functions $f$.  If we take $\delta(z - z_0) \equiv \frac 1 {2 \pi i} \frac 1 {z - z_0}$ and the integral to be a contour integral on a contour around $z_0$, the property in \eqref{eq:delta_def} is satisfied for all the functions $f$ which are holomorphic inside the integration contour.

This interpretation is compatible with the usual properties of Fourier integrals if we define the Fourier integral to be a complex integral along a contour from zero to infinity, chosen in such a way to insure the convergence.  For example, in the case of real $z$, the Fourier transform of the identity is defined as follows
\begin{equation}
  \int_0^{+i \infty} \frac {d k}{2 \pi} e^{i k z} = - \frac 1 {2 \pi
    i} \frac 1 z = - \delta(z).
\end{equation}  A similar contour, from zero to infinity was
already used in ref.~\cite{Cachazo:2004kj} in a heuristic discussion of the twistor-space propagator.

This interpretation is fully compatible with the delta function manipulations in ref.~\cite{Roiban:2004yf}.  For example, we have
\begin{equation}
  \int \delta(f(z)) = \frac 1 {2 \pi i} \oint \frac {d z}{f(z)} =
  \sum_{z_i \in \{z| f(z)=0\}} \frac 1 {f'(z_i)}.
\end{equation}

Note also that multiple roots of $f(z) = 0$ do not contribute and the result is obtained by using the Jacobian instead of the absolute value of the Jacobian.  This is indeed what is required to obtain a correct answer in the connected prescription computation~\cite{Roiban:2004yf}.  The choice of contour is such that all the poles at finite distance are included inside the contour.

The arguments for factorisation put forward in ref.~\cite{Vergu:2006np} use this interpretation of the integrals. The main idea of the argument is to integrate over the configurations where the vertex operators are widely separated (presumably, only these configurations contribute to the residue of the pole where an internal momentum goes on-shell).  In the limit of widely-separated vertex operators the moduli of the algebraic curves and the positions of the vertex operators can be reparametrised in such a way as to make possible the identification of a product of tree amplitudes as in eq.~\eqref{eq:fact}.

A related approach was used by Gukov, Motl and Neitzke in ref.~\cite{Gukov:2004ei}, to prove the equivalence between the connected and disconnected prescription and also to propose some mixed prescriptions.

Let us describe in more detail the approach of ref.~\cite{Vergu:2006np} for
the factorisation of the connected prescription.  Consider an $n$-point
tree-level amplitude with $q$ negative helicity and $n-q$ positive helicity
external particles.  We will study the factorisation of this $n$-point
amplitude into two tree-level amplitudes (a left and a right one) with $n_l
+ 1$ and $n_r + 1$ legs ($n_l + n_r = n$) having $q_l$ respectively $q_r$
negative helicity legs ($q_l + q_r = q + 1$).  The degrees of the algebraic
curves describing these tree amplitudes in the connected prescription of
Roiban, Spradlin and Volovich, are $d_l = q_l - 1$ and $d_r = q_r - 1$.

The physical intuition behind the factorisation is that the internal line
going on-shell allows for a propagation of the internal state over long
distances.  Even if what distance means in the twistor string language is
not completely obvious, we will take it to be given roughly by the
difference of the $\sigma$ coordinates.  The strategy for proving
factorisation is then, to restrict to a region of the integration region
where the coordinates $\sigma$ (which are also the positions of the vertex
operators in Berkovits' version of the twistor string~\cite{Berkovits:2004hg}) are widely separated in two clusters.

Taking $L$ to be the (large) scale of the separation, we restrict therefore
to a region in the integration domain where
\begin{align}
  a_k & =
  \begin{cases}
    \hat{a}_{d_l - k} L^{d_r}, \quad \text{if $0 \leq k \leq d_l$},\\
    \bar{a}_{k - d_l} L^{d - k}, \quad \text{if $d_l \leq k \leq d$},
  \end{cases}\\
  \sigma_i & =
  \begin{cases}
    \frac 1 {\hat{\sigma}_i}, \quad \text{if $i$ is on the left},\\
    \bar{\sigma}_i, \quad \text{if $i$ is on the right},
  \end{cases}\\
  \xi_i & =
  \begin{cases}
    \hat{\xi}_i \hat{\sigma}_i^{d_l} L^{-d_r}, \quad \text{if $i$ is on
      the left},\\
    \bar{\xi}_i L^{-d} \bar{\sigma}_i^{- d_l}, \quad \text{if $i$ is on
      the right}
  \end{cases}.
\end{align}

There are several constraints that we need to impose in order for the
proposed scaling with $L$ to work.  First, all the hatted and barred
variables need to be of order one and second all the $\hat{\sigma}$ and
$\bar{\sigma}$ must be non-zero.  Note also that we have $\hat{a}_0 =
\bar{a}_0$.

Using the notations introduced above, we can write the polynomials
\begin{equation}
  \xi_i Z^I(\sigma_i) = \xi_i \sum_{k=0}^d a_k^I \sigma_i^k,
\end{equation} as
\begin{equation}
  \xi_i Z^I(\sigma_i) = \hat{\xi}_i \sum_{k=0}^{d_l} \hat{a}_k^I
  \hat{\sigma}_i^k + \hat{\xi}_i \sum_{k=1}^{d_r} \bar{a}_k^I
  \hat{\sigma}_i^{-k} L^{-k},
\end{equation} if $i$ is on the left side, and
\begin{equation}
  \xi_i Z^I(\sigma_i) = \bar{\xi}_i \sum_{k=0}^{d_r} \bar{a}_k^I
  \bar{\sigma}_i^k + \bar{\xi}_i \sum_{k=1}^{d_l} \hat{a}_k^I
  \bar{\sigma}_i^{-k} L^{-k},
\end{equation} if $i$ is on the right side.

In the limit $L \to \infty$ the hatted and barred moduli separate, except
for the identification $\hat{a}_0 = \bar{a}_0$.  It is easy to establish
that $L \to \infty$ corresponds to an internal line going on-shell.  Take
the momenta that are on the left side, $p_i^{\alpha \dot{\alpha}} =
\lambda_i^\alpha \tilde{\lambda}_i^{\dot{\alpha}}$.  Then, by using the
formulae for the $\lambda_i$ that result from the delta functions in the
connected prescription, we get
\begin{equation}
  P^{\alpha \dot{\alpha}} \rightarrow \hat{a}_0^\alpha \sum_{i \in L}
  \hat{\xi}_i \tilde{\lambda}_i^{\dot{\alpha}} + \mathcal{O}(L^{-1}).
\end{equation}

The connected prescription integrand also contains a product of differences
of adjacent $\sigma_i$, which can be rewritten as follows \footnote{We have
  introduced a $0$, anticipating the fact that the internal line will be
  attached at $\sigma=0$.  See below.}
\begin{multline}
  \prod_{i=1}^n \frac {d \sigma_i}{\sigma_i - \sigma_{i + 1}} = \frac
  {\prod_{i=1}^{n_l} d \sigma_i} {(\sigma_1 - \sigma_2) \cdots
    (\sigma_{n_l} - 0) (0 - \sigma_1)} \\ \times \frac {\prod_{i=n_l+1}^n d
    \sigma_i} {(\sigma_{n_l + 1} - \sigma_{n_l + 2}) \cdots (\sigma_n - 0)
    (0 - \sigma_{n_l + 1})} \times \frac {\sigma_{n_l} (- \sigma_1)
    \sigma_n (- \sigma_{n_l + 1})}{(\sigma_{n_l} - \sigma_{n_l + 1})
    (\sigma_n - \sigma_1)} =\\= \frac 1 L \times \frac {\prod_{i=1}^{n_l} d
    \hat{\sigma}_i} {(\hat{\sigma}_1 - \hat{\sigma}_2) \cdots} \times \frac
  {\prod_{i=n_l+1}^n d \bar{\sigma}_i}{(\bar{\sigma}_{n_l + 1} -
    \bar{\sigma}_{n_l + 2}) \cdots} \left(1 + \mathcal{O}(L^{-1})\right).
\end{multline} Therefore, this part also factorises at the leading order in
$L$.

In order to prove factorisation, we need to introduce two more vertex
operators corresponding to the internal line going on-shell in the
factorisation limit.  We introduce the notation:
\begin{equation}
  \Psi_{\pi, \bar{\pi}, \eta}(\lambda, \mu, \psi) = \int \frac {d \xi} \xi \delta^2(\pi - \xi \lambda) \exp(i \xi \left[\mu, \bar{\pi}\right]) \exp(i \xi \psi^A \eta_A).
\end{equation}  For these wave functions one can prove orthonormality and
completeness relations.

The orthonormality relation is\footnote{The complex conjugation does not
  act on the fermionic variables.}
\begin{multline}
  \int \frac {d^2 \lambda d^2\mu d^4 \psi}{(2 \pi)^2\text{Gl}(1)}
  \Psi^*_{\pi, \bar{\pi}, \eta}(\lambda, \mu, \psi) \Psi_{\pi', \bar{\pi}',
    \eta'}(\lambda, \mu, \psi) =\\= \int \frac {d \xi} \xi \delta^2(\xi \pi
  - \pi') \delta^2(\bar{\pi} - \xi \bar{\pi}') \delta^4(\eta - \xi \eta')
  \equiv \delta_{\pi, \bar{\pi}, \eta; \pi', \bar{\pi}', \eta'},
\end{multline} where the $\Gl(1)$ group comes from the following symmetry 
\begin{equation}
  \lambda \rightarrow t \lambda,\quad
  \mu \rightarrow t \mu,\quad
  \psi \rightarrow t \psi,\quad
  (\xi, \xi') \rightarrow t^{-1} (\xi, \xi').
\end{equation}

The formula above can be easily proven by gauge-fixing $\lambda^1 = \pi^1$,
which introduces a Jacobian equal to $\lambda^1 = \pi^1$.  Then the
integral over $\xi$ sets $\xi = 1$ and cancels the Jacobian $\pi^1$ and
finally the integral over $\lambda^2$ sets $\lambda^2 = \pi^2$ (then, the
result above is obtained by renaming $\xi'$ to $\xi$).

But what we need to insert in order to achieve factorisation is a
completeness relation.
\begin{multline}
  \int \frac {d^2 \pi d^2 \bar{\pi} d^4 \eta}{(2 \pi)^2 \text{Gl}(1)}
  \Psi^*_{\pi, \bar{\pi}, \eta}(\lambda, \mu, \psi) \Psi_{\pi, \bar{\pi},
    \eta}(\lambda', \mu', \psi') =\\= \int \frac {d \xi} \xi
  \delta^2(\lambda - \xi \lambda') \delta^2(\mu - \xi \mu') \delta^4(\psi -
  \xi \psi') \equiv \delta_{\lambda, \mu, \psi; \lambda', \mu', \psi'},
\end{multline} where the $\Gl(1)$ group comes from the following symmetry
of the integral
\begin{align}
  \pi \rightarrow t \pi,\quad
  \bar{\pi} \rightarrow t^{-1} \bar{\pi},\quad
  \eta \rightarrow t^{-1} \eta,\quad
  (\xi, \xi') \rightarrow t (\xi, \xi').
\end{align}

The above formulae can be used to separate the integrals over moduli
(remember that we still have the constraint $\hat{a}_0 = \bar{a}_0$).
Then, schematically
\begin{multline}
  \int d^{4(d+1)|4(d+1)} a \cdots = \int d^{4(d_l+1)|4(d_l+1)} \hat{a}
  d^{4(d_r+1)|4(d_r+1)} \bar{a} \delta^{4|4}(\hat{a}_0 - \bar{a}_0) \cdots
  =\\= \int \frac {d^2 \pi d^2 \bar{\pi} d^4 \eta}{(2 \pi)^2 \Gl(1)} \int
  \frac {d^{4(d_l+1)|4(d_l+1)} \hat{a} d^{4(d_r+1)|4(d_r+1)} \bar{a}}
  {\Gl(1)} \Psi^*_{\pi, \bar{\pi}, \eta}(\hat{a}_0) \Psi_{\pi, \bar{\pi},
    \eta}(\bar{a}_0) \cdots,
\end{multline} where we have inserted the delta function from the
completeness relation and $\Gl(1)$ acts projectively on $\bar{a}$ or
$\hat{a}$ moduli.  The dots in the above formula stand for a function of
the moduli $a$ which is invariant under a scaling of $\bar{a}$ and
$\hat{a}$ separately.  At the dominant order in $L$ this property is
satisfied by the integrands we consider.  The delta function really stands
for $\delta^{4|4}(\hat{Z}(\sigma)-\bar{Z}(\sigma'))$, understood to be
evaluated at $\sigma = \sigma' = 0$.

Now that we have managed to factorise the integrand to the dominant order
in $L$, we come to the issue of gauge fixing.  We can fix the gauge in
several different ways.  If we gauge-fix one component of $\hat{a}_0$, say
$\hat{a}_0^1$, together with $\hat{\sigma}_i$, $\hat{\sigma}_j$ and
$\bar{\sigma}_p$ we get a Jacobian
\begin{equation}
  \label{eq:Jacobian_J}
  J = - \frac 1 L \hat{a}_0^1 (\hat{\sigma}_i - \hat{\sigma}_j)(-1 + L
  \bar{\sigma}_p \hat{\sigma}_i)(-1 + L \bar{\sigma}_p \hat{\sigma}_j).
\end{equation} The $\Gl(1)$ gauge invariance at the right can be
gauge-fixed independently and gives a Jacobian $\bar{a}_0^1$.
Note that if anyone of $\hat{\sigma}_i$, $\hat{\sigma}_j$ or
$\bar{\sigma}_p$ is zero, the Jacobian is $J \sim L^{-1}$ and this will not
contribute in the factorisation limit (in order to get a contribution in
the limit $L \rightarrow \infty$ it has to cancel the factor in $\frac 1 L$
from the product of $\sigma$).  This is consistent with the
interpretation we gave that the internal line has $\hat{\sigma} =
\bar{\sigma} =0$ so, it already is gauge-fixed at zero.

When $\hat{\sigma}_i$, $\hat{\sigma}_j$ and $\bar{\sigma}_p$ are different
from zero we have
\begin{equation}
  J = - L \hat{a}_0^1 \bar{\sigma}_p^2 \hat{\sigma}_i
  \hat{\sigma}_j (\hat{\sigma}_i - \hat{\sigma}_j) + \mathcal{O}(1).
\end{equation} We want to compare this with the case when the left- and
right-hand sides are completely gauge-fixed, however on the right-hand side
there are only two $\bar{\sigma}$'s and a modulus fixed.  We use the fact
that $\bar{a}_0^1 \bar{\sigma}_p \bar{\sigma}_q (\bar{\sigma}_p -
\bar{\sigma}_q)$ times the right hand integrals where we don't integrate
over $\bar{a}_0^1$, $\bar{\sigma}_p$, $\bar{\sigma}_q$ is independent of
$\bar{\sigma}_q$ and can be taken out of the integral over
$\bar{\sigma}_q$.

Dividing the full Jacobian $J$ by the Jacobians needed to recombine the
left and right parts into gauge invariant amplitudes gives
\begin{equation}
  \frac J {L J_l J_r} = \frac {\bar{\sigma}_p}{\bar{\sigma}_q
    (\bar{\sigma}_p - \bar{\sigma}_q)} = \frac 1 {\bar{\sigma}_q} + \frac 1
  {\bar{\sigma}_p - \bar{\sigma}_q},
\end{equation} where $J_l = \hat{\sigma}_i \hat{\sigma}_j (\hat{\sigma}_i -
\hat{\sigma}_j)$ and $J_r = \bar{\sigma}_p \bar{\sigma}_q (\bar{\sigma}_p -
\bar{\sigma}_q)$.  We are left with the integral over $\bar{\sigma}_q$
\begin{equation}
  \oint d \bar{\sigma}_q \left(\frac 1 {\bar{\sigma}_q} -
    \frac 1 {\bar{\sigma}_q - \bar{\sigma}_p}\right).
\end{equation} This integral is zero if we interpret it in the most naive
way possible, by taking a contour around $0$ and $\bar{\sigma}_p$ in the
$\bar{\sigma}_q$ plane.  However, we have to recall that the region in the
neighbourhood of $\bar{\sigma}_q = 0$ is special and is not included in our
integration domain (at any rate, a contour around zero which is included in
our integration domain cannot be shrunk to $\bar{\sigma} = 0$ while
staying inside the integration domain).  Therefore, we propose to do the
next less naive thing possible and take a contour which does not go around
$\bar{\sigma} = 0$.  The result of the integration is then $-2 \pi i$.

We can test this prescription for the contour by gauge-fixing the linear
combination $\bar{\sigma}_p + \zeta \bar{\sigma}_q$.  This has a Jacobian
$J_\zeta$ which is such that
\begin{equation}
  \frac {J_\zeta}{L J_l J_r} = \frac 1 {\bar{\sigma}_q} -
  \frac {1 + \zeta}{\bar{\sigma}_p - \bar{\sigma}_q} - \frac \zeta
  {\bar{\sigma}_p} + \mathcal{O}(L^{-1}).
\end{equation} Now suppose $\bar{\sigma}_p + \zeta \bar{\sigma}_q$ is
gauge-fixed to a value $\tau$.  This is implemented by introducing a delta
function $\delta(\bar{\sigma}_p + \zeta \bar{\sigma}_q - \tau)$ and the
Jacobian $J_\zeta$ in the integral.  After integrating over
$\bar{\sigma}_p$ we are left with the following integral over
$\bar{\sigma}_q$
\begin{equation}
  \oint d \bar{\sigma}_q \left(\frac 1 {\bar{\sigma}_q} -
    \frac 1 {\bar{\sigma}_q - \frac \tau {1 + \zeta}} + \frac \zeta {\zeta
      \bar{\sigma}_q - \tau}\right).
\end{equation} In the integrand, the first and the last term correspond to
$\bar{\sigma}_q = 0$ and $\bar{\sigma}_p = 0$ respectively.  Therefore, as
before, we argue that the choice of contour is such that they don't
contribute.  The remaining term yields $-2 \pi i$.

There is however a problem for $\zeta = -1$ and, in this particular case,
our prescription does not work.  It does work however, for the whole family
of gauge-fixing conditions where $\zeta \neq -1$.  (Note that when $\zeta
\rightarrow -1$ the pole which contributes to the integral is sent to
infinity and also out of our domain of integration.)

Now we can perform the integration $\int \frac {d^2 \pi d^2 \bar{\pi} d^4
\eta}{(2 \pi)^2 \Gl(1)}$.  The fermionic part of the integral imposes that
the helicities on the two sides of the internal line are opposite.

Let us concentrate on the bosonic part.  After gauge-fixing $\pi^1$ the
measure is $\pi^1 d \pi^2 d^2 \bar{\pi}$.  The right and left part each
contain a momentum conserving delta function.  Denote $P$ the total
momentum at left and $Q$ the total momentum at right
\begin{multline}
  \int \pi^1 d \pi^2 d^2 \bar{\pi}^{\dot{\alpha}} \delta^4(P^{\alpha
    \dot{\alpha}} - \pi^\alpha \bar{\pi}^{\dot{\alpha}}) \delta^4(Q^{\alpha
    \dot{\alpha}} - \pi^\alpha \bar{\pi}^{\dot{\alpha}}) =\\= \delta^4(P -
  Q) \int \pi^1 d \pi^2 d^2 \bar{\pi}^{\dot{\alpha}} \delta^4(P^{\alpha
    \dot{\alpha}} - \pi^\alpha \bar{\pi}^{\dot{\alpha}}).
\end{multline} The integral above can be computed straightforwardly and the
result is $\delta(P^2)$.

We interpret this as a holomorphic delta function $\delta(z) \equiv \frac 1
{2 \pi i} \frac 1 z$.  Granted this interpretation, we obtain the pole we
were looking for
\begin{equation}
  \delta(P^2) \equiv \frac 1 {2 \pi i} \frac 1 {P^2}.
\end{equation}

\chapter{Berkovits twistor string}

Berkovits has proposed an alternative twistor string theory in ref.~\cite{Berkovits:2004hg} (see ref.~\cite{Siegel:2004dj} for yet another proposal which we will not discuss).  In describing this theory we will use the notation of ref.~\cite{Dolan:2007vv} rather than those of the original paper.  One difference is that ref.~\cite{Dolan:2007vv} use a complex twistor space as target space, while Berkovits uses a real twistor space as target space.

This is an open string with the action
\begin{equation}
  S = \int d^2 z \left(Y^{\overline{z}} \cdot \nabla_{\overline{z}} Z + \tilde{Y}^z \cdot \nabla_z \tilde{Z}\right) + S_c,
\end{equation} where $S_c$ is the action for a current algebra, $Z$ are the coordinates on the twistor space $Z = (\lambda, \mu, \psi)$ and $\tilde{Z}$ are the complex conjugate coordinates, $\tilde{Y}^{\overline{z}}$ and $Y^z$ are the conjugated momenta, the $\cdot$ stands for summation over twistor space indices and
\begin{equation}
  \nabla_{\overline{z}} = \partial_{\overline{z}} - A_{\overline{z}}, \quad \nabla_z = \partial_z - A_z,
\end{equation} where $A$ is a $\Gl(1, \mathbb{C})$ worldsheet gauge field.  Under this $\Gl(1, \mathbb{C})$ gauge symmetry the fields $Z$ and $\tilde{Z}$ have charge $+1$ and $Y^{\overline{z}}$ and $\tilde{Z}^z$ have charge $-1$.

This is a conformal theory where $Z$, $\tilde{Z}$ have dimension zero and $Y^{\overline{z}}$ has dimension $(1, 0)$ and $\tilde{Y}^z$ has dimension $(0, 1)$.

The $b$--$c$ ghost system contributes $-26$ to the central charge, the $u$--$v$ $\Gl(1, \mathbb{C})$ ghosts contribute $-2$ and the $Y$--$Z$ system contributes $0$ because of a cancellation between the bosons and fermions.  In order to cancel the central charge the current algebra with action $S_c$ must contribute $28$.

The equations of motion are
\begin{equation}
  \nabla_{\overline{z}} Z = 0, \quad \left(\partial_{\overline{z}} + A_{\overline{z}}\right) Y^{\overline{z}} = 0,
\end{equation} the constraints
\begin{equation}
  Y^{\overline{z}} \cdot Z = 0, \quad \tilde{Y}^z \cdot \tilde{Z} = 0
\end{equation} and the boundary conditions
\begin{equation}
  n_{\overline{z}} Y^{\overline{z}} \cdot \delta Z + n_z \tilde{Y}^z \cdot \delta \tilde{Z} = 0.
\end{equation}

We search solutions for the boundary conditions of the form $\tilde{Z} = U Z$.  The condition $\tilde{Z} = \overline{Z}$ implies that $|U| = 1$ on the boundary.

The $\Gl(1, \mathbb{C})$ gauge group acts on the fields as
\begin{alignat}{3}
  Z & \rightarrow g Z, & \quad Y^{\overline{z}} & \rightarrow g^{-1} Y^{\overline{z}}, & \quad A_{\overline{z}}' & \rightarrow A_{\overline{z}} - g \partial_{\overline{z}} g^{-1},\\
  \tilde{Z} & \rightarrow \overline{g} \tilde{Z}, & \quad \tilde{Y}^z & \rightarrow \overline{g}^{-1} \tilde{Y}, & \quad  A_z & \rightarrow A_z - \overline{g} \partial_z \overline{g}^{-1}.
\end{alignat}

We can then set $A_z$ and $A_{\overline{z}}$ to zero and still make gauge transformations where $g(z)$ is holomorphic (and $\overline{g}(\overline{z})$ anti-holomorphic).

Let us now discuss the boundary conditions for the disk, which is relevant for computing tree amplitudes (in ref.~\cite{Dolan:2007vv} Dolan and Goddard also discuss the case of the annulus, which is relevant for loop amplitudes).

The phase of $U(z)$ changes by $- 2 \pi n$ as $z$ goes around the unit circle.  Then, the function $\ln \left( z^n U(z)\right)$ is analytic in a neighbourhood of the unit circle.  By using its expansion in Laurent series, we see that it can be written as a sum of two functions, one holomorphic in the unit disk and the other holomorphic outside of the unit disk.
\begin{equation}
  \ln \left( z^n U(z)\right) = f^<(z) + f^>(z),
\end{equation} where
\begin{equation}
  f^<(z) = \sum_{n \geq 0} c_n^< z^n, \quad f^>(z) = \sum_{n \geq 0} c_n^> z^{-n}.
\end{equation}

The function $\ln \left( z^n U(z)\right)$ has a zero real part on the unit circle.  This implies that the coefficients $c_n^<$ and $c_n^>$ are linked by $c_n^< + c_n^> = 0$.  This in turn implies
\begin{equation}
  \label{eq:fgt_flt}
  f^<(z) = - \overline{f^> \left(\frac 1 {\overline{z}}\right)}.
\end{equation}

Under a gauge transformation the function $U$ transforms as
\begin{equation}
  U \rightarrow \overline{g}^{-1} U g.
\end{equation}  By making a holomorphic gauge transformation $g(z) = \exp (- f^<(z))$ and using eq.~\eqref{eq:fgt_flt} and the fact that $f^>\left(\frac 1 {\overline{z}}\right) - f^>(z) = 0$ on the unit circle we get that $U(z) = z^{-n}$ on the unit circle.

Because $A_z$ and $A_{\overline{z}}$ are zero the equations of motion impose that $Z$ is holomorphic and $\tilde{Z}$ is anti-holomorphic in the unit disk
\begin{equation}
  Z(z) = \sum_{k \geq 0} Z_k z^k, \quad \tilde{Z}(\overline{z}) = \sum_{k \geq 0} \overline{Z}_k \overline{z}^k,
\end{equation} which satisfy the reality condition $\tilde{Z}(\overline{z}) = \overline{Z(z)}$.  The boundary condition $\tilde{Z}(\overline{z}) = z^{-n} Z(z)$ implies $Z_m = \overline{Z}_{n-m}$.  This has solutions only for $n \geq 0$ and then\footnote{This is for the classical theory.  In the quantum theory the modes of $Y$ and $Z$ have canonical (anti-)commutators and therefore cannot vanish.  Rather, these expressions can be used inside a vacuum expectation value, as the other modes yield zero when acting on the left or right vacua (we have $Z_m \ket{0} = 0$ and $\bra{0} Z_{n-m} = 0$ if $m < 0$).  This differs from the usual CFT conventions where positive modes annihilate the vacuum.}
\begin{equation}
  Z(z) = \sum_{k = 0}^n Z_k z^k, \quad \tilde{Z}(\overline{z}) = \sum_{k = 0}^n \overline{Z}_k \overline{z}^k.
\end{equation}

Another way to understand this is to use the `doubling trick' and identify the theory with holomorphic and anti-holomorphic fields on the disk with a theory of holomorphic fields on the sphere.  Then, the boundary conditions for the fields $Z$ are encoded in the topology, more precisely the instanton number, of the worldsheet gauge field.  This worldsheet gauge field can also be interpreted as a worldsheet Levi-Civita connection corresponding to a shifted conformal dimension of the fields $Y$ and $Z$ (see ref.~\cite{Berkovits:2004tx} for a more detailed explanation).  A related way to see this is to consider the definition of the adjoint in a CFT (see ref.\cite{Ginsparg:1988ui}, for example).  For a holomorphic field with conformal dimension $h$ the definition of the adjoint is
\begin{equation}
  Z^\dagger(z) = \frac 1 {\overline{z}^{2 h}} Z\left(\frac 1 {\overline{z}}\right).
\end{equation}  This, together with the boundary condition $\tilde{Z}(\overline{z}) = z^{-n} Z(z)$ on $|z| = 1$ yields $h = - \frac n 2$.

The current algebra contains holomorphic currents $j^a$ with the OPE
\begin{equation}
  j^a(y) j^b(z) \sim \frac {k \delta^{a b}}{(y - z)^2} + \frac {f^{a b}_{\hphantom{a b} c}}{y - z}.
\end{equation}  These currents can be used to make Yang-Mills vertex operators
\begin{equation}
  V^a(z) = j^a \phi(Z(z)),
\end{equation} where $\phi(Z)$ is any function on $\mathbb{CP}^{3|4}$.  By taking
\begin{equation}
  \phi(Z(z)) = \delta\left(\frac {\lambda^2(z)}{\lambda^1(z)} - \frac {\lambda^2}{\lambda^1}\right) \delta^2\left(\frac {\mu^{\dot{\alpha}}(z)}{\lambda^1(z)} - \frac {\mu^{\dot{\alpha}}}{\lambda^1}\right) \delta^4\left(\frac {\psi^A(z)}{\lambda^1(z)} - \frac {\psi^A}{\lambda^1}\right),
\end{equation} the tree-level computation is identical to that of the connected prescription of Roiban, Spradlin and Volovich.  Note that the colour factors come from the current algebra and not from Chan-Paton factors.  This is very close in spirit to the original proposal of Nair (see ref.~\cite{Nair:1988bq}).

In ref.~\cite{Berkovits:2004jj} Witten and Berkovits studied the conformal supergravity sector present in the twistor string theories, both in the twistor string theory proposed by Witten~\cite{Witten:2003nn} and in the twistor string theory proposed by Berkovits~\cite{Berkovits:2004hg}.  See ref.\cite{Fradkin:1985am} for a review of conformal supergravity.

The analysis is simpler in Berkovits' twistor string theory.  In this theory, the conformal supergravity states are described by the vertex operators
\begin{equation}
  V_f = Y_I^{\overline{z}} f^I(Z), \quad V_g = g_I(Z) \partial Z^I,
\end{equation} where $f^I$ has $\Gl(1, \mathbb{C})$ charge $1$ and $g_I$ has $\Gl(1, \mathbb{C})$ charge zero.

The conditions that $V_f$ and $V_g$ be primary fields are
\begin{equation}
  \partial_I f^I = 0, \quad Z^I g_I = 0
\end{equation} and they have the gauge invariance
\begin{equation}
  \delta f^I = Z^I \Lambda, \quad \delta g_I = \partial_I \chi.
\end{equation}

It has been proposed that usual (i.e. non-conformal) supergravity theories arise from gaugings of Berkovits's twistor string in ref.~\cite{AbouZeid:2006wu}.  In this reference the $++-$ three-graviton amplitude has been computed and shown to agree with the three graviton amplitude in Einstein (not conformal) gravity.  More recently, however, Nair gave some arguments for the vanishing of the $--+$ amplitude in ref.~\cite{Nair:2007md}.  This means that the (super-)gravity theories constructed in ref.~\cite{AbouZeid:2006wu} are chiral versions of (super-)gravity theories.

There are several issues that are still not properly understood in this area:
\begin{itemize}
\item how to obtain non-chiral supergravity from twistor-string theories?
\item how to separate $\mathcal{N}=4$ loop amplitudes from the twistor-string loop amplitudes?  (At tree level the separation can be achieved by restricting to single trace amplitudes.)
\item what are the supplementary consistency constraints that limit the choice of gauge group?  (In a theory of conformal supergravity coupled to $\mathcal{N}=4$ super-Yang-Mills the $SU(4)_R$ $R$-symmetry group is gauged and this gauge symmetry is in general anomalous because the helicity $+\tfrac 1 2$ fermions transform as $\overline{\mathbf{4}}$ while the helicity $-\tfrac 1 2$ fermions transform as $\mathbf{4}$ of this group.  Anomaly cancellation imposes that the dimension of the $\mathcal{N}=4$ gauge group $G$ be equal to four, so $G = SU(2) \times U(1)$ or $G = U(1)^4$.)
\item what is the link between Witten and Berkovits twistor string theories?  The two do not seem to equivalent as there is a supplementary parameter, the level of the current algebra $k$, in Berkovits' twistor string theory.
\item is it possible to reproduce the one-loop twistor string scattering amplitudes computed by Dolan and Goddard in ref.~\cite{Dolan:2007vv} from a field theory computation?  What about infrared regularisation?
\end{itemize}


\chapter{Conclusion}

There are several problems that remain to be solved.
\begin{itemize}
\item So far, all the twistor strings contain conformal supergravity states.  Is it possible to formulate a twistor string theory that does not contain conformal supergravity?  A related question is if it is possible to compute Yang-Mills loop amplitudes by using the twistor string and somehow decouple the conformal supergravity contributions.  At tree level this can be achieved by restricting to single-trace amplitudes but at loop level this does not work.
\item Is it possible to formulate twistor-string theories for supergravity and for gauge and gravity theories with less than maximal or no supersymmetry?  The scattering amplitudes in these theories also have a relatively simple twistor space structure which suggests that such constructions might be possible.  If these generalisations exist, they will most likely be restricted to UV-finite theories, since the twistor string theories should be finite also.  First steps towards constructing twistor strings for theories with less than maximal or no supersymmetry were taken in refs.~\cite{Park:2004bw, Giombi:2004xv, Gao:2006mw}.
\item A complete understanding of the structure of IR divergences in the twistor string theories is lacking.  In Witten's initial paper it was observed that, in some cases, the IR divergences arise from transformation from twistor space to space-time while the twistor space amplitudes are IR finite.  Is this general?
\item As also remarked above, it is not clear how some constraints on the gauge group in $\mathcal{N}=4$ super-Yang-Mills coupled to conformal supergravity arise.
\item If a twistor string theory describing $\mathcal{N}=8$ supergravity exists, it might shed some light on the issue of UV finiteness of this theory.
\item In the light of the evidence for a scattering amplitudes--Wilson loop duality to be discussed in part~\ref{part:scattering}, it might be interesting to find the twistor space representation of Wilson loops and their geometrical interpretation.
\end{itemize}

\part{Scattering Amplitudes in SCFTs}
\label{part:scattering}

\chapter{Unitarity Method}
\label{ch:unitarity}

The implications of unitarity for field theory scattering amplitudes were found by Cutkosky in ref.~\cite{Cutkosky:1960sp}.  He formulated some rules, called Cutkosky rules, that compute the imaginary (absorptive) part of loop amplitudes by taking products of on-shell tree amplitudes.  In some cases, the full amplitude can be reconstructed by using dispersion relations.

By using the unitarity method~\cite{Bern:1994cg, Bern:1997sc}, one can compute the full amplitude, not only the absorptive parts, by considering the discontinuities across the branch cuts in different channels.

\begin{figure}
\centering
\label{fig:cut_example}
\beginpgfgraphicnamed{4}
\begin{tikzpicture}[baseline=0]
  \draw [propagator, ->] (-1,.7) -- ++(135:1)
        node [above left]{$k_2$};
  \draw (-1,.7) -- (1,.7);
        node[pos=0.5]{\tikz\draw[->] (-1pt,0pt) -- (0pt,0pt);}
        node[pos=0.5, above]{$l_1$};
  \draw [propagator, ->] (1,.7) -- ++(45:1)
        node [above right]{$k_3$};

  \draw [propagator, ->] (-1,-0.7) -- +(-135:1)
        node [below left]{$k_1$};
  \draw (-1,-0.7) -- (1,-0.7)
        node[pos=0.5]{\tikz\draw[->] (-1pt,0pt) -- (0pt,0pt);}
        node[pos=0.5, below]{$l_2$};
  \draw [propagator, ->] (1,-0.7) -- ++(-45:1)
        node [below right]{$k_4$};

  \draw [cut] (-.5,-1) -- (.5,1);
  \draw [cut'] (-.5,-1) -- (.5,1);
  \filldraw [blob] (-1,0) ellipse (.3 and 1); 
  \filldraw [blob] (1,0) ellipse (.3 and 1); 
  \draw (0,-1.4) node[below]{($a$)};
\end{tikzpicture}
\endpgfgraphicnamed \hspace{5em}%
\beginpgfgraphicnamed{5}
\begin{tikzpicture}[baseline=0]
  \draw [propagator, ->] (-.7,1) -- ++(135:1)
        node [above left]{$k_2$};
  \draw (-.7,-1) -- (-.7,1)
        node[pos=0.5, sloped]{\tikz\draw[<-] (-1pt,0pt) -- (0pt,0pt);}
        node[pos=0.5, left]{$l_1$};
  \draw [propagator, ->] (-.7,-1) -- ++(-135:1)
        node [below left]{$k_1$};

  \draw [propagator, ->] (.7,1) -- +(45:1)
        node [above right]{$k_3$};
  \draw (.7,1) -- (.7,-1)
        node[pos=0.5, sloped]{\tikz\draw[->] (-1pt,0pt) -- (0pt,0pt);}
        node[pos=0.5, right]{$l_2$};
  \draw [propagator, ->] (.7,-1) -- ++(-45:1)
        node [below right]{$k_4$};

  \draw [cut] (-1,-.5) -- (1,.5);
  \draw [cut'] (-1,-.5) -- (1,.5);
  \filldraw [blob] (0,-1) ellipse (1 and .3); 
  \filldraw [blob] (0,1) ellipse (1 and .3); 
  \draw (0,-1.8) node[below]{($b$)};
\end{tikzpicture}
\endpgfgraphicnamed
\caption{The $s$ and $t$ cuts for a four-point, one-loop amplitude.}
\end{figure}
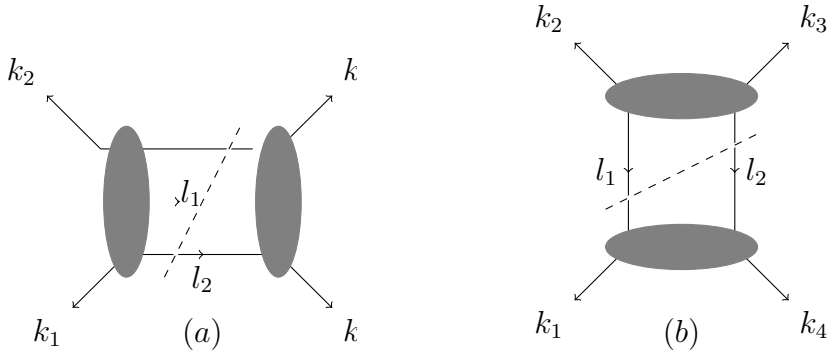

Take the example of a one-loop four gluon amplitude $A(1^-, 2^-, 3^+, 4^+)$ (it is easy to see that the colour algebra factorises so we will work with partial amplitudes).  Then, the discontinuity across the branch cut in the kinematical invariant $s$ is given by (see Appendix~\ref{ch:cuts_and_disc})
\begin{multline}
  \left. A^{\text{1-loop}}(1^-, 2^-, 3^+, 4^+)\right|_{s-\text{cut}} =  \int \frac {d^4 l_1}{(2 \pi)^4} 2 \pi \delta^{(+)}(l_1^2) 2 \pi \delta^{(+)}(l_2^2) \\A^{\text{tree}}(1^-, 2^-, l_1^+, l_2^+) A^{\text{tree}}(-l_2^-, -l_1^-, 3^+, 4^+),
\end{multline} where $\delta^{(+)}(p^2) = \theta(p_0) \delta(p^2)$ and we use a clockwise ordering for the legs of the tree amplitudes, as necessary for the colour ordered amplitudes.  Note that the loop integral is restricted to the phase space of the two exchanged on-shell particles.

However, if we replace the $\delta^{(+)}$ functions with propagators, the $s$-cut is still the same
\begin{multline}
  \label{eq:cut_equality}
  \left. A^{\text{1-loop}}(1^-, 2^-, 3^+, 4^+)\right|_{s-\text{cut}} = \\ \int \frac {d^4 l_1}{(2 \pi)^4} \frac i {l_1^2} A^{\text{tree}}(1^-, 2^-, l_1^+, l_2^+) \left.\frac i {l_2^2} A^{\text{tree}}(-l_2^-, -l_1^-, 3^+, 4^+)\right|_{s-\text{cut}}.
\end{multline}

Several observations are in order.
\begin{itemize}
\item Initially, the expressions for the tree-level amplitudes in equation eq.~\eqref{eq:cut_equality} are defined only when $l_1$ and $l_2$ are on-shell.  The integrals in eq.~\eqref{eq:cut_equality} are not restricted to the phase space of the two exchanged on-shell particles anymore, so we need a way to extend the integrand off-shell.  In order to extend the integrand off-shell, we need to pass from spinor to vector language (we will show how to do that in an example below).
\item The integral in the right-hand side of eq.~\eqref{eq:cut_equality} has the same $s$-cut as the one-loop amplitude, but its $t$-cut might be different.
\item The amplitude might be a sum of contributions some of which do not have any branch cuts.  These terms will not be visible in any of the cuts.
\end{itemize}

We will give an example computation for the MHV amplitude considered above (see ref.~\cite{Bern:1996je} for more details).  The tree-level amplitudes are
\begin{align}
  A^{\text{tree}}(1^-, 2^-, l_1^+, l_2^+) & = i \frac {\langle 1\ 2\rangle^3}{\langle 2\ l_1\rangle \langle l_1\ l_2\rangle \langle l_2\  1\rangle},\\
  A^{\text{tree}}(-l_2^-, -l_1^-, 3^+, 4^+) & = i \frac {\langle (-l_2)\  (-l_1)\rangle^3}{\langle (-l_1)\ 3\rangle \langle 3\ 4\rangle \langle 4\ (-l_2) \rangle}.
\end{align}

When computing MHV amplitudes one can pull out a factor of the tree-level MHV amplitude in front of the integral
\begin{multline}
  \left. A^{\text{1-loop}}(1^-, 2^-, 3^+, 4^+)\right|_{s-\text{cut}} = \\ -i A^{\text{tree}} \left.\int \frac {d^4 l_1}{(2 \pi)^4} \frac i {l_1^2} \frac i {l_2^2} \frac {\langle l_1\ l_2\rangle^2 \langle 2\ 3\rangle \langle 4\ 1\rangle} {\langle 2\ l_1\rangle \langle l_2\ 1\rangle \langle l_1\ 3\rangle \langle 4\ l_2\rangle}\right|_{s-\text{cut}},
\end{multline} where we have cancelled some phases arising from spinors $|(-l_1)\rangle$ and $|(-l_2)\rangle$.

Now, still using the on-shell conditions, we can `rationalise' the denominators
\begin{equation}
  \frac 1 {\langle 2\ l_1\rangle} = \frac {[l_1\ 2]}{2 k_2 \cdot l_1} = \frac {[l_1\ 2]}{(k_2 + l_1)^2},
\end{equation} and the same for all the other denominators.

After this `rationalisation' the denominators look like ordinary Feynman propagators but some of them are squared.  Omitting the cut propagators, the integrand is
\begin{equation}
  - \frac {[l_2\ 4] \langle 4\ 1\rangle [1\ l_2] \langle l_2\ l_1\rangle [l_1\ 2] \langle 2\ 3\rangle [3\ l_1] \langle l_1\ l_2\rangle}{\left[(k_2 + l_1)^2\right]^2 \left[(k_3 - l_1)^2\right]^2}.
\end{equation}  The numerator of this equation can be written as a trace
\begin{equation}
  [l_2\ 4] \langle 4\ 1\rangle [1\ l_2] \langle l_2\ l_1\rangle [l_1\ 2] \langle 2\ 3\rangle [3\ l_1] \langle l_1\ l_2\rangle = \tr_+(l_2 4 1 l_2 l_1 2 3 l_1),
\end{equation} where
\begin{equation}
  \tr_+(a b \cdots) = \frac 1 2 \tr\left((1 + \gamma^5) \slashed{a} \slashed{b} \cdots\right).
\end{equation}

Let us make some comments on the computation of these traces.  When computing these traces one can find odd terms.  For example
\begin{equation}
  \tr_+(a b c d) = 2 (a \cdot d) (b \cdot c) - 2 (a \cdot c) (b \cdot d) + 2 (a \cdot b) (c \cdot d) - 2 i \epsilon(a, b, c, d),
\end{equation} where $\epsilon(a, b, c, d) = \epsilon_{\mu \nu \rho \sigma} a^\mu b^\nu c^\rho d^\sigma$.  By using momentum conservation we see that, in the four-point case, the only independent odd terms we can write are $\epsilon(k_1, k_2, k_3, k_4)$ and $\epsilon(k_i, k_j, k_l, l_1)$.  The first is zero by momentum conservation and the second is zero after integrating over $l_1$.  The idea is that an integral with an $l_1^\mu$ numerator can be decomposed over a basis of external momenta.  By using the fact that the $\epsilon$ with two identical arguments vanishes and that $\epsilon(k_1, k_2, k_3, k_4) = 0$ the result follows.  So in the four-point case we can neglect the odd parts as they do not contribute in the final result.\footnote{In some cases it is profitable to form multiple traces in the numerator.  Then, the product of two odd parts can contribute to the even part.}  Starting at five points, one has non-vanishing odd parts.

The trace can be computed (still using the on-shell condition for the cut propagators) and the final result is:
\begin{multline}
  \left. A^{\text{1-loop}}(1^-, 2^-, 3^+, 4^+)\right|_{s-\text{cut}} = i s t A^{\text{tree}}(1^-, 2^-, 3^+, 4^+)\\ =\left.\int \frac {d^4 l_1}{(2 \pi)^4} \frac 1 {l_1^2 (k_1 + k_2 + l_1)^2 (k_2 + l_1)^2 (k_3 - l_1)^2}\right|_{s-\text{cut}} = \\
i s t A^{\text{tree}}(1^-, 2^-, 3^+, 4^+) \left(
\beginpgfgraphicnamed{6}
\begin{tikzpicture}[baseline=0]
  \def\r{1cm};
  \def\l{3mm};
  \draw (-\r/2,-\r/2) rectangle (\r/2, \r/2)
  (-\r/2,-\r/2) -- +(-135:\l)
  (-\r/2,\r/2) -- +(135:\l)
  (\r/2, \r/2) -- +(45:\l)
  (\r/2, -\r/2) -- +(-45:\l);
\end{tikzpicture}
\endpgfgraphicnamed\right)_{s-\text{cut}}.
\end{multline}  Note that the result for the trace reduced the propagators squared to ordinary propagators.

This integral is actually infrared divergent so some kind of regularisation has to be used.  One can try to regularise the final answer but it is not obvious this is the right thing to do.  On the other hand, if we use a regulator like dimensional regularisation in the initial stages of the computation, we will not be able to use the powerful spinor techniques we used above.\footnote{These spinor techniques are restricted to four dimensions.}  We will come back to these questions below, where we will see how to turn this apparent disadvantage into an advantage.

Let us now compute the cut in the $t$-channel.  The result is the same as the one for the $s$-cut, but the direct computation is more difficult since one has to sum over all the states of the $\mathcal{N}=4$ super-multiplet that can be exchanged.  In the case of the $s$-cut the only states that could be exchanged were gluons with helicities plus, as all the other amplitudes vanish by the supersymmetry Ward identities.

Fortunately, there is a shortcut.  In a theory with $\mathcal{N}=4$ supersymmetry, the supersymmetry Ward identities imply that for MHV amplitudes, the ratio of the one-loop and the tree amplitudes does not depend on the position of helicity minus gluons.  In our case,
\begin{equation}
  \label{eq:loop-tree_ratio}
  \frac {A^{\text{1-loop}}(1^-, 2^-, 3^+, 4^+)}{A^{\text{tree}}(1^-, 2^-, 3^+, 4^+)} = \frac {A^{\text{1-loop}}(1^+, 2^-, 3^-, 4^+)}{A^{\text{tree}}(1^+, 2^-, 3^-, 4^+)} = \frac {A^{\text{1-loop}}(2^-, 3^-, 4^+, 1^+)}{A^{\text{tree}}(2^-, 3^-, 4^+, 1^+)},
\end{equation} where we have also used the cyclic symmetry.  But this is the same computation up to relabelling of external lines and the exchange $s \leftrightarrow t$.

In conclusion, we have that the $s$ and $t$ cuts of the box integral coincide (up to some factors) with the $s$ and $t$ cuts of the one-loop amplitude.  So we have computed the one-loop four-point amplitude up to possible additive rational (cut-free) contributions.

Once we have computed the $--++$ amplitude we can find the result for the $-+-+$ amplitude without any further computation, by just using the result in eq.~\eqref{eq:loop-tree_ratio}.  We see here that big simplifications come from using the supersymmetry Ward identities.

Let us now return to the issues of regularisation and of rational terms.  It turns out that both these problems can be solved simultaneously.

Let us focus on the case of an amplitude $A$ in a massless theory regularised by dimensional regularisation.  Suppose that the coupling constant in the regularised theory is $g \mu^\epsilon$, where $\mu$ is the dimensional regularisation scale.   So in dimensional regularisation the coupling constant is dimensionful.  The amplitude is a dimensionless quantity which means that the general form of the amplitude at order $2 k$ in the coupling constant $g$ is
\begin{equation}
  A = g^{2 k} \sum_{K_i} \left(- \frac {K_i} {\mu^2}\right)^{- k \epsilon} f_i,
\end{equation} where $K_i$ are kinematic invariants and the $f_i$ are dimensionless functions of kinematic invariants and possibly $\epsilon$.  The point is that, when expanding around $\epsilon = 0$, these terms produce branch cuts at order $\epsilon$
\begin{equation}
  A = g^{2 k} \sum_{K_i} \left(1 - k \epsilon \ln \left(- \frac {K_i} {\mu^2}\right)\right) f_i.
\end{equation}

So, even if the quantities $f_i$ do not have branch cuts, they can be found when computing the cuts to one higher order in $\epsilon$ than the four-dimensional cuts.\footnote{If the contribution at this higher order in $\epsilon$ has a finite, non-zero limit when $\epsilon \rightarrow 0$, then it contributes to the rational part of the amplitude.}  So we have solved both of our problems, that of regularisation and that of missing rational parts.  In the case of $\mathcal{N}=4$ theory in four dimensions, the computations can be done by using the dimensional reduction variant of dimensional regularisation, which is compatible with supersymmetry, and doing all the manipulations in the $\mathcal{N}=1$ super Yang-Mills theory in ten dimensions (see ref.~\cite{Brink:1976bc} for the original construction of $\mathcal{N}=4$ theory by dimensional reduction of the $\mathcal{N}=1$ super Yang-Mills theory in ten dimensions).

How about the computation we presented above, where we computed the cuts in four dimensions?  Is the box integral the full answer, or there are further rational contributions?  It turns out that the result obtained there is the complete result, (the one-loop result for the four-point amplitude in $\mathcal{N}=4$ theory was first obtained in ref.~\cite{Green:1982sw} by taking the zero slope limit of string theory) by using a power counting theorem proved in ref.~\cite{Bern:1994cg}.  This theorem is very effective at one loop in supersymmetric theories.  At more than one loop, no general argument is known.

A very powerful technique, which is very useful in higher-loop computations or for amplitudes with large numbers of external legs, is generalised unitarity (see ref.~\cite{Bern:1997sc}).  In the generalised unitarity method, one uses several unitarity cuts simultaneously, thus isolating a smaller set of integrals.  In ref.~\cite{Britto:2004nc}, Britto, Cachazo and Feng used the generalised unitarity method together with complex on-shell momenta.  They were able to reproduce the results for the one-loop amplitudes in $\mathcal{N}=4$ super-Yang-Mills by cutting four internal lines.

After this general discussion let us discuss a two loop example: a four-point two loop amplitude in $\mathcal{N}=4$ theory at planar level\footnote{This means that we only keep single trace contributions.} (this was first studied in ref.~\cite{Bern:1997nh}).  There are several types of cuts one can consider: three-particle cuts and double two-particle cuts in both $s$ and $t$ channels.

\begin{figure}
\centering
\label{fig:2loop_cuts_example}
\beginpgfgraphicnamed{7}
\begin{tikzpicture}[baseline=0]
  \draw [propagator, ->] (-1,.7) -- ++(135:1)
        node [above left]{$k_2$};
  \draw (-1,0.7) -- (1,0.7)
        node[pos=0.3]{\tikz\draw[->] (-1pt,0pt) -- (0pt,0pt);}
        node[pos=0.3, above]{$l_1$};
  \draw [propagator, ->] (1,0.7) -- ++(45:1)
        node [above right]{$k_3$};

  \draw (-1,0) -- (1,0)
        node[pos=0.3]{\tikz\draw[->] (-1pt,0pt) -- (0pt,0pt);}
        node[pos=0.3, above]{$l_2$};

  \draw [propagator, ->] (-1,-0.7) -- +(-135:1)
        node [below left]{$k_1$};
  \draw (-1,-0.7) -- (1,-0.7)
        node[pos=0.5]{\tikz\draw[->] (-1pt,0pt) -- (0pt,0pt);}
        node[pos=0.5, below]{$l_3$};
  \draw [propagator, ->] (1,-0.7) -- ++(-45:1)
        node [below right]{$k_4$};

  \draw [cut] (-.5,-1) -- (.5,1);
  \draw [cut'] (-.5,-1) -- (.5,1);
  \filldraw [blob] (-1,0) ellipse (.3 and 1); 
  \filldraw [blob] (1,0) ellipse (.3 and 1); 
  \draw (0,-1.4) node[below]{($a$)};
\end{tikzpicture}
\endpgfgraphicnamed \hspace{5em}%
\beginpgfgraphicnamed{8}
\begin{tikzpicture}[baseline=0]
  \draw [propagator, ->] (-2, .7) -- ++(135:1)
        node [above left]{$k_2$};
  \draw [propagator, ->] (-2, -.7) -- ++(-135:1)
        node [below left]{$k_1$};
  \draw [propagator, ->] (2, .7) -- ++(45:1)
        node [above right]{$k_3$};
  \draw [propagator, ->] (2, -.7) -- ++(-45:1)
        node [below right]{$k_4$};

  \draw [propagator] (-2, .7) -- (2, .7)
     node[pos=.25]{\tikz\draw[->] (-1pt,0pt) -- (0pt,0pt);}
     node[pos=.25,above]{$l_1$}
     node[pos=.75]{\tikz\draw[<-] (-1pt,0pt) -- (0pt,0pt);}
     node[pos=.75,above]{$l_4$};

  \draw [propagator] (-2, -.7) -- (2, -.7)
     node[pos=.25]{\tikz\draw[->] (-1pt,0pt) -- (0pt,0pt);}
     node[pos=.25,below]{$l_2$}
     node[pos=.75]{\tikz\draw[<-] (-1pt,0pt) -- (0pt,0pt);}
     node[pos=.75,below]{$l_3$};

  \draw [cut] (-1.5,-1) -- (-.5,1);
  \draw [cut'] (-1.5,-1) -- (-.5,1);
  \draw [cut] (.5,-1) -- (1.5,1);
  \draw [cut'] (.5,-1) -- (1.5,1);

  \filldraw [blob] (-2,0)ellipse (.3 and 1); 
  \filldraw [blob] (0,0) ellipse (.3 and 1); 
  \filldraw [blob] (2,0) ellipse (.3 and 1); 
  \draw (0,-1.8) node[below]{($b$)};
\end{tikzpicture}
\endpgfgraphicnamed
\caption{The three particle $s$-cut $(a)$ and the double two-particle $s$-cut for a four-point, two-loop amplitude $(b)$.}
\end{figure}
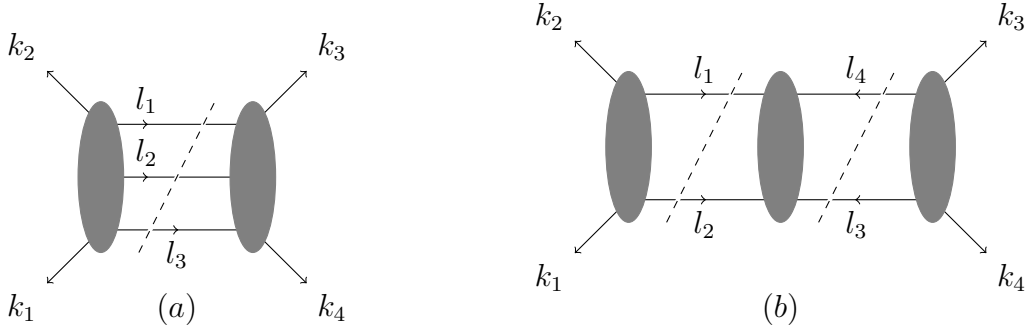

It is very easy to compute the iterated double particle cut because one can reuse the results for the one-loop amplitude.  The two particle cut on the left yields a cut of the box integral times the tree amplitude and this tree amplitude is sewn onto the right tree amplitude and again yields a box amplitude.  The final result for the ratio of two-loop four point amplitude to the tree amplitude is:
\begin{equation}
  \label{eq:N=4_2loop_result}
  s^2 t %
\beginpgfgraphicnamed{9}
\begin{tikzpicture}[baseline=-1ex]
  \def\r{1cm};
  \def\l{3mm};
  \draw (-\r,-\r/2) rectangle (\r, \r/2)
  (0,\r/2) -- (0,-\r/2)
  (-\r,-\r/2) -- +(-135:\l)
  (-\r,\r/2) -- +(135:\l)
  (\r, \r/2) -- +(45:\l)
  (\r, -\r/2) -- +(-45:\l);
\end{tikzpicture}
\endpgfgraphicnamed + s t^2
\beginpgfgraphicnamed{10}
\begin{tikzpicture}[baseline=-1ex]
  \def\r{1cm};
  \def\l{3mm};
  \draw (-\r/2,-\r) rectangle (\r/2, \r)
  (-\r/2,0) -- (\r/2,0)
  (-\r/2,-\r) -- +(-135:\l)
  (-\r/2,\r) -- +(135:\l)
  (\r/2, \r) -- +(45:\l)
  (\r/2, -\r) -- +(-45:\l);
\end{tikzpicture}
\endpgfgraphicnamed.
\end{equation}

Let us emphasize at this point that the diagrams we drawn above are a graphical representation of the final result as scalar integrals and should not be thought as Feynman diagrams.  The number of Feynman diagrams contributing to this two-loop amplitude is much larger than two!

The three particle cut is a bit more difficult to compute but it agrees with the double two-particle cut result.  (This is a general feature of the unitarity method: one given integral can be detected in several cuts and the coefficients obtained in the two cases have to be identical.  This provides a check on the computation.  Also, the coefficients are subjected to different symmetry properties.  For example, in the two-loop computation presented above the horizontal and vertical double boxes must have the same coefficient when replacing $s \leftrightarrow t$.)

Another feature worth emphasis is the fact that the same integral can appear several times in a cut.  For example, the horizontal double box appears twice in the $s$-channel three-particle cut (see fig.~\ref{fig:horizontal_double_box_s-cut}).\footnote{If an integral appears twice in an given cut, it must do so with identical coefficients.}  In the final result the integral appears only once.

\begin{figure}
  \centering
\beginpgfgraphicnamed{11}
\begin{tikzpicture}[baseline=0]
  \draw (-1,-1/2) rectangle (1, 1/2)
  (0,1/2) -- (0,-1/2)
  (-1,-1/2) -- +(-135:.3)
  (-1,1/2) -- +(135:.3)
  (1, 1/2) -- +(45:.3)
  (1, -1/2) -- +(-45:.3);
  \draw [cut] (-.3,-.8) -- (.3,.8);
  \draw [cut'] (-.3,-.8) -- (.3,.8);
\end{tikzpicture}
\endpgfgraphicnamed \qquad
\beginpgfgraphicnamed{12}
\begin{tikzpicture}[baseline=0]
  \draw (-1,-1/2) rectangle (1, 1/2)
  (0,1/2) -- (0,-1/2)
  (-1,-1/2) -- +(-135:.3)
  (-1,1/2) -- +(135:.3)
  (1, 1/2) -- +(45:.3)
  (1, -1/2) -- +(-45:.3);
  \draw [cut] (.3,-.8) -- (-.3,.8);
  \draw [cut'] (.3,-.8) -- (-.3,.8);
\end{tikzpicture}
\endpgfgraphicnamed
  \caption{The horizontal double box is detected twice by the $s$-channel three-particle cut.}
  \label{fig:horizontal_double_box_s-cut}
\end{figure}

Before ending this chapter, let us make some remarks on the fermionic signs that some readers might be worrying about.  There fermionic signs arise in two places: a minus sign for each fermionic loop and a relative sign when adding some Feynman diagrams.  It turns out that these two contributions conspire to cancel.

In the double two-particle cut of a four-point two loop amplitude, the middle four-fermion tree amplitude receives contributions from two kinds of diagrams (we only present the contribution with gluon exchange; the contributions with scalar exchange can be analysed similarly)
\begin{equation}
  \beginpgfgraphicnamed{13}
  \begin{tikzpicture}[baseline=-1ex]
     \draw (-.5, -.5) -- (.5, -.5)
           (-.5, .5) -- (.5, .5);
     \filldraw [blob] (0,0) ellipse (.3 and .7);
  \end{tikzpicture}
  \endpgfgraphicnamed \qquad \ni \quad %
  \beginpgfgraphicnamed{14}
  \begin{tikzpicture}[baseline=-1ex]
     \draw [gluon, decorate] (-.5,0) -- (.5,0);
     \draw (-1,-.5) arc (-90:90:.5);
     \draw (1, .5) arc (90:270:.5);
  \end{tikzpicture}
  \endpgfgraphicnamed \quad - \quad%
  \beginpgfgraphicnamed{15}
  \begin{tikzpicture}[baseline=-1ex]
     \draw [gluon, decorate] (0,-.5) -- (0,.5);
     \draw (-.5, -.5) -- (.5, -.5)
           (-.5, .5) -- (.5, .5);
  \end{tikzpicture}
  \endpgfgraphicnamed.
\end{equation}  Note the relative minus between the two contributions.  When the left and the right fermion lines are sewn onto the middle tree, there has to be a sign difference between the two contributions because one has two fermion loops and the other only has one fermion loop.

\begin{equation}
  \beginpgfgraphicnamed{16}
  \begin{tikzpicture}[baseline=-1ex]
     \draw (-1.1,-.5) arc (270:90:.5);
     \draw [cut] (-1,-.6) -- (-1,.6);
     \draw [cut'] (-1,-.6) -- (-1,.6);     
     \draw (1.1, .5) arc (90:-90:.5);
     \draw [gluon, decorate] (-.5,0) -- (.5,0);
     \draw (-1,-.5) arc (-90:90:.5);
     \draw (1, .5) arc (90:270:.5);
     \draw [cut] (1,-.6) -- (1,.6);
     \draw [cut'] (1,-.6) -- (1,.6);     
  \end{tikzpicture}
  \endpgfgraphicnamed \quad - \quad %
  \beginpgfgraphicnamed{17}
  \begin{tikzpicture}[baseline=-1ex]
     \draw (-.6,-.5) arc (270:90:.5);
     \draw [cut] (-.6,-.6) -- (-.6,.6);
     \draw [cut'] (-.6,-.6) -- (-.6,.6);     
     \draw (.6, .5) arc (90:-90:.5);
     \draw [cut] (.6,-.6) -- (.6,.6);
     \draw [cut'] (.6,-.6) -- (.6,.6);     
     \draw [gluon, decorate] (0,-.5) -- (0,.5);
     \draw (-.5, -.5) -- (.5, -.5)
           (-.5, .5) -- (.5, .5);
  \end{tikzpicture}
  \endpgfgraphicnamed
\end{equation}

This sign difference is provided by the relative minus between the two contributions to the four-fermion tree amplitude.  Note that when considering the relative minus signs one has to compare amplitudes that have the same colour structure.  For example, there also is a relative sign between the following diagrams
\begin{equation}
  \beginpgfgraphicnamed{18}
  \begin{tikzpicture}[baseline=-1ex]
     \draw [gluon, decorate] (0,-.5) -- (0,.5);
     \draw (-.5, -.5) -- (.5, -.5)
     (-.5, .5) -- (.5, .5);
  \end{tikzpicture}
  \endpgfgraphicnamed \qquad - \qquad
  \beginpgfgraphicnamed{19}
  \begin{tikzpicture}[baseline=-1ex]
     \draw [gluon, decorate] (0,-.5) -- (0,.5);
     \draw (-.5, -.5) -- (0, -.5) -- (.7, .5)
           (-.5, .5) -- (0, .5);
           \draw [cut] (0, .5) -- (.7,-.5);
           \draw (0, .5) -- (.7,-.5);
  \end{tikzpicture}
  \endpgfgraphicnamed,
\end{equation} but they have a different colour structure.

Therefore, there is no need to separately keep track of the fermion signs when computing the cuts.

\chapter{\texorpdfstring{One-loop MHV amplitudes in $\mathcal{N}=4$}{One-loop MHV amplitudes in N=4}}

The one-loop planar MHV amplitudes in $\mathcal{N}=4$ where computed for \emph{arbitrary} number of external legs in ref.~\cite{Bern:1994zx}.  Their twistor space structure was analysed in refs.~\cite{Witten:2003nn, Cachazo:2004zb, Cachazo:2004by}.  These one-loop amplitudes will play a central role in the formulation of the ABDK iteration relation (proposed by Anastasiou, Bern, Dixon and Kosower in ref.~\cite{Anastasiou:2003kj}) and of the all orders BDS ansatz (proposed by Bern, Dixon and Smirnov in ref.~\cite{Bern:2005iz}).

The leading colour (single trace) $n$-point $l$-loop amplitude can be written
\begin{equation}
  \mathcal{A}_n^{(l)} = g^{n-2} \left[(4 \pi e^{-\gamma})^\epsilon \frac \lambda {8 \pi^2}\right]^l \sum_{\rho \in \mathcal{S}/\mathbb{Z}_n} \tr (T^{a_{\rho(1)}} \cdots T^{a_{\rho(n)}}) A_n^{(l)}(\rho(1), \ldots \rho(n)),
\end{equation} where $\lambda = g^2 N_c$ is the 't Hooft coupling and the sum is over non-cyclic permutations of external legs.

For MHV amplitudes it is convenient to present the results for the ratio,
\begin{equation}
  M_n^{(l)} = \frac {A_n^{(l)}}{A_n^{\text{tree}}},
\end{equation} where $A_n^{(l)}$ and $A_n^{\text{tree}}$ are colour ordered amplitudes with the same distribution of negative helicity gluons.  The ratio $M_n^{(l)}$ does not depend of the position of negative helicity gluons.

It is also convenient to subtract the divergent part, i.e. the part that diverges in the limit $\epsilon \rightarrow 0$, when the infrared regulator is removed.

We can split the $n$-point one-loop amplitude divided by the $n$-point tree amplitude into a divergent and a finite part
\begin{equation}
  M_n^{(1)}(\epsilon) = \text{div}_n^{(1)}(\epsilon) + F_n^{(1)}(\epsilon),
\end{equation} where
\begin{equation}
  F_n^{(1)}(\epsilon) \xrightarrow{\epsilon \rightarrow 0} F_n^{(1)}(0)
\end{equation} and $F_n^{(1)}(0)$ is finite.

The divergent part is universal and is given by
\begin{equation}
  \text{div}_n^{(1)}(\epsilon) = -\frac 1 2 \frac 1 {\epsilon^2} \sum_{i=1}^n \left(\frac {\mu^2}{- s_{i, i+1}}\right)^\epsilon,
\end{equation} where $\mu$ is the scale introduced by the dimensional regularisation and $s_{i,i+1} = (k_i + k_{i+1})^2$.  Here and below all the summation over the indices of external particles is understood modulo $n$, i.e. $k_{n+1} \equiv k_1$.

The finite terms have the form
\begin{equation}
F_n^{(1)}(0) = \frac 1 2 \sum_{i=1}^n g_{n,i},
\end{equation} where
\begin{equation}
g_{n,i} =
-\sum_{r=2}^{\lfloor n/2 \rfloor -1} \ln \left(\frac {-s_{i\cdots(i+r-1)}} {- s_{i\cdots(i+r)}}\right) \ln \left(\frac { -s_{(i+1)\cdots(i+r)}} {-s_{i\cdots(i+r)}}\right) +
D_{n,i} + L_{n,i} + \frac 3 2 \zeta_2,
\end{equation} in which $\lfloor n \rfloor$ is the greatest integer less than or equal to $n$ and $s_{i \cdots j} = (k_i + \cdots + k_j)$ when $j > i$.

The form of $D_{n, i}$ and $L_{n, i}$ depends on $n$.  If $n = 2 m$ and $m > 2$
\begin{align*}
D_{2m,i} & = -\sum_{r=2}^{m-2}
\Li_2 \left( 1 - \frac {s_{i\cdots(i+r-1)} s_{(i-1)\cdots(i+r)}} {s_{i\cdots(i+r)} s_{(i-1)\cdots(i+r-1)}} \right)
- \frac 1 2 \Li_2 \left( 1- \frac {s_{i\cdots(i+m-2)} s_{(i-1)\cdots(i+m-1)}} {s_{i\cdots(i+m-1)} s_{(i-1)\cdots(i+m-2)}} \right)\\
L_{2m,i} & = \frac 1 4 \ln^2 \left(\frac {-s_{i\cdots(i+m-1)}} {-s_{(i+1)\cdots(i+m)}} \right).
\end{align*}

If $n = 2 m + 1$
\begin{align*}
D_{2m+1,i} & = - \sum_{r=2}^{m-1} \Li_2 \left( 1 - \frac {s_{i\cdots(i+r-1)} s_{(i-1)\cdots(i+r)}}{s_{i\cdots(i+r)} s_{(i-1)\cdots(i+r-1)}} \right),\\
L_{2m+1,i} & = - \frac 1 2 \ln \left( \frac {-s_{i\cdots(i+m-1)}} {-s_{i\cdots(i+m)}} \right) \ln \left(\frac {-s_{(i+1)\cdots(i+m)}} {-s_{(i-1)\cdots(i+m-1)}} \right).
\end{align*}

The case $n = 4$ is special and the result in this case is
\begin{equation}
  F_4^{(1)}(0) = \frac 1 2 \ln^2 \left( \frac s t \right) + 4 \zeta_2.
\end{equation}

\chapter{ABDK/BDS ansatz}
\label{ch:abdk-bds}

Scattering amplitudes in gauge theories have a known structure of infrared singularities that constrains their form.  Indeed, the infrared divergences of the scattering amplitudes have to cancel in physical observables.  They cancel among the real and virtual contributions in jet observables, for example.  Also, the infrared-divergent part should cancel against parton distribution functions when computing scattering of colourless states.  Because of this, one should expect some universal structure that allows these cancellations to take place.

Based on an iteration relation for the one- and two-loop splitting functions, Anastasiou, Bern, Dixon and Kosower (ABDK) proposed (see ref.~\cite{Anastasiou:2003kj}) an ansatz for the iteration of all MHV amplitudes to two-loop order.  Below we review the collinear factorisation that was an essential ingredient in ABDK proposal and then we review the ABDK proposal itself.  We end with a review of the all-order BDS (Bern, Dixon and Smirnov) ansatz for the MHV amplitudes.

A colour ordered $L$-loop amplitude has an universal factorisation property when two adjacent external legs become collinear.  This is called collinear factorisation.
\begin{equation}
  \label{eq:collinear_factorisation}
  A_n^{(L)} (\ldots, a^{h_a}, b^{h_b}, \ldots) \xrightarrow{k_a \parallel k_b} \sum_{l=0}^L \sum_{h=\pm} \text{Split}_{-h}^{(l)}(z, a^{h_a}, b^{h_b}) A_{n-1}^{(L-l)}(\ldots, P^h, \ldots),
\end{equation} where $P = k_a + k_b$ and $k_a \rightarrow z P$, $k_b \rightarrow (1 - z) P$.  This is an extension of the tree-level collinear factorisation
  discussed in Appendix~\ref{ch:ir_divergencies} to loop level.

The collinear factorisation can be represented graphically in the following way (this is restricted to two loops, but at higher loops it works similarly).
\begin{multline}
\beginpgfgraphicnamed{20}
\begin{tikzpicture}[scale=1/4,baseline=0]
  \draw (0,0) -- (-135:4) node[left]{\ensuremath{a}}
  (0,0) -- (135:4) node[left]{\ensuremath{b}}
  (0,0) -- (45:4)
  (0,0) -- (-45:4);
  \fill[black] (0,0) ++ (30:3) circle (2pt)
  (0,0) ++ (0:3) circle (2pt)
  (0,0) ++ (-30:3) circle (2pt);
  \filldraw[fill=gray, draw=black] (0,0) circle (2);
  \filldraw[fill=white, draw=black] (-1,0) ellipse (1/2 and 1);
  \filldraw[fill=white, draw=black] (1,0) ellipse (1/2 and 1);
\end{tikzpicture}
\endpgfgraphicnamed \quad \xrightarrow{k_a \parallel k_b} \quad
\beginpgfgraphicnamed{21}
\begin{tikzpicture}[scale=1/4,baseline=0]
  \draw (0,0) -- (-135:4) node[left]{\ensuremath{a}}
  (0,0) -- (135:4) node[left]{\ensuremath{b}}
  (0,0) -- (4,0);
  \filldraw[fill=gray, draw=black] (0,0) circle (2);
\end{tikzpicture}
\endpgfgraphicnamed \times %
\beginpgfgraphicnamed{22}
\begin{tikzpicture}[scale=1/4,baseline=0]
  \draw (0,0) -- (-4,0)
  (0,0) -- (45:4)
  (0,0) -- (-45:4);
  \fill[black] (0,0) ++ (30:3) circle (2pt)
  (0,0) ++ (0:3) circle (2pt)
  (0,0) ++ (-30:3) circle (2pt);
  \filldraw[fill=gray, draw=black] (0,0) circle (2);
  \filldraw[fill=white, draw=black] (-1,0) ellipse (1/2 and 1);
  \filldraw[fill=white, draw=black] (1,0) ellipse (1/2 and 1);
\end{tikzpicture}
\endpgfgraphicnamed \quad + \quad %
\beginpgfgraphicnamed{23}
\begin{tikzpicture}[scale=1/4,baseline=0]
  \draw (0,0) -- (-135:4) node[left]{\ensuremath{a}}
  (0,0) -- (135:4) node[left]{\ensuremath{b}}
  (0,0) -- (4,0);
  \filldraw[fill=gray, draw=black] (0,0) circle (2);
  \filldraw[fill=white, draw=black] (0,0) circle (1);
\end{tikzpicture}
\endpgfgraphicnamed \times %
\beginpgfgraphicnamed{24}
\begin{tikzpicture}[scale=1/4,baseline=0]
  \draw (0,0) -- (-4,0)
  (0,0) -- (45:4)
  (0,0) -- (-45:4);
  \fill[black] (0,0) ++ (30:3) circle (2pt)
  (0,0) ++ (0:3) circle (2pt)
  (0,0) ++ (-30:3) circle (2pt);
  \filldraw[fill=gray, draw=black] (0,0) circle (2);
  \filldraw[fill=white, draw=black] (0,0) circle (1);
\end{tikzpicture}
\endpgfgraphicnamed \quad +\\
\quad + \quad %
\beginpgfgraphicnamed{25}
\begin{tikzpicture}[scale=1/4,baseline=0]
  \draw (0,0) -- (-135:4) node[left]{\ensuremath{a}}
  (0,0) -- (135:4) node[left]{\ensuremath{b}}
  (0,0) -- (4,0);
  \filldraw[fill=gray, draw=black] (0,0) circle (2);
  \filldraw[fill=white, draw=black] (-1,0) ellipse (1/2 and 1);
  \filldraw[fill=white, draw=black] (1,0) ellipse (1/2 and 1);
\end{tikzpicture}
\endpgfgraphicnamed \times %
\beginpgfgraphicnamed{26}
\begin{tikzpicture}[scale=1/4,baseline=0]
  \draw (0,0) -- (-4,0)
  (0,0) -- (45:4)
  (0,0) -- (-45:4);
  \fill[black] (0,0) ++ (30:3) circle (2pt)
  (0,0) ++ (0:3) circle (2pt)
  (0,0) ++ (-30:3) circle (2pt);
  \filldraw[fill=gray, draw=black] (0,0) circle (2);
\end{tikzpicture}
\endpgfgraphicnamed
\end{multline}

The supersymmetry Ward identities imply that the ratio of the splitting functions at loop and tree-level is independent of the parton helicities.  Therefore, it is convenient to work with their ratio
\begin{equation}
  \label{eq:rs_def}
  r_s^{(L)}(\epsilon, z, s=(k_1 + k_2)^2 ) = \frac {\text{Split}_{-h}^{(L)}(\epsilon, z, 1^{h_1}, 2^{h_2})} {\text{Split}_{-h}^{(0)}(\epsilon, z, 1^{h_1}, 2^{h_2})}.
\end{equation}

By using eq.~\eqref{eq:collinear_factorisation} and eq.~\eqref{eq:rs_def} for $L=2$ we have that, in any given collinear limit
\begin{align}
  M_n^{(1)} & \rightarrow M_{n-1}^{(1)} + r_S^{(1)},\\
  M_n^{(2)} & \rightarrow M_{n-1}^{(2)} + r_S^{(1)} M_{n-1}^{(1)} + r_S^{(2)}.
\end{align}

By using the methods developed in ref.~\cite{Kosower:1999xi}, ABDK computed the two loop splitting functions with the result
\begin{equation}
  \label{eq:rs_2loop_iteration}
  r_S^{(2)}(\epsilon, z, s) = \frac 1 2 \left(r_s^{(1)}(\epsilon, z, s)\right)^2 + f(\epsilon) r_s^{(1)}(2 \epsilon, z, s),
\end{equation} where
\begin{equation}
  f(\epsilon) = \frac {\psi(1 - \epsilon) - \psi(1)} \epsilon = -(\zeta_2 + \epsilon \zeta_3 + \epsilon^2 \zeta_4 + \cdots),
\end{equation} where $\psi(1) = -\gamma$, $\psi(z) = \tfrac d {d z} \Gamma(z)$.

This is a relation that gives the two-loop splitting function in terms of the the one-loop splitting function and a function $f$ that has no dependence on the kinematics.  Note that in the right-hand side of eq.~\eqref{eq:rs_2loop_iteration} one of the $r_s$ is evaluated for $2 \epsilon$ and the other for $\epsilon$.

Then, ABDK formulated an ansatz for two-loop MHV amplitudes with arbitrary number of legs and tested it for the case of four-point amplitudes.  Their ansatz is
\begin{equation}
  \label{eq:ABDK_ansatz}
  M_n^{(2)}(\epsilon) = \frac 1 2 \left(M_n^{(1)}\right)^2 + f^{(2)}(\epsilon) M_n^{(1)}(2 \epsilon) - \frac 5 4 \zeta_4.
\end{equation}  It is very easy to see that this ansatz is consistent with the two-loop iteration relation for the splitting function in eq.~\ref{eq:rs_2loop_iteration} (when taking a collinear limit in the left-hand side and right-hand side of eq.~\ref{eq:ABDK_ansatz} we get eq.~\ref{eq:ABDK_ansatz} back with $n \rightarrow n -1$).  This implies that if the ABDK ansatz is true for $n$-point amplitudes it will automatically be true for lower point amplitudes.

The ABDK ansatz is the simplest ansatz that is consistent with both the collinear limits and the iteration relations in eq.~\ref{eq:rs_2loop_iteration} for the splitting function.  In principle, one can add to the right-hand side of eq.~\ref{eq:ABDK_ansatz} any function that vanishes in all collinear limits.  The ABDK ansatz was constructed in such a way that it works for four-point amplitudes.

Following an initial guess (see ref.~\cite{Bern:1997it}) of Bern, Rozowsky and Yan for what the five-point two-loop amplitude should be in planar $\mathcal{N}=4$ Yang-Mills and the proof by Cachazo, Spradlin and Volovich that the even part of this ansatz satisfies the required iteration relation, the expectation that the ABDK ansatz works for five-point amplitudes was confirmed in ref.~\cite{Bern:2006vw} by Bern, Czakon, Kosower, Roiban and Smirnov.

The fact that the ABDK ansatz works also for the odd part of the five-point amplitude is non-trivial as the odd part vanishes in the collinear limits.  A more stringent test would come at three-loops where the square of the odd part can yield contributions to the finite part of even part of the amplitude.

Before going to the all-order BDS ansatz, let us quickly resume here what the ABDK ansatz is.  It is described by the following three formulae
\begin{align}
  M_n^{(2)}(\epsilon) & = \frac 1 2 \left(M_n^{(1)}\right)^2 +   f(\epsilon)^{(2)} M_n^{(1)}(2 \epsilon) + C^{(2)},\\
  f(\epsilon) & = \frac {\psi(1 - \epsilon) - \psi(1)} \epsilon =   -(\zeta_2 + \epsilon \zeta_3 + \epsilon^2 \zeta_4 + \cdots),\\
  C^{(2)} & = - \frac 5 4 \zeta_4.
\end{align}

Bern, Dixon and Smirnov did a three-loop four-point computation in ref.~\cite{Bern:2005iz} and discovered that there is an extension of the ABDK ansatz to three loops
\begin{align}
  M_4^{(3)}(\epsilon) & = - \frac 1 3 \left(M_4^{(1)}(\epsilon)\right)^3 + M_4^{(1)}(\epsilon) M_4^{(2)}(\epsilon) + f^{(3)}(\epsilon) M_4^{(1)}(3 \epsilon) + C^{(3)} + \mathcal{O}(\epsilon),\\
  f^{(3)}(\epsilon) & = \frac {11} 2 \zeta_4 + \epsilon \left(6 \zeta_5 + 5 \zeta_2 \zeta_3\right) + \epsilon^2 \left(c_1 \zeta_6 + c_2 \zeta_3^2\right),\\
  C^{(3)} & = \left(\frac {341}{216} + \frac 2 9 c_1\right) \zeta_6 + \left(- \frac {17} 9 + \frac 2 9 c_2\right) \zeta_3^2.
\end{align}

The constants $c_1$ and $c_2$ cancel in the three-loop iteration relation for four-point amplitudes.  As the coefficients of the expansion of $f$ in powers of epsilon and also $C^{(3)}$ are expected to be of uniform transcendentality, $c_1$ and $c_2$ are expected to be rational numbers.  They cannot be computed from the four-point calculation since they cancel in the final result, but they might contribute to the iteration relations for amplitudes with five or more external legs (remember that the \emph{same} functions $f^{(3)}$ and $C^{(3)}$ are expected to contribute to the three-loop iteration of amplitudes with arbitrary number of external legs).

Then, BDS proposed an all-orders ansatz.
\begin{equation}
\begin{split}
  \mathcal{M}_n(\epsilon) & = 1 + \sum_{l = 1}^\infty a^l M_n^{(l)}(\epsilon) \\ & = \exp \left(\sum_{l=1}^\infty a^l \left(f^{(l)}(\epsilon) M_n^{(1)}(l \epsilon) + C^{(l)} + E_n^{(l)}(\epsilon)\right)\right),
\end{split}
\end{equation} where $M_n^{(1)}(l \epsilon)$ is the one-loop amplitude evaluated in $D = 4 - 2 l \epsilon$ dimensions,
\begin{equation}
  f^{(l)}(\epsilon) = f_0^{(l)} + \epsilon f_1^{(l)} + \epsilon^2 f_2^{(l)}
\end{equation} and $E_n^{(l)}$ is an $\mathcal{O}(\epsilon)$ contribution.  Note that $f_k^{(l)}$ and $C^{(l)}$ do not depend on the number of external legs and on the kinematics.  They are expected to be polynomials in Riemann zeta values $\zeta_m$ with uniform degree of transcendentality.

In order to check this conjecture to low loop orders it is necessary to be able to write an extension to the ABDK and to the three-loop BDS ansatz.  In order to do this, we define a remainder $X_n^{(l)}(\epsilon)$ as
\begin{equation}
  M_n^{(l)} - \left(f^{(l)}(\epsilon) M_n^{(1)}(l \epsilon) + C^{(l)} + E_n^{(l)}\right) \equiv X_n^{(l)}(\epsilon).
\end{equation}  Then, the all-order BDS ansatz is
\begin{equation}
  1 + \sum_{l = 1}^\infty a^l M_n^{(l)}(\epsilon) = \exp \left(\sum_{l=1}^\infty a^l \left(M_n^{(l)} - X_n^{(l)}\right)\right).
\end{equation}  Expanding in $a$ and identifying the coefficients, we get
\begin{equation}
  f^{(L)}(\epsilon) M_n^{(1)}(L \epsilon) + C^{(L)} + E_n^{(L)} = \left. \ln\left(1 + \sum_{l=1}^\infty a^l M_n^{(l)}(\epsilon)\right)\right|_{\text{coefficient of $a^L$}}.
\end{equation}  It is obvious that the $L$-loop quantity $M_n^{(L)}(\epsilon)$ is expressible in terms of lower loop quantities ($M_n^{(l)}(\epsilon)$ with $l < L$).

\chapter{Pseudo-conformal integrals}
\label{ch:pseudo-conformal}

Besides the surprising fact that the $\mathcal{N}=4$ super-Yang-Mills loop scattering amplitudes can be expressed as sums of a small number of integrals, it turns out that the integrals themselves have a surprising property of conformal invariance.

This property has been proposed by Drummond, Henn, Smirnov and Sokatchev in ref.~\cite{Drummond:2006rz}, by analysing the results of the four-point computations.

Let us recall the results for the scattering amplitudes up to three loops.  The one-loop computation is given in terms of the box integral with a coefficient $s t$
\begin{equation}
  s t
\beginpgfgraphicnamed{27}
\begin{tikzpicture}[baseline=0]
  \def\r{1cm};
  \def\l{3mm};
  \draw (-\r/2,-\r/2) rectangle (\r/2, \r/2)
  (-\r/2,-\r/2) -- +(-135:\l)
  (-\r/2,\r/2) -- +(135:\l)
  (\r/2, \r/2) -- +(45:\l)
  (\r/2, -\r/2) -- +(-45:\l);
\end{tikzpicture}
\endpgfgraphicnamed.
\end{equation}

The two-loop amplitude is given in terms of a horizontal and vertical double box.
\begin{equation}
  s^2 t
\beginpgfgraphicnamed{28}
\begin{tikzpicture}[baseline=0]
  \def\r{1cm};
  \def\l{3mm};
  \draw (-\r,-\r/2) rectangle (\r, \r/2)
  (0,\r/2) -- (0,-\r/2)
  (-\r,-\r/2) -- +(-135:\l)
  (-\r,\r/2) -- +(135:\l)
  (\r, \r/2) -- +(45:\l)
  (\r, -\r/2) -- +(-45:\l);
\end{tikzpicture}
\endpgfgraphicnamed, \qquad s t^2
\beginpgfgraphicnamed{29}
\begin{tikzpicture}[baseline=0]
  \def\r{1cm};
  \def\l{3mm};
  \draw (-\r/2,-\r) rectangle (\r/2, \r)
  (-\r/2,0) -- (\r/2,0)
  (-\r/2,-\r) -- +(-135:\l)
  (-\r/2,\r) -- +(135:\l)
  (\r/2, \r) -- +(45:\l)
  (\r/2, -\r) -- +(-45:\l);
\end{tikzpicture}
\endpgfgraphicnamed.
\end{equation}

Finally, the tree-loop amplitude is given in terms of a three-loop ladder integral and the so-called `tennis court' integral.
\begin{equation}
  s^3 t
\beginpgfgraphicnamed{30}
\begin{tikzpicture}[baseline=0]
  \def\r{1cm};
  \def\l{3mm};
  \draw (-3*\r/2,-\r/2) rectangle (3*\r/2, \r/2)
  (-\r/2,\r/2) -- (-\r/2,-\r/2)
  (\r/2,\r/2) -- (\r/2,-\r/2)
  (-3*\r/2,-\r/2) -- +(-135:\l)
  (-3*\r/2,\r/2) -- +(135:\l)
  (3*\r/2, \r/2) -- +(45:\l)
  (3*\r/2, -\r/2) -- +(-45:\l);
\end{tikzpicture}
\endpgfgraphicnamed, \qquad s t^2 (l_1 + l_2)^2
\beginpgfgraphicnamed{31}
\begin{tikzpicture}[baseline=0]
  \def\r{1cm};
  \def\l{3mm};
  \draw (\r,-\r) -- (-\r,-\r) -- (-\r, \r) -- (\r,\r)
  (0,-\r) -- (0,0) node[midway,right]{$l_1$} node[midway,sloped]{\tikz\draw[>=stealth,->] (-1pt,0pt) -- (0pt,0pt);} -- (0,\r)
  (\r,-\r) -- (\r,0) node[midway,left]{$l_2$} node[midway,sloped]{\tikz\draw[>=stealth,->] (-1pt,0pt) -- (0pt,0pt);} -- (\r,\r)
  (0,0) -- (\r,0)
  (-\r,-\r) -- +(-135:\l)
  (-\r,\r) -- +(135:\l)
  (\r, \r) -- +(45:\l)
  (\r, -\r) -- +(-45:\l);
\end{tikzpicture}
\endpgfgraphicnamed
\end{equation}

Here we have just listed the integrals that contribute but, in order to get the full amplitude one has to sum over all circular permutations of the external legs and divide by corresponding symmetry factors of the integrals.  We emphasize again that these diagrams are not Feynman diagrams; they are just a useful representation of the integrals, each line corresponding to a factor in the denominator that is the same as a propagator.

It turns out that the numerator factors of these integrals, including the odd-looking numerator for the tennis court integral are precisely the factors that make the integrals conformal in a sense we detail below.

\begin{figure}
  \centering
\beginpgfgraphicnamed{32}
\begin{tikzpicture}[baseline=0]
  \def\r{1cm};
  \def\l{3mm};
  \draw (-\r,-\r) rectangle (\r,\r)
  (0,-\r) -- (0,\r)
  (0,0) -- (\r,0)
  (-\r,-\r) -- +(-135:\l) node[below left]{$k_1$}
  (-\r,\r) -- +(135:\l) node[above left]{$k_2$}
  (\r, \r) -- +(45:\l) node[above right]{$k_3$}
  (\r, -\r) -- +(-45:\l) node[below right]{$k_4$};
  \path (0,-\r-\l) coordinate (x1)
  (-\r-\l,0) coordinate (x2)
  (0,\r+\l) coordinate (x3)
  (\r+\l,0) coordinate (x4)
  (-\r/2,0) coordinate (x5)
  (\r/2,\r/2) coordinate (x6)
  (\r/2,-\r/2) coordinate (x7);
  \fill (x1) circle (1pt) node[below] {$x_1$}
  (x2) circle (1pt) node[left] {$x_2$}
  (x3) circle (1pt) node[above] {$x_3$}
  (x4) circle (1pt) node[right] {$x_4$}
  (x5) circle (1pt) node[below] {$x_5$}
  (x6) circle (1pt) node[above] {$x_6$}
  (x7) circle (1pt) node[below] {$x_7$};
  \draw (x1) ++ (0,-2*\l) node[below] {$(a)$};
\end{tikzpicture}
\endpgfgraphicnamed\qquad
\beginpgfgraphicnamed{33}
\begin{tikzpicture}[baseline=0]
  \def\r{1cm};
  \def\l{3mm};
  \draw (-\r,-\r) rectangle (\r,\r)
  (0,-\r) -- (0,\r)
  (0,0) -- (\r,0)
  (-\r,-\r) -- +(-135:\l)
  (-\r,\r) -- +(135:\l)
  (\r, \r) -- +(45:\l)
  (\r, -\r) -- +(-45:\l);
  \path (0,-\r-\l) coordinate (x1)
  (-\r-\l,0) coordinate (x2)
  (0,\r+\l) coordinate (x3)
  (\r+\l,0) coordinate (x4)
  (-\r/2,0) coordinate (x5)
  (\r/2,\r/2) coordinate (x6)
  (\r/2,-\r/2) coordinate (x7);
  \draw [blue] (x1) -- (x5) (x2) -- (x5) (x3) -- (x5) (x3) -- (x6) (x4) -- (x6) (x4) -- (x7) (x1) -- (x7) (x5) -- (x6) (x5) -- (x7) (x6) -- (x7);
  \path[draw=red, style=dashed] 
           (x1) edge[bend right] (x3)
           (x1) edge[bend left] (x3)
           (x4) edge[bend right] (x5)
           (x2) edge[bend right] (x4);
  \draw (x1) ++ (0,-2*\l) node[below] {$(b)$};
\end{tikzpicture}
\endpgfgraphicnamed
\caption{The points $x_i$ in $(a)$ form the vertices of the dual graph, which is drawn with continuous lines in $(b)$.  The numerator factors are represented by dashed lines in $(b)$.  To each solid line joining vertices $x_i$ and $x_j$ in the dual graph we associate a factor of $\tfrac 1 {x_{i j}^2}$ and for each dashed line joining vertices $x_i$ and $x_j$ we associate a numerator factor of $x_{i j}^2$.}
  \label{fig:tennis_court_conformal}
\end{figure}
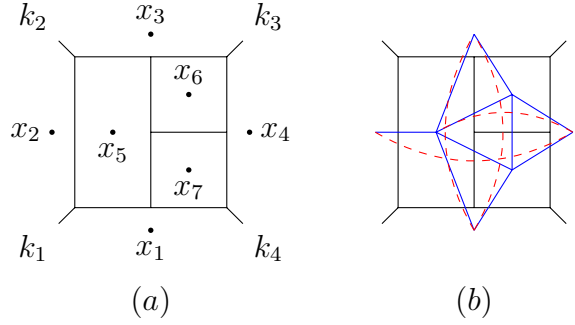

In order to illustrate this conformal symmetry we will use the example of the tennis court integral (see  fig.~\ref{fig:tennis_court_conformal}).

For every planar graph one can construct the dual graph by drawing a vertex inside every loop (it is convenient for this purpose to think of the the external lines as joining at infinity in a single point) and to each edge that is common to two loops we associate an edge joining the vertices inside these loops.  In fig.~\ref{fig:tennis_court_conformal}~$(a)$ the vertices of the dual graph are denoted by $x_i$, with $i=1, \ldots, 7$.

It turns out that the vertices of the dual graph can also be thought as an alternative description of the momenta flowing through the lines of the initial graph.  Through each line in the initial graph flows a momentum and also to each line in the initial graph corresponds a line in the dual graph.  One can encode the information about the momentum that flows through a given line in the difference of coordinates of the corresponding dual graph (after choosing an orientation in the plane).  For example, in fig.~\ref{fig:tennis_court_conformal}~$(a)$ we take $k_1 = x_2 - x_1$, $k_2 = x_3 - x_2$, etc.  Momentum conservation is automatically satisfied.  Note that the coordinates $x_i$ in the `dual space' are only defined up to an additive constant quadrivector.

Using these dual space coordinates, the tennis court integral, including the numerator factor can be written as follows
\begin{equation}
  \int d^4 x_5 d^4 x_6 d^4 x_7 \frac {x_{2 4}^2 \left(x_{1 3}^2\right)^2 x_{4 5}^2}{x_{1 5}^2 x_{2 5}^2 x_{3 5}^2 x_{3 6}^2 x_{4 6}^2 x_{4 7}^2 x_{1 7}^2 x_{5 6}^2 x_{5 7}^2 x_{6 7}^2}.
\end{equation}

Following Drummond, Henn, Smirnov and Sokatchev (see ref.~\cite{Drummond:2006rz}) we observe that these integrals are conformal invariant in the dual space parametrised by the coordinates $x$.  To see this it is enough to verify the invariance under inversions because the integrals are manifestly translation and rotation invariant and the invariance under inversion and translation implies the invariance under special conformal transformations.  The inversion acts on all the $x_i$ as follows,
\begin{equation}
  x^\mu \rightarrow \frac {x^\mu}{x^2}.
\end{equation}  Under inversion we also have
\begin{equation}
  x_{i j}^2 \rightarrow \frac {x_{i j}^2}{x_i^2 x_j^2}, \qquad d^4 x_k \rightarrow \frac {d^4 x_k}{(x_k^2)^4}.
\end{equation}

Putting this information together, it is easy to see that the necessary and sufficient condition for conformal invariance is to have weight zero at every vertex of the dual graph, where the weight is computed by taking the difference between the number of solid lines and the number of dotted lines incident with this vertex.  In particular, this implies that there can be no integrals that contain loops with the form of triangles.\footnote{For triangle sub-integrals, there are three lines in the dual graph meeting at the point dual to the triangle loop.  This is not enough to cancel the transformation of the integration measure over the dual coordinate, which has weight four under inversion.}  Even though there is no construction related to dual conformal symmetry for $\mathcal{N}=8$ supergravity, it has been conjectured in refs.~\cite{Bern:2005bb, BjerrumBohr:2006yw} that triangles do not appear in $\mathcal{N}=8$ supergravity amplitudes either.  This is the so-called `no-triangle hypothesis.'  The fact that there are no triangle subintegrals, hints that some unexpected cancellations are taking place (see ref.~\cite{BjerrumBohr:2008ji} for a recent paper on the cancellation of triangles in $\mathcal{N}=8$ supergravity).

It is important to stress that this dual conformal invariance is \emph{not} the usual conformal invariance of the $\mathcal{N}=4$ theory.  Also, note that these integrals are dual conformal invariant only in four dimensions, and dimensional regularisation explicitly breaks the dual conformal invariance.

An alternative way to regularise these integrals is to put the external legs off-shell, i.e. $k_1^2 = x_{1 2}^2 \neq 0$, etc.  (Not all the integrals that are conformal according to the counting described above can be regularised in this way).  Moreover, this way of regularising the integrals is not a regularisation in the usual sense of quantum field theory.  The difference is that a regulator in field theory is \emph{unchanged} by the symmetry operations in the theory, whereas in this case the virtuality of external lines $x_{1 2}^2$, etc is changed by the dual conformal transformations.  See ref.~\cite{Nguyen:2007ya} for a recent study of the conformal integrals with off-shell regulator.

Also, one should not conclude that the off-shell scattering amplitude can be obtained by simply taking the on-shell amplitude with the external momenta off-shell and removing the dimensional regularisation.

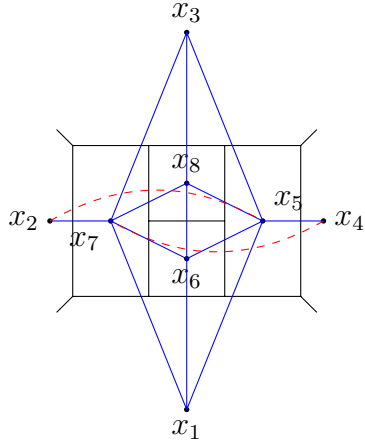
\begin{figure}
  \centering
\beginpgfgraphicnamed{34}
\begin{tikzpicture}
  \def\r{1cm};
  \def\l{3mm};
  \draw (-3*\r/2,-\r) rectangle (3*\r/2,\r)
  (-\r/2,-\r) -- (-\r/2,\r)
  (\r/2,-\r) -- (\r/2,\r)
  (-\r/2,0) -- (\r/2,0)
  (-3*\r/2,-\r) -- +(-135:\l)
  (-3*\r/2,\r) -- +(135:\l)
  (3*\r/2, \r) -- +(45:\l)
  (3*\r/2, -\r) -- +(-45:\l);
  \path (0,-5*\r/2) coordinate (x1)
  (-3*\r/2-\l,0) coordinate (x2)
  (0,5*\r/2) coordinate (x3)
  (3*\r/2+\l,0) coordinate (x4)
  (\r,0) coordinate (x5)
  (0,-\r/2) coordinate (x6)
  (-\r,0) coordinate (x7)
  (0,\r/2) coordinate (x8);
  \fill (x1) circle (1pt) node[below] {$x_1$}
  (x2) circle (1pt) node[left] {$x_2$}
  (x3) circle (1pt) node[above] {$x_3$}
  (x4) circle (1pt) node[right] {$x_4$}
  (x5) circle (1pt) node[above right] {$x_5$}
  (x6) circle (1pt) node[below] {$x_6$}
  (x7) circle (1pt) node[below left] {$x_7$}
  (x8) circle (1pt) node[above] {$x_8$};
\draw [blue] (x1) -- (x5) (x1) -- (x7) (x2) -- (x7) (x3) -- (x7) (x3) -- (x5) (x4) -- (x5) (x1) -- (x6) (x3) -- (x8) (x6) -- (x8) (x7) -- (x8) (x5) -- (x8) (x7) -- (x6) (x5) -- (x6);
   \path[draw=red, style=dashed] 
            (x2) edge[bend left] (x5)
            (x4) edge[bend left] (x7);
\end{tikzpicture}
\endpgfgraphicnamed
  \caption{Four-loop conformal integral with vanishing coefficient.}
  \label{fig:four-loop_vanishing}
\end{figure}

For four-point amplitudes, the hypothesis that only dual conformal integral appear has been checked through five loops (see refs.~\cite{Bern:1997nh, Bern:2005iz, Bern:2006ew, Bern:2007ct}).  It is striking that the conformal integrals appear with a coefficient of zero, or plus and minus one.  In fig.~\ref{fig:four-loop_vanishing} we present an example of four-loop integral that appears with coefficient zero.

In ref.~\cite{Drummond:2007aua} an argument was presented for the vanishing of the coefficient of this integral.  If we continue this integral to Euclidean signature and consider the region in the integration space where $x_{3 i}^2 \sim \rho \rightarrow 0$ for $i = 5, 6, 7, 8$ and we also have (because we are in Euclidean signature) $x_{5 6}^2 \sim \rho$, $x_{5 7}^2 \sim \rho$, etc (i.e. we take the region in the integration domain where the points $x_5$, $x_6$, $x_7$ and $x_8$ approach the point $x_3$), one can then prove that the integral is logarithmically divergent.\footnote{The measure of integration $d^4 x_5 d^4 x_6 d^4 x_7 d^4 x_8$ scales as $\rho^{15} d \rho$ for small $\rho$ and the denominator scales as $\tfrac 1 {\rho^{16}}$.}

This argument works at four and five loops for the four-point amplitudes.  For example, out of the $59$ dual conformal integrals that one can draw at five loops only $34$ actually contribute to the amplitude (see ref.~\cite{Bern:2007ct}) and these are precisely the ones that are finite when continued off-shell, as argued by Drummond, Korchemsky and Sokatchev in ref.~\cite{Drummond:2007aua}.  We emphasize that these conformal integrals that are off-shell divergent can be regularised in dimensional regularisation.

From this four-point discussion it seems that we have a very precise prescription for what an amplitude at arbitrary loop order is, up to some integer coefficients of $\pm 1$ (The prescription is: enumerate all the conformal integrals, eliminate the ones that are divergent off-shell and then the only unknowns are the integer coefficients of the remaining integrals).  There is as yet no first-principles understanding of this `experimental' observation, but Cachazo and Skinner have taken the first steps towards such an explanation in ref.~\cite{Cachazo:2008dx}.

Beyond four points, the amplitudes\footnote{We will restrict to MHV amplitudes when discussing the dual conformal properties.  The question of the possible extension to non-MHV amplitudes is interesting, but not much is known about these amplitudes beyond one-loop.} have both even and odd parts (the odd part contains factors like $\epsilon_{\mu \nu \rho \sigma} k_1^\mu k_2^\nu k_3^\rho k_4^\sigma$).  The coefficients of the odd part (see ref.~\cite{Bern:2006vw} for the five-point example) are considerably more complicated and their dual conformal properties (if any) are not clear.  Recent work~\cite{Cachazo:2008vp} by Cachazo lead to a more democratic treatment of the even and odd parts of the five-point two-loop amplitude but, in the representation he gives for the final answer, the conformal properties are obscured.  It is probably necessary to find an action of the dual conformal symmetry on the spinors that appear in the decomposition of the on-shell external momenta.  However, in order for this dual conformal symmetry to work, one has to put the external legs off-shell and, by doing that, one loses the possibility to decompose the momentum as a product of spinors.

It turns out that the scattering amplitudes with more than four external legs, cannot be constructed uniquely from four dimensional cuts.  When computing $D$-dimensional cuts, one finds integrals like the hexabox and a double-pentagon with numerator given by a scalar product of $(-2 \epsilon)$-dimensional components of the loop momenta (see ref.~\cite{Bern:2008ap} and sec.~\ref{ch:WL-SA_duality}).  These integrals are not conformal but they cancel when taking the logarithm of the amplitude against contributions from the square of the one-loop result.

\chapter{Integrability and AdS/CFT correspondence}

Because the AdS/CFT correspondence \cite{Maldacena:1997re, Gubser:1998bc, Witten:1998qj} is a weak--strong duality, checking it is very difficult.  Integrability provides a handle on both weak and strong-coupling regimes and allows for non-trivial tests of the correspondence.

On the CFT side, the natural objects to consider are gauge-invariant local operators with well-defined scaling dimensions.  In order to get gauge-invariant operators one should take traces of products of fields.  Among these the simplest are single trace operators.  Single trace operators can be interpreted in the dual $AdS$ picture as single particle states (or fundamental fields), while multiple trace operators which can be formed from the product of single trace operators, are interpreted as bound states of single trace operators.  The spectrum of scaling dimensions in the CFT is related to the spectrum of masses in the $AdS$ dual.

The $\mathcal{N}=4$ super-Yang-Mills has six scalars which can be described by three complex scalar fields $X_i$, with $i=1,2,3$.  One usually studies the super-conformal primaries, that is the operators of the lowest dimension in a representation of the super-conformal algebra, so the simplest objects are the single trace operators that only contain scalar fields.  From the primary operators we can get the other members of the super-conformal multiplet by acting with the supercharges $Q$ which have dimension $\tfrac 1 2$.

Operators of type $\tr \left(X_1^n\right)$ are BPS operators and their dimension is not renormalised.  The next simplest type of single trace super-conformal primary operator is one that contains only $X_1$ and $X_2$ scalar fields
\begin{equation}
  \tr\left(X_1^{n_1} X_2^{n_2}\right),
\end{equation} where we have indicated inside the trace only the number of operators of each kind; they can appear in all possible orderings up to cyclic permutations.  The operators of this kind form the $SU(2)$ sector.

In ref.~\cite{Minahan:2002ve}, Minahan and Zarembo mapped the problem of finding the spectrum of anomalous dimensions in this $SU(2)$ sector to the problem of finding the spectrum of an integrable spin-chain.

The states of the spin-chain are spins up and down on each site (up for an appearance of $X_1$ and down for an appearance of $X_2$, for example) and the Hamiltonian is the one-loop dilatation operator in the planar limit.  This model is integrable by means of Bethe ansatz and the complicated mixing problem for the renormalisation of the operators can be solved by integrability techniques.  The integrability at higher loops of the dilatation operator has been also proved and the methods were extended to sectors larger than $SU(2)$.  Integrability is also known to play an important role at strong coupling (see ref.~\cite{Bena:2003wd}).

A more complicated sector is the $SL(2)$ sector, which contains operators of the type
\begin{equation}
  \tr\left(\cdots X_1 \mathcal{D}^{+} X_1 \cdots\right),
\end{equation} where $\mathcal{D}^{+}$ is a covariant derivative in a light-cone direction.  Here one finds an ubiquitous and important quantity, the cusp anomalous dimension $f(\lambda)$, where $\lambda$ is the 't Hooft coupling (see Appendix~\ref{ch:wilson_loops}).  Historically, the anomalous dimension first appeared in studies of the renormalisation properties of Wilson loops with cusps (hence the name).  It also appears in the anomalous dimension of the large spin twist-two operators and in the infrared behaviour form-factors and of scattering amplitudes.  An example of twist-two operator in a theory with scalars transforming in the adjoint representation of the gauge group is
\begin{equation}
  \mathcal{O}_S = \tr\left(X_1 (\mathcal{D}^+)^S X_1\right).
\end{equation}

In the large spin limit, the dimensions of these operators have the following behaviour
\begin{equation}
  \Delta_S \sim S + f(\lambda) \ln S.
\end{equation} See ref.~\cite{Alday:2007mf} for an argument of this scaling with the spin that holds in all conformal field theories.  The $AdS$ dual of these twist-two operators is a folded closed string spinning around its center (see ref.~\cite{Gubser:2002tv}).

The integrability assumption for the Hamiltonian corresponding to the all-orders dilatation operator led to an all-orders proposal for the cusp anomalous dimension in the form of an integral equation (see ref.~\cite{Eden:2006rx}). (This integral equation depends on a $2 \rightarrow 2$ magnon $S$-matrix which is fixed by the symmetries up to a multiplicative phase, the dressing factor.)

It soon became clear that the ansatz proposed by Eden and Staudacher in ref.~\cite{Eden:2006rx} is in disagreement with the four-loop computation~\cite{Bern:2006ew} by Bern, Czakon, Dixon, Kosower and Smirnov, where the fourth order coefficient in the weak coupling expansion of the cusp anomalous dimension was extracted from a four-point, four-loop scattering amplitude computation.  The numerical results in ref.~\cite{Bern:2006ew} were refined by Cachazo, Spradlin and Volovich in ref.~\cite{Cachazo:2006az}.

Simultaneously with Bern \emph{et al.}~\cite{Bern:2006ew} Beisert, Eden and Staudacher proposed in ref.~\cite{Beisert:2006ez} a modified integral equation (BES equation) that incorporated a non-trivial dressing factor and which was in agreement with the fourth order computation by Bern \emph{et al.}~\cite{Bern:2006ew}.

The BES equation was studied at strong coupling first numerically in ref.~\cite{Benna:2006nd} and then analytically in ref.~\cite{Basso:2007wd}.  The strong-coupling expansion agrees with the strong-coupling perturbative computations in $\tfrac 1 {\sqrt \lambda}$ in refs.~\cite{Gubser:2002tv, Frolov:2002av, Roiban:2007jf, Roiban:2007dq}.

We give below all the coefficients of the cusp anomalous dimension that have been computed so far at weak and strong coupling.  They are in agreement with the weak- and strong-coupling expansions of the BES equation.
\begin{align}
f(\lambda) & = \frac{\lambda}{2\pi^2}\left( 1-\frac{\lambda}{48} + \frac{11\,\lambda^2}{11520} - \left(\frac{73}{1290240} + \frac{\zeta_3^2}{512 \pi^6}\right)\lambda^3 + \cdots \right),\qquad \text{when $\lambda \rightarrow 0$},\\
f(\lambda) & = \frac{\sqrt{\lambda}}{\pi}\left(1 - \frac{3 \ln
2}{\sqrt{\lambda}} - \frac{\mathrm{K}}{\lambda} + \cdots\right), \qquad \text{when $\lambda \rightarrow \infty$}.
\end{align}  Here $K$ is the Catalan's constant
\begin{equation}
   K = \sum_{n=0}^\infty \frac {(-1)^n}{(2 n + 1)^2} \approx 0.91596559.
\end{equation}

\chapter{Scattering at Strong Coupling}

Recently, Alday and Maldacena (see ref.~\cite{Alday:2007hr}) provided a prescription for computing on-shell scattering amplitudes at strong coupling by using the AdS/CFT correspondence.  They also computed the four-point scattering amplitude at the leading order in $\sqrt{\lambda}$ and showed that its functional form is identical to the BDS ansatz and the value of the cusp anomalous dimension appearing in their strong coupling computation agrees with the previously known expression.  This computation also had the by-product of computing the strong coupling limit of the collinear anomalous dimension which was previously unknown (the collinear anomalous dimension characterises the sub-leading IR divergences; see \eqref{eq:cusp_coll_def} in Appendix~\ref{ch:ir_scattering}).  In fact, the only known definition of the collinear anomalous dimension at strong coupling is by scattering amplitudes.

Before going to the Alday and Maldacena construction, let us review the high-energy fixed-angle scattering in string theory in flat space (see refs.~\cite{Gross:1987ar, Gross:1987kza}).  We will do a naive analysis based on the tree-level Koba-Nielsen formula (Gross and Mende also considered the contributions from higher genus).

Let us consider the scattering of four photons.  The tree-level amplitude in string theory is
\begin{equation}
  A_4 = \int_{z_1 < z_2 < z_3 < z_4} d z_2 \mu_{KN} \exp \sum_{i \neq j} \left(\frac 1 2 k_i \cdot k_j \ln |z_i - z_j| + \frac 1 2 \frac {\epsilon_i \cdot \epsilon_j}{(z_i - z_j)^2} + \frac {k_i \cdot \epsilon_j}{z_i - z_j}\right),
\end{equation} where we have set $\alpha' = \tfrac 1 2$ and used the signature $-++\cdots$ as is usual in string theory literature.  Here the factors containing the polarisations $\epsilon_i$ have to be expanded and only the terms containing one polarisation tensor for each external leg should be kept.

For high energy, fixed angle scattering, the dominant contribution comes from the products $k_i \cdot k_j$ in the exponential, which can be treated by a saddle point method.  We gauge-fix $z_1 = 0$, $z_3 = 1$ and $z_4 = \infty$ so the exponent becomes
\begin{equation}
  - \frac s 2 \ln (z_2) - \frac t 2 \ln (1 - z_2),
\end{equation} where $s = - (k_1 + k_2)^2 = -2 k_1 \cdot k_2$ and $t = - (k_2 + k_3)^2 = -2 k_2 \cdot k_3$.  The consistency conditions imply that the extremum of the above expression in $z_2$ should be a maximum and also that $0 < \tfrac s {s + t} < 1$ ($z_2 = \tfrac s {s + t}$ is the value where the exponent is extremal and this is the condition that $z_2$ should be inside the integration region).  These conditions imply space-like kinematics $s < 0$ and $t < 0$.

Keeping only the dominant contribution, we have the following behaviour at high energy scattering
\begin{equation}
  A_4 \sim \exp \left[- \frac s 2 \ln \left(\frac s {s + t}\right) - \frac t 2 \ln \left(\frac t {s + t}\right)\right].
\end{equation}  This asymptotic behaviour was already found by Veneziano in his famous paper~\cite{Veneziano:1968yb}.  The important point in the above analysis is that the asymptotic behaviour of an amplitude in some particular kinematic configuration is obtained by a saddle point analysis.  Also, this asymptotic behaviour does not depend on the states that are scattered, but only on their momentum.  The information about what states are scattered enters in sub-leading terms.

We now turn to a discussion of Alday and Maldacena prescription for computing $\mathcal{N}=4$ scattering amplitudes in the planar limit at strong coupling.  An essential characteristic of on-shell scattering amplitudes is that they are IR divergent so they have to be regularised.  In weak-coupling computations a popular regularisation is dimensional regularisation, or, when one wants to preserve supersymmetry, dimensional reduction (see sec.~\ref{sec:regularisation_schemes} for more details).

A regularisation that is more natural in the dual string theory description is to give a mass to the scattered gauge bosons by going to the Coulomb branch where the gauge group is broken from $U(N+1)$ to $U(N) \times U(1)$.  This can be done by separating a $D$-brane (which we will call the IR brane in the following) from the stack of $N + 1$ $D$-branes at $z=0$ and placing it at $z_{IR}$.  We will scatter the massive gauge bosons represented by strings stretched between the branes sitting at $z=0$ and $z=z_{IR}$.

Now, in order to use the AdS/CFT correspondence, we take the Maldacena limit, which amounts to replacing the stack of $N$ $D$-branes at $z=0$ by an $AdS_5 \times \mathbb{S}^5$ background.\footnote{In a perturbative string theory set-up this interaction corresponds to summing over all worldsheets that connect the IR brane to the stack of $N$ $D$-branes.}  The open strings stretched between the branes at $z=0$ and $z=z_{IR}$ are replaced by strings starting and ending on the brane at $z=z_{IR}$ but interacting with the $AdS_5 \times \mathbb{S}^5$ background.

In a Poincar\' e patch, the resulting $AdS$ metric is\footnote{We do not write the $\mathbb{S}^5$ metric since it will not play any role in the following.}
\begin{equation}
  d s^2 = R^2 \frac {d z^2 + d x_{3 + 1}^2}{z^2}.
\end{equation}  Here $z = 0$ is the boundary of the $AdS$ space and $z = \infty$ is the horizon of the stack of $N$ $D$-branes.\footnote{The $AdS$ metric can also be written as
\begin{equation}
  d s^2 = \frac {r^2}{R^2} d x_\mu d x^\mu + \frac {R^2}{r^2} d r^2,
\end{equation} where $r$ is the coordinate distance to the horizon.  In these coordinates the horizon is at $r = 0$ and the boundary is at $r = \infty$.  The relation between the two coordinate systems is $z = \tfrac {R^2} r$ so in the $z$ coordinate the boundary is at $z = 0$ and the horizon at $z = \infty$.}

At this point we should ask what does it mean to remove the IR regulator.  In the language of $D$-branes placed in flat space this means that we should let the the IR brane approach the stack of $N$ $D$-branes.  When this happens, the massive gauge bosons become massless and the gauge symmetry is restored.  This means that we should let $r_{IR} \rightarrow 0$ or, equivalently, $z_{IR} \rightarrow \infty$.\footnote{It might seem that, when taking $z_{IR} \rightarrow \infty$ the asymptotic states are defined off the boundary of the $AdS$ space which is contrary to the spirit of the AdS/CFT correspondence which states that good observables are defined on the boundary of the $AdS$ space.  In fact, the surface $z_{IR} \rightarrow \infty$ intersects the boundary when $x^\mu \rightarrow \infty$, as can be seen by describing the $AdS$ space in global coordinates.  The solution found by Alday and Maldacena has this property.}

When removing the IR cutoff, we should be careful to keep the momentum of the scattered gauge bosons fixed.  The metric has an isometry which is a translation symmetry in the $x^\mu$ directions so there is a conserved momentum $p_\mu = - i \partial_\mu$.  The momentum in the directions $x^\mu$ for an observer at $z$ in a local inertial frame is $\tfrac z R k_{(R)}$, where $k_{(R)}$ is the momentum as seen by an observer at $z = R$.  Then, it is easy to see that when one removes the IR cutoff $z_{IR} \rightarrow \infty$, the proper momentum becomes very high and the scattering problem reduces to the problem Gross and Mende solved long ago in refs.~\cite{Gross:1987ar, Gross:1987kza} for the flat space case.  That is, it reduces to finding the classical action for a string in $AdS$, with appropriate boundary conditions.

Let us discuss the boundary conditions.  The action for the string sigma model in this background is\footnote{We only write the bosonic $AdS$ part of the action.  The fermions and the $\mathbb{S}^5$ part of the action will not play any role in what follows.}
\begin{equation}
  S = \frac {\sqrt{\lambda}}{4 \pi} \int d \tau d \sigma \frac 1 {Z^2} \left(\partial_\alpha X \cdot \partial^\alpha X + \partial_\alpha Z \partial^\alpha Z\right) + \int d \tau \sum_{i=1}^n k_i \cdot X(0, \tau) \delta(\tau - \tau_i),
\end{equation} where we have included the contributions of $n$ vertex operators representing particles with momenta $k_i$ inserted on the boundary of the upper half-plane at $\tau_i$.  The first integral is over the upper half-plane $\sigma > 0$ and the second is over the real axis parametrised by $\tau$.  For the field $Z$ we have Dirichlet boundary conditions $Z(0, \tau) = z_{IR}$.

By variation of the action we find the following boundary conditions for the fields $X$
\begin{equation}
  - \frac {\sqrt{\lambda}}{2 \pi} \frac 1 {z_{IR}^2} \partial_\sigma X^\mu(0, \tau) + \sum_{i=1}^n k_i^\mu \delta(\tau - \tau_i) = 0.
\end{equation}

In order to impose these boundary conditions, it is useful to pass to $T$-dual coordinates ($T$-dualise in all space-time directions).  For a warped metric
\begin{equation}
  d s^2 = w^2(z) d x_\mu d x^\mu,
\end{equation} the $T$-dual fields $Y$ are defined by
\begin{equation}
  \partial_\alpha Y^\mu = i w^2(z) \epsilon_{\alpha \beta} \partial_\beta X^\mu.
\end{equation}  In our case, $w = \tfrac R z$.  The action in terms of these $T$-dual fields is also an action in an $AdS$ background with metric
\begin{equation}
  d s^2 = R^2 \frac {d y_\mu d y^\mu + d r^2}{r^2}, \quad \text{where $r = \frac {R^2} z$}.
\end{equation}

Using the formula
\begin{equation}
  \frac 1 {Z^2} \partial_\sigma X^\mu = - \frac i {R^2} \partial_\tau Y^\mu
\end{equation}for the dual field $Y$ in the boundary conditions for $X$, we find, after integrating the differential equation for $Y^\mu$,
\begin{equation}
  Y^\mu(0, \tau) = 2 \pi i \frac {R^2}{\sqrt{\lambda}} \sum_{i=1}^n k_i^\mu \theta(\tau - \tau_i) + \text{const}.
\end{equation}

The boundary conditions are such that for $\tau_{i-1} < \tau < \tau_i$, $Y^\mu(0, \tau)$ is a constant, and when $\tau = \tau_i$ it jumps by a quantity proportional to $k_i$.  So the worldsheet has Dirichlet boundary conditions in $T$-dual coordinates and the boundary is constructed out of the light-like momenta $k_i$, which form a closed contour because of momentum conservation $\sum k_i = 0$.

Let us summarise the prescription: in order to compute the strong coupling scattering amplitude of states with momenta $k_i$, we need to find the saddle point action for a worldsheet with the topology of a disk\footnote{Since we are working in the large $N$ limit, contributions from handles are suppressed by $\tfrac 1 {N^2}$.} with Dirichlet boundary conditions on a polygonal line made out of the light-like momenta of the scattered particles and with the $z$ coordinate equal to $z_{IR}$.  This is the same as computing a Wilson line on the polygonal contour made out of the light-like momenta $k_i$ (see refs.~\cite{Maldacena:1998im, Rey:1998ik, Drukker:1999zq}).

The problem of finding the minimal surfaces in $AdS$ with these boundary conditions is very difficult.  Only the solution for four points is known explicitly and it was found in the original paper~\cite{Alday:2007hr}.  Alday and Maldacena used a solution for a light-like cusp found earlier by Kruczenski in ref.~\cite{Kruczenski:2002fb}.  This solution has four cusps in global coordinates, but in Poincar\'e coordinates, which cover only part of the $AdS$ space only one cusp is visible.  Alday and Maldacena showed that one can bring all the cusps to finite distance by using $SO(2, 4)$ transformations.  The solution for the minimal surface is incredibly simple and it can be given by three relations among the five coordinates $y^\mu$ for $\mu=0, \ldots 3$ and $r$.  In the case $s = t$ it is given by
\begin{equation}
  y_0 = y_1 y_2, \qquad y_3 = 0, \qquad r = \sqrt{(1 - y_1^2)(1 - y_2^2)}.
\end{equation}

In order to make contact with the perturbation theory results obtained at small $\lambda$ and because the computations turned out to be easier, Alday and Maldacena used a kind of dimensional regularisation which amounts to consider the theory on $D_p$ branes with $p = 3 - 2 \epsilon$.  The dual of the theories living on these $D_p$ branes is a string theory with the background metric
\begin{equation}
  d s^2 = f^{-\frac 1 2} d x_D^2 + f^{\frac 1 2} \left(d r^2 + r^2 d \Omega_{9 - D}^2\right),
\end{equation} where $D = 4 - 2 \epsilon$ and
\begin{align}
  f & = \frac {2^{4 \epsilon} \pi^{3 \epsilon} \Gamma(2 + \epsilon)     \lambda_D}{r^{8 - D}},\\
  \lambda_D & = \frac {\lambda \mu^{2 \epsilon}}{(4 \pi e^{-\gamma})^\epsilon}, \qquad \gamma = - \Gamma'(1).
\end{align}  Most of the discussion above in the off-shell regularisation carries through to this regularisation: one can pass to the $T$-dual variables with $w^2 = f^{-\frac 1 2}$, etc.  There are only two subtle points: one is that we cannot trust the gravity description in the region where $r \rightarrow 0$ because the curvature becomes important there.  The other difference is that this  regularisation modifies the metric and therefore modifies the solution for the minimal surface.

Using this, Alday and Maldacena obtain the following result for the four-point function
\begin{equation}
  A_4 \sim \exp \left(2 i S_s^{div} + 2 i S_t^{div} + i S^{fin}\right),
\end{equation} where $S_s^{div}$ and $S_t^{div}$ are the IR-divergent pieces associated with the cusps of the adjacent gluons in the $s$ and $t$ channels (and each appears twice because there are a total of four cusps) and $S^{fin}$ is the finite part.  These are given by
\begin{align}
  i S_s^{div} & = - \frac 1 {\epsilon^2} \frac 1 {2 \pi} \sqrt{\frac {\lambda \mu^{2 \epsilon}}{(-s)^\epsilon}} - \frac 1 \epsilon \frac {1 - \ln 2}{4 \pi} \sqrt{\frac {\lambda \mu^{2 \epsilon}}{(-s)^\epsilon}},\\
  S^{fin} & = \frac {\sqrt{\lambda}}{8 \pi} \ln^2 \frac s t + \frac {\sqrt{\lambda}}{4 \pi} \left(\frac {\pi^2} 3 + 2 \ln 2 - \ln^2 2\right).
\end{align}

The BDS ansatz is
\begin{equation}
  A_4 = A_4^{\text{tree}} \left(A_s^{div}\right)^2 \left(A_t^{div}\right)^2 \exp \left(\frac {f(\lambda)} 8 \left(\ln^2 \frac s t + \frac {4 \pi^2} 3\right) + C(\lambda)\right),
\end{equation} where, because of the known structure of the IR divergences
\begin{equation}
  A_s^{div} = \exp\left(- \frac 1 {8 \epsilon^2} f^{(-2)}\left(\frac {\lambda \mu^{2 \epsilon}}{(-s)^\epsilon}\right) - \frac 1 {4 \epsilon} g^{(-1)}\left(\frac {\lambda \mu^{2 \epsilon}}{(-s)^\epsilon}\right)\right),
\end{equation} and
\begin{align}
  \left(\lambda \frac d {d \lambda}\right)^2 f^{(-2)}(\lambda) & =   f(\lambda),\\
  \left(\lambda \frac d {d \lambda}\right) g^{(-1)}(\lambda) & =   g(\lambda).
\end{align}  Here $f$ is the cusp anomalous dimension and $g$ is the collinear anomalous dimension.

Using the above expressions computed at strong coupling, we have, in the strong coupling limit,
\begin{align}
  f(\lambda) & \sim \frac {\sqrt{\lambda}} \pi,\\
  g(\lambda) & \sim \frac {1 - \ln 2} {2 \pi} \sqrt{\lambda}.
\end{align}  Note that the above strong coupling computation is IR consistent in the sense that $f^{(-2)}$ appearing in $S^{div}$ and $f$ appearing in $S^{fin}$ are linked by $\left(\lambda \frac d {d \lambda}\right)^2 f^{(-2)}(\lambda) = f(\lambda)$ as they should be.

The Alday and Maldacena prescription was further considered from several points of view.  In ref.~\cite{Abel:2007mw} the dependence on helicities of external states was discussed, the minimal surfaces for more than four points were studied in refs.~\cite{Mironov:2007qq, Astefanesei:2007bk}, quark scattering amplitudes were considered in refs.~\cite{McGreevy:2007kt, Komargodski:2007er}, finite temperature scattering in ref.~\cite{Ito:2007zy}, infrared~\cite{Buchbinder:2007hm} and collinear~\cite{Komargodski:2008wa} divergences and scattering in beta-deformed $N=4$ super-Yang-Mills was studied in ref.~\cite{Oz:2007qr}.  Higher order corrections in $\frac 1 {\sqrt{\lambda}}$ were considered in ref.~\cite{Kruczenski:2007cy} where some problems with dimensional regularisation were found at one-loop level.

\chapter{MHV scattering amplitude--Wilson loop duality}
\label{ch:WL-SA_duality}

One important conclusion of the last section was that the strong coupling computation of the scattering amplitude in the large $N$ limit was identical to the computation of a light-like Wilson loop in $T$-dual coordinates.  This Wilson loop is defined on a polygonal contour made from the momenta of the scattered particles (the contour closes by momentum conservation).  This strong coupling computation also agrees with the strong coupling limit of the BDS ansatz in the case of four-point amplitudes.

This was the case at strong coupling, but how about the weak coupling?  This question was addressed in ref.~\cite{Drummond:2007aua} for one-loop four-point case and in ref.~\cite{Brandhuber:2007yx} for one-loop and an arbitrary number of points.  Then Drummond, Henn, Korchemsky and Sokatchev computed the two-loop corrections at four~\cite{Drummond:2007cf}, five~\cite{Drummond:2007au} and six~\cite{Drummond:2007bm, Drummond:2008aq} point Wilson loops.

In ref.~\cite{Kruczenski:2007cy}, Kruczenski, Roiban, Tirziu and Tseytlin attempted to compute sub-leading corrections in $\tfrac 1 {\sqrt{\lambda}}$ at strong coupling, but they encountered some difficulties with the dimensional regularisation used by Alday and Maldacena.

It was already known that there is a link between the IR divergences of the scattering amplitudes and the Wilson lines built from the momenta of the scattered particles (see Appendix~\ref{ch:ir_scattering} for a more in-depth discussion and references).

The duality between scattering amplitudes and Wilson loops is an UV--IR duality and, in particular, links the IR divergences of the scattering amplitudes to the UV divergences of the Wilson loop.

In ref.~\cite{Drummond:2007au} an anomalous Ward identity was shown to hold for the polygonal Wilson loops with light-like edges.  This anomalous Ward identity is the consequence of the conformal invariance of the $\mathcal{N}=4$ theory. (The Wilson loop with cusps is divergent so it has to be regularised.  The regularisation explicitly breaks conformal invariance and the conformal Ward identity is anomalous.  See ref.~\cite{Drummond:2007au} for more details.)

If the vertices of the polygon for which the Wilson loop is defined have coordinates $x_i$, with $i = 1, \ldots, n$, then the anomalous Ward identity from ref.~\cite{Drummond:2007au} reads
\begin{equation}
  \label{eq:conformal_ward_id}
  \sum_{i=1}^n (2 x_i^\nu x_i \cdot \partial_i - x_i^2 \partial_i^\nu) \ln F_n = \frac  1 2 \Gamma_{\text{cusp}}(a) \sum_{i=1}^n \ln \frac {x_{i, i+2}^2}{x_{i-1,i+1}^2} x_{i, i+1}^\nu,
\end{equation} where $F_n$ is the finite\footnote{Here by finite we mean the part of the Wilson loop after the subtraction of the UV divergences, evaluated for $\epsilon=0$.  So the finite part does not have an $\epsilon$ dependence.} part of the Wilson loop.

These equations can be solved and have an unique solution for $n=4$ and $n=5$
\begin{align}
  \ln F_4 & = \frac 1 4 \Gamma_{\text{cusp}}(a) \ln^2 \left(\frac     {x_{13}^2}{x_{24}^2}\right) + \text{const},\\
  \ln F_5 & = \frac 1 8 \Gamma_{\text{cusp}}(a) \sum_{i=1}^5 \ln \left(\frac {x_{i,i+2}^2}{x_{i, i+3}^2}\right) \ln \left(\frac {x_{i+1,i+3}^2}{x_{i+2, i+4}^2}\right) + \text{const},
\end{align} where we have used the notation
\begin{equation}
  \label{eq:xp_translation}
  x_{i, i+j}^2 = (x_i - x_{i+j})^2 = (p_i + \cdots + p_{i+j-1})^2.
\end{equation}  This form of the finite part of the Wilson loop is the same as the finite form of the BDS ansatz for the four- and five-point amplitudes.  The value of the constant in the logarithm of the finite part is not fixed by the Ward identity.

For amplitudes beyond five points, the conformal Ward identity is not powerful enough to completely determine the finite part of the Wilson loop.  At six points, one can form three conformal cross-rations
\begin{equation}
  u_1 = \frac {x_{13}^2 x_{46}^2}{x_{14}^2 x_{36}^2}, \qquad u_2 = \frac {x_{24}^2 x_{15}^2}{x_{25}^2 x_{14}^2}, \qquad u_3 = \frac {x_{35}^2 x_{26}^2}{x_{36}^2 x_{25}^2},
\end{equation} and any function of these conformal cross-ratios satisfies the conformal Ward identity.

The BDS ansatz for the finite part of the amplitude for more that six point is also a solution of the conformal Ward identity but the Wilson loop result may differ from the BDS proposal by an arbitrary\footnote{This remainder function is not entirely arbitrary.  For example, it is constrained by the circular permutation symmetry acting on the vertices and a flip symmetry arising from the reality of the Wilson loop (see ref.~\cite{Drummond:2007au}).  In the case of the hexagon Wilson loop this means that the function should be completely symmetric in the three conformal cross-ratios.  There  are also constraints that come from multiple collinear limits.} function of conformal cross-ratios.

Given the possibility of having a non-zero remainder function starting at six points, a non-trivial test of the Wilson loop--scattering amplitudes duality would be to compare the computations of six-point Wilson loops and MHV scattering amplitudes at two loops.  (The agreement at one-loop and an arbitrary number of points was shown in ref.~\cite{Brandhuber:2007yx}).  The computation of the six-cusps two-loops Wilson loop was reported in refs.~\cite{Drummond:2007bm, Drummond:2007bm} and the computation of the six-point, two-loops MHV amplitude was reported in ref.~\cite{Bern:2008ap}.  In ref.~\cite{Bern:2008ap} it was also shown that the BDS ansatz breaks down at six-point two-loops.  The amplitude and the Wilson loop were evaluated numerically at several kinematic points and the results were found to agree within the numerical errors.

We will briefly describe below the computation of the scattering amplitude.  The computation was done using the unitarity method~\cite{Bern:1994cg, Bern:1994zx,  Bern:1997sc}, using a set of four- and $D$-dimensional cuts.  The $D$-dimensional cuts were necessary because the result contains some contributions which do not have any four-dimensional cuts.

\begin{figure}
\label{fig:all-cuts}
\centering
\beginpgfgraphicnamed{35}
\begin{tikzpicture}
  \matrix[row sep = 3ex, column sep=2em]
  {
\begin{scope}
  \draw [propagator,<->] (-3,.7) -- (3,.7);
  \draw [propagator,<-] (-3,0) -- (-2,0);
  \draw [propagator,<-] (3,0) -- (2,0);
  \draw [propagator,<->] (-3,-.7) -- (3,-.7);
  \draw [cut] (-1.5,-1) -- (-.5,1);
  \draw [cut'] (-1.5,-1) -- (-.5,1);
  \draw [cut] (.5,-1) -- (1.5,1);
  \draw [cut'] (.5,-1) -- (1.5,1);
  \filldraw [blob] (-2,0) ellipse (.3 and 1); 
  \filldraw [blob] (0,0) ellipse (.3 and 1); 
  \filldraw [blob] (2,0) ellipse (.3 and 1); 
  \draw (0,-1.5) node[below]{$(a)$};
\end{scope} &%
\begin{scope}
  \draw [propagator,<->] (-3,.7) -- (3,.7);
  \draw [propagator,->] (0,0) -- (0,1.5);
  \draw [propagator,<-] (3,0) -- (2,0);
  \draw [propagator,<->] (-3,-.7) -- (3,-.7);
  \draw [cut] (-1.5,-1) -- (-.5,1);
  \draw [cut'] (-1.5,-1) -- (-.5,1);
  \draw [cut] (.5,-1) -- (1.5,1);
  \draw [cut'] (.5,-1) -- (1.5,1);
  \filldraw [blob] (-2,0) ellipse (.3 and 1); 
  \filldraw [blob] (0,0) ellipse (.3 and 1); 
  \filldraw [blob] (2,0) ellipse (.3 and 1); 
  \draw (0,-1.5) node[below]{$(b)$};
\end{scope}\\
\begin{scope}
  \draw [propagator,<->] (-3,.7) -- (3,.7);
  \draw [propagator,->] (2,0) -- (2,1.5);
  \draw [propagator,->] (2,0) -- (2,-1.5);
  \draw [propagator,<->] (-3,-.7) -- (3,-.7);
  \draw [cut] (-1.5,-1) -- (-.5,1);
  \draw [cut'] (-1.5,-1) -- (-.5,1);
  \draw [cut] (.5,-1) -- (1.5,1);
  \draw [cut'] (.5,-1) -- (1.5,1);
  \filldraw [blob] (-2,0) ellipse (.3 and 1); 
  \filldraw [blob] (0,0) ellipse (.3 and 1); 
  \filldraw [blob] (2,0) ellipse (.3 and 1); 
  \draw (0,-1.5) node[below]{$(c)$};
\end{scope} &%
\begin{scope}
  \draw [propagator,<->] (-3,.7) -- (3,.7);
  \draw [propagator,->] (0,0) -- (0,1.5);
  \draw [propagator,->] (0,0) -- (0,-1.5);
  \draw [propagator,<->] (-3,-.7) -- (3,-.7);
  \draw [cut] (-1.5,-1) -- (-.5,1);
  \draw [cut'] (-1.5,-1) -- (-.5,1);
  \draw [cut] (.5,-1) -- (1.5,1);
  \draw [cut'] (.5,-1) -- (1.5,1);
  \filldraw [blob] (-2,0) ellipse (.3 and 1); 
  \filldraw [blob] (0,0) ellipse (.3 and 1); 
  \filldraw [blob] (2,0) ellipse (.3 and 1); 
  \draw (0,-1.5) node[below]{$(d)$};
\end{scope}\\
\begin{scope}
  \draw [propagator,<->] (-3,.7) -- (3,.7);
  \draw [propagator,->] (0,1) -- +(60:.5);
  \draw [propagator,->] (0,1) -- +(120:.5);
  \draw [propagator,<->] (-3,-.7) -- (3,-.7);
  \draw [cut] (-1.5,-1) -- (-.5,1);
  \draw [cut'] (-1.5,-1) -- (-.5,1);
  \draw [cut] (.5,-1) -- (1.5,1);
  \draw [cut'] (.5,-1) -- (1.5,1);
  \filldraw [blob] (-2,0) ellipse (.3 and 1); 
  \filldraw [blob] (0,0) ellipse (.3 and 1);  
  \filldraw [blob] (2,0) ellipse (.3 and 1); 
  \draw (0,-1.5) node[below]{$(e)$};
\end{scope} \\
  };
\end{tikzpicture}
\endpgfgraphicnamed
\caption{The double two-particle cuts used to determine the integrand.}
\end{figure}
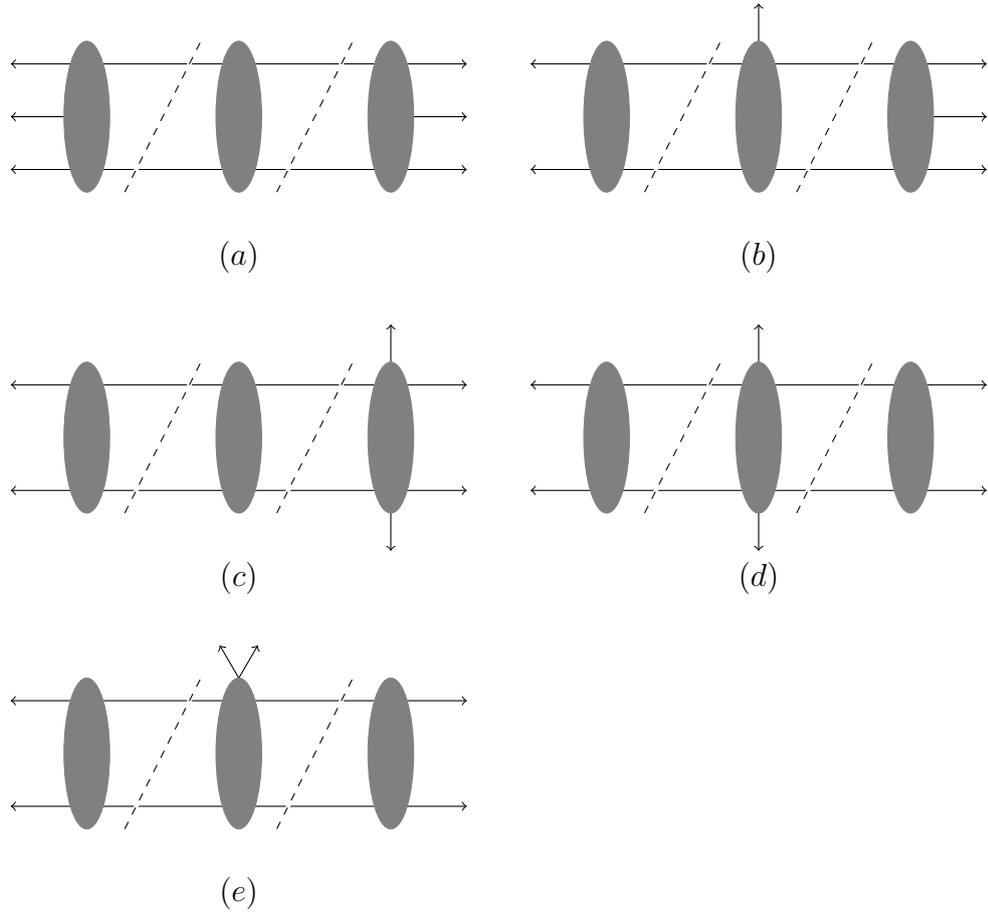

There are several kinds of unitarity cuts one can consider, but the three-particle cuts are difficult to compute because they involve NMHV tree amplitudes that are more complicated.  We therefore consider the set of double two-particle cuts represented in fig.~\ref{fig:all-cuts}.  These cuts have the advantage that they can be computed using only MHV amplitudes.

Given the no-triangle constraint, these double two-particle cuts are sufficient to determine the amplitude.  The no-triangle constraint is the conjecture that the result of $\mathcal{N}=4$ amplitude computations can always be written as a sum of integrals which do not contain any triangle subintegral.\footnote{In fact, the integrals do not contain any bubble subintegrals either, which is reasonable since bubble integrals are UV divergent in four dimensions.}  We also remark that the no-triangle constraint follows from the hypothesis of dual conformal invariance (see sec.~\ref{ch:pseudo-conformal}) but is a weaker assumption than the assumption that only pseudo-conformal integrals contribute.

\begin{figure}
  \centering
\includegraphics{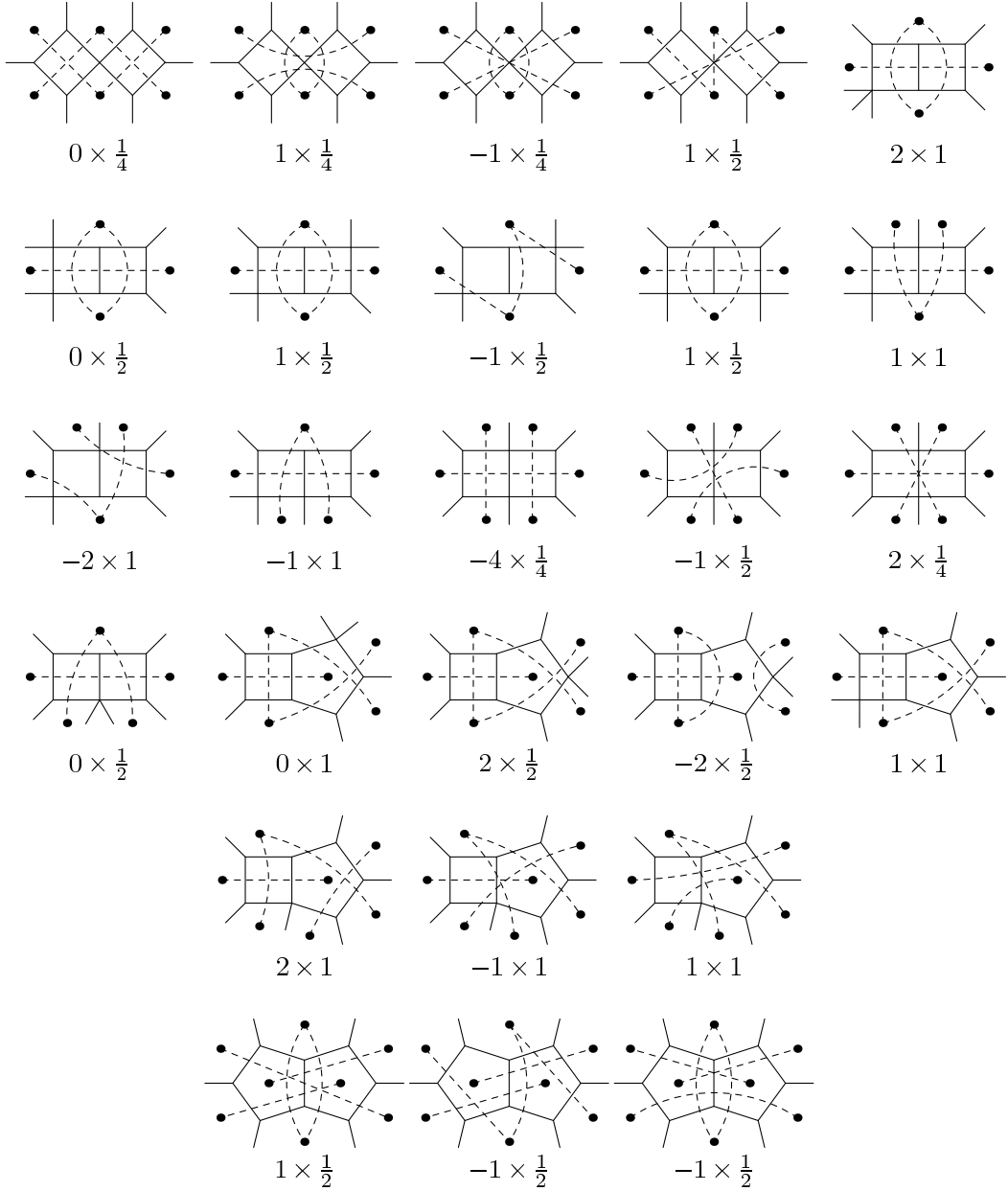}
\caption{The 26 dual conformal integrals.  Beneath each diagram is the   coefficient with which the corresponding integral enters the result constructed from four-dimensional cuts.  An overall factor of $\tfrac 1 {16}$ is suppressed and it is understood that one should sum over the $12$ cyclic and reflection permutations of the external legs.  In each coefficient, the second factor is a symmetry factor that accounts for overcounting in this sum.  This figure is taken from ref.~\cite{Bern:2008ap}.}
  \label{fig:motherfigure}
\end{figure}

\begin{figure}
\centering
\includegraphics{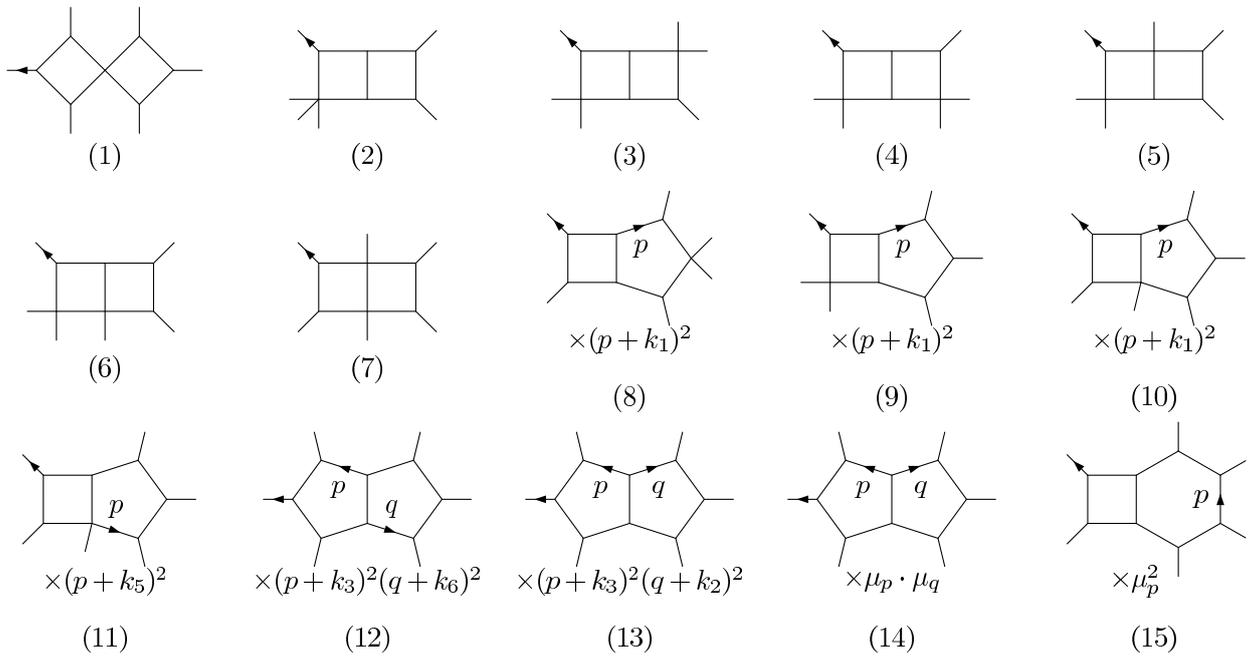}
\caption{The 15 independent integrals which contribute to the even part of the six-point MHV amplitude at two loops.  The external momenta are labelled clockwise with $k_1$ denoted by an arrow.  Integrals (8)--(15) are defined to include the indicated numerator factors involving the loop momenta.  In the last two integrals, $\mu_p$ denotes the $(-2 \epsilon)$-dimensional component of the loop momentum $p$.  This figure is taken from ref.~\cite{Bern:2008ap}.}
\label{fig:ContributingIntegrals}
\end{figure}

\begin{figure}
\centering
\includegraphics{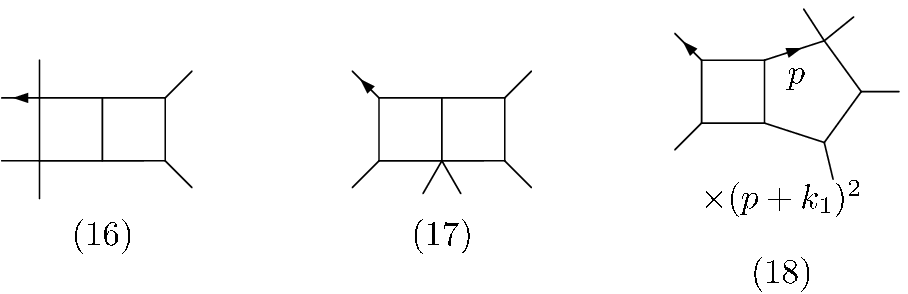}
\caption{The three independent two-loop diagrams which can be made pseudo-conformal by including appropriate numerators but which do not contribute to the amplitude ({\it i.e.}, they enter with zero coefficient).  This figure is taken from ref.~\cite{Bern:2008ap}.}
\label{fig:NonContributingIntegrals}
\end{figure}

The result of the computation for the even part of the amplitude is that the part which can be detected by the four-dimensional cuts can be written as a sum of conformal integrals.  The conformal integrals that can appear are listed in fig.~\ref{fig:motherfigure}, together with the numerator factors, the symmetry factors and the integer coefficients.  The contributing integrals are also listed in fig.~\ref{fig:ContributingIntegrals} where they are also numbered and the integrals that are conformal but appear with coefficient zero are listed in fig.~\ref{fig:NonContributingIntegrals}.

Let us present the results before describing in some detail the computation of one of the cuts.  In the MHV case, it is convenient to compute the ratio $M_6^{(2)}$ of the two-loop amplitude to the tree amplitude, since this ratio does not depend on the helicity distribution.  This ratio, in turn, can be separated into an even and an odd part, where the odd part contains Levi-Civita tensors in combinations like
\begin{equation}
  \epsilon_{\mu \nu \rho \sigma} k_i^\mu k_j^\nu k_l^\rho k_m^\sigma,
\end{equation} with $k_i, k_j, k_l, k_m$ external momenta.  We have not computed the odd part since the Wilson loop results do not contain an odd part.  Very recently, Cachazo, Spradlin and Volovich computed the odd part of the six-point MHV amplitude and found that the ABDK iteration relation holds (see ref.~\cite{Cachazo:2008hp}).  This computation used the leading singularity method~\cite{Cachazo:2008vp} of Cachazo.

The results are
\begin{align}
    M_6^{(2),D=4}(\epsilon) & = 
  \begin{aligned}[t]
    \frac{1}{16} \sum_{12~\text{perms.}} \Bigg[&
       \frac{1}{4} c_1 I^{(1)}(\epsilon) + c_2 I^{(2)}(\epsilon) + \frac{1}{2} c_3 I^{(3)}(\epsilon) + \frac{1}{2} c_4 I^{(4)}(\epsilon) + c_5 I^{(5)}(\epsilon)\\
      & + c_6 I^{(6)}(\epsilon) + \frac{1}{4} c_7 I^{(7)}(\epsilon) + \frac{1}{2} c_8 I^{(8)}(\epsilon) + c_9 I^{(9)}(\epsilon) + c_{10} I^{(10)}(\epsilon)\\
      & + c_{11} I^{(11)}(\epsilon) + \frac{1}{2} c_{12} I^{(12)}(\epsilon) + \frac{1}{2} c_{13} I^{(13)}(\epsilon)
\Bigg],
  \end{aligned}\label{eq:two-loop_assembly4}\\
    M_6^{(2),\mu}(\epsilon) & = \frac{1}{16} \sum_{12~\text{perms.}}
\Bigg[
  \frac{1}{4} c_{14} I^{(14)}(\epsilon)
+ \frac{1}{2} c_{15} I^{(15)}(\epsilon)
\Bigg].\label{TwoLoopAssemblyD}
\end{align}

Here $M_6^{(2),D=4}(\epsilon)$ is the part which can be detected by four-dimensional cuts and $M_6^{(2),\mu}(\epsilon)$ is the part that can only be detected by $D$-dimensional cuts.  The numerical factors are symmetry factors of the integrals and are cancelled when summing over the permutations.  The twelve permutations over which we sum are
\begin{equation}
  \label{eq:permutations}
  \begin{aligned}
& (1,2,3,4,5,6), \quad (2,3,4,5,6,1), \quad (3,4,5,6,1,2), \quad (4,5,6,1,2,3),\\
& (5,6,1,2,3,4), \quad (6,1,2,3,4,5), \quad (6,5,4,3,2,1), \quad (1,6,5,4,3,2),\\
& (2,1,6,5,4,3), \quad (3,2,1,6,5,4), \quad (4,3,2,1,6,5), \quad (5,4,3,2,1,6).
  \end{aligned}
\end{equation}

For the permutation $(1,2,3,4,5,6)$ the factors are
\begin{equation}
  \label{eq:6p2l-coeffs}
  \begin{aligned}
    c_1 & = s_{61} s_{34} s_{123} s_{345} + s_{12} s_{45} s_{234} s_{345} + s_{345}^2 (s_{23} s_{56} - s_{123} s_{234}),\\
    c_2 & = 2 s_{12} s_{23}^2,\\
    c_3 & = s_{234} (s_{123} s_{234} - s_{23} s_{56}),\\
    c_4 & = s_{12} s_{234}^2,\\
    c_5 & = s_{34} (s_{123} s_{234} - 2 s_{23} s_{56}),\\
    c_6 & = - s_{12} s_{23} s_{234},\\
    c_7 & = 2 s_{123} s_{234} s_{345} - 4 s_{61} s_{34} s_{123} - s_{12} s_{45} s_{234} - s_{23} s_{56} s_{345},\\
    c_8 & = 2 s_{61} (s_{234} s_{345} - s_{61} s_{34}),\\
    c_9 & = s_{23} s_{34} s_{234},\\
    c_{10} & = s_{23} (2 s_{61} s_{34} - s_{234} s_{345}),\\
    c_{11} & = s_{12} s_{23} s_{234},\\
    c_{12} & = s_{345} (s_{234} s_{345} - s_{61} s_{34}),\\
    c_{13} & = - s_{345}^2 s_{56},\\
    c_{14} & = -2 s_{126} (s_{123} s_{234} s_{345} - s_{61} s_{34} s_{123} - s_{12} s_{45} s_{234} - s_{23} s_{56} s_{345}),\\
    c_{15} & = 2 s_{61} ( s_{123} s_{234} s_{345} - s_{61} s_{34} s_{123} - s_{12} s_{45} s_{234} - s_{23} s_{56} s_{345}).
  \end{aligned}
\end{equation}

\begin{figure}
  \centering
  \beginpgfgraphicnamed{36}
  \begin{tikzpicture}
  \draw [propagator,<->] (-3,.7) -- (3,.7)
     node[pos=0,left]{$k_3$}
     node[pos=.3]{\tikz\draw[->] (-1pt,0pt) -- (0pt,0pt);}
     node[pos=.3,above]{$l_1$}
     node[pos=.6]{\tikz\draw[<-] (-1pt,0pt) -- (0pt,0pt);}
     node[pos=.6,above]{$l_4$}
     node[pos=1,right]{$k_4$};
  \draw [propagator,<-] (-3,0) -- (-2,0)
     node[pos=0,left]{$k_2$};
  \draw [propagator,<-] (3,0) -- (2,0)
     node[pos=0,right]{$k_5$};
  \draw [propagator,<->] (-3,-.7) -- (3,-.7)
     node[pos=.4]{\tikz\draw[->] (-1pt,0pt) -- (0pt,0pt);}
     node[pos=.4,below]{$l_2$}
     node[pos=.7]{\tikz\draw[<-] (-1pt,0pt) -- (0pt,0pt);}
     node[pos=.7,below]{$l_3$}
     node[pos=0,left]{$k_1$}
     node[pos=1,right]{$k_6$};
  \draw [cut] (-1.5,-1) -- (-.5,1);
  \draw [cut'] (-1.5,-1) -- (-.5,1);
  \draw [cut] (.5,-1) -- (1.5,1);
  \draw [cut'] (.5,-1) -- (1.5,1);
  \filldraw [blob] (-2,0) ellipse (.3 and 1); 
  \filldraw [blob] (0,0) ellipse (.3 and 1); 
  \filldraw [blob] (2,0) ellipse (.3 and 1); 
  \end{tikzpicture}
  \endpgfgraphicnamed
  \caption{Labelling for cut $(a)$.}
  \label{fig:a-detail}
\end{figure}
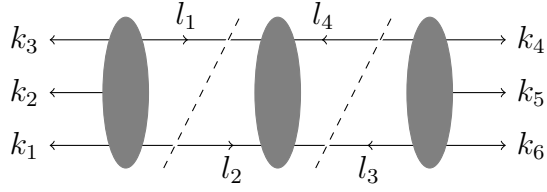

Let us now describe in some detail the computation of cut $(a)$ in fig.~\ref{fig:all-cuts}.  We use the labelling in fig.~\ref{fig:a-detail}.

The product of the three tree amplitudes corresponding to cut $(a)$ is
\begin{equation}
i\frac{\langle 1\ 2\rangle^3}{\langle 2\ 3\rangle \langle 3\ l_1\rangle \langle l_1\ l_2\rangle \langle l_2\ 1\rangle} 
\times i\frac{\langle (-l_2)\ (-l_1)\rangle^3}{\langle (-l_1)\ (-l_4)\rangle \langle(-l_4)\ (-l_3)\rangle \langle (-l_3)\ (-l_2)\rangle} \times i\frac{\langle l_3\ l_4\rangle^3}{\langle l_4\ 4\rangle \langle 4\ 5\rangle \langle 5\ 6\rangle \langle 6\ l_3\rangle}.
\label{eq:Cut_a}
\end{equation} 

After dividing by the tree amplitude we obtain
\begin{equation}
  \frac{\left\langle 1\ 6\right\rangle 
   \left\langle 3\ 4\right\rangle 
   \left\langle l_1\ l_2\right\rangle ^2
   \left\langle l_3\ l_4\right\rangle
   ^2}{\left\langle 1\ l_2\right\rangle 
   \left\langle 3\ l_1\right\rangle 
   \left\langle 4\ l_4\right\rangle 
   \left\langle 6\ l_3\right\rangle 
   \left\langle l_1\ l_4\right\rangle 
   \left\langle l_2\ l_3\right\rangle }.
\end{equation}

In the next step we `rationalise' the denominators
\begin{equation}
  \frac 1 {\langle a, b\rangle} = \frac {[b, a]}{2 p_a \cdot p_b},
\end{equation} where $p_a$ and $p_b$ are light-like momenta corresponding to the spinors $a$ and $b$.  This formula is a consequence of the identity $2 p_a \cdot p_b = \langle a, b\rangle [b, a]$.

After rationalising denominators to Lorentz products, we find that the numerator can be written (see also ch.~\ref{ch:unitarity}),
\begin{align*}
  N & = \left\langle 1\ 6\right\rangle \left\langle 3\ 4\right\rangle \left\langle l_1\ l_2\right\rangle^2 \left\langle l_3\ l_4\right\rangle^2 \left[l_2\ 1\right] \left[l_1\ 3\right] \left[l_4\ 4\right] \left[l_3\ 6\right] \left[l_4, l_1\right] \left[l_3, l_2\right] \\
    & = \left(\left[l_2\ 1\right] \langle 1\ 6\rangle \left[6\ l_3\right] \langle l_3\ l_4\rangle \left[l_4\ 4\right] \langle 4\ 3\rangle \left[3\ l_1\right] \langle l_1\ l_2\rangle\right) \left(\left[l_2\ l_3\right] \langle l_3\ l_4\rangle \left[l_4\ l_1\right] \langle l_1\ l_2\rangle\right) \\
    & = \tr_+\left[l_2 k_1 k_6 l_3 l_4 k_4 k_3 l_1\right] \tr_+\left[l_2 l_3 l_4 l_1\right],
\end{align*} where $\tr_+[\cdots] = \frac 1 2 \tr[(1 + \gamma_5) \cdots]$.

When expanded, each trace has an even and an odd part (the origin of the odd terms lies in the presence of the $\gamma_5$ matrix inside the traces).  The product of two epsilon tensors would yield an even term, but only the longer trace here can actually produce an epsilon tensor, as $\epsilon(l_1, l_2, l_3, l_4)$ vanishes because of momentum conservation.

In order to identify the coefficients of the integrals in \ref{fig:ContributingIntegrals}, we use momentum conservation to re-express all Lorentz invariants in terms of independent invariants.  The required simplifications can be done analytically, but in some cases (for example cut $(d)$) it is easier to do them numerically, by matching to a target expression.

Doing so, we obtain for the final result of cut $(a)$, in the $(3,4,5,6,1,2)$ permutation with respect to fig.~\ref{fig:ContributingIntegrals},
\begin{equation}
\begin{aligned}
 \frac 1 4 \Biggl[&
    \frac {s_{123}^2 s_{34} s_{61} - s_{123}^2 s_{234} s_{345} 
           + s_{123} s_{234} s_{12} s_{45} + s_{123} s_{345} s_{23} s_{56}}
          {(k_1 + l_2)^2 (k_3 + l_1)^2 (k_4 + l_4)^2 (k_6 + l_3)^2} \\ 
  &+ \frac {s_{123}^2 s_{345} - s_{123} s_{12} s_{45}}
           {(k_3 + l_1)^2 (l_2 + l_3)^2 (k_6 + l_3)^2}  
   + \frac {s_{123}^2 s_{234} - s_{123} s_{23} s_{56}}
           {(k_1 + l_2)^2 (l_2 + l_3)^2 (k_4 + l_4)^2}  \\ 
  &+ \frac {s_{123}^2 s_{34}}
           {(k_3 + l_1)^2 (l_2 + l_3)^2 (k_4 +l_4)^2}  
   + \frac {s_{123}^2 s_{61}}
	   {(k_1 + l_2)^2 (l_2 + l_3)^2 (k_6 +l_3)^2}  \\
  &+ \frac {s_{123} s_{12} s_{23} (k_6 - l_2)^2}
           {(k_1 + l_2)^2 (k_3 + l_1)^2 (l_2 + l_3)^2 (k_6 + l_3)^2} 
   + \frac {s_{123} s_{12} s_{23} (k_4 - l_1)^2}
           {(k_1 + l_2)^2 (k_3 + l_1)^2 (l_2 + l_3)^2 (k_4 + l_4)^2} \\ 
  &+ \frac {s_{123} s_{45} s_{56} (k_3 - l_4)^2}
           {(k_3 + l_1)^2 (l_2 + l_3)^2 (k_4 + l_4)^2 (k_6 + l_3)^2} 
   + \frac {s_{123} s_{45} s_{56} (k_1 - l_3)^2}
           {(k_1 + l_2)^2 (l_2 + l_3)^2 (k_4 + l_4)^2 (k_6 + l_3)^2} \\ 
  &+ \frac {1}
           {(k_1 + l_2)^2 (k_3 + l_1)^2 (l_2 + l_3)^2 (k_4 + l_4)^2 (k_6 + l_3)^2}  \times \\ 
  & \hspace{.2\textwidth}
    \begin{aligned}
    \Bigl(&- s_{123}^2 s_{61} (k_3 - l_4)^2 (k_4 - l_1)^2 - s_{123}^2 s_{34} (k_1 - l_3)^2 (k_6 - l_2)^2  \\
          &+ s_{123} (s_{123} s_{234} - s_{23} s_{56})  (k_3 - l_4)^2 (k_6 - l_2)^2  \\
          &+ s_{123} (s_{123} s_{345} - s_{12} s_{45}) (k_1 - l_3)^2 (k_4 - l_1)^2
    \Bigr)  \Biggr]\,.
    \end{aligned}
\end{aligned}\label{eq:cutabig}
\end{equation}

One can then read off the coefficients of the integrals detected by the four-dimensional cut $(a)$ on the result in eq.~\eqref{eq:cutabig}.  The coefficients of the remaining integrals can be detected by at least one of the remaining cuts in fig.~\ref{fig:all-cuts}.

We cannot be sure, however, that the four-dimensional cuts are sufficient to construct the amplitude as there can be contributions that do not have four-particle cuts but still contribute to the finite or divergent parts in $\epsilon$.

These contributions must be computed using $D$-dimensional cuts.  In ref.~\cite{Bern:2008ap} the cuts in fig.~\ref{fig:all-cuts}~$(a)$ and $(c)$ were computed in $D$ dimensions.  These cuts determine the coefficients of integrals~(14) and~(15), respectively, in fig.~\ref{fig:ContributingIntegrals}.  The calculations were done by taking advantage of the equivalence between the $\mathcal{N}=4$ theory and ten-dimensional $\mathcal{N}=1$ super-Yang-Mills theory compactified on a torus.  The cuts are computed with the spin algebra performed in the ten-dimensional theory, keeping loop momenta in $D$ dimensions. (External momenta can be taken to be four-dimensional.)  The ten-dimensional gluon corresponds to a four-dimensional gluon and six real scalar degrees of freedom, while the ten-dimensional Majorana-Weyl fermions correspond to four flavours of gluinos.

The integrals are evaluated numerically by using the Mellin-Barnes representation.  The Mellin-Barnes representation is well suited for numerical (and sometimes analytical) computations of complicated integrals of Feynman type. See ref.~\cite{Smirnov:2004ym} for an in-depth discussion.

It relies on the following identity
\begin{equation}
  \label{eq:MB_fund_id}
  \frac 1 {(A + B)^\nu} = \frac 1 {\Gamma(\nu)} \frac 1 {2 \pi i} \int_{-i
    \infty}^{i \infty} d z \frac {A^z}{B^{z + \nu}} \Gamma(-z) \Gamma(\nu +
  z),
\end{equation} where $|A| > |B|$ and the contour of integration separates the poles of $\Gamma(-z)$ and of $\Gamma(\nu + z)$ (recall that the function $\Gamma(z)$ has poles at $z=-k$ for $k=0, 1, \ldots$ with residues $\frac {(-1)^k}{k !}$).

\begin{figure}
  \centering
  \beginpgfgraphicnamed{37}
  \begin{tikzpicture}
  \draw[thick,>=stealth,->] (0, -2) -- (0, 4);
  \draw[thick,>=stealth,->] (-4, 0) -- (6, 0);
  \filldraw (0,0) circle (2pt) node[below right]{$0$}
            (1,0) circle (2pt) node[below]{$1$}
            (2,0) circle (2pt) node[below]{$2$}
            (1.5,1) circle (2pt) node[above]{$-\nu$}
            (0.5,1) circle (2pt) node[above]{$-\nu-1$}
            (-0.5,1) circle (2pt) node[above]{$-\nu-2$};

  \draw {[rounded corners=10pt] (-.2,-2) -- (-.2, 0.75) -- (1.90, 0.75) --
         (1.90, 1.5) -- (-.2, 1.5) -- (-.2, 4)};
  \end{tikzpicture}
  \endpgfgraphicnamed
  \caption{The integration contour for the complex integral in
    eq.~\eqref{eq:MB_fund_id}}
  \label{fig:MB_contour}
\end{figure}
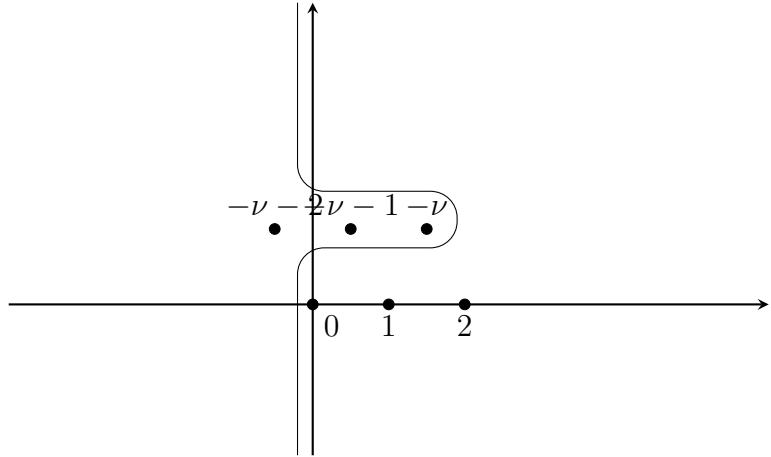

There are several ways to use this identity.  One way is to transform massive propagators into massless ones (in general, massless integrals are easier to compute than massive ones).

The formula~\eqref{eq:MB_fund_id} can also be used in a parametric representation of the Feynman integrals and it results in a multiple contour integral.  Some of these contour integrals can be computed by using contour deformation and Cauchy's theorem or by using some lemmas due to Barnes
\begin{gather}
  \int_{-i \infty}^{i \infty} d z\, \Gamma(a+z) \Gamma(b+z) \Gamma(c-z)
  \Gamma(d-z) = \frac {\Gamma(a+c) \Gamma(a+d) \Gamma(b+c)
    \Gamma(b+d)}{\Gamma(a+b+c+d)},\\
  \begin{split}
    \int_{-i \infty}^{i \infty} d z\, \frac{\Gamma(a+z) \Gamma(b+z) \Gamma(c+z)
      \Gamma(d-z) \Gamma(e-z)}{\Gamma(a+b+c+d+e+z)} =\\ \frac {\Gamma(a+d)
      \Gamma(a+e) \Gamma(b+d) \Gamma(b+e) \Gamma(c+d)
      \Gamma(c+e)}{\Gamma(a+b+d+e) \Gamma(a+c+d+e) \Gamma(b+c+d+e)}
  \end{split}
\end{gather}

The contours for integrals in Barnes lemmas are such that they separate the poles of $\Gamma$ functions with negative $z$ arguments from the poles of $\Gamma$ functions with positive $z$ arguments.

The choice of integration contours, the simplification using Barnes lemmas and numerical evaluation of integrals can be automated by using Czakon's \texttt{MB} package for \texttt{Mathematica}, documented in ref.~\cite{Czakon:2005rk} and the \texttt{CUBA} package for multidimensional numerical integration, documented in ref.~\cite{Hahn:2004fe}.

In ref.~\cite{Bern:2008ap} it was shown that the BDS ansatz fails at two loops for six-point amplitudes.  Before this, the behaviour of the BDS ansatz in several Regge limits was analysed in refs.~\cite{Bartels:2008ce, Brower:2008nm}.  The BDS ansatz was tested at strong coupling in ref.~\cite{Alday:2007he} and it was found to fail in the limit of infinite number of gluons.

It is interesting to find the remainder of the six-point amplitude from the BDS ansatz.  Following ref.~\cite{Bern:2008ap} we denote this function by $R_A$
\begin{equation}
  R_A = M_6^{(2)} - M_6^{\text{BDS}}.
\end{equation}  A priori, this is an arbitrary function of the coupling and of the kinematics, but in ref.~\cite{Bern:2008ap} some numerical evaluations suggest that $R_A$ only depends on the conformal cross-ratios.

We present below the numerical results from ref.~\cite{Bern:2008ap} and the comparison with the Wilson loop computations~\cite{Drummond:2007bm, Drummond:2008aq}.  The amplitude has been evaluated numerically for the kinematic points in eq.~\eqref{eq:KinematicPoints}.\footnote{If the external momenta are four-dimensional, then any five or more momenta are linearly dependent.  The linear dependence is encoded in the vanishing of the Gram determinant of any five (or more) momenta.  For example, the Gram determinant of momenta $k_1, \ldots, k_5$ is defined by
\begin{equation}
  \det (k_i \cdot k_j)_{1 \leq i, j \leq 5},
\end{equation} and should be zero.  This imposes additional constraints on the kinematic variables $s_{i, i+1}$ and $s_{i-1, i, i+1}$.  In the list of our kinematical points, the points $K^{(0)}, K^{(1)}, K^{(2)}, K^{(3)}$ satisfy the Gram determinant constraint, whereas $K^{(4)}$ and $K^{(5)}$ do not.}

\begin{subequations}
\label{eq:KinematicPoints}
\begin{align}
K^{(0)} &:  s_{i,i+1} = -1, \quad s_{i,i+1,i+2} = -2, \\
K^{(1)} &:
\begin{aligned}[t]
  & s_{12} = -0.723 6200, \quad s_{23} = -0.921 3500, \quad s_{34} = -0.272 3200, \\
  & s_{45} = -0.358 2300, \quad s_{56} = -0.423 5500, \quad s_{61} = -0.321 8573, \\
  & s_{123} = -2.148 6192, \quad s_{234} = -0.726 4904, \quad s_{345} =  -0.482 5841, 
\end{aligned}\\
K^{(2)} &:
\begin{aligned}[t]
  & s_{12} =-0.322 3100, \quad s_{23} = -0.2323220, \quad s_{34} = -0.523 8300, \\
  & s_{45} = -0.823 7640, \quad s_{56} = -0.532 3200, \quad s_{61} = -0.923 7600, \\
  & s_{123} = -0.732 2000, \quad s_{234} = -0.828 6700, \quad s_{345} = -0.662 6116,
\end{aligned}\\
K^{(3)} &: s_{i,i+1} = -1, \quad s_{123} = -1/2, \quad s_{234} = -5/8, \quad s_{345} = -17/14,\\
K^{(4)} &: s_{i,i+1} = -1, \quad s_{i,i+1,i+2} = -3,\\
K^{(5)} &:  s_{i,i+1} = -1, \quad s_{i,i+1,i+2} = -9/2.
\end{align}
\end{subequations}

In table~\ref{tab:RemainderTable} we present the results for the remainder function $R_A$.  We also present the values of the conformal cross-ratios
\begin{equation}
  \label{eq:SixPtConformalCrossRatios}
  u_1 = \frac {s_{1 2} s_{4 5}}{s_{1 2 3} s_{3 4 5}}, \quad
  u_2 = \frac {s_{2 3} s_{5 6}}{s_{2 3 4} s_{1 2 3}}, \quad
  u_3 = \frac {s_{3 4} s_{6 1}}{s_{3 4 5} s_{2 3 4}}.
\end{equation}

Observe that the kinematical points $K^{(0)}$ and $K^{(1)}$ have the same conformal cross-ratios and the corresponding values of the remainder function $R_A$ are equal, within the numerical uncertainties.  This provides some support for the conjecture that the remainder function only depends on the conformal cross-ratios.

\begin{table}
\label{tab:RemainderTable}
\caption{The numerical remainder compared with the ABDK/BDS ansatz for the kinematic points in eq.~\eqref{eq:KinematicPoints}. The second column gives the conformal cross ratios defined in eq.~\eqref{eq:SixPtConformalCrossRatios}.  This table is taken from ref.~\cite{Bern:2008ap}.}

\begin{tabular}{||c|c||c||}
\hline
\hline
kinematic point & $(u_1, u_2, u_3)$ & $R_A$    \\
\hline
\hline
$K^{(0)}$ & $(1/4, 1/4, 1/4)$  & $ 1.0937 \pm  0.0057$   \\
\hline
$K^{(1)}$ & $(1/4, 1/4, 1/4)$  & $  1.076 \pm 0.022$  \\
\hline
$K^{(2)}$ &  $\, (0.547253,\, 0.203822,\, 0.881270) \,$ 
                            & $ -1.659 \pm  0.014$   \\
\hline
$K^{(3)}$ & $(28/17, 16/5, 112/85)$  & $ -3.6508 \pm 0.0032\,$  \\
\hline
$K^{(4)}$ & $(1/9, 1/9, 1/9)$ & $  5.21  \pm   0.10$  \\
\hline
$K^{(5)}$ & $(4/81, 4/81, 4/81)$ &  $ 11.09 \pm 0.50 $ \\
\hline
\end{tabular}
\end{table}

Let us now discuss the comparison with the Wilson loop computations.  For the Wilson loop, one can use an analog of the ABDK/BDS ansatz and define a remainder function at two loops by
\begin{equation}
  R_W = W_6^{(2)} - W_6^{\text{BDS}}.
\end{equation}  The fact that $R_W$ is non-zero was discovered in ref.~\cite{Drummond:2007bm}.  The two remainder functions $R_A$ and $R_W$ differ by an inessential constant.  In the collinear limit $R_A$ vanishes since the collinear singularities are correctly taken into account by the ABDK/BDS ansatz (see sec.~\ref{ch:abdk-bds}).  This constant can be determined by taking a collinear limit in the Wilson loop result, but it turns out to be better from the point of view of numerical errors to compare the differences of $R_A$ and $R_W$ at two different kinematic points.  The results are presented in table~\ref{tab:ComparisonTable}.  The agreement between the third and fourth column provides strong numerical evidence for the equality of the finite parts of the scattering amplitudes and Wilson loops.

\begin{table}
\label{tab:ComparisonTable}
\caption{The comparison between the remainder functions $R_A$ and $R_W$ for the MHV amplitude and the Wilson loop.  To account for various constants of the kinematics, we subtract from the remainders their values at the standard kinematic point $K^{(0)}$, denoted by $R_A^0$ and $R_W^0$.  The third column contains the difference of remainders for the amplitude, while the fourth column has the corresponding difference for the Wilson loop.  The numerical agreement between the third and fourth columns provides strong evidence that the finite remainder for the Wilson loop is identical to that for the MHV amplitude.  This table is taken from ref.~\cite{Bern:2008ap}.}

\begin{tabular}{||c|c||c|c||}
\hline
\hline
kinematic point & $(u_1, u_2, u_3)$ & $R_A - R_A^0$ & $R_W - R_W^0$ \\
\hline
\hline
$K^{(1)}$ &$(1/4, 1/4, 1/4)$ & $-0.018 \pm  0.023 $ &  $ <10^{-5}$  \\
\hline
$K^{(2)}$ & $\,(0.547253,\, 0.203822,\, 0.881270)\,$  & $-2.753 \pm 0.015$ 
 & $ -2.7553$ \\
\hline
$K^{(3)}$ & $(28/17, 16/5, 112/85)$ & $\, -4.7445 \pm  0.0075\, $ & $ -4.7446$ \\
\hline
$K^{(4)}$ & $(1/9, 1/9, 1/9)$ & $4.12 \pm 0.10$ & $4.0914$  \\
\hline
$K^{(5)}$ & $(4/81, 4/81, 4/81)$ & $10.00 \pm 0.50$ & $9.7255$  \\
\hline
\end{tabular}
\end{table}

\chapter{Conclusion}

In conclusion, the $\mathcal{N}=4$ scattering amplitudes were shown to exhibit very interesting properties.  These amplitudes seem to satisfy a dual conformal symmetry whose origin remains mysterious.  Their even part has the interesting property of being expressible as a sum of conformal integrals, where the conformal transformations act in momentum space.

The duality between scattering amplitudes and Wilson loops is supported by weak- and strong-coupling arguments, and also by general arguments about the structure of IR divergences of scattering amplitudes and UV divergences of Wilson loops.

The BDS ansatz was shown to fail for six-point amplitudes and a remainder function was defined.  This remainder function seems to be dual-conformal invariant, i.e. it depends only on conformal cross-rations.

There are some remaining issues that need to be addressed
\begin{itemize}
\item There is by now fairly convincing numerical evidence for the equality at weak coupling of scattering amplitudes and Wilson loops.  It would be better to have analytical expressions for the remainder function, which is not constrained by dual conformal symmetry.  This function can be computed from triply collinear splitting functions as detailed in ref.~\cite{Bern:2008ap}.  The integrals appearing in the Wilson loop computation are much simpler and here a direct approach, combined with a matching to an ansatz of uniform transcendentality might work.
\item Once the remainder function is found, it would be interesting to try to correct the BDS ansatz.
\item If there really is a scattering amplitudes--Wilson loops duality, is there an underlying symmetry, beyond the dual conformal symmetry, from which this duality follows?  Obviously, this can't be a symmetry of the underlying Lagrangian because it only holds in the planar limit.  The appearance of the cusp anomalous dimension in the BDS ansatz and in the Wilson loop computations hints that integrability might play a role.
\item Is there a weak coupling analog of the $T$-duality used by Alday and Maldacena?  This weak coupling $T$-duality would map the computation of a scattering amplitude to the computation of a Wilson loop constructed from the on-shell momenta of the scattered particles.  It would also help to understand the origin of the dual conformal symmetry of the scattering amplitudes.  See ref.~\cite{McGreevy:2008zy} for a recent discussion of the link between Wilson loops and scattering amplitudes.
\item It would be useful to have analytic or at least numerical information at strong coupling beyond four-point amplitudes.
\item Is it possible to prove the equivalence between Wilson loops and scattering amplitudes at strong coupling to all orders in $\frac 1 {\sqrt{\lambda}}$?
\item It would be interesting to find an operator definition of the collinear anomalous dimension and a BES-like equation that would allow weak and strong coupling computations.  The existence of an operator interpretation is not certain because the collinear anomalous dimension depends on the regularisation scheme.  See ref.~\cite{Dixon:2008gr} for a very recent discussion on the collinear anomalous dimension.
\item Until now, the search for an iteration relation was restricted to MHV amplitudes.  Does an iteration relation hold for non-MHV amplitudes?  The first step towards answering this question is a computation of two-loop six-point NMHV amplitudes.
\end{itemize}

\appendix
\part{Appendix}

\chapter{Spinor Conventions}
\label{ch:spinor_conventions}

There are a large number of different conventions for representing Weyl spinors.  We will use the conventions of ref.~\cite{Sohnius:1985qm}.

The metric signature is $(+---)$ and $\epsilon_{0 1 2 3} = 1$.  We use the following rules for raising indices
\begin{gather}
  \psi^\alpha = \epsilon^{\alpha \beta} \psi_\beta,\\
  \overline{\psi}^{\dot{\alpha}} = \overline{\psi}_{\dot{\beta}}
  \epsilon^{\dot{\beta} \dot{\alpha}},
\end{gather} where $\alpha$, $\beta$ are two dimensional spinor
indices and $\epsilon^{\alpha \beta}$ is an antisymmetric tensor,
$\epsilon^{1 2} = -\epsilon^{2 1} = 1$.

Then, we lower indices via,
\begin{gather}
  \psi_\beta = \psi^\alpha \epsilon_{\alpha \beta},\\
  \overline{\psi}_{\dot{\beta}} = \epsilon_{\dot{\beta} \dot{\alpha}}
  \overline{\psi}^{\dot{\alpha}},
\end{gather} where $\epsilon_{1 2} = -\epsilon_{2 1} = 1$.

The rules above are consistent because
\begin{gather}
  \epsilon^{\alpha \beta} = \epsilon^{\alpha \gamma} \epsilon^{\beta
    \delta} \epsilon_{\gamma \delta},\\
  \epsilon^{\dot{\alpha} \dot{\beta}} = \epsilon_{\dot{\gamma}
    \dot{\delta}} \epsilon^{\dot{\gamma} \dot{\alpha}}
  \epsilon^{\dot{\delta} \dot{\beta}}.
\end{gather}

We also define complex conjugation by
\begin{gather}
  \left(\psi^\alpha\right)^* = \overline{\psi}^{\dot{\alpha}},\\
  \left(\psi_\alpha\right)^* = \overline{\psi}_{\dot{\alpha}}.
\end{gather}

In order for raising and lowering of indices to be compatible with complex conjugation, we need to have $\left(\epsilon^{\alpha \beta}\right)^* = \epsilon^{\dot{\beta} \dot{\alpha}}$.  This implies that $\epsilon_{\dot{1} \dot{2}} = \epsilon_{\dot{2} \dot{1}}
= -1$.

The products $\psi^\alpha \chi_\beta$ are $SL(2, \mathbb{C})$ (and therefore Lorentz) invariant.  Let us introduce some notation for them
\begin{gather}
  \langle \psi \chi\rangle  = \psi^\alpha \chi_\alpha,\\
  \left[\overline{\psi} \overline{\chi}\right] =
  \overline{\psi}_{\dot{\alpha}} \overline{\chi}^{\dot{\alpha}}.
\end{gather}  Note that the these spinor products are antisymmetric $\langle \psi \chi\rangle = - \langle \chi \psi\rangle$ and in particular $\langle \psi \psi\rangle = 0$.

The Schouten identity is a important identity that can sometimes be used to dramatically simplify expressions involving spinor products.  It can be proven by observing that there is no rank-three completely antisymmetric tensor.  Therefore, $\psi_{[\alpha} \chi_\beta \rho_{\gamma]} = 0$.  It is easy to see that this implies that
\begin{equation}
  \label{eq:schouten}
  \langle \psi \chi\rangle \langle \rho \mu\rangle + \langle \psi \rho\rangle \langle \mu \chi\rangle + \langle \psi \mu\rangle \langle \chi \rho\rangle = 0.
\end{equation}  An easy way to remember this identity is to observe that the first spinor remains the same while the last three are permuted circularly.  A similar identity exists for the $[\,]$ spinor products.

We define a matrix-valued vector, $\sigma^\mu = (\mathbf{1}, \vec{\sigma})$, where the three-dimensional vector $\vec{\sigma}$ has as components the Pauli matrices.  More precisely, this defines the $\sigma$ matrices with the index structure $\sigma^\mu_{\alpha \dot{\alpha}}$.  Then, one can define the matrices $\overline{\sigma}$ by $\left(\bar{\sigma}^\mu\right)^{\dot{\alpha} \beta} = \left(\sigma^\mu\right)^{\beta \dot{\alpha}}$.  Note that here the matrices $\sigma^\mu$ have raised indices (they are obtained from $\sigma^\mu = (\mathbf{1}, \vec{\sigma})$ by using the rules of index raising).  It is easy to prove that $\overline{\sigma}^\mu = (\mathbf{1}, -\vec{\sigma})$, where the index structure of the matrix $\overline{\sigma}^\mu$ is $\left(\overline{\sigma}^\mu\right)^{\dot{\alpha} \beta}$.

Here are some more identities involving the matrices $\sigma^\mu$ and $\overline{\sigma}^\nu$,
\begin{align}
  (\sigma^\mu \overline{\sigma}^\nu + \sigma^\nu
  \overline{\sigma}^\mu)_\alpha^{~\beta} =& 2 \eta^{\mu \nu}
  \delta_\alpha^\beta,\\
  (\overline{\sigma}^\mu \sigma^\nu + \overline{\sigma}^\nu
  \sigma^\mu)^{\dot{\alpha}}_{~\dot{\beta}} =& 2 \eta^{\mu \nu}
  \delta_{\dot{\alpha}}^{\dot{\beta}},\\
  \tr{\sigma^\mu \overline{\sigma}^\nu} =& 2 \eta^{\mu \nu},\\
  \sigma^\mu_{\alpha \dot{\alpha}} \overline{\sigma}_\mu^{\dot{\beta}
    \beta} =& 2 \delta_\alpha^\beta
  \delta_{\dot{\alpha}}^{\dot{\beta}}.
\end{align}  The last two identities are sometimes called completeness
relations.  They are used to translate back and forth between the two
languages (spinor and vector)
\begin{equation}
  \label{eq:spinor-vector_translation}
  v_{\alpha \dot{\alpha}} = \sigma^\mu_{\alpha \dot{\alpha}} v_\mu, \quad
  v^\mu = \frac 1 2 (\overline{\sigma}^\mu)^{\dot{\alpha} \alpha} v_{\alpha
    \dot{\alpha}}.
\end{equation}

Starting from the transformation of a vector in spinor language, we
define the transformation of the tensor $\eta_{\mu \nu}$ in spinor
language
\begin{equation}
  \eta_{\alpha \dot{\alpha} \beta \dot{\beta}} = \sigma^\mu_{\alpha
    \dot{\alpha}} \sigma^\nu_{\beta \dot{\beta}} \eta_{\mu \nu}.
\end{equation}  Of course, $\eta$ is an invariant tensor so it must be
expressible in terms of the invariant tensors $\epsilon_{\alpha
  \beta}$ and $\epsilon_{\dot{\alpha} \dot{\beta}}$.  An ansatz that
has the right index structure and symmetries is $\eta_{\alpha
  \dot{\alpha} \beta \dot{\beta}} \propto \epsilon_{\alpha
  \beta} \epsilon_{\dot{\alpha} \dot{\beta}}$.  The proportionality
constant can be found by contracting with $\epsilon_{\alpha
  \beta} \epsilon_{\dot{\alpha} \dot{\beta}}$ and it turns out to be
equal to minus two.  Thus,
\begin{equation}
  \eta_{\alpha \dot{\alpha} \beta \dot{\beta}} = - 2 \epsilon_{\alpha
    \beta} \epsilon_{\dot{\alpha} \dot{\beta}}.
\end{equation}  Similarly, the $\eta$ tensor with upper indices turns
out to be
\begin{equation}
  \eta^{\alpha \dot{\alpha} \beta \dot{\beta}} = - \frac 1 2 \epsilon^{\alpha
    \beta} \epsilon^{\dot{\alpha} \dot{\beta}}.
\end{equation}  One minor annoyance of this convention is that the
raising of spinor and vector indices are not compatible upon
translation from vector to spinor language.  In order to have this,
one should define the invariant tensors $\epsilon_{\alpha \beta}$ and
$\epsilon_{\dot{\alpha} \dot{\beta}}$ to include a factor of
$\sqrt{2}$ and the corresponding upper index tensors to include a
factor of $\tfrac 1 {\sqrt{2}}$.  Alternatively, one can choose to use
a $\tfrac 1 {\sqrt{2}}$ in front of both equations in
eq.~\ref{eq:spinor-vector_translation}.

Consider now two light-like vectors $p_{\alpha \dot{\alpha}} =
\lambda_\alpha \overline{\lambda}_{\dot{\alpha}}$ and $q_{\beta
  \dot{\beta}} = \mu_\beta \overline{\mu}_{\dot{\beta}}$.  Compute
their dot product using the formulae above for the $\eta$ tensor in
spinor language
\begin{equation}
  \label{eq:dot_to_spinor}
  p \cdot q = \eta^{\mu \nu} p_\mu q_\nu = \eta^{\alpha \dot{\alpha}
    \beta \dot{\beta}} \lambda_\alpha
  \overline{\lambda}_{\dot{\alpha}} \mu_\beta
  \overline{\mu}_{\dot{\beta}} = \frac 1 2 \langle \lambda \mu\rangle
  \left[\overline{\mu} \overline{\lambda}\right].
\end{equation}

Let us now introduce the expressions for the polarisation tensors in
spinor language.  They are
\begin{gather}
  \epsilon^-_{\alpha \dot{\alpha}}(p;q) = \sqrt{2} \frac {\lambda_\alpha \overline{\mu}_{\dot{\alpha}}}{\left[\overline{\lambda} \overline{\mu}\right]},\\
  \epsilon^+_{\alpha \dot{\alpha}}(p;q) = \sqrt{2} \frac {\mu_\alpha \overline{\lambda}_{\dot{\alpha}}}{\langle \mu \lambda \rangle},
\end{gather} where $\lambda$, $\overline{\lambda}$ are the spinor
corresponding to the momentum of the photon (gluon) and $q$ is a
light-like reference vector with corresponding spinors $\mu$ and
$\overline{\mu}$.

It is obvious that the polarisation vectors $\epsilon^\pm$ are
light-like.  Using the formula in eq.~\ref{eq:dot_to_spinor} it is
easy to prove that $\epsilon^+(p;q) \cdot \epsilon^-(p;q') = -1$.

It is obvious that
\begin{gather}
  p \cdot \epsilon^\pm(p;q) = 0,\\
  q \cdot \epsilon^\pm(p;q) = 0,\\
  \epsilon^\pm(k; q) \cdot \epsilon^\pm(p; q) = 0,\\
  \left(\epsilon^\pm_{\alpha \dot{\beta}}\right)^* =
  \epsilon^\mp_{\beta \dot{\alpha}}.
\end{gather}

It is interesting to also compute the the sum $\epsilon_\mu^+(p; q)
\epsilon_\nu^-(p; q) + \epsilon_\nu^-(p; q) \epsilon_\mu^+(p; q)$,
which is the projector on the physical states in a certain gauge.  In
spinor language this is
\begin{align*}
  & \frac 2 {\langle \mu \lambda \rangle \left[\overline{\lambda}
      \overline{\mu}\right]} \left(\mu_\alpha
    \overline{\lambda}_{\dot{\alpha}} \lambda_\beta
    \overline{\mu}_{\dot{\beta}} + {}^{\alpha \leftrightarrow
      \beta}_{\dot{\alpha} \leftrightarrow \dot{\beta}}\right) =\\
  & = \frac 2 {\langle \mu \lambda \rangle \left[\overline{\lambda}
      \overline{\mu}\right]} \left( (\mu_\alpha \lambda_\beta -
    \mu_\beta \lambda_\alpha) (\overline{\lambda}_{\dot{\alpha}}
    \overline{\mu}_{\dot{\beta}} - \overline{\lambda}_{\dot{\beta}}
    \overline{\mu}_{\dot{\alpha}}) + \mu_\alpha \lambda_\beta
    \overline{\lambda}_{\dot{\beta}} \overline{\mu}_{\dot{\alpha}} +
    \mu_\beta \lambda_\alpha \overline{\lambda}_{\dot{\alpha}}
    \overline{\mu}_{\dot{\beta}}\right) =\\
  & = - 2 \epsilon_{\alpha \beta} \epsilon_{\dot{\alpha} \dot{\beta}}
  + 2 \frac {\lambda_\beta \overline{\lambda}_{\dot{\beta}} \mu_\alpha
    \overline{\mu}_{\dot{\alpha}} + \lambda_\alpha
    \overline{\lambda}_{\dot{\alpha}} \mu_\beta
    \overline{\mu}_{\dot{\beta}}}{\langle \mu \lambda \rangle
    \left[\overline{\lambda} \overline{\mu}\right]},
\end{align*} where we have used
\begin{gather}
  \lambda_\alpha \mu_\beta - \lambda_\beta \mu_\alpha = - \langle
  \lambda \mu\rangle \epsilon_{\alpha \beta},\\
  \overline{\lambda}_{\dot{\alpha}} \overline{\mu}_{\dot{\beta}} -
  \overline{\lambda}_{\dot{\beta}} \overline{\mu}_{\dot{\alpha}} =
  \left[\overline{\lambda} \overline{\mu}\right]
  \epsilon_{\dot{\alpha} \dot{\beta}}.
\end{gather}

If we now translate back to the vector language we obtain
\begin{equation}
  \epsilon_\mu^+(p; q) \epsilon_\nu^-(p; q) + \epsilon_\nu^-(p; q)
  \epsilon_\mu^+(p; q) = - \eta_{\mu \nu} + \frac {p_\mu q_\nu + p_\nu
  q_\mu}{p \cdot q},
\end{equation} which corresponds to the light-cone gauge with light-cone vector $q$.

\chapter{Wavefunctions}
\label{ch:wavefunctions}

When computing amplitudes by using traditional Feynman diagram approach one needs to use wavefunctions for external lines.  In this section we will describe the wavefunctions for particles of spin-$\tfrac 1 2$ through spin-two.

The spin $\tfrac 1 2$ wavefunctions are the simplest: for a on-shell momentum $p$ such that $p_{\alpha \dot{\alpha}} = \lambda_\alpha \overline{\lambda}_{\dot{\alpha}}$, the helicity $-\tfrac 1 2$ wavefunction in momentum space is\footnote{Here and below we leave out labels such as colour, etc.} $\lambda_\alpha$, while the helicity $\tfrac 1 2$ wavefunction is $\overline{\lambda}_{\dot{\alpha}}$.  It is easy to prove that the Weyl equations for each chirality are satisfied by the above wavefunctions.

The wavefunctions for spin one particles are more complicated because of gauge structure.  We need to choose a light-like reference vector $q_{\alpha \dot{\alpha}} = \mu_\alpha \overline{\mu}_{\dot{\alpha}}$ and then the polarisation vectors are
\begin{equation}
  \epsilon_{\alpha \dot{\alpha}}(p, -; q) = \sqrt{2} \frac {\lambda_\alpha \overline{\mu}_{\dot{\alpha}}}{\left[\overline{\lambda}\  \overline{\mu}\right]}, \qquad \epsilon_{\alpha \dot{\alpha}}(p, +; q) = \sqrt{2} \frac {\mu_\alpha \overline{\lambda}_{\dot{\alpha}}}{\langle \mu\ \lambda\rangle}.
\end{equation}

Transformations of $q$, or equivalently of $\mu$ and $\overline{\mu}$ produce gauge transformations on spin one wavefunctions.  In fact, $\mu_\alpha$ and $\lambda_\alpha$ are a basis in the vector space of two-dimensional spinors (they are independent since otherwise $\langle\lambda\ \mu\rangle = 0$).  Therefore, $\delta \mu_\alpha = A \mu_\alpha + B \lambda_\alpha$.  Plugging this into the formula for $\epsilon_{\alpha \dot{\alpha}}(p, +; \mu)$ yields $\delta \epsilon_\mu(p, +) \propto p_\mu$ which is the usual gauge transformation for spin one particles.  The same analysis can be done for $\epsilon_{\alpha \dot{\alpha}}(p, -; \mu)$.

For spin $\tfrac 3 2$ wavefunctions we take
\begin{equation}
  \psi_{\mu \alpha}\left(p, -\frac 3 2\right) = \epsilon_\mu(p, -) \lambda_\alpha, \qquad \psi_{\mu \dot{\alpha}}\left(p, +\frac 3 2\right) = \epsilon_\mu(p, +) \overline{\lambda}_{\dot{\alpha}}.
\end{equation}  Note that these wavefunctions are irreducible because, contracting the free spinor index with a spinor index of the same type inside the polarisation vector $\epsilon$ yields zero.  The other possible choices $\psi_{\mu \alpha}(p, +\frac 1 2) = \epsilon_\mu(p, +) \lambda_\alpha$ and $\psi_{\mu \dot{\alpha}}(p, -\frac 1 2) = \epsilon_\mu(p, -) \overline{\lambda}_{\dot{\alpha}}$ are not irreducible, so they are not part of the spin $\tfrac 3 2$ wavefunction.  In this language, the constraints one imposes on a Rarita-Schwinger field are very easy to understand.

The spin-$\tfrac 3 2$ wavefunctions also have a gauge invariance arising from that of the spin-one component (the polarisation vector).

Finally, for spin two we choose $h_{\mu \nu}(p, \pm 2) = \epsilon_\mu(p, \pm) \epsilon_\nu(p, \pm)$.  This is symmetric and traceless since $\epsilon(p, \pm) \cdot \epsilon(p, \pm) = 0$.  The fact that the graviton wavefunction is the `square' of the photon wavefunction is an important observation that also carries over to the level of the graviton amplitudes.  This is embodied in the Kawai-Lewellen-Tye relations from string theory.

\chapter{IR Divergences}
\label{ch:ir_divergencies}

As already alluded above, the on-shell scattering amplitudes in massless theories are IR divergent.  However, the structure of the IR divergences is tightly constrained and it has proven to be an useful guide to checking the computations and as a source of conjectures.

So, even if the study of IR divergences is not a computational tool in itself, it is remarkably effective at checking the correctness of a computation as it imposes highly non-trivial constraints on the results.

Let us start with a discussion of the IR singularities at tree level.  The
discussion in this section is inspired by the ref.~\cite{Mangano:1990by}.  We will restrict to gluon amplitudes in the following.

There are two kinds of IR singularities: soft and collinear.  For the soft singularities the momentum of one of the gluons goes to zero (in the sense that all its components go to zero at the same speed).  For the collinear singularities, the momenta of two neighbouring gluons become parallel.

It turns out that one of the most direct ways to compute these IR singularities is to use the Koba-Nielsen representation of the amplitudes, which is derived in string theory.  If the string worldsheet is mapped onto the upper half-plane and the vertex operators corresponding to momenta and polarisations $k_i$ and $\epsilon_i$, with $i = 1, \ldots, n$ are placed on the boundary (the real axis) at points $z_i$ such that $z_1 < z_2 \cdots < z_n$, then the scattering amplitude is
\begin{equation}
  \label{eq:koba_nielsen}
  A^{\text{tree}} = \int_{z_1 < z_2 < \cdots < z_n} \prod_{i=3}^{n-1} d z_i \mu_{KN} \prod_{1 \leq j < i \leq n} (z_i - z_j)^{k_i \cdot k_j} \exp \sum_{i \neq j} \left( \frac 1 2 \frac {\epsilon_i \cdot \epsilon_j}{(z_i - z_j)^2} + \frac {k_i \cdot \epsilon_j}{z_i - z_j}\right),
\end{equation} where $\mu_{KN} = (z_2 - z_1)(z_n - z_1)(z_n - z_2)$ is the Koba-Nielsen Jacobian arising from fixing the positions of the vertex operators corresponding to particles $1$, $2$ and $n$.  Above we have omitted the factors of $\alpha'$ by setting $\alpha' = \tfrac 1 2$; these factors are necessary when taking the zero slope limit but can be reinstated by dimensional analysis (for example, one should do the following replacements above $k_i \cdot k_j \rightarrow 2 \alpha' k_i \cdot k_j$, $\epsilon_i \cdot k_j \rightarrow \sqrt{2 \alpha'} \epsilon_i \cdot k_j$).  Also, the exponential in the above expression is to be expanded and only terms which are multilinear in the polarisations vectors should be kept.\footnote{This arises as follows: one needs to compute the correlation function of vector vertex operators $V(\epsilon, k, z) = \epsilon \cdot \dot{X}(z) \exp \left(i k \cdot X(z)\right)$ (where $k^2 = \epsilon \cdot k = 0$ and $z$ is a coordinate on the boundary of the string worldsheet) on the upper half-plane and after computing this correlation function one needs to integrate over the positions of these vertex operators while imposing a cyclic ordering.  By fixing the position of three vertex operators one fixes the residual $SL(2, \mathbb{R})$ gauge invariance which yields a Faddeev-Popov Jacobian equal to $\mu_{KN}$.

When computing the correlation function of the vertex operators it is convenient to work with modified vertex operators $\tilde{V} =  \exp \left(i k \cdot X(z) + \epsilon \cdot \dot{X}(z)\right)$ because, in this case, one can compute the correlation functions using the formula
\begin{equation*}
  \left \langle e^{\mathcal{O}_1} e^{\mathcal{O}_2} \cdots e^{\mathcal{O}_n} \right \rangle = \exp \left(\sum_{i < j} \langle \mathcal{O}_i \mathcal{O}_j\rangle\right),
\end{equation*} for operators $\mathcal{O}_i$ which are linear in $X$.  The result we seek is then obtained by expanding `to first order in $\epsilon$.'}

Let us discuss the soft gluon singularities using the above formalism (see~\cite{Mangano:1990by}).  Gauge fix $z_1 = 0$, $z_2 = 1$ and $z_n = \infty$.  Let the soft gluon have momentum $p$ and polarisation $\zeta$ and insert it at $w$, such that $0 < w < 1$.  Then, the Koba-Nielsen formula becomes
\begin{equation}
\begin{aligned}
  \label{eq:koba_nielsen_soft}
  \int_{0 < w < 1 < z_3 < \cdots} d w & \prod_{i=3}^{n-1} d z_i \mu_{KN} \prod_{1 \leq j < i \leq n} (z_i - z_j)^{k_i \cdot k_j} \cancel{\prod_{2 < i \leq n} (z_i - w)^{k_i \cdot p}} w^{k_1 \cdot p} (1 - w)^{k_2 \cdot p}\\ \exp \Bigg( & \sum_{i \neq j} \bigg(\frac 1 2 \frac {\epsilon_i \cdot \epsilon_j}{(z_i - z_j)^2} + \frac {k_i \cdot \epsilon_j}{z_i - z_j}\bigg) + \\ & \sum_{i = 1}^n \bigg( \frac {\epsilon_i \cdot \zeta}{(z_i - w)^2} + \frac {k_i \cdot \zeta}{z_i - w} - \cancel{\frac {p \cdot \epsilon_i}{z_i - w}} \bigg)\Bigg),
\end{aligned} \end{equation} where we crossed out the terms that do not contribute to the singularity when $p \rightarrow 0$.

Also, it is obvious that the only source of singularities are the zones in the integration region where two or more $z$ coordinates come close together.  The zones where more than two $z$ come close together yield multiparticle poles, so if we want to separate the IR singularities from the multiparticle poles we need to consider only the cases where two $z$ coordinates come close together.  Then, the cyclic ordering imposes that the only possible sources of singularities are from the regions where $w \sim 0$ and $w \sim 1$.

This implies that the last line of (eq.~\eqref{eq:koba_nielsen_soft}) can be replaced by
\begin{equation*}
  \frac {\epsilon_1 \cdot \zeta}{w^2} + \frac {\epsilon_2 \cdot \zeta}{(1 - w)^2} - \frac {k_1 \cdot \zeta} w + \frac {k_2 \cdot \zeta}{1 - w}
\end{equation*} if we are concerned only with soft singularities.

It is then easy to see that, in the $p \rightarrow 0$ limit, the initial $n+1$ point amplitude factorises into an $n$ point amplitude and an singular factor we will now compute.  The integrals over $w$ have the general form
\begin{equation*}
  B(k_1 \cdot p + a + 1, k_2 \cdot p + b + 1) = \int_0^1 d w w^{k_1 \cdot p + a} (1 - w)^{k_2 \cdot p + b} \sim_{p \rightarrow 0} B(a + 1, b + 1),
\end{equation*} where $B(x,y) = \frac {\Gamma(x) \Gamma(y)}{\Gamma(x+y)}$ is Euler's beta function.  Recalling that for $k \in \mathbb{N}$
\begin{equation*}
  \Gamma(-k + z) \sim_{z \rightarrow 0} \frac {(-1)^k}{k !} \frac 1 z,
\end{equation*} we finally obtain
\begin{equation}
  \label{eq:ir_soft}
  A_{n+1}^{\text{tree}}(1; p, \zeta; 2; \cdots n) \sim_{p \rightarrow 0} \left(\frac {\zeta \cdot k_2}{p \cdot k_2} - \frac {\zeta \cdot k_1}{p \cdot k_1}\right) A_n^{\text{tree}}(1; \cdots n).
\end{equation}

This can be expressed using spinor language (see ch.~\ref{ch:spinor_conventions} and sec.~\ref{sec:spinor_helicity}).  If the soft momentum $p$ can be expressed in spinor language using spinors $\lambda$ and $\overline{\lambda}$ and momenta $p_i$ can be described using spinors $\lambda^i$ and $\overline{\lambda}^i$ then, for a helicity plus soft gluon, 
\begin{equation}
  \left(\frac {\zeta \cdot k_2}{p \cdot k_2} - \frac {\zeta \cdot       k_1}{p \cdot k_1}\right) = \sqrt{2} \frac {\langle 1 2\rangle}{\langle 1 \lambda\rangle \langle \lambda 2\rangle}.
\end{equation}  This can be proven by writing out the expressions for the polarisation vectors, using the rules for index contraction and, finally, using the Schouten identity, eq.~\eqref{eq:schouten}.  Another simpler method is to choose the reference momentum in the polarisation tensor in order to get one of the two terms to cancel.

The final result is 
\begin{equation}
  \label{eq:ir_soft_spinors}
  A_{n+1}^{\text{tree}}(1; p, h; 2; \cdots n) \sim_{p \rightarrow 0}
  \begin{cases}
    \sqrt{2} \frac {\langle 1 2\rangle}{\langle 1 \lambda\rangle \langle \lambda 2\rangle} A_n^{\text{tree}}(1; \cdots n), \quad \text{if}\quad h=+1 \\
    - \sqrt{2} \frac {[1 2]}{[1 \tilde{\lambda}] [ \tilde{\lambda} 2 ]} A_n^{\text{tree}}(1; \cdots n), \quad     \text{if}\quad h = -1.
  \end{cases}
\end{equation}

Let us now turn to collinear singularities, where two momenta become parallel.  Here also we will use the Koba-Nielsen language.  Let momenta $p_1$ and $p_2$ become parallel and $P = p_1 + p_2$; denote the corresponding polarisations by $\zeta_1$ and $\zeta_2$.  Insert
the vertex operator of the first particle at $w_1 = 0$ and of the second at $w_2 = w$ and let the cyclic ordering be defined by $0 < w < z_1 = 1 < z_2 < \cdots < z_n = \infty$.  Then, the Koba-Nielsen amplitude is
\begin{equation}
  \begin{aligned}
  \label{eq:koba-nielsen_collinear}
  \int_{0 < w < z_1 = 1 < z_2 < \cdots < z_n} & d w \prod_{i=2}^{n-1} d z_i \mu_{KN} \prod_{1 \leq j < i \leq n} (z_i - z_j)^{k_i \cdot k_j} w^{p_1 \cdot p_2} \prod_{i=1}^n z_i^{p_i \cdot p_1} (z_i - w)^{p_i \cdot p_2}\\ \exp \Bigg( \sum_{i \neq j}& \bigg(\frac 1 2 \frac {\epsilon_i \cdot \epsilon_j}{(z_i - z_j)^2} + \frac {k_i \cdot \epsilon_j}{z_i - z_j}\bigg) + \frac {\zeta_1 \cdot \zeta_2}{w^2} + \frac {p_2 \cdot \zeta_1 - p_1 \cdot \zeta_2} w + \\ \sum_{i=1}^n & \bigg( \frac {\epsilon_i \cdot \zeta_1} {z_i^2} + \frac {k_i \cdot \zeta_1 - p_1 \cdot \epsilon_i}{z_i} \bigg) + \sum_{i=1}^n \bigg( \frac {\epsilon_i \cdot \zeta_2} {(z_i - w)^2} + \frac {k_i \cdot   \zeta_2 - p_2 \cdot \epsilon_i}{z_i - w} \bigg)\Bigg).
  \end{aligned}
\end{equation}

Now we want to expand the exponent above in Laurent series in $w$.  We also need to remember that, when expanding the exponential, the quadratic and higher order terms in each of the polarisations should be neglected.  Also, we won't touch the first sum in the exponent since that will be used to form the remaining amplitude after the collinear limit and we will neglect the terms of order $\alpha'$ such as $(p_1 \cdot \zeta_2) (p_2 \cdot \zeta_1)$.

First we we make the transformations
\begin{equation}
  \label{eq:collinear_transf1}
  \prod_{i=1}^n z_i^{p_i \cdot p_1} (z_i - w)^{p_i \cdot p_2} = \prod_{i=1}^n z_i^{P \cdot p_i} \exp \left(- w \sum_{i=1}^n \frac {p_2 \cdot k_i}{z_i} + \mathcal{O}(w^2)\right).
\end{equation}

Then, the exponent in (eq.~\eqref{eq:koba-nielsen_collinear}) except the first sum can be rewritten
\begin{align*}
  \label{eq:collinear_transf2}
  \frac {\zeta_1 \cdot \zeta_2}{w^2} + \frac {p_2 \cdot \zeta_1 - p_1 \cdot \zeta_2} w + \\ (\zeta_1 + \zeta_2) \cdot \sum_{i=1}^n & \bigg( \frac {\epsilon_i} {z_i^2} + \frac {k_i}{z_i} \bigg) - P \cdot \sum_{i=1}^n \frac {\epsilon_i} {z_i} - w \sum_{i=1}^n \frac {p_2 \cdot \epsilon_i}{z_i^2}.
\end{align*}

If to this equation we add the exponent in (eq.~\eqref{eq:collinear_transf1}), after the expansion of exponential we can rewrite all the $w$ dependence as (for $w \sim 0$)
\begin{equation*}
  w^{p_1 \cdot p_2} \left[\frac {\zeta_1 \cdot \zeta_2}{w^2} - \frac 1 w \left( (\zeta_1 \cdot \zeta_2) p_2^\mu + (p_1 \cdot \zeta_2) \zeta_1^\mu - (p_2 \cdot \zeta_1) \zeta_2^\mu\right) \sum_{i=1}^n \left(\frac {(\epsilon_i)_\mu}{z_i^2} + \frac {(k_i)_\mu}{z_i}\right)\right].
\end{equation*}

The expression above can only be trusted when $w \sim 0$ and we will only integrate it in a neighbourhood of zero.  This will be sufficient to compute the singularity.  There are two integrals to be computed.  The first is
\begin{equation*}
  \int_0 d w\ w^{p_1 \cdot p_2 - 2}.
\end{equation*} This integral does not always converge.  We choose to define it for values of the kinematical invariant $p_1 \cdot p_2$ for which it is convergent in $w \rightarrow 0$ and then to analytically continue in the kinematical invariant to values for which the integral is divergent.  When defined in this way, the integral above will actually yield a pole corresponding to the exchange of a tachyon coming from a vector-vector-tachyon interaction.  However, this pole is not in the limit we are interested in (where $p_1$ and $p_2$ are on-shell and parallel) and is clearly non-physical so we will neglect this contribution.

The second integral
\begin{equation*}
  \int_0 d w\ w^{p_1 \cdot p_2 - 1} \equiv - \frac 1 {p_1 \cdot p_2}
\end{equation*} yields a result which is divergent in the collinear limit.  Then,
\begin{equation}
  \begin{aligned}
    \label{eq:collinear_general}
  A_{n+2}^{\text{tree}}(p_1,p_2, k_1, \ldots, k_n) \xrightarrow{1 \parallel 2}&\\ \frac 1 {p_1 \cdot p_2} \Big[ (\zeta_1 \cdot \zeta_2) Q^\mu +& (p_1 \cdot \zeta_2) \zeta_1^\mu - (p_2 \cdot \zeta_1) \zeta_2^\mu \Big] \frac \partial {\partial \zeta^\mu} A_{n+1}^{\text{tree}}(P, k_1, \ldots, k_n),
  \end{aligned}
\end{equation} where $Q = \tfrac 1 2 (p_2 - p_1)$ and we have written $p_2 = \tfrac P 2 + Q$ and we have used the Ward identity
\begin{equation*}
  P^\mu \frac \partial {\partial \zeta^\mu} A_{n+1}^{\text{tree}}(P, k_1, \ldots, k_n) = 0,
\end{equation*} which is correct in the collinear limit since then $P$ is on-shell.

This can also be translated to spinor language.  Let $p_1 \sim z P$, $p_2 \sim (1 - z) P$. There are several possibilities for the helicities
\begin{align}
  A_{n+2}^{\text{tree}}(1^+, 2^+, \ldots) &\stackrel{1 \parallel     2}{\rightarrow} \frac 1 {\sqrt{z (1 - z)}} \frac 1 {\langle 1     2\rangle} A_n^{\text{tree}}(P^+, \ldots),\\
  A_{n+2}^{\text{tree}}(1^+, 2^-, \ldots) &\stackrel{1 \parallel     2}{\rightarrow} - \frac {z^2} {\sqrt{z (1 - z)}} \frac 1 {[1 2]}   A_n^{\text{tree}}(P^+, \ldots) + \frac {(1 - z)^2} {\sqrt{z (1 -       z)}} \frac 1 {\langle 1 2\rangle} A_n^{\text{tree}}(P^-, \ldots),\\
  A_{n+2}^{\text{tree}}(1^-, 2^-, \ldots) &\stackrel{1 \parallel     2}{\rightarrow} - \frac 1 {\sqrt{z (1 - z)}} \frac 1 {[1 2]}   A_n^{\text{tree}}(P^-, \ldots).
\end{align}

\chapter{Penrose transform}
\label{ch:penrose_transform}

One useful way to understand twistors is by using them to generate solutions to massless field equations in four dimensions.  The idea is to make use of the power of complex analysis.

Let us start with a simple example\footnote{The discussion in this section is partly inspired by ref.\cite{Nair:2005wh}.} of the wave equation in two dimensions.  In this case, it is useful to pass to complex coordinates $z = x_1 + i x_2$, where the Laplacian is $\partial \overline{\partial}$.  The wave equation
\begin{equation}
  \partial \overline{\partial} \phi (x) = 0,
\end{equation} has solutions,
\begin{equation}
  f(x) = g(z) + h(\overline{z}),
\end{equation} where $g$ is a holomorphic function and $h$ is an anti-holomorphic function.

This strategy does not apply in four dimensions because in this case, unlike in two dimensions, there are many choices of complex structure.  There are $O(4)/U(2) = \mathbb{CP}^1$ inequivalent complex structures, compatible with the flat space constraint. (To define a complex structure on a flat space one needs to give a linear mapping of $\mathbb{R}^4$ to $\mathbb{C}^2$ at each point.  An $U(2)$ transformation acting on $\mathbb{C}^2$ does not change the complex structure.  Therefore, the space of complex structures can be locally described as the coset above.)  This is similar in spirit to harmonic superspace constructions (see ref.~\cite{Galperin:2001uw}).

We see here that each particular choice of complex structure breaks the $O(4)$ symmetry down to $U(2)$.  In order to preserve the symmetry one should consider all the complex structures at the same time.  The way to do this is to consider a $\mathbb{CP}^1$ fiber bundle over $\mathbb{R}^4$, whose sections are in one to one correspondence with complex structures on $\mathbb{R}^4$.

Let us give an example.  Take the coordinates $x^\mu$ and define $x_{\alpha \dot{\alpha}} = \sigma_{\alpha \dot{\alpha}}^\mu x_\mu$.  Then, by choosing an element of $\mathbb{CP}^1$ given by the homogeneous coordinates $\lambda^\alpha$ with $\alpha = 1,2$ we can define two complex coordinates $\mu_{\dot{\alpha}}$, $\dot{\alpha} = \dot{1}, \dot{2}$ by
\begin{equation}
  \mu_{\dot{\alpha}} = - \lambda^\alpha x_{\alpha \dot{\alpha}}.
\end{equation}  This is the same as the twistor equation that first appeared in sec.~\ref{ch:super-twistor_space} but here we arrived at this result from another point of view.

Let us now see how to generate solutions to massless wave equations in four dimensions by using twistors.  We start with negative helicity fields, $h = - \tfrac n 2$, with $n \geq 0$.  Consider $g(\lambda, \mu)$ a function of degree of homogeneity $-n-2$ and then take the integral
\begin{equation}
  \phi_{\alpha_1 \cdots \alpha_n} = \frac 1 {2 \pi i} \oint_{\mathcal{C}} \langle \lambda\ d \lambda\rangle \lambda_{a_1} \cdots \lambda_{a_n} g(\lambda, - \lambda^\alpha x_{\alpha \dot{\alpha}}),
\end{equation} where $\mathcal{C}$ is an arbitrary integration contour.  Note also that the integrand has degree of homogeneity zero.

It is easy to see that, if $n \neq 0$ this satisfies the massless wave equation\footnote{If $n=0$ this is a scalar field and it must satisfy the Klein-Gordon equation $\partial_{\alpha \dot{\alpha}} \partial^{\alpha \dot{\alpha}} \phi =0$.  One can also check that it does satisfy this equation.}
\begin{equation}
  \partial_{\alpha \dot{\alpha}} \phi^\alpha_{~ \alpha_1 \cdots \alpha_{n-1}} = 0.
\end{equation}  The wavefunction $\phi_{\alpha_1 \cdots \alpha_n}$ is also completely symmetric and therefore transforms in an irreducible representation of the Lorentz group.  It corresponds to a massless particle with helicity $h = -\tfrac n 2$ (see also Appendix~\ref{ch:wavefunctions}).

For positive helicity consider a homogeneous function $g$ defined on the twistor space with degree of homogeneity $n-2$.  Then compute the contour integral
\begin{equation}
  \phi_{\dot{\alpha}_1 \cdots \dot{\alpha}_n} = \frac 1 {2 \pi i} \oint_{\mathcal{C}} \langle \lambda\ d \lambda\rangle \left.\frac \partial {\partial \mu^{\dot{a_1}}} \cdots \frac \partial {\partial \mu^{\dot{a_n}}} g(\lambda, \mu)\right|_{\mu_{\dot{\alpha}} = - \lambda^\alpha x_{\alpha \dot{\alpha}}}.
\end{equation}

Similarly, this wavefunction is completely symmetric and satisfies the wave equation
\begin{equation}
  \partial_{\alpha \dot{\alpha}} \phi^{\dot{\alpha}}_{~\dot{\alpha}_1 \cdots \dot{\alpha}_{n-1}} = 0.
\end{equation}  It corresponds to a state with helicity $h = \tfrac n 2$.

We have shown that to each homogeneity $-n-2$ function on the twistor space we can associate a solution of the equations of motion of a massless particle of helicity $-\tfrac n 2$ and to each homogeneity $n-2$ function on the twistor space we can associate a solution of the equations of motion of a massless particle of helicity $\tfrac n 2$.

However, this correspondence is not one-to-one.  The twistor space wavefunctions and/or the integration contour can be modified while still getting the same solutions for the equations of motion.  The integration contour separates the $\mathbb{CP}^1$ into two parts.  For example, by transforming the twistor space function $g$
\begin{equation}
  \label{eq:twistor-space-freedom}
  g \to g + h - \tilde{h},
\end{equation} where $h$ is holomorphic on one side and $\tilde{h}$ is holomorphic on the other side of the contour, the result of the integration is unchanged.  The twistor space objects that are in one-to-one correspondence with solutions of massless field equations for helicity $h$ are Cech cohomology classes of sheafs of homogeneous functions with degree of homogeneity $2 h - 2$.  Though this formulation might seem intimidating at first, it is basically the statement that the corresponding homogeneous functions in twistor space are equivalent under the transformations in eq.~\eqref{eq:twistor-space-freedom}.  See ref.~\cite{Huggett:1986fs} for more details.  The original papers where this construction was carried out are~\cite{Penrose:1968me, Penrose:1969ae}.

For early attempts to generalise twistor constructions to ten dimensions see ref.~\cite{Witten:1985nt} and to general even dimensions see refs.~\cite{Hughston}.  For recent work establishing links between higher dimensional twistors and pure spinors see ref.~\cite{Berkovits:2004bw}.

It appears that pure spinors are the closest analog of twistors in higher-dimensional spaces.  The pure spinors in $d = 2 n$ dimensions are complex commuting spinors $\lambda^a$ which satisfy the constraints
\begin{equation}
  \lambda^a \sigma_{a b}^{\mu_1 \cdots \mu_j} \lambda^b = 0,
\end{equation} for $0 \leq j < n$ where $\sigma^{\mu_1 \cdots \mu_j}$ is the antisymmetrised product of the higher dimensional Pauli matrices.  Equivalently, the pure spinor constraints can be written
\begin{equation}
  \lambda^a \lambda^b = \frac 1 {n! 2^n} \sigma_{\mu_1 \cdots \mu_n}^{a b} \left(\lambda^c \sigma_{c d}^{\mu_1 \cdots \mu_n} \lambda^d\right).
\end{equation}  It can be shown that the projective pure spinors in $d = 2 n$ dimensional Euclidean space parametrise the coset $SO(2 n)/U(n)$ and this is also the space of complex structures on $\mathbb{R}^{2 n}$, compatible with the flat space metric.

Starting in eight dimensions the pure spinor constraints are non-trivial and non-linear.  In these cases, the construction of solutions of massless field equations becomes more difficult and it is not clear how to generalise the Penrose transform.

\chapter{Short Introduction to Topological Field and String Theories}
\label{ch:topological}

There are many review articles on topological field and string theories; this appendix is based on refs.~\cite{Vonk:2005yv} and~\cite[Chap.~4]{Cachazo:2005ga}.

The topological field theories are field theories that do not depend (more precisely their observables do not depend) on the the choice of background metric on the manifold on which they are defined.  Note that sometimes the observables are allowed to depend parameters that are not topological in nature (for example, the correlation functions sometimes depend on the choice of the complex structure).  They are not allowed to depend on the metric, however.

In the case of a theory which is invariant under diffeomorphisms, the metric independence has the consequence that the correlation functions of local operators cannot depend on the insertion points of those local operators.  The argument goes as follows: under a diffeomorphism the coordinates of the insertion points of local operators get transformed together with the metric.  However, by independence of the metric, we can change back to the initial metric without changing the correlation function, so the net effect is a displacement of the insertion points.

There are several ways of constructing topological field theories.  The first and most natural way is to construct an action that does not depend on the metric.  An example of this kind is the Chern-Simons theory, whose Lagrangian is
\begin{equation}
  \label{eq:Chern-Simons_Lagrangian}
  L = \tr \left( A \wedge d A - \frac 2 3 A \wedge A \wedge A\right),
\end{equation} where $A$ is a connection (gauge field) on a vector bundle over a three-dimensional base space $\mathcal{M}$.

There is a group of gauge transformations acting on $A$ as
\begin{equation}
  A \rightarrow g A g^{-1} - g d g^{-1}.
\end{equation}

If the manifold $\mathcal{M}$ on which the theory is defined doesn't have a boundary, the Lagrangian $L$ is invariant under small gauge transformations (gauge transformations that can be continuously connected to the identity).  Under large gauge transformations, the Lagrangian is not invariant, but picks up a contribution that integrates to $8 \pi^2$ times an integer.  So if we want to have an theory that is invariant also under large gauge transformations, we must take
\begin{equation}
  S = \frac k {4 \pi} \int L,
\end{equation} where $k$ is an integer.  With this choice, $e^{i S}$ is invariant.

In this approach of constructing topological theories, the metric independence is obvious, but the gauge-fixing procedure can introduce a metric dependence.  It can be proven however that after quantisation the theory remains topological.

One can construct correlation functions of Wilson loops, and compute topological invariants of the embedding of these loops inside $\mathcal{M}$.  One of the simplest topological invariants is the Jones polynomial, and it can be reproduced (together with other more complex invariants) by the above construction (see ref.~\cite{Witten:1988hf}).

Another method of constructing topological field theories rests on the existence of a fermionic operator $Q$ such that $Q^2 = 0$.  The theories constructed in this way are called cohomological field theories because of the resemblance of this construction with the cohomology theory.

The physical operators are $Q$-closed ($O$ is a physical operator if $[Q,O]_\pm = 0$, where the commutator is used if $O$ is bosonic and the anti-commutator is used if $O$ is fermionic) and two operators differing by a $Q$-exact quantity are equivalent ($O \sim O + [Q, \lambda]_\pm$).

There are some more conditions to be satisfied in order to have a topological theory.  The vacuum must be invariant under the operator $Q$, $Q \ket{0} = 0$ and the energy-momentum tensor must be $Q$-exact, $T_{\alpha \beta} = \{Q, G_{\alpha \beta}\}$.  The conditions imposed on the energy-momentum tensor are stronger than the conditions imposed on the other physical operators; the fact that it is $Q$-closed can be proven from the fact that it is $Q$-exact.

This stronger condition is necessary to insure the metric independence of the correlation functions.
\begin{equation}
  \frac \delta {\delta h^{\alpha \beta}} \langle O_1 \cdots O_n\rangle = i \langle O_1 \cdots O_n \frac {\delta S}{\delta h^{\alpha \beta}}\rangle \propto i \langle O_1 \cdots O_n T_{\alpha \beta}\rangle = i \langle O_1 \cdots O_n \{Q,G_{\alpha \beta}\}\rangle = 0,
\end{equation} where the operators $O_i$ are physical, so they (anti-)commute with $Q$.

A practical way to construct a cohomological field theory is to have a $Q$-exact Lagrangian, $L = \{Q, V\}$, so
\begin{equation}
  S = \left\{Q, \int_{\mathcal{M}} V\right\}.
\end{equation}

The examples we will be interested in can be constructed by a procedure called `twisting', starting with $\mathcal{N} = (2, 2)$ supersymmetric theories in two dimensions.

Let us first briefly review the $\mathcal{N} = (2, 2)$ supersymmetric theories in two dimensions.  In two dimensional space of Euclidean signature, the Lorentz group is $SO(2) \simeq U(1)$.  This is an abelian group so its irreducible representations are one-dimensional.  It is therefore useful to pass to complex coordinates $(z, \overline{z})$ which transform in irreducible representations of $U(1)$:
\begin{equation}
  z \rightarrow e^{i \alpha} z, \qquad \overline{z} \rightarrow e^{-i \alpha} \overline{z}.
\end{equation}

In order to construct a supersymmetric theory we need to introduce spinors.  In two dimensions with Euclidean signature one can introduce left and right chirality complex spinors, $\theta^\pm$ with transformation under Lorentz group
\begin{equation}
  \theta^\pm \rightarrow e^{\pm i \alpha/2} \theta^\pm.
\end{equation}  The complex conjugated spinors are defined by $\overline{\theta}^\pm = \left(\theta^\mp\right)^*$ and transform according to their index
\begin{equation}
  \overline{\theta}^\pm \rightarrow e^{\pm i \alpha/2} \overline{\theta}^\pm.
\end{equation}  In order to indicate their charges under the $U(1)$ Lorentz group, sometimes one writes $\partial_{++}$ instead of $\partial_z \equiv \partial$ and $\partial_{--}$ instead of $\partial_{\overline{z}} \equiv \overline{\partial}$.

Introduce the supercharges and the covariant derivatives
\begin{alignat}{2}
  Q_\pm & = \frac \partial {\partial \theta^\pm} + i \overline{\theta}^\pm \partial_\pm, & \qquad \overline{Q}_\pm & = - \frac \partial {\partial \overline{\theta}^\pm} - i \theta^\pm \partial_\pm,\\
  D_\pm & = \frac \partial {\partial \theta^\pm} - i \overline{\theta}^\pm \partial_\pm, & \qquad \overline{D}_\pm & = - \frac \partial {\partial \overline{\theta}^\pm} + i \theta^\pm \partial_\pm.
\end{alignat}

The supersymmetry algebra is
\begin{alignat}{2}
  \left\{Q_\pm, \overline{Q}_\pm\right\} & = P \pm H, & \qquad   \left\{Q_\pm, \overline{Q}_\mp\right\} & = 0,\\
  Q_\pm^2 & = 0, & \qquad \overline{Q}_\pm^2 & = 0,
\end{alignat} where $P$ is the momentum and $H$ is the Hamiltonian.

A general superfield
\begin{equation}
  \Phi(z, \overline{z}, \theta^\pm, \overline{\theta}^\pm) = \phi(z, \overline{z}) + \psi_+ \theta^+ + \psi_- \theta^- + \cdots,
\end{equation} contains 16 ordinary fields in its expansion in the odd coordinates.  These 16 fields transform in a reducible representation of the supersymmetry.

In order to obtain a reducible representation, one has to impose further constraints on the general superfield.  The simplest possibility leads to chiral superfields.  A left chiral superfield $\Phi$ satisfies the constraints $\overline{D}_\pm \Phi = 0$.

For chiral superfields one can construct an action manifestly invariant under supersymmetry.
\begin{equation}
  S = \int d^2 z d^4 \theta K(\Phi^i, \overline{\Phi}^i) + \left(\int d^2 z d^2 \theta \left.W(\Phi^i)\right|_{\overline{\theta} = 0} + \text{cc}\right),
\end{equation} where $K$ is the K\"ahler potential and $W$ is the superpotential.

This theory has a vector and an axial $R$-symmetry.
\begin{equation}
  \text{vector:}
  \left\{
    \begin{aligned}
    \theta^\pm & \rightarrow e^{i \beta} \theta^\pm,\\
    \overline{\theta}^\pm & \rightarrow e^{- i \beta} \overline{\theta}^\pm,
    \end{aligned}
  \right.
\end{equation}
\begin{equation}
  \text{axial:}
  \left\{
    \begin{aligned}
      \theta^\pm & \rightarrow e^{\pm i \beta} \theta^\pm,\\
      \overline{\theta}^\pm & \rightarrow e^{\mp i \beta} \overline{\theta}^\pm.
    \end{aligned}\right.
\end{equation}

The vector $R$-symmetry is non-anomalous for any K\"ahler target space if the the $R$-charges of the scalar components of the chiral superfields are chosen to vanish.  However, the axial $R$-symmetry is anomalous in general; it is non-anomalous if the target space is a Calabi-Yau manifold.\footnote{If $\Sigma$ is the two dimensional space on which we define our topological theory (the worldsheet) and $\mathcal{M}$ is the target space, a necessary condition for the absence of anomalies, consequence of the Atiyah-Singer index theorem, is
\[
\int_{\phi(\Sigma)} c_1(\mathcal{M}) = 0,
\] where $\phi(\Sigma)$ is the image of the worldsheet inside the target space and $c_1(\mathcal{M})$ is the first Chern class of the target space (equivalently, one can consider the integral on $\Sigma$ of the pullback of the Chern class $c_1(\mathcal{M})$).  If the Chern class is zero, the axial $R$-symmetry is non-anomalous on any worldsheet.  A Calabi-Yau manifold is a Ricci flat K\"ahler manifold, whose first Chern class is therefore necessarily zero.  This implies that the axial $R$-symmetry is non-anomalous for every choice of worldsheet $\Sigma$ if the target space is a Calabi-Yau manifold.  While a field theory can be defined for a fixed $\Sigma$, a string theory sums over all possible worldsheets, so the stronger condition of consistency on for all possible choices of $\Sigma$ is necessary in string theory.}  For a more in-depth presentation of the anomaly computation, see ref.~\cite{Vonk:2005yv}.

Now, in order to pursue the construction of topological field theories we need to find a fermionic operator $Q$ such that $Q^2 = 0$ and such that $P$ and $H$ are $Q$-exact.  Obviously, $Q$ will have to be made out of the supercharges $Q_\pm$ and $\overline{Q}_\pm$.  The following combinations
\begin{align}
  Q_A & = \overline{Q}_+ + Q_-,\\
  Q_B & = \overline{Q}_+ + \overline{Q}_-,
\end{align} satisfy all the requirements.  We have that $Q_A^2 = Q_B^2 = 0$ and also that
\begin{align}
  \left\{Q_A, Q_+\right\} & = P + H,\\
  \left\{Q_A, \overline{Q}_-\right\} & = P - H,
\end{align} and
\begin{align}
  \left\{Q_B, Q_+\right\} & = P + H,\\
  \left\{Q_B, Q_-\right\} & = P - H.
\end{align}  Then, we can conclude that $P$ and $H$ are $Q_A$ and $Q_B$-exact.

This does not mean that we have succeeded in formulating a cohomological theory because we first need to be able to formulate the theory on a curved space.  This is done by covariantising the derivatives and contracting the indices with the worldsheet metric.  However, the covariantisation procedure is not compatible with the global supersymmetry we had.  In order to preserve the global supersymmetry, one must have that the parameters of the supersymmetry transformations (the spinors multiplying the supercharges) are covariantly constant.  But they must be covariantly constant with respect to an arbitrary metric, which is not possible.

A related difficulty is the following: ultimately we would like the use the operator $Q$ as a BRST operator.  However, $Q_A$ and $Q_B$ are not Lorentz invariant, so the theory constructed with this BRST operator is not guaranteed to be Lorentz invariant.  This is where the twisting procedure comes into play.  One finds a new Lorentz group under which some of the supercharges transform with spin zero (but remain anti-commuting).  The generators of this new Lorentz algebra are constructed out the the generators of the old Lorentz algebra and the generators of the vector and axial $R$-symmetries (this is why we emphasized the $R$-symmetries and discussed their anomalies).

Let $M$ be the generator of the old Lorentz symmetry and $R_V$, $R_A$ the generators of the vector and axial $R$-symmetries.  The transformations of the supercharges can be inferred from the transformations of the odd coordinates $\theta$, $\overline{\theta}$.

It turns out that with respect to the new Lorentz generator $M_A = M + \tfrac 1 2 R_V$, the supercharges $\overline{Q}_+$ and $Q_-$ are scalars so $Q_A$ is also a scalar.  The remaining supercharges $Q_+$ and $\overline{Q}_-$ transform with spin $+1$ and $-1$ respectively.

One can similarly define $M_B = M + \tfrac 1 2 R_A$.  With respect to this Lorentz group $\overline{Q}_\pm$ (and therefore also $Q_B$) are scalars, while $Q_+$, $Q_-$ have charges $+1$, respectively $-1$.

Now, when going from the theory formulated on a flat space to the theory formulated on a curved space, we replace the derivatives with covariant derivatives with respect to the transformation properties under the new Lorentz group.  The supersymmetry parameters are still anti-commuting but they are scalars with respect to the new Lorentz group.  Therefore, there is no problem keeping the global supersymmetry on a curved space.

It is now possible to write down the action for the $\mathcal{N}=(2,2)$ theories and covariantise it using the new Lorentz transformations.  The theory which is $M_A$ invariant is called the $A$-model and the one invariant under $M_B$ is called the $B$-model.  Moreover, the energy-momentum tensor for these theories can be computed and it can be shown to be $Q_{A/B}$ exact.

\chapter{Landau equations}

The purpose of Landau equations is to characterise the position of singularities that can appear in Feynman integrals.  When dealing with Feynman integrals it proves very fruitful to continue them analytically in the complex plane, as functions of kinematical invariants.  The singularities arise when one encounters obstructions to analytic continuation.

In the end we will be interested in studying multiple integrals which, after analytic continuation become hyper-contour integrals.  We will start with a one-dimensional example since it is simpler and it already exhibits some features that generalise to the more complicated case of multiple integrals.

Define a function $f$ by
\begin{equation}
  f(z) = \int_{\mathcal{C}} d w g(z, w),
\end{equation} where the singularities of the function $g$ in the $(z,
w)$ variables can be described by $w_r = w_r(z)$, with $r$ a discrete
index.  We consider the contour to be compact, for simplicity; if the
contour goes to infinity, some further singularities can appear.

The function $f$ is analytic at a point $z_0$ if all the singularities $w_r(z_0)$ are away from the contour $\mathcal{C}$.  If we now start to continue analytically starting from $z_0$, the singularities $w_r(z)$ move in the $w$-plane in a complicated way.  As long as it is possible to move the contour in such a way as to avoid collision with the contour $\mathcal{C}$, one can analytically continue the function $f$.

Where one cannot deform the contour anymore, a singularity occurs.  There are several situations where the contour cannot be deformed: they are called pinch and end-point singularities.

In the case of pinch singularities, two singularities of the function $g$, say $w_i(z)$ and $w_j(z)$ approach the contour $\mathcal{C}$ from different sides when $z \rightarrow z_{\text{sing}}$.  In this case, the contour is pinched between these singularities, so it cannot be deformed anymore.

The other possibility arises when the contour $\mathcal{C}$ is open. If one of the singularities $w_r$ approaches one of the end-points, obviously the contour can't be deformed to avoid it.

It is not difficult to argue that the singularities found in this way are usually branch-points.  The discussion can be extended to the case of contours going to infinity.  In that case, one must also verify if `pinching' happens at infinity.  This is most easily done by mapping the infinity to zero.

For two external variables, but still one integration,
\begin{equation}
  f(z_1, z_2) = \int_{\mathcal{C}} d w g(z_1, z_2, w),
\end{equation} $g$ will have some singularity surfaces defined by $w_r
= w_r(z_1, z_2)$ and $\mathcal{C}$ is contour as above.  In this case
as well, the singularities are of pinch and end-point type.

Pinch singularities can arise at $(z_1, z_2)$ if the contour $\mathcal{C}$ is pinched between two points $w_i(z_1, z_2) = w_j(z_1, z_2)$.  End-point singularities can appear for $(z_1 ,z_2)$ for which one $w_i(z_1, z_2)$ is equal to an end-point of the open contour $\mathcal{C}$.  These conditions are realised in general on varieties of complex dimension one in the space of $(z_1, z_2)$ of complex dimension two.  The conditions above are however only necessary but not sufficient conditions.  It might well happen that one has $w_i(z_1, z_2) = w_j(z_1, z_2)$, but the two points do not lie on the contour $\mathcal{C}$ so no pinching can occur.  The end-point singularities separate between the pinching and non-pinching, so they are at the boundary which separates the singular and non-singular regions on the surface $w_i(z_1, z_2) = w_j(z_1, z_2)$.

It can also happen that $w_i(z_1, z_2) = w_j(z_1, z_2)$ but there is no pinching (the two points approach the contour from the same side).

Let us finally discuss the case of one external complex variable and multiple integrals.  Here we will only list the results without much discussion; some of them can be understood by analogy with the case of one-dimensional integrals, but the hyper-contour deformations are of course much more difficult to visualise.

Suppose we have a function
\begin{equation}
  f(z) = \int_{\mathcal{H}} \prod_{i=1}^n d w_i g(z, w_i),
\end{equation} where $\mathcal{H}$ is an $n$-dimensional
hyper-contour.  The singularities of $g$ are given by the implicit
equations $S_r(z, w_i) = 0$, where $r$ is a discrete index.  For fixed
$z$, the equations $S_r = 0$ describe an $n - 1$ dimensional variety
inside the $n$-dimensional space parametrised by $w_i$.

When we vary $z$, the surfaces $S_r = 0$ move inside the $n$-dimensional space parametrised by $w_i$, where the hyper-contour $\mathcal{H}$ lives.

When one singularity surface $S_k$ advances towards the hyper-contour $\mathcal{H}$, this contour can be deformed in the direction normal to $S_k$, in such a way as to avoid collision.

The pinch singularities arise when two surfaces $S_k$ and $S_l$ approach the hyper-contour from opposite sides and the directions of the normals coincide.  Then, the hyper-contour is pinched and a singularity develops for the value of $z$ where
\begin{gather}
  S_k(z, w_i) = S_l(z, w_i) = 0,\\
  \alpha_k \frac {\partial S_k}{\partial w_j}(z, w_i) + \alpha_l \frac
  {\partial S_l}{\partial w_j}(z, w_i) = 0, \quad \text{for i = 1,
    \ldots, n},
\end{gather} where the last equation expresses the identity of the tangent spaces (and therefore of the normals) to surfaces $S_k$ and $S_l$.

In case several surfaces $S_k$, $S_l$, $S_m, \ldots$ contribute to the pinching one can write the same kind of equations but adding further constants $\alpha_m, \ldots$

One more complicated possibility is when the hyper-contour
$\mathcal{H}$ is pinched at the singularity of a single surface.  This
is the case for conical singularities.  They are described by
\begin{equation}
  S_k = \frac {\partial S_k}{w_i} = 0, \quad \text{i = 1, \ldots, n}.
\end{equation}

The `end-point' singularities can also be analysed by specifying the boundary of $\mathcal{H}$ by equations $\tilde{S}_r$.  The boundary must remain fixed so in particular it cannot move in the orthogonal direction to $\tilde{S}_r$.  These surfaces can then be treated just like the surfaces $S_i$ above.

Finally, the conditions for the existence of singularities can be assembled as follows, by introducing variables $\alpha_i$ and $\tilde{\alpha}_r$
\begin{align}
  \alpha_i S_i = 0, \quad \forall i,\\
  \intertext{which implies that ether $\alpha_i$ or $S_i$ are zero}
  \tilde{\alpha}_r \tilde{S}_r = 0, \quad \forall r,\\
  \frac \partial {\partial w_i} \left(\sum_i \alpha_i S_i + \sum_r
    \tilde{\alpha}_r \tilde{S}_r\right) = 0, \quad \forall w_i.
\end{align}

After this rather long introduction, we are ready to study the Landau equations which are just the equations for singularities derived above, applied to the case where the integrals are of Feynman type.

There are several different ways of writing the Landau equation depending of the way the integral under study is written (one can use the form obtained straightforwardly from Feynman rules, or the form using Feynman parameters after the integration of loop momenta, etc).

Consider for example an integral like
\begin{equation}
  I = \int \frac {d^4 k_1 \cdots d^4 k_l}{\prod_{i=1}^N (q_i^2 - m_i^2)}.
\end{equation}  There are no $\tilde{S}$ and $S_i = q_i^2 - m_i^2$.
Then, the Landau equations are
\begin{gather}
  q_i^2 = m_i^2, \quad \text{or}\ \alpha_i = 0,\\
  \frac \partial {\partial k_j} \sum_i \alpha_i (q_i^2 - m_i^2) = 0.
\end{gather} Using the fact that the momenta $q_i$ are linear combinations of $k_i$ (the loop momenta) and of external momenta, the last equation above can be written
\begin{equation}
  \sum_{\text{loop} j} \alpha_i q_i = 0,
\end{equation} where $\sum_{\text{loop} j}$ is the sum over the edges
of loop $j$.

One can perform the following construction: start with the initial graph and attribute to each $i$ edge a momentum $q_i$ and a constant $\alpha_i$.  The Landau equations then have the following interpretation: find solutions for the momenta $q_i$ and the constants $\alpha_i$ such that for each edge $i$ either $q_i$ is on-shell ($q_i^2 = m_i^2$), or $\alpha_i = 0$ and the momenta $q_i$ satisfy momentum conservation.\footnote{In case $\alpha_j = 0$ one should   assign a momentum equal to zero to that edge.  The corresponding   graph is drawn with the edge $j$ collapsed to a point.}

If for a graph one has solutions for the Landau equations where none of the $\alpha_i$ are equal to zero, the corresponding singularity is called leading singularity.  The singularities where one of the $\alpha_i$ is zero are identical to singularities of graphs obtained by collapsing the edge $j$, corresponding to $\alpha_j = 0$.  These are called lower-order singularities.

\chapter{Cuts and Discontinuities}
\label{ch:cuts_and_disc}

In the previous section we discussed how one should find the location of singularities for a Feynman integral.  We will show in the following that the singularities given by the Landau equations are branch points, and we compute the discontinuities across the corresponding branch cuts.

We will find that the expressions for the discontinuities across branch cuts have simple form and a physical interpretation.  This is inspired by the discussion in ref.~\cite{Olive:1967}.

Start with a Feynman integral
\begin{equation}
  I(z) = \int \frac {d^4 k_1 \cdots d^4 k_l}{\prod_{i=1}^N (q_i^2 - m_i^2)}
\end{equation} and study the discontinuity associated with the
singularity corresponding to $r$ internal lines going on-shell (by
Landau equations, the remaining $N-r$ internal lines have $\alpha =
0$).  We denoted by $z$ all the dependence on the external momenta.
Choose a notation for the momenta $q$ such that $q_i^2 = m_i^2$ for $i
= 1, \ldots, r$.

One can make a change of variables such that $r$ of the $4 l$ integration variables are $q_i^2$ for $i = 1, \ldots, r$.  In these variable we write the integration measure as $\prod_{i=1}^r d q_i^2 d^{4 l - r} \xi$, where $d^{4 l - r} \xi$ contains the remaining integration variables and also the Jacobian that was generated by the change of variables.

The integral now reads
\begin{equation}
  I(z) = \int_{l_1}^{u_1} d q_1^2 \cdots \int_{l_r}^{u_r} d q_r^2 \int
  \frac {d^{4 l - r} \xi}{\prod_{i=1}^N (q_i^2 - m_i^2)}.
\end{equation}  The integration limits $l_i$, $u_i$ for the integral
over $q_i^2$ can be obtained by fixing all the values of $q_j^2$ with
$j < i$ and extremising the value of $q_i^2$ subject to the conditions
of momentum conservation around the loops containing momenta $q_j$
with $j < i$.

\begin{figure}
  \label{fig:pinching_disc}
\beginpgfgraphicnamed{38}
\begin{tikzpicture}[baseline=0]
  \coordinate (a1) at (-3,0);
  \coordinate (b1) at (3,0);
  \coordinate (m) at (0,-.4);
  \coordinate (q) at (0,.4);
    \draw (a1) .. controls +(-30:2) and +(150:2) .. (b1) node[midway,sloped]{\tikz\draw[>=stealth,very thick,->] (-1pt,0pt) -- (0pt,0pt);}
    {[fill] (m) circle (1pt) node[below]{$m_1^2$}}
    {[fill] (q) circle (1pt) node[above]{$\tilde{q}_1^2$}}
    (a1) node[left]{$l_1$}
    (b1) node[right]{$u_1$};
\end{tikzpicture}
\endpgfgraphicnamed\hspace{3em}
\beginpgfgraphicnamed{39}
\begin{tikzpicture}[baseline=0]
  \coordinate (a1) at (-3,0);
  \coordinate (b1) at (3,0);
  \coordinate (m) at (0,-.4);
  \coordinate (q) at (0,.4);
    \draw (a1) .. controls +(-40:3) and +(-140:3) .. (b1) node[midway,sloped]{\tikz\draw[>=stealth,very thick,->] (-1pt,0pt) -- (0pt,0pt);}
    {[fill] (m) circle (1pt) node[above]{$m_1^2$}}
    {[fill] (q) circle (1pt) node[above]{$\tilde{q}_1^2$}}
    (m) circle (.7)
    (0,-1.1) node{\tikz\draw[>=stealth,very thick,<-] (-1pt,0pt) --       (0pt,0pt);}
    (a1) node[left]{$l_1$}
    (b1) node[right]{$u_1$};
\end{tikzpicture}
\endpgfgraphicnamed
   \caption{The deformation of the $q_1^2$ integration contour.  The integrand has a singularity when $q_1^2 = \tilde{q}_1^2$ and when $q_1^2 = m_1^2$.}
\end{figure}
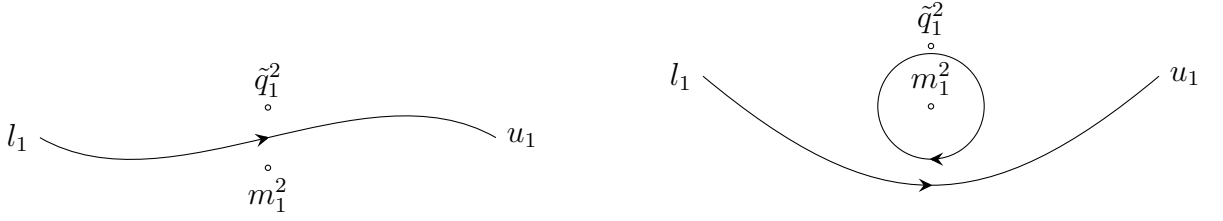

So the general form of the integral is
\begin{equation}
  I(z) = \int_{l_1}^{u_1} d q_1^2 \frac {I_1 (q_1^2, z)}{q_1^2 - m_1^2}.
\end{equation}  The singularity we are interested in arises for $q_1^2
= m_1^2$.  An important point is that $l_1$ and $u_1$ do not depend on
$m_1^2$ and, more generally, $l_i$ and $u_i$ do not depend on
$m_i^2$.  So the singularity is not of end-point type.  The only way
to obtain a singularity is to have a pinch of the integration contour (see fig.~\ref{fig:pinching_disc}).

The integral over the open contour after the deformation (see the right side of fig.~\ref{fig:pinching_disc}) is not pinched so it doesn't contain any singularity.  The singularity comes entirely from the integral over the circle contour (which is pinched).  But this integral can be computed exactly and the result is
\begin{equation}
  - 2 \pi i I_1(m_1^2, z).
\end{equation}

Now we can repeat the argument above using the fact that
\begin{equation}
  I_1(q_1^2, z) = \int_{l_2}^{u_2} d q_2^2 \frac {I_2(q_2^2; q_1^2, z)}{q_2^2 - m_2^2}.
\end{equation}  In the end, the discontinuity in the external kinematical invariants across the branch cut determined by the internal momenta $q_1^2, \ldots q_r^2$ going on-shell is
\begin{equation}
  \text{disc} I(z) = (-2 \pi i)^r \int \delta^{(+)}(q_1^2 - m_1^2) \cdots \delta^{(+)}(q_r^2 - m_r^2) \frac {d^4 k_1 \cdots d^4 k_l}{\prod_{i=r+1}^N (q_i^2 - m_i^2)}  
\end{equation}

\chapter{Wilson loops}
\label{ch:wilson_loops}

In this part we will restrict our attention to Euclidean signature, unless specified otherwise.  Let us define a path ordered exponential by the following formula
\begin{multline}
  P \exp \left(\int d x f(x) \right) = 1 + \int d x_1 f(x_1) + \int d x_1 d x_2 f(x_1) f(x_2) \theta(x_1, x_2) + \cdots \\+ \int d x_1 d x_2 \cdots d x_n f(x_1) f(x_2) \cdots f(x_n) \theta(x_1, x_2, \ldots, x_n) + \cdots,
\end{multline} where\footnote{This formula works when the range of integration is not a loop.  If, as we will see below, we integrate over a loop, there is no well defined ordering globally: for example, for two points $x$ and $y$ on a loop we can interpret them as having both $x > y$ and $x < y$.  Therefore, once we have chosen an orientation for the curve we can have $x_1 > x_2 > \cdots > x_n$ but also $x_2 > x_3 > \cdots > x_n > x_1$ and all the circular permutations.  This is the origin of factors of $\tfrac 1 n$ that sometimes appear in the literature.}
\begin{equation}
  \theta(x_1, x_2, \ldots, x_n) =
  \begin{cases}
    1, & \text{if} \quad x_1 > x_2 > \cdots > x_n,\\
    0, & \text{otherwise}
  \end{cases}.
\end{equation}

The path ordered exponential
\begin{equation}
  U(P_{x, y}) = P \exp \left(i g \int_{P_{x, y}} A_\mu d x^\mu\right),
\end{equation} where $P_{x,y}$ is a path from $x$ to $y$ has simple properties with respect to gauge transformations
\begin{equation}
  U(P_{x, y}) \rightarrow U^g(P_{x, y}) = g(x) U(P_{x, y}) g^{-1}(y).
\end{equation}  Such path ordered exponentials are called Wilson lines.

The trace of a Wilson line whose beginning and ending points are identical is gauge invariant and it is called a Wilson loop.  The Wilson loops are very important quantities because they provide a criterion for confinement (the area law).  We will denote the expectation value of the Wilson loop over the loop $\mathcal{C}$ by $W(\mathcal{C})$.

Expanding the exponential, we have\footnote{The factor $\tfrac 1 n$ here has the same origin as the one in the previous footnote.}
\begin{equation}
  \label{eq:wilson_expanded}
  W(\mathcal{C}) = 1 + \sum_{n=2}^\infty \frac {(i g)^n} n \oint_{\mathcal{C}} d x_1^{\mu_1} \cdots d x_n^{\mu_n} \theta(x_1, x_2, \ldots, x_n) \langle A_{\mu_1} (x_1) \cdots A_{\mu_n} (x_n) \rangle.
\end{equation}

It is easy to see that there will be divergences coming from the Green functions $\langle A_{\mu_1} (x_1) \cdots A_{\mu_n} (x_n) \rangle$.  The divergences in these Green functions are renormalised in the usual way by charge and wavefunction renormalisation.  There is a further source of divergences from the integrations over the contour $\mathcal{C}$ (the contour is compact so the divergences are only possible if there are singularities in the integrand, i.e. if the Green function is singular when two or more coordinates $x_i$ collapse to a single point).  These are short distance (UV) divergences.

The first study of the renormalisation properties of Wilson loops was done by Polyakov in ref.~\cite{Polyakov:1980ca}.  He observed that the renormalisation properties of Wilson loops in Euclidean signature depend essentially on the smoothness of the loop.

Let us first consider the leading perturbative correction for the case of a smooth loop.
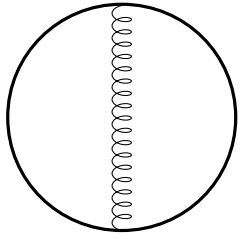
\begin{figure}
  \centering
  \beginpgfgraphicnamed{40}
  \begin{tikzpicture}
    \draw[very thick] (0,0) circle (1.5);
    \draw[gluon,decorate] (0,-1.5) -- (0,1.5);
  \end{tikzpicture}
  \endpgfgraphicnamed
  \caption{First correction in perturbation theory to the Wilson loop.}
  \label{fig:wl_first-correction}
\end{figure}

This first correction is given by the exchange of a gluon between two points on the loop.  In position space and in Feynman gauge ($\alpha = 1$) the gluon propagator is
\begin{equation}
  \label{eq:gluon_propagator}
  \frac 1 {(2 \pi)^2} \frac {\delta_{a_1, a_2} \delta_{\mu_1, \mu_2}} {(x_1 - x_2)^2},
\end{equation} where $a_1$, $a_2$ are colour factors.

At this point we consider a pure gauge theory (no fermions and no scalars).  The ghost propagator has the same position space behaviour
but we only need it if we want to compute the Green functions entering the integrand at loop level.

It is easy to compute the first correction and see that it is divergent.

Following Polyakov, \cite{Polyakov:1980ca} we have\footnote{In this case we have cancelled the factor of $\tfrac 1 2$ and we let both integrations (over $x$ and over $y$) run over the contour $\mathcal{C}$ unrestricted.}
\begin{equation}
  \label{eq:wilson_loop_first_correction}
  W^{(1)}(\mathcal{C}) \propto \oint_{\mathcal{C}} \oint_{\mathcal{C}} d x_\mu d y_\mu \frac 1 {(x - y)^2}.
\end{equation}  As we are in Euclidean space, we don't bother to raise or lower indices.  This integral is divergent.  Regularising by introducing a small gluon mass $a$, we get
\begin{equation}
  \label{eq:first_correction}
  \oint_{\mathcal{C}} \oint_{\mathcal{C}} \frac {d x_\mu d
    y_\mu}{(x - y)^2 + a^2} = \int \frac {\dot{x}(s) \cdot \dot{x}(s +
    t) d s\ d t}{(x(s+t) - x(s))^2 + a^2}.
\end{equation}  Choosing a loop parametrisation such that $\dot{x}^2 = \text{const}$ and therefore $\dot{x} \cdot \ddot{x} = 0$ and keeping the dominant contribution in $t$ we get
\begin{align}
  W^{(1)}(\mathcal{C}) &\propto \int d s \dot{x}^2(s) \int \frac {d t}{t^2 \dot{x}^2 + a^2} + \text{finite}\\
  &= \frac {\pi} a \int d s \sqrt{\dot{x}^2} + \text{finite} = \frac
  {\pi L_{\mathcal{C}}} a + \text{finite},
\end{align} were $L_{\mathcal{C}}$ is the length of the curve $\mathcal{C}$.

This kind of divergence can be thought as a contribution to the mass renormalisation of a heavy coloured test particle moving on a loop $\mathcal{C}$ and interacting with its own radiation field.

The renormalisation properties to all orders in perturbation theory were studied by Dotsenko and Vergeles and by Brandt, Neri and Sato in refs.~\cite{Dotsenko:1979wb,Brandt:1981kf}.  They proved that for a smooth Wilson loop the only kind of divergences that appear can be eliminated by a ``mass renormalisation'' discussed above.  So one can define the renormalised Wilson loop by
\begin{equation}
  \label{eq:smooth_wilson_renormalisation}
  W(\mathcal{C}) = \exp (- K(a) L_{\mathcal{C}}) W_\text{ren}(\mathcal{C}),
\end{equation} were $W_\text{ren}(\mathcal{C})$ is finite and $K(a)$ is a divergent factor (in the limit where the cutoff is removed $a \rightarrow 0$).

\begin{figure}
  \centering
  \beginpgfgraphicnamed{41}
  \begin{tikzpicture}
  \draw[very thick] (0,0) .. controls +(135:6) and +(45:6) .. (0,0);
  \draw[dashed] (0,0) -- ++(-135:1.3)
                (0,0) -- ++(135:1.3);
  \draw (0,0)++(135:1) arc (135:225:1);
  \draw (-1,0) node[left]{$\gamma$};
  \end{tikzpicture}
  \endpgfgraphicnamed
  \caption{A Wilson loop with cusp}
  \label{fig:wilson_cusp}
\end{figure}
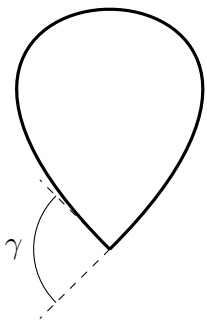

In the case where the loop is not smooth, there are further divergences possible.  They have also been computed by Polyakov to first order in perturbation theory (see ref.~\cite{Polyakov:1980ca}).  The computations are too long to be included here, but the final result is (the notation is depicted in fig.~\ref{fig:wilson_cusp})
\begin{equation}
  \label{eq:cusp_divergence}
  W^{(1)}(\mathcal{C}) \propto \frac {\pi L_{\mathcal{C}}} a + (\gamma \cot \gamma - 1) \ln\left(\frac {L_{\mathcal{C}}} a\right) + \text{finite}.
\end{equation}

This supplementary divergence can also be interpreted in terms of a heavy coloured test particle.  Its origin is in the violent Bremsstrahlung due to the infinite acceleration the particle has at the cusp.

The origin of this divergence is then local so the reasonable expectation is that if there are several cusps, each contributes a factor as in eq.~\eqref{eq:cusp_divergence}.  In ref.~\cite{Brandt:1981kf} it has been proven that this is indeed true and the results above were also extended to self-intersecting Wilson loops.

Then, the renormalisation properties of smooth Wilson loops presented in eq.~\eqref{eq:smooth_wilson_renormalisation} can be extended to the case of Wilson loops with cusps
\begin{equation}
  \label{eq:cusp_wilson_renormalisation}
  W(\mathcal{C}) = \exp (- K(a) L_{\mathcal{C}}) Z(\gamma, g; a) W_\text{ren}(\mathcal{C}).
\end{equation}

In the case of the Wilson loop with a cusp, we can define an anomalous dimension by
\begin{equation}
  \label{eq:cusp_anomalous_dim}
  \Gamma_{\text{cusp}}(g, \gamma) = \left.\frac {\partial \ln Z(g, \gamma; a)}{\partial \ln a}\right|_{a \rightarrow 0}.
\end{equation}  To first order in perturbation theory the cusp anomalous dimension is
\begin{equation}
  \label{eq:cusp_anomalous_dim_1st}
  \Gamma_{\text{cusp}}^{(1)}(g, \gamma) = - (\gamma \cot \gamma - 1) \frac {g^2}{4 \pi^2} C_1,
\end{equation} where $C_1$ is a colour factor $C_1 = T^a T^a$.

One important result, called ``non-abelian exponentiation'' was proved for expectation values of Wilson loops (see ref.~\cite{Dotsenko:1979wb,Gatheral:1983cz,Frenkel:1984pz}).  It states that the expectation value $W(\mathcal{C})$ of a Wilson loop on a contour $\mathcal{C}$ can be written in a natural way as
\begin{equation}
  \label{eq:exponentiation_th}
  W(\mathcal{C}) = \exp (w(\mathcal{C})).
\end{equation}  This statement does not have any real content in itself, but it is important in the sense that far fewer diagrams contribute to $w(\mathcal{C})$ than to $W(\mathcal{C})$.  We also need to specify what diagrams one should sum over in $w(\mathcal{C})$.

When discussing the exponentiation proposed in eq.~\ref{eq:exponentiation_th} it is important to keep in mind that the Green functions appearing in the expansion of the Wilson loop can be decomposed in sums of products of \emph{connected} Green functions.  This makes it reasonable to expect an exponentiation theorem in agreement with the link between the generating functions of Green functions and of connected Green functions.

However, the Green functions in the expansion of the Wilson loop also contain the gauge algebra generators and their arguments on the loop are restricted by inequalities.

Let us start with an example (see ref.~\cite{Dotsenko:1979wb} for a related discussion) to see how the perturbation theory can be reorganised.

\begin{figure}
  \centering
\beginpgfgraphicnamed{42}
\begin{tikzpicture}
\matrix[row sep=3ex, column sep=3em]{
  \begin{scope}
  \draw[very thick] (0,0) circle (1.5);
  \path (0,0)++(135:1.5) coordinate (A)
        (0,0)++(-135:1.5) coordinate (B)
        (0,0)++(45:1.5) coordinate (C)
        (0,0)++(-45:1.5) coordinate (D);
  \draw[gluon, decorate] (A) -- (C)
               (B) -- (D);
\end{scope} & node{a} 
\begin{scope}
  \draw[very thick] (0,0) circle (1.5);
  \path (0,0)++(135:1.5) coordinate (A)
        (0,0)++(-135:1.5) coordinate (B)
        (0,0)++(45:1.5) coordinate (C)
        (0,0)++(-45:1.5) coordinate (D);
  \draw[gluon, decorate] (A) -- (B)
               (C) -- (D);
\end{scope} & node {b}
\begin{scope}
  \draw[very thick] (0,0) circle (1.5);
  \path (0,0)++(135:1.5) coordinate (A)
        (0,0)++(135:.2) coordinate (A')
        (0,0)++(-135:1.5) coordinate (B)
        (0,0)++(45:1.5) coordinate (C)
        (0,0)++(-45:1.5) coordinate (D)
        (0,0)++(-45:.2) coordinate (D');
  \draw[gluon, decorate] (A) -- (A');
  \draw[gluon, decorate] (D') -- (D);
  \draw[gluon, decorate] (B) -- (C);
\end{scope} \\
\begin{scope}
  \draw[very thick] (0,0) circle (1.5);
  \path (0,0)++(90:1.5) coordinate (A)
        (0,0)++(210:1.5) coordinate (B)
        (0,0)++(-30:1.5) coordinate (C);
  \draw[gluon, decorate] (0,0) -- (A);
  \draw[gluon, decorate] (0,0) -- (B);
  \draw[gluon, decorate] (0,0) -- (C);
\end{scope} &
\begin{scope}
  \draw[very thick] (0,0) circle (1.5);
  \draw[gluon, decorate] (0,-1.5) -- (0,-.25)(0,.25) --(0,1.5);
  \draw[fill, fill opacity=0.2] (0,0) circle (.25);
\end{scope} \\
};
\end{tikzpicture}
\endpgfgraphicnamed
  \caption{Second order contribution in perturbation theory to the     Wilson loop.  The shaded blob contains the one-loop correction to the gluon propagator.}
  \label{fig:wl_second-correction}
\end{figure}
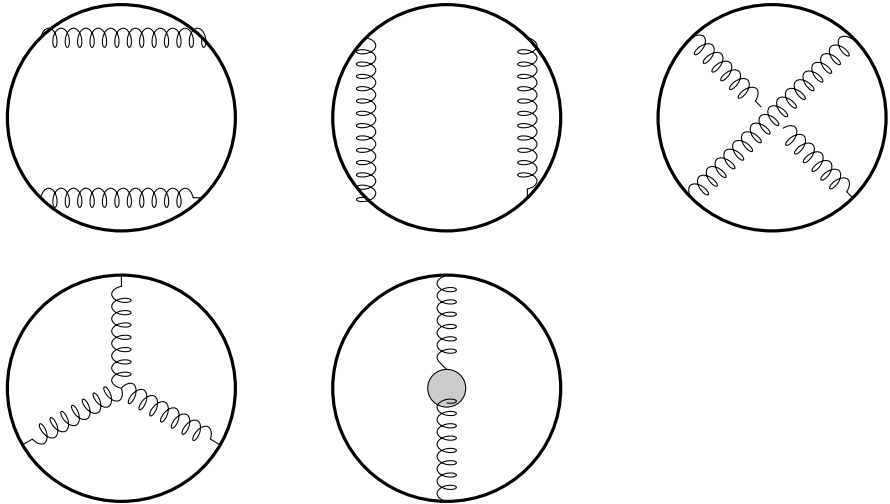

When expanding the exponential of the first correction (see fig.~\ref{fig:wl_first-correction}) one obtains all the orderings of the external legs as in the first three diagrams in fig.~\ref{fig:wl_second-correction}, but the colour factors are different.  For the first correction the colour factor is $C_1 = T^a T^a$ and for the first two diagrams in fig.~\ref{fig:wl_second-correction} it is $C_1^2 = T^a T^a T^b T^b$.  For the third diagram in fig.~\ref{fig:wl_second-correction} the colour factor is
\begin{equation}
  T^a T^b T^a T^b = T^a \left[T^b, T^a\right] T^b + T^a T^a T^b T^b = i f^{b a c} T^a T^c T^b + C_1^2 = \frac 1 2 C_2 C_1^2 + C_1^2,
\end{equation} where we used
\begin{align}
  T^a T^a =& C_1,\\
  i f^{a b c} T^b T^c =& \frac 1 2 C_2 T^a.
\end{align}

Now, the $C_1^2$ term in the colour decomposition of the third diagram combines with the first two diagrams to yield the first order in the expansion of the exponential of the first order correction to the Wilson loop (the three diagrams correspond to the summation over the relative positions four the four points on the loop).

The construction proceeds recursively.  If a diagram is two-particle reducible with respect to the Wilson loop (meaning that it is possible to break it in two parts by cutting the Wilson loop twice), it appears in the expansion of the exponential if its components appear in the exponent.\footnote{That one can choose the coefficients of the diagrams in the exponent such that each diagram in the Wilson loop expansion appears with the right coefficient is a not very illuminating combinatorial exercise.}  So we are lead to the conclusion that one should include in the exponent only 2PI (two-particle irreducible) diagrams with respect to the Wilson loop. (Note that these diagrams can be disconnected when considered as contributions to the Green functions in eq.~\eqref{eq:wilson_expanded}.  They are connected by the Wilson loop.)

The final prescription is as follows: include in the exponent all the 2PI diagrams with respect to the Wilson loop, with a certain colour factor that can be found following the steps we presented for the second order correction (this is \emph{not} the naive colour factor assigned to the diagram by Feynman rules).

The exponentiation theorem implies that for QED (without electrons), which is a free theory and whose colour structure is trivial, the Wilson loop is given by the exponential of the first order correction.  This result is just the fact that, for a stochastic variable $X$ with a Gaussian distribution
\begin{equation}
  \langle e^{i X}\rangle = e^{-\frac 1 2 \langle X^2\rangle}.
\end{equation}  The exponentiation theorem is the extension of this result to an interacting theory.

The exponentiation property implies that, while $W(\mathcal{C})$ renormalises multiplicatively, $w(\mathcal{C})$ renormalises additively.  As $w(\mathcal{C})$ is easier to compute (fewer diagrams), it is preferable to compute the additive renormalisation of $w(\mathcal{C})$ and compute the cusp anomalous dimension using it.

The cusp anomalous dimension has been computed to two loops in perturbation theory in ref.~\cite{Korchemsky:1987wg}.  At two loops, the effect of adding fermions has been computed in ref.~\cite{Korchemsky:1988si}.

Let us now briefly discuss the Wilson loops in Minkowski signature.  In this case, the expectation values should be replaced with vacuum $T$-ordered matrix elements.  Note that after the expansion of the exponential one has both a path ordering and a time ordering for the arguments of the vector potential.  The prescription to deal with potential ambiguities in the two orderings is to compute the $T$-ordered Green functions first and then to compute the path ordered integrals on the Wilson loop.

In Minkowski signature the results are unchanged, except when part of the Wilson loop is a light-like line.  These kinds of Wilson loops have been studied in ref.~\cite{Korchemskaya:1992je}.

\begin{figure}
  \centering
  \beginpgfgraphicnamed{43}
  \begin{tikzpicture}
    \draw[very thick] (0,0) coordinate (A) -- ++(60:2.5) coordinate (B) node[midway,right]{$x$} node[midway,sloped]{\tikz\draw[>=stealth,very thick,->] (-1pt,0pt) -- (0pt,0pt);} -- ++(0:4) node[midway,below]{$y$} node[midway,sloped]{\tikz\draw[>=stealth,very thick,->] (-1pt,0pt) -- (0pt,0pt);} -- ++(-120:2.5) node[midway,sloped]{\tikz\draw[>=stealth,very thick,-<] (-1pt,0pt) -- (0pt,0pt);} -- ++(180:4) node[midway,sloped]{\tikz\draw[>=stealth,very thick,-<] (-1pt,0pt) -- (0pt,0pt);} -- cycle;
    \draw[dashed] (A) -- ++(180:1);
    \draw (A) ++(60:.5) arc (60:180:.5);
    \draw (A) ++(120:.5) node[above]{$\gamma(x,-y)$};
    \draw[dashed] (B) -- ++(60:1);
    \draw (B) ++(0:.5) arc (0:60:.5);
    \draw (B) ++(30:.5) node[right]{$\gamma(x,y)$};
  \end{tikzpicture}
  \endpgfgraphicnamed
  \caption{A Wilson loop with four cusps.}
  \label{fig:four_cusps}
\end{figure}
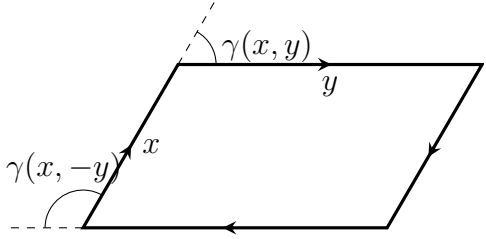

In Minkowski signature\footnote{With signature is $(+,-,-,-)$.} and for time-like four-vectors $x$ and $y$ such that $x \cdot y > 0$ one can define an angle formed by $x$ and $y$ by
\begin{equation}
  \cosh \gamma(x, y) = \frac {x \cdot y}{\sqrt{x^2 y^2}}.
\end{equation}  This can be obtained by a Wick rotation from the Euclidean signature result.

It is obvious that when one of the four-vectors is light-like one can't use the same formula to define an angle.  However, much can be learned by just taking the limit $x^2 \rightarrow 0$ in the Wick rotated results for the Wilson loop.  In this limit $\gamma \rightarrow \infty$.

By finding a differential equation for the Wilson loop where it is safe to set $x^2$ to zero, Korchemskaya and Korchemsky proved in ref.~\cite{Korchemskaya:1992je} that the property of multiplicative renormalisation for Wilson loops with light-like lines is lost.

They showed that the Euclidean signature renormalisation-group evolution for a Wilson loop on a polygonal contour with $N$ vertices at $0$, $x_1$, $x_1 + x_2$, ..., $x_1 + \cdots x_{N-1}$,
\begin{equation}
  \label{eq:wl_renormalisation}
  \left(\mu \frac \partial {\partial \mu} + \beta(g) \frac \partial {\partial g}\right) \ln W(\mu^2 x_i \cdot x_j, \mu^2 x_k^2; \mu) = - \sum_{i=1}^{N-1} \Gamma_{\text{cusp}}(\gamma(x_i, x_{i+1}), g)
\end{equation} should be modified when one of the $x_i$ is light-like.
The modifications are the following
\begin{itemize}
\item make the following replacements in the right-hand side
  \begin{equation}
    \Gamma_{\text{cusp}}(\gamma(x_i, x_{i+1}), g) \rightarrow
    \begin{cases}
      \Gamma_{\text{cusp}}(g) \ln(\mu^2 (x_i \cdot x_{i+1} - 0 i)),&       \quad \text{if}\ x_i^2 = x_{i+1}^2 = 0,\\
      \tfrac 1 2 \Gamma_{\text{cusp}}(g) \ln\left(\frac {\mu^2 (x_i \cdot x_{i+1} - 0 i)}{x_i^2 - 0 i}\right),& \quad \text{if}\ x_i^2 \neq 0, x_{i+1}^2 = 0
    \end{cases}.
  \end{equation}
\item add a new anomalous dimension $\Gamma(g)$ whose origin lies in the light-cone singularities.  In ref.~\cite{Korchemskaya:1992je} it appears as an integration constant.
\end{itemize}

Note that in the case of Wilson loops with light-like lines there appears a new quantity $\Gamma_{\text{cusp}}$, without dependence on the angle (since one can't define an angle where one side is on the light-cone).  This quantity is also called ``cusp anomalous dimension.''

The large $\gamma$ asymptotics of $\Gamma_{\text{cusp}}(\gamma,g)$ are known (see ref.~\cite{Korchemsky:1987wg}) to all orders in perturbation theory by a power counting argument.  There, it was proven that, to all orders in $g$,
\begin{equation}
  \frac {\Gamma_{\text{cusp}}(\gamma,g)} \gamma \sim_{\gamma \rightarrow \infty} \mathcal{O}(1).
\end{equation}  The proportionality factor in the large $\gamma$ limit is $\Gamma_{\text{cusp}}(g)$.

\chapter{\texorpdfstring{Cusp anomalous dimension and IR divergences of scattering amplitudes\footnotemark}{Cusp anomalous dimension and IR divergences of scattering amplitudes}}
\footnotetext{I thank Gregory Korchemsky for patiently explaining to me many of the things discussed here.}
\label{ch:ir_scattering}

The cusp anomalous dimension is the main character in the BES equation, obtained by integrability techniques.  It is a striking example of highly non-trivial quantity for which we have both weak- and strong-coupling expansions.  The BES equation actually provides an exact expression for the cusp anomalous dimension in the form of an integral equation.

The cusp anomalous dimension appears in many seemingly unrelated places: the IR divergences of form-factors and scattering amplitudes, the anomalous dimension of twist two operators and the UV divergences of Wilson loops with cusps.  See refs.~\cite{Sudakov:1954sw, Sen:1981sd, Ivanov:1985np, Korchemsky:1988si, Korchemsky:1992xv, Collins:1989bt, Catani:1998bh, Magnea:1990zb, Sterman:2002qn} for a discussion of IR divergences.

In this section we will discuss how the cusp anomalous dimension appears in the IR divergences of form-factors and of scattering amplitudes.  For a discussion of Wilson loops see Appendix~\ref{ch:wilson_loops}.  We will not discuss twist-two operators here, but see ref.~\cite{Kruczenski:2002fb} for a pedagogical discussion.

We will first study the IR divergences of the quark form-factor.  It has been argued in refs.\cite{Korchemsky:1988si, Korchemsky:1992xv} that IR divergences can be captured exactly in the eikonal approximation.  Consider the tree-level interaction of an on-shell incoming quark with a soft gluon
\begin{wrapfigure}{l}{5cm}
\beginpgfgraphicnamed{44}
\begin{tikzpicture}[baseline=0]
  \draw (-1,-1) -- (1,1)
  node[pos=0, below left]{$p$}
  node[pos=.3, sloped]{\tikz\draw[->] (-1pt,0pt) -- (0pt,0pt);};
  \draw[soft gluon, decorate] (0,0) -- (1.5,0) node[above right]{$a, \mu$};
\end{tikzpicture}
\endpgfgraphicnamed
\end{wrapfigure}
which has the following expression,
\begin{equation}
  \frac {i (\slashed{p} + \slashed{k} + m)}{(p + k)^2 - m^2 + 0 i} i g \gamma^\mu t^a u(p).
\end{equation}

In the eikonal approximation we make the replacements
\begin{align}
  \frac 1 {(p + k)^2 - m^2 + 0 i} & \sim \frac 1 {2 p \cdot k},\\
  (\slashed{p} + \slashed{k} + m) \gamma^\mu u(p) & \sim 2 p^\mu u(p).
\end{align}

In the eikonal approximation, the expression for the tree level interaction above simplifies to
\begin{equation}
  i g \frac {i p^\mu}{p \cdot k} t^a u(p).
\end{equation}  This construction can be iterated and, for the interaction with $n$ soft gluons we have, in the eikonal approximation,
\begin{equation}
  \label{eq:eikonal_product}
  (i g)^n \frac {i p^{\mu_n} t^{a_n}}{p \cdot (k_1 + \cdots + k_n)} \frac {i p^{\mu_{n-1}} t^{a_{n-1}}}{p \cdot (k_1 + \cdots + k_{n-1})} \cdots \frac {i p^{\mu_1} t^{a_1}}{p \cdot k_1} u(p).
\end{equation}  Note that the eikonal approximation drastically modifies the UV behaviour of the propagator.

Let us now return to the quark form factor.  It is defined by
\begin{equation}
  F^\mu(q) = \langle p_1 | \mathbf{J}^\mu(q) | p_2\rangle,
\end{equation} where $q$ is the momentum transfer $q = p_1 - p_2$ and $\mathbf{J}^\mu$ is the current operator, $\mathbf{J}^\mu = \overline{\Psi} \gamma^\mu \Psi$.  Gauge invariance requires current conservation $q_\mu F^\mu = 0$, so we have that
\begin{equation}
  F^\mu(q) = \bar{v}(p_2) \gamma^\mu u(p_1) F(q^2).
\end{equation}

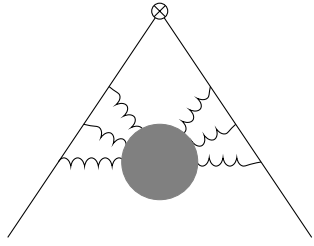
\begin{figure}
\centering
\beginpgfgraphicnamed{45}
\begin{tikzpicture}
  \draw (-2,-2) -- (0,1) -- (2,-2);
  \draw[soft gluon, decorate] (-4/3,-1) -- (0,-1)
                              (-1,-1/2) -- (0,-1)
                              (-2/3,0) -- (0,-1)
                              (4/3,-1) -- (0,-1)
                              (1,-1/2) -- (0,-1)
                              (2/3,0) -- (0,-1);
  \draw[blob] (0,-1) circle (1/2);
  \filldraw[fill=white, draw=black] (0, 1) circle (3pt);
  \draw (0,1) +(45:3pt) -- +(-135:3pt)
              +(135:3pt) -- +(-45:3pt);
\end{tikzpicture}
\endpgfgraphicnamed
  \caption{The general diagrammatic structure of a correction to the form factor.  The shaded blob is a (possibly disconnected) Green function.}
  \label{fig:form_factor_corr}
\end{figure}

The corrections to the form factor have the general diagrammatic structure indicated in fig.~\ref{fig:form_factor_corr}, where the shaded blob is a (possibly disconnected) Green function.

Suppose at first that $n$ gluon legs only attach to the quark line at the left.  Then, by using the following formula in eq.~\eqref{eq:eikonal_product}
\begin{equation}
  \int_0^\infty d s e^{i s (p \cdot k + 0 i)} = \frac i {p \cdot k + 0 i}
\end{equation} and also inserting the Green function for the gluons, represented by the shaded blob, we have
\begin{multline}
  (i g)^n \int \prod_{i=1}^n \frac {d^4 k_i}{(2 \pi)^4} d s_i \exp (i s_1 p_1 \cdot k_1 + i s_2 p_2 \cdot (k_1 + k_2) + \cdots + i s_n p_1 \cdot (k_1 + \cdots + k_n))\\ t^{a_n} \cdots t^{a_1} u(p_1) \langle p_1 \cdot A^{a_1}(k_1) \cdots p_1 \cdot A^{a_n}(k_n)\rangle.
\end{multline}  Observe now that the integrals over $k$ can be done exactly because they are Fourier transforms and they yield the gluon field in configuration space.  By changing the variables to $t_i = s_i + s_{i+1} + \cdots + s_n$ and integrating over the region $t_1 \geq t_2 \geq \cdots t_n$ we obtain the result
\begin{equation}
  P \exp\left(i g \int_{-\infty}^0 d t p_1 \cdot \tilde{A}(p_1 t)\right) u(p_1),
\end{equation} where $P$ stands for path ordering.  It is now easy to see that, if there are gluon legs on both the incoming and outgoing quark lines, we get a similar result with the contour of integration going from $-\infty$ to zero for the incoming quark line and from zero to $\infty$ for the outgoing quark line.

We have to remember that, in deriving this result, we integrated over all values for $k$ but the eikonal approximation can only be trusted in part of this integration region.  It is easy to see that the small $k$ divergences are now given by large $t$ divergences, which is in agreement with the intuition that the IR divergences are related to gluon propagation over large distances.

It is important to mention at this point that the Wilson contour does not depend on the quark mass separately, because the momenta $p_1$ and $p_2$ describing the contour can be rescaled without changing the result.  The Wilson contour will only depend on the `angle' $\gamma$ defined by
\begin{equation}
  \cosh \gamma = \frac {p_1 \cdot p_2}{\sqrt{p_1^2 p_2^2}} = \frac {p_1 \cdot p_2} {m^2}.
\end{equation}  When the mass of the quark goes to zero, $\gamma \rightarrow \infty$ and, in this limit, the cusp anomalous dimension scales linearly with $\gamma$ (see Appendix~\ref{ch:wilson_loops} for more detailed discussion and references to the relevant literature)
\begin{equation}
  \Gamma_{\text{cusp}}(g, \gamma) = \Gamma_{\text{cusp}}(g) \gamma + \mathcal{O}(\gamma^0).
\end{equation}  If we now define $Q^2 = - (p_2 - p_1)^2$, we can easily see that, in the limit when the quarks become massless,
\begin{equation}
  \frac d {d \ln Q^2} \Gamma_{\text{cusp}}(g, \gamma) = \Gamma_{\text{cusp}}(g).
\end{equation}

So we have proved that the Wilson contour described above correctly captures the IR divergences but it does not describe correctly the UV ones.  In fact, the Wilson contour introduces spurious UV divergences that have to be cancelled.  Because the Wilson contour has both IR and UV divergences and, when it is made out of two light-like lines it does not depend on any other mass or length scale, it has to depend on the ratio of the UV and IR regulators, $W = W(g, \gamma, \frac {\mu_{UV}}{\mu_{IR}})$.

The Wilson loop with light-like segments satisfies the following renormalisation group equation (see eq.~\eqref{eq:wl_renormalisation} and the discussion following it)
\begin{equation}
  \frac d {d \ln Q^2} \left(\mu_{UV} \frac \partial {\partial \mu_{UV}} + \beta(g) \frac \partial {\partial g}\right) \ln W = - \Gamma_{\text{cusp}}(g).
\end{equation}

Now, by a factorisation argument, the form factor can be written as
\begin{equation}
  F\left(\frac {Q^2}{\mu_{IR}^2}\right) = H\left(\frac {Q^2}{\mu_{UV}^2}\right) W\left(g, \gamma, \frac {\mu_{UV}}{\mu_{IR}}\right).
\end{equation}

For a conformal theory, where $\beta(g) = 0$, we have
\begin{align}
  \frac \partial {\partial \ln Q^2} & \frac \partial {\partial \ln \mu_{IR}} \ln F\left(\frac {Q^2}{\mu_{IR}^2}\right) = \frac \partial {\partial \ln Q^2} \frac \partial {\partial \ln \mu_{IR}} \ln W\left(g, \gamma, \frac {\mu_{UV}}{\mu_{IR}}\right) =\\& - \frac \partial {\partial \ln Q^2} \frac \partial {\partial \ln \mu_{UV}} \ln W\left(g, \gamma, \frac {\mu_{UV}}{\mu_{IR}}\right) = \Gamma_{\text{cusp}}(g).
\end{align}  The sign in the second line comes from using
\begin{equation}
  \frac \partial {\partial \mu_{\text{UV}}} = - \frac \partial {\partial \mu_{\text{IR}}},
\end{equation} when acting on a function of the ratio $\frac {\mu_{\text{UV}}} {\mu_{\text{IR}}}$.  Also, $\ln W$ depends on $Q^2$ only though the `angle' $\gamma$.

We also have
\begin{equation}
  \frac \partial {\partial \ln Q^2} \frac \partial {\partial \ln \mu_{IR}} \ln F\left(\frac {Q^2}{\mu_{IR}^2}\right) = - 2 \frac {\partial^2}{\partial \ln\left(\frac {Q^2}{\mu_{IR}^2}\right)^2} \ln F\left(\frac {Q^2}{\mu_{IR}^2}\right)
\end{equation} and, together with the previous equation, we have
\begin{equation}
  \frac {\partial^2}{\partial \ln\left(\frac {Q^2}{\mu_{IR}^2}\right)^2} \ln F\left(\frac {Q^2}{\mu_{IR}^2}\right) = - \frac 1 2 \Gamma_{\text{cusp}}(g).
\end{equation}  This equation can be integrated with the result
\begin{equation}
  \ln F\left(\frac {Q^2}{\mu_{IR}^2}\right) = - \frac 1 4 \Gamma_{\text{cusp}}(g) \ln^2 \left(\frac {Q^2}{\mu_{IR}^2}\right) - \frac 1 2 \Gamma(g) \ln \left(\frac {Q^2}{\mu_{IR}^2}\right) - \frac 1 2 C(g).
\end{equation}

In dimensional regularisation, the form factor can be written as\footnote{Sometimes this is written using the ratio $\frac {\mu_{IR}^2}{Q^2}$ rather than $\frac {Q^2}{\mu_{IR}^2}$ and also using the definition $a = \tfrac {g N^2}{8 \pi^2} (4 \pi e^{-\gamma})^\epsilon$.}
\begin{equation}
  \ln F\left(\frac {Q^2}{\mu_{IR}^2}\right) = - \frac 1 2 \sum_{l=1}^\infty \left\{a^l \left(\frac {\Gamma_{\text{cusp}}^l}{(l \epsilon)^2} + \frac {\Gamma^l}{l \epsilon} + C^l\right) \left(\frac {Q^2}{\mu_{IR}^2}\right)^{l \epsilon}\right\},
\end{equation} where $a = \tfrac {g N^2}{8 \pi^2}$ and
\begin{equation}
  \label{eq:cusp_coll_def}
  \Gamma_{\text{cusp}} = \sum_{l=1}^\infty a^l   \Gamma_{\text{cusp}}^l,\qquad
  \Gamma = \sum_{l=1}^\infty a^l \Gamma^l,\qquad
  C = \sum_{l=1}^\infty a^l C^l.
\end{equation}  The quantity $\Gamma_{\text{cusp}}$ is called cusp anomalous dimension while $\Gamma$ is called collinear anomalous dimension.

Let us now discuss scattering amplitudes of particles in a conformal gauge theory, in the adjoint representation of the gauge group.  We will restrict to the large $N$ (planar) limit.

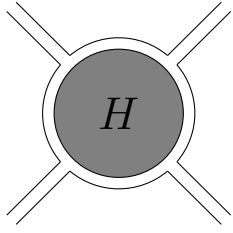
\begin{figure}
  \centering
  \beginpgfgraphicnamed{46}
  \begin{tikzpicture}[scale=1/2]
    \draw (-40:2) arc (-40:40:2)
    (50:2) arc (50:130:2)
    (140:2) arc (140:220:2)
    (230:2) arc (230:310:2);
    \draw (310:2) -- +(-45:2)
    (-40:2) -- +(-45:2)
    (40:2) -- +(45:2)
    (50:2) -- +(45:2)
    (130:2) -- +(135:2)
    (140:2) -- +(135:2)
    (220:2) -- +(-135:2)
    (230:2) -- +(-135:2);
    \draw[fill=gray] (0,0) node{\Large{$H$}} circle (1.7);
  \end{tikzpicture}
  \endpgfgraphicnamed
  \caption{The colour flow for a four-point scattering amplitude. The hard part of the interaction is denoted by $H$.}
  \label{fig:four-point_colour}
\end{figure}

Here, the treatment of the IR divergences is the same in spirit.  One important difference is that the Wilson loops are to be computed in the adjoint representation.  The colour flow for a four-point scattering amplitude is as shown in fig.~\ref{fig:four-point_colour}. 

In the large $N$ limit, the IR divergences appear from the exchange of soft gluons between adjacent external legs.  The analysis is identical to the one for the form factor: the amplitude factorises in a product of a hard part and a product of contributions from exchanges between all the adjacent external lines.

Summing all the contributions, the IR divergent part of an $n$-point scattering amplitude is
\begin{equation}
  \text{Div} = -\frac 1 4 \sum_{l=1}^\infty a^l \left(\frac {\Gamma_{\text{cusp}}^l}{(l \epsilon)^2} + \frac {\Gamma^l}{l \epsilon}\right) \sum_{i=1}^n \left(\frac {- s_{i, i+1}}{\mu_{IR}^2}\right)^{l \epsilon}.
\end{equation}

\end{document}